\documentclass[paper,toc,nohyper]{JHEP3}
\usepackage{epsfig,cite}
\usepackage{amsbsy}

\def\doubletilde#1{\widetilde{\vphantom{\raise 1.5pt \hbox{#1}}\smash{\kern -2pt\widetilde{#1}}}}
\def\MA#1#2{{\cal M}^{#1}_{A,#2}}
\def\MB#1#2{{\cal M}^{#1}_{B,#2}}

\def\NF{N_F}
\def\Flavour{F}
\def\Poles{{\cal P}oles}
\def\Finite{{\cal F}inite}
\def\Re{\mbox{Re}}
\def\bom#1{{\mbox{\boldmath $#1$}}}
\def\MSbar{$\overline{{\rm MS}}$}

\def\JET{J}

\def\Sab{s_{12}}
\def\Sac{s_{13}}
\def\Sad{s_{14}}
\def\Sbc{s_{23}}
\def\Sbd{s_{24}}
\def\Scd{s_{34}}
\def\Sae{s_{15}}

\def\Sce{s_{35}}
\def\Sde{s_{45}}
\def\Sabc{s_{123}}
\def\Sabd{s_{124}}
\def\Sacd{s_{134}}
\def\Sbcd{s_{234}}

\def\Sace{s_{135}}
\def\Sade{s_{145}}

\def\Scde{s_{345}}

\def\e{\epsilon}
\def\d{\hbox{d}}

\def\Li{\hbox{Li}}
\def\ln{\hbox{ln}}

\title{\boldmath 
Antenna Subtraction at NNLO
}
\author{
A.~Gehrmann--De Ridder\\
Institute for Theoretical Physics, ETH, CH-8093 Z\"urich,
Switzerland\\ 
E-mail: \email{gehra@phys.ethz.ch}}
\author{
T.~Gehrmann\\
Institut f\"ur Theoretische Physik, Universit\"at Z\"urich,
Winterthurerstrasse 190,\\ CH-8057 Z\"urich, Switzerland\\
E-mail: \email{thomas.gehrmann@physik.unizh.ch}}
\author{E.W.N.~Glover\\
Institute of Particle Physics Phenomenology, 
        Department of Physics,\\
        University of Durham, Durham, DH1 3LE, UK\\
	E-mail: \email{e.w.n.glover@durham.ac.uk}}

\abstract{
The computation of exclusive QCD jet observables at higher orders
requires a method for the subtraction of infrared singular
configurations arising from multiple radiation of real partons.
We present a subtraction scheme relevant for NNLO perturbative
calculations in $e^+e^- \to $~jets.   The building blocks of the scheme are
antenna functions derived from the matrix elements for 
tree-level $1\to 3$ and  $1 \to 4$ 
and one-loop $1 \to 3$ processes.   
By construction, these building blocks have
the correct infrared behaviour when one or two particles are unresolved.   At
the same time, their integral over the
antenna  phase space is straightforward.
As an example of how to use the scheme we compute  the NNLO contributions to
the
subleading colour QED-like contribution to $e^+e^- \to 3$~jets. 
 To 
illustrate the application  of NNLO antenna subtraction for  different colour 
structures, we construct the integrated forms of the subtraction terms 
needed for the five-parton and four-parton contributions to 
$e^+e^- \to 3$~jets at NNLO
in all colour factors, and show that their infrared poles 
cancel analytically with the infrared poles of the two-loop virtual correction
to this observable.
}

\preprint{{ZU-TH 07/05}, {IPPP/05/18}, {hep-ph/0505111}}

\begin{document}

\section{Introduction}

Experimental measurements of jet production observables are among the 
most sensitive tests of the theory of Quantum Chromodynamics (QCD), 
and yield very accurate determinations of QCD parameters~\cite{dissertori},
especially of the strong coupling constant $\alpha_s$. At present, 
the precision of many of these determinations is limited not by the 
quality of the experimental data, but 
by the error on the theoretical (next-to-leading order, NLO) 
calculations used for the extraction of 
the QCD parameters. To 
improve upon this situation, an extension of the theoretical calculations to 
next-to-next-to-leading order (NNLO) is therefore mandatory~\cite{glover}.

In the recent past, many ingredients to NNLO calculations of collider
 observables have been derived,
including the universal three-loop QCD splitting 
functions~\cite{mvv} which
govern the evolution of parton distribution functions at NNLO. 
The massless
two-loop $2\to 2$ and $1 \to 3$ matrix elements relevant to 
NNLO jet production 
have been computed~\cite{twol,3jme} using several 
innovative methods~\cite{twolmeth},
and are now available for many processes of phenomenological relevance.
The one-loop corrections to $2\to 3$ and $1\to 4$ matrix elements 
have been known for longer~\cite{onel-3,onel-4} and form part of NLO calculations of 
the respective multi-jet observables~\cite{nlomult,cullen}. 
These NLO matrix elements naturally
contribute to NNLO jet observables of lower multiplicity if one of the 
partons involved becomes unresolved (soft or collinear)~\cite{onelstr}. 
In these cases, the infrared 
singular parts of the matrix elements need to be extracted and integrated 
over the phase space appropriate to the unresolved configuration 
to make the infrared pole structure explicit. Methods for the 
extraction of soft and collinear limits of one-loop matrix elements are  
worked out in detail in the 
literature~\cite{onelstr,oneloopsoft,onelstr1,onelstr2,twolstr,babisdy}. 
As a final ingredient, the tree level $2\to 4$ and $1\to 5$ processes also
contribute to ($2\to 2$)- and ($1 \to 3$)-type jet observables at NNLO. 
These 
contain double real radiation singularities corresponding to two 
partons becoming simultaneously soft and/or 
collinear~\cite{audenigel,campbell,cg,campbellandother}. 
To determine the
contribution to NNLO jet observables from these configurations, one has to 
find two-parton subtraction terms which coincide with the full matrix element 
and are still sufficiently simple to be integrated analytically in order 
to cancel  their  infrared pole structure with the two-loop virtual and 
the one-loop single-unresolved contributions. In the past, such configurations 
were only dealt with on a case-by-case basis in the context of specific 
calculations~\cite{audenigel,uwer,fg,daleo,adamson}, while no general method 
was available. Several methods have been 
proposed recently to accomplish this 
task~\cite{nnlosub1,nnlosub2,nnlosub3,nnlosub4,nnlosub5}. 
Up to now, only 
one method has been fully worked through for observables of physical
interest:the sector decomposition algorithm~\cite{secdec,ggh}.
In this method,  
 both phase space and loop integrals are analytically decomposed into their 
Laurent expansion in dimensional regularisation, and the coefficients 
of the expansion are numerically integrated.
Results have been  
obtained for $e^+e^- \to 2j$~\cite{babis2j}, $pp \to H+X$~\cite{babishiggs} 
and most recently for muon decay~\cite{babismu}
at NNLO. In contrast to all other
approaches, in the  sector decomposition method one does not have to 
integrate the subtraction term analytically.

Following a number of pioneering calculations~\cite{ERT,ert2,kn},
infrared subtraction of real radiation singularities is well understood 
at NLO, where several generic process independent methods 
exist~\cite{singleun,cs,ant}. One of these methods is the so-called 
antenna subtraction~\cite{cullen,ant}, which derives NLO subtraction terms 
from three-parton tree-level
matrix elements which naturally encapsulate all singular limits due to
unresolved  
emission of a single parton between two colour-connected~\cite{colord,ddm}
 hard partons (radiators). 
In~\cite{our2j}, we described the construction of NNLO subtraction terms 
for $e^+e^- \to 2j$ based on full four-parton tree-level 
and three-parton one-loop matrix elements, which can be integrated
analytically over the appropriate phase spaces~\cite{ggh}. If 
normalised appropriately, these 
full four-parton tree-level and three-parton one-loop matrix elements 
can be interpreted as antenna functions at NNLO. 
These NNLO antenna subtraction terms
 derived from four-parton matrix elements with a hard quark-antiquark
pair 
in~\cite{our2j} were used subsequently~\cite{our3j} 
to compute the $\alpha_s^3 C_F^3$--correction to $e^+e^- \to 3j$ 
at NNLO. 

The calculation 
presented in~\cite{our2j} included only antenna functions involving 
radiation off a hard quark-antiquark pair. We derived 
colour-ordered matrix elements (and thus NNLO antenna functions) 
for the two other partonic configurations (hard quark-gluon radiators and 
hard gluon-gluon radiators) by using appropriate effective Lagrangian 
densities coupling an external current to a gluino--gluon~\cite{chi} or 
gluon-gluon~\cite{h} system. 

In the present paper, we describe the antenna subtraction
method at NNLO in detail and provide an algorithm for the construction of 
antenna subtraction terms at NNLO from the basic three-parton and four-parton 
antenna functions. For the sake of clarity, we restrict ourselves to the 
kinematical situation of a colour-neutral particle decaying into coloured 
final state partons, as relevant to $e^+e^- \to m$~jets at NNLO. The 
basic structure of the NNLO antenna subtraction can however be carried 
over to configurations with partons in the initial state. 

The paper is structured as follows. In Section~\ref{sec:method} we describe 
the antenna subtraction method at NLO and NNLO, and explain the 
construction of the subtraction functions for tree-level and one-loop 
real radiation matrix elements up to NNLO accuracy from the basic three- and 
four-parton antenna functions. All antenna functions are derived and 
integrated over their appropriate final state
phase spaces in the following sections.
Section \ref{sec:notation} establishes the notation, and lists the 
different possible antenna configurations. In Section \ref{sec:operators},
we introduce the colour-ordered
infrared singularity operators~\cite{catani} appearing in the 
integrated NLO and NNLO antenna functions. The quark-antiquark, 
quark-gluon and gluon-gluon antenna functions in both unintegrated and 
integrated forms are then derived in 
Sections \ref{sec:qq}--\ref{sec:gg}. Their behaviour in all single and double 
unresolved limits is summarised in Section~\ref{sec:limits}. As a 
first non-trivial application of our method, we document the 
calculation of the 
subleading colour NNLO QCD correction to the
$e^+e^- \to 3j$ cross section (which was already reported briefly 
in~\cite{our3j})
in Section~\ref{sec:3j}. To illustrate the generality of the method, and to 
outline future applications of it, we reconstruct the infrared pole structure 
of the integrated NNLO subtraction terms for $e^+e^- \to 3j$
in all colour factors, and show that these cancel the explicit infrared poles
of the two-loop virtual corrections to this observable in 
Section~\ref{sec:3jir}.
Finally, 
Section~\ref{sec:conc} contains our conclusions and an outlook on 
applications and extensions of the method presented here.

\section{Infrared subtraction terms}
\label{sec:method}

To obtain the perturbative corrections to a jet observable at a given  order,
all partonic multiplicity channels contributing to that order  have to be
summed. In general, each partonic channel contains both ultraviolet and
infrared (soft and collinear) singularities.  The ultraviolet poles are removed
by renormalisation, however the soft and collinear infrared poles cancel
among each other when all partonic channels are summed over~\cite{kln}.

While infrared singularities from purely virtual corrections are obtained 
immediately after integration over the loop momenta, their extraction is  more
involved for real emission (or mixed real-virtual) contributions. Here, the
infrared singularities only become explicit after integrating  the real
radiation matrix elements over the phase space appropriate to  the jet
observable under consideration. In general, this integration  involves the
(often iterative) definition of the jet observable, such that  an analytic
integration is not feasible (and also not appropriate). Instead,   one would
like to have a flexible method that can be easily adapted to  different jet
observables or jet definitions. Therefore, the infrared singularities  of the real radiation
contributions should be extracted using  infrared subtraction  terms.  The
crucial points that all  subtraction terms must satisfy are  that (a) they
approximate the full  real radiation matrix elements in all singular limits and
(b) are still   sufficiently simple to be integrated analytically over a
section of  phase space that encompasses all regions corresponding to singular
configurations. Note that the subtraction terms should also be local and
should fully account for the limit they are aimed at without introducing 
spurious infrared singularities in other limits. 

In this paper, we shall restrict ourselves to the kinematical situation of
a massive, colour-neutral particle decaying into massless coloured partons,
such as multijet production in electron--position annihilation:
\begin{displaymath}
e^+e^- \to m~\mbox{jets} \;.
\end{displaymath} 
In this situation, infrared singularities appear only due to final state
radiation. In the more general
partonic scattering kinematics, coloured partons are also present in the 
initial state, yielding initial state and fixed initial/final state
singularities. The method presented here can, like the antenna factorisation 
method at NLO~\cite{ant}, be extended to these situations. This 
extension is beyond the scope of this paper, and will only be commented on 
briefly in Section~\ref{sec:conc} below.

To specify the notation, we define 
the tree-level $n$-parton contribution to the $m$-jet cross 
section (for tree
level cross sections $n = m$; we leave $n \neq m$ for later
reference) 
in $d$ dimensions by,
\begin{equation}
{\rm d} \sigma^{B}={\cal N}\,\sum_{{n}}{\rm d}\Phi_{n}(p_{1},\ldots,p_{n};q)
\frac{1}{S_{{n}}}\,|{\cal M}_{n}(p_{1},\ldots,p_{n})|^{2}\; 
\JET_{m}^{(n)}(p_{1},\ldots,p_{n}).
\label{eq:sigm}
\end{equation}
the normalisation factor 
${\cal N}$ includes all QCD-independent factors as well as the 
dependence on the renormalised QCD coupling constant $\alpha_s$,
$\sum_{n}$ denotes the sum over all configurations with $n$ partons,
${\rm d}\Phi_{n}$ is the phase space for an $n$-parton final state with total
four-momentum $q^{\mu}$ in
$d=4-2\e$ space-time dimensions,
\begin{equation}
\d \Phi_n(p_{1},\ldots,p_{n};q) 
= \frac{\d^{d-1} p_1}{2E_1 (2\pi)^{d-1}}\; \ldots \;
\frac{\d^{d-1} p_n}{2E_n (2\pi)^{d-1}}\; (2\pi)^{d} \;
\delta^d (q - p_1 - \ldots - p_n) \,,
\end{equation}
while $S_{n}$ is a
symmetry factor for identical partons in the final state.
$|{\cal M}_{n}|^2$ denotes a 
squared, colour-ordered tree-level $n$-parton matrix element, where
particle 1 is colour connected to particle 2 which is colour connected to
particle 3 and so on as illustrated in Figure~\ref{fig:connect}.
\FIGURE[h!]{ 
\epsfig{file=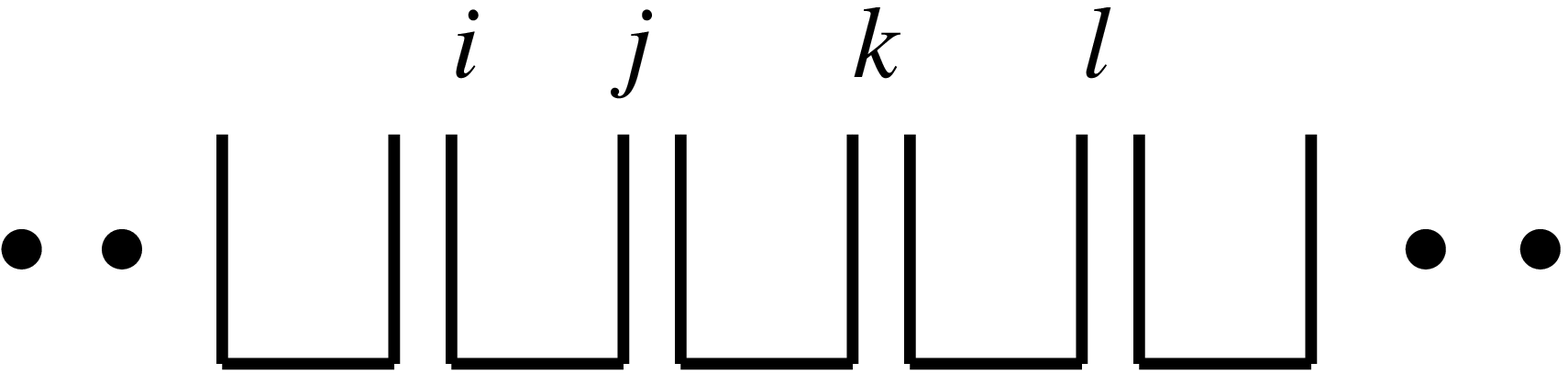,height=1.2cm}
\label{fig:connect}
\caption{Illustration of colour connection.}
}
Contributions to the squared matrix element which are subleading in the number 
of colours can equally be treated in the same context, noting that 
these subleading terms yield configurations where a 
certain number of essentially non-interacting particles
are emitted between a pair of hard 
radiators. By carrying 
out the colour algebra, it becomes evident that non-ordered gluon emission 
inside a colour-ordered system is equivalent to photon emission off the 
outside legs of the system~\cite{campbell,colord}. For simplicity,
these subleading colour contributions are also denoted as squared matrix 
elements $|{\cal M}_m|^2$, 
although they often correspond purely to interference
terms between different amplitudes.

The precise definition depends on the number and types
of particles involved in the process.   
However, all colour orderings are summed
over in $\sum_{m}$ with the appropriate colour weighting.
The jet function $ \JET_{m}^{(n)}$ defines the procedure for 
building $m$ jets out of $n$ partons.
The main property of $\JET_{m}^{(n)}$ is that the jet observable defined
above is collinear and infrared safe as explained in \cite{irsafe,cs}.
In general $\JET_{m}^{(n)}$ contains $\theta$ and $\delta$-functions.  
 $\JET_{m}^{(n)}$ can also represent the definition of the $n$-parton 
contribution to an event shape observable related to $m$-jet final states. 

From (\ref{eq:sigm}), one obtains the leading order approximation to the 
$m$-jet cross section by integration over the appropriate phase space. 
\begin{equation}
{\rm d}\sigma_{LO}=\int_{{\rm d}\Phi_{m}}{\rm d}\sigma^{B} \;.
\end{equation}
Depending on the jet function used, 
this cross section can still be differential in certain kinematical
quantities. 

\subsection{NLO infrared subtraction terms}
\label{sec:nloant}

At NLO, we consider the following $m$-jet cross section,
\begin{equation}
{\rm d}\sigma_{NLO}=\int_{{\rm d}\Phi_{m+1}}\left({\rm d}\sigma^{R}_{NLO}
-{\rm d}\sigma^{S}_{NLO}\right) +\left [\int_{{\rm d}\Phi_{m+1}}
{\rm d}\sigma^{S}_{NLO}+\int_{{\rm d}\Phi_{m}}{\rm d}\sigma^{V}_{NLO}\right].
\end{equation}
The cross section ${\rm d}\sigma^{R}_{NLO}$ has the same expression as the 
Born cross section ${\rm d}\sigma^{B}_{NLO}$ (\ref{eq:sigm}) above
except that $m \to m+1$, while 
${\rm d}\sigma^{V}_{NLO}$ is the one-loop virtual correction to the 
$m$-parton Born cross section ${\rm d}\sigma^{B}$.
The cross section ${\rm d}\sigma^{S}_{NLO}$ is a 
(preferably local) counter-term for  
 ${\rm d}\sigma^{R}_{NLO}$. It has the same unintegrated
singular behaviour as ${\rm d}\sigma^{R}_{NLO}$ in all appropriate limits.
Their difference is free of divergences 
and can be integrated over the $(m+1)$-parton phase space numerically.
The subtraction term  ${\rm d}\sigma^{S}_{NLO}$ has 
to be integrated analytically over all singular regions of the 
$(m+1)$-parton phase space. 
The resulting cross section added to the virtual contribution 
yields an infrared finite result.

A systematic procedure for finding NLO infrared subtraction terms is the
antenna formalism 
introduced in~\cite{cullen,ant}.
The antenna subtraction terms are obtained as 
 sum of antennae:
\begin{eqnarray}
\lefteqn{{\rm d}\sigma_{NLO}^{S} =
{\cal N}\,\sum_{m+1}{\rm d}\Phi_{m+1}(p_{1},\ldots,p_{m+1};q)
\frac{1}{S_{{m+1}}} }\nonumber \\ 
&\times &\sum_{j}\;X^0_{ijk}\,
|{\cal M}_{m}(p_{1},\ldots,\tilde{p}_{I},\tilde{p}_{K},\ldots,p_{m+1})|^2\,
\JET_{m}^{(m)}(p_{1},\ldots,\tilde{p}_{I},\tilde{p}_{K},\ldots,p_{m+1})
\;,
\label{eq:sigmasNLO}
\end{eqnarray}
such that,
\begin{eqnarray}
\lefteqn{
{\rm d}\sigma_{NLO}^{R}-{\rm d}\sigma_{NLO}^{S}
= {\cal N}\,\sum_{m+1}{\rm d}\Phi_{m+1}(p_{1},\ldots,p_{m+1};q)
\frac{1}{S_{{m+1}}} }\nonumber \\
&\times& \,\Bigg [|{\cal M}_{m+1}(p_{1},\ldots,p_{m+1})|^{2}\;
\JET_{m}^{(m+1)}(p_{1},\ldots,p_{m+1}) \nonumber \\
&&-\sum_{j}\;X^0_{ijk}\,
|{\cal M}_{m}(p_{1},\ldots,\tilde{p}_{I},\tilde{p}_{K},\ldots,p_{m+1})|^2\,
\JET_{m}^{(m)}(p_{1},\ldots,\tilde{p}_{I},\tilde{p}_{K},\ldots,p_{m+1})\;\Bigg
].\nonumber \\
\label{eq:sub1}
\end{eqnarray}
\FIGURE[h!]{ 
\epsfig{file=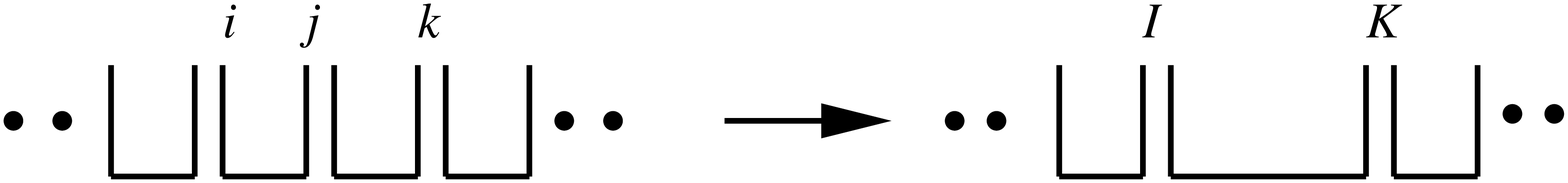,height=1.2cm}
\label{fig:combine}
\caption{Colour connection of the partons showing the parent and daughter
partons for the single unresolved antenna.}
}

The subtraction term involves the $m$-parton amplitude 
depending only on the redefined on-shell momenta
$p_{1},\ldots,\tilde{p}_{I},\tilde{p}_{K},\ldots,p_{m+1}$ 
where $\tilde{p}_{I},\tilde{p}_{K}$
 are linear combinations of $p_{i},p_{j},{p}_{k}$
 while the tree antenna function $X^0_{ijk}$ depends only on 
$p_{i},p_{j},{p}_{k}$.
$X^0_{ijk}$ describes all of the configurations (for this colour-ordered
amplitude)
where parton $j$
 is unresolved.  This occurs because of the particular factorisation
 properties of colour-ordered amplitudes.  In particular, when particle $j$ 
 is a gluon and becomes
 soft, the colour-ordered
 matrix element undergoes QED-like factorisation so that, 
\begin{displaymath}
|{\cal M}_{m+1}(p_{1},\ldots,p_i,p_j,p_k,\ldots,p_{m+1})|^{2}
\to \frac{2s_{ik}}{s_{ij}s_{jk}} 
|{\cal M}_{m}(p_{1},\ldots,\tilde{p}_{I},\tilde{p}_{K},\ldots,p_{m+1})|^{2}.
\end{displaymath}
In this limit $\tilde{p}_{I} \equiv p_i$ and $\tilde{p}_{K} \equiv p_k$.
Together particles $I$ and $K$ form a colour connected 
hard antenna that radiates particle $j$. In doing so, 
the momenta of 
the radiators change 
to form particles $i$ and $k$. The type of particle may also change.
For example when particle $j$ is a gluon, 
the colour factor is modified at the amplitude level by, 
$$T^{a_j}_{ik}  \leftrightarrow   \delta_{iI}  \delta_{IK} \delta_{Kk},$$
as illustrated in Figure~\ref{fig:combine}. The antenna factorisation of
squared matrix element and phase space can be illustrated pictorially, as 
displayed in Figure~\ref{fig:nloant}.

The antenna approach described here 
is closely related to the commonly used dipole
factorisation formalism  derived by Catani and Seymour~\cite{cs}. 
Here, the corresponding
subtraction term for the same colour-ordered 
squared matrix element is given by
\begin{eqnarray}
{\rm d}\sigma_{NLO}^{S}
&= & {\cal N}\,\sum_{m+1}{\rm d}\Phi_{m+1}(p_{1},\ldots,p_{m+1};q)
\frac{1}{S_{{m+1}}} \,\Bigg [\;
\nonumber \\
&&\;\sum_{j}{\cal D}_{ij,k}\,
|{\cal M}_{m}((p_{1},\ldots,\tilde{p}_{ij},\tilde{p}_{k},\ldots,p_{m+1})|^2\,
\JET_{m}^{(m)}(p_{1},\ldots,\tilde{p}_{ij},\tilde{p}_{k},\ldots,p_{m+1})\; 
\nonumber \\
&+&\;\sum_{j}{\cal D}_{kj,i}\,
|{\cal M}_{m}((p_{1},\ldots,\tilde{p}_{i},\tilde{p}_{kj},\ldots,p_{m+1})|^2\,
\JET_{m}^{(m)}(p_{1},\ldots,\tilde{p}_{i},\tilde{p}_{kj},\ldots,p_{m+1})\;\Bigg
].\nonumber \\
\end{eqnarray}
In the first term, 
the dipole contribution  involves the $m$-parton amplitude 
which only depends on the redefined on-shell momenta
$p_{1},\ldots,\tilde{p}_{ij},\tilde{p}_{k},\ldots,p_{m+1}$ 
and the dipole function ${\cal D}_{ij,k}$ which depends on 
$p_{i},p_{j},{p}_{k}$.
The momenta ${p}_{i}$, $p_j$ and ${p}_{k}$ are respectively 
the emitter, unresolved parton  and the spectator momenta 
corresponding to a single dipole term.
In the second term, the role of emitter and spectator are exchanged.
The redefined on-shell momenta 
$\tilde{p}_{ij},\tilde{p}_{k}$ ($\tilde{p}_{kj},\tilde{p}_{i} $)
are different linear combinations of $p_i$, $p_j$ and $p_l$ for each dipole.
In the antenna approach, the momentum mapping would be 
the same for each dipole
contribution and the two terms combine to form the 
tree antenna, $X^0_{ijk}$. The two dipoles combining to an antenna have a
common unresolved parton, and contain the two possible emitter/spectator
combinations. In the antenna language, emitter and spectator act as 
radiators. 
Note that we can always choose to divide the antenna and use 
different momentum maps for the two parts.
\FIGURE[h!]{ 
\epsfig{file=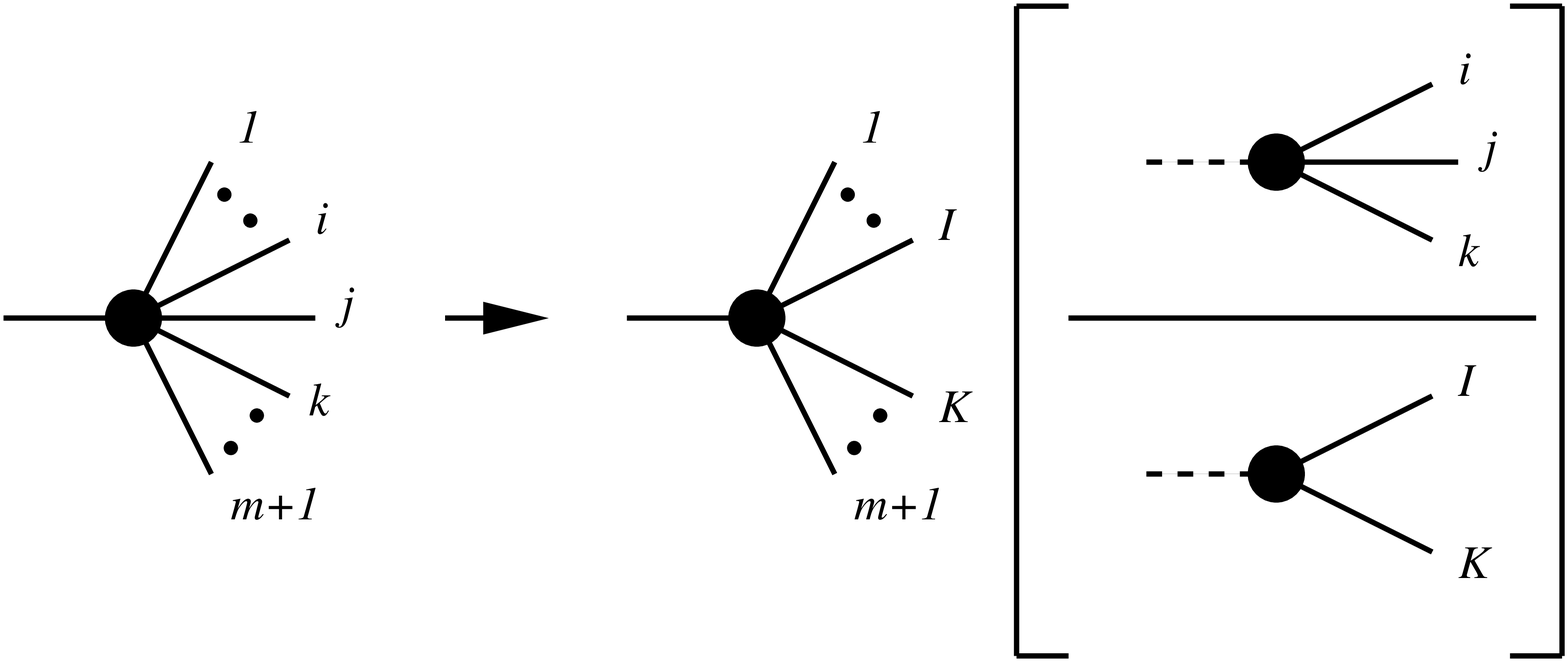,width=10cm}
\label{fig:nloant}
\caption{Illustration of NLO antenna factorisation representing the
factorisation of both the squared matrix elements and the $(m+1)$-particle phase
space. The term in square brackets
represents both the antenna function $X^0_{ijk}$ and the antenna phase space
${\rm d}\Phi_{X_{ijk}}$.}}

The jet function $\JET^{(m)}_m$ in (\ref{eq:sub1}) does not depend on the 
individual momenta ${p}_{i}$, $p_j$ and ${p}_{k}$, but only on 
$\tilde{p}_{I},\tilde{p}_{K}$. One can therefore carry out the integration
over the unresolved dipole phase space appropriate to 
   ${p}_{i}$, $p_j$ and ${p}_{k}$ analytically, exploiting the 
factorisation of the phase space,
\begin{equation}
\label{eq:psx3}
\d \Phi_{m+1}(p_{1},\ldots,p_{m+1};q)  = 
\d \Phi_{m}(p_{1},\ldots,\tilde{p}_{I},\tilde{p}_{K},\ldots,p_{m+1};q)\cdot 
\d \Phi_{X_{ijk}} (p_i,p_j,p_k;\tilde{p}_{I}+\tilde{p}_{K})\;.
\end{equation}
The NLO antenna phase space $\d \Phi_{X_{ijk}}$ is proportional to the three-particle phase space,
as can be seen by using $m=2$ in the above formula and exploiting 
the fact that the two-particle phase space is a constant,
\begin{equation}
P_2 = \int \d \Phi_2 = 2^{-3+2\epsilon}\pi^{-1+\epsilon}
\frac{\Gamma(1-\epsilon)}{\Gamma(2-2\epsilon)}\left(q^2\right)^{-\epsilon}\;,
\end{equation}
such that
\begin{equation}
{\rm d}\Phi_{3}= P_2\; {\rm d}\Phi_{X_{ijk}}\;.
\end{equation}
 
For the analytic integration, 
we can use (\ref{eq:psx3}) to rewrite
each of the subtraction terms  in the form, 
\begin{displaymath}
|{\cal M}_{m}|^2\,
\JET_{m}^{(m)}\; 
{\rm d}\Phi_{m}
\int {\rm d} \Phi_{X_{ijk}}\;X^0_{ijk},
\end{displaymath}
where $|{\cal M}_{m}|^2$, $\JET_{m}^{(m)}$ and ${\rm d}\Phi_{m}$ depend only on
$p_1,,\ldots,\tilde{p}_{I},\tilde{p}_{K},\ldots,p_{m+1}$
and ${\rm d} \Phi_{X_{ijk}}$ and $X^0_{ijk}$ depend only on $p_i, p_j, p_k$.
The analytic integral of the subtraction term is 
therefore defined as the antenna function
integrated over the fully inclusive 
antenna phase space, normalised appropriately,
\begin{equation}
\label{eq:x3int}
{\cal X}^0_{ijk}(s_{ijk}) = \left(8\pi^2 \,(4\pi)^{-\e}\, e^{\e\gamma}\right) 
\int {\rm d} \Phi_{X_{ijk}}\;X^0_{ijk}.
\end{equation}
This integration is performed analytically in $d$ dimensions 
to make the infrared singularities explicit and
added directly to the one-loop $m$-particle contributions.
The factor $\left(8\pi^2 \,(4\pi)^{-\e}\, e^{\e\gamma}\right)$ in the 
above equation is related to the normalisation of the renormalised 
coupling constant, and its relation to the bare coupling parameter 
$g=\sqrt{4\pi \alpha_0}$ appearing 
in the QCD Lagrangian density:
\begin{equation}
\alpha_0\mu_0^{2\e} S_\e = \alpha_s \mu^{2\e}\left[
1- \frac{\beta_0}{\e}\left(\frac{\alpha_s}{2\pi}\right) 
+\left(\frac{\beta_0^2}{\e^2}-\frac{\beta_1}{2\e}\right)
\left(\frac{\alpha_s}{2\pi}\right)^2+{\cal O}(\alpha_s^3) \right]\; ,
\end{equation}
where
\begin{displaymath}
S_\e =(4\pi)^\e e^{-\e\gamma}\qquad \mbox{with Euler constant }
\gamma = 0.5772\ldots
\end{displaymath}
and $\mu_0^2$ is the mass parameter introduced 
in dimensional regularisation  to maintain a 
dimensionless coupling 
in the bare QCD Lagrangian density; $\beta_0$ and $\beta_1$ are the first 
two coefficients of the QCD $\beta$-function:
\begin{equation}
\beta_0 = \frac{11 N - 2 N_F}{6}\;, \qquad \beta_1 =
 \frac{34 N^3 - 13 N^2  N_F+   3 N_F}{12N}\;,
\label{eq:qcdbeta}
\end{equation}
with $N=3$ colours and $N_F$ massless quark flavours.

\subsection{NNLO infrared subtraction terms}
At NNLO, the $m$-jet production is induced by final states containing up to
$(m+2)$ partons, including the one-loop virtual corrections to $(m+1)$-parton final 
states. As at NLO, one has to introduce subtraction terms for the 
$(m+1)$- and $(m+2)$-parton contributions. 
Schematically the NNLO $m$-jet cross section reads,
\begin{eqnarray}
{\rm d}\sigma_{NNLO}&=&\int_{{\rm d}\Phi_{m+2}}\left({\rm d}\sigma^{R}_{NNLO}
-{\rm d}\sigma^{S}_{NNLO}\right) + \int_{{\rm d}\Phi_{m+2}}
{\rm d}\sigma^{S}_{NNLO}\nonumber \\ 
&&+\int_{{\rm d}\Phi_{m+1}}\left({\rm d}\sigma^{V,1}_{NNLO}
-{\rm d}\sigma^{VS,1}_{NNLO}\right)
+\int_{{\rm d}\Phi_{m+1}}{\rm d}\sigma^{VS,1}_{NNLO}  
\nonumber \\&&
+ \int_{{\rm d}\Phi_{m}}{\rm d}\sigma^{V,2}_{NNLO}\;,
\end{eqnarray}
where $\d \sigma^{S}_{NNLO}$ denotes the real radiation subtraction term 
coinciding with the $(m+2)$-parton tree level cross section 
 $\d \sigma^{R}_{NNLO}$ in all singular limits. 
Likewise, $\d \sigma^{VS,1}_{NNLO}$
is the one-loop virtual subtraction term 
coinciding with the one-loop $(m+1)$-parton cross section 
 $\d \sigma^{V,1}_{NNLO}$ in all singular limits. 
Finally, the two-loop correction 
to the $m$-parton cross section is denoted by ${\rm d}\sigma^{V,2}_{NNLO}$.

\subsection{Tree-level double real radiation subtraction terms}

Let us first consider the construction of the subtraction terms 
for the double radiation
contribution ${\rm d}\sigma^{S}_{NNLO}$, which shall correctly subtract all 
single and double unresolved singularities contained in the $(m+2)$-parton 
real 
radiation contribution to $m$-jet final states, 
\begin{eqnarray}
\lefteqn{{\rm d}\sigma^{R}_{NNLO}
= {\cal N}\,\sum_{m+2}{\rm d}\Phi_{m+2}(p_{1},\ldots,p_{m+2};q)
\frac{1}{S_{{m+2}}} }\nonumber \\ &\times&
|{\cal M}_{m+2}(p_{1},\ldots,p_{m+2})|^{2}\;
\JET_{m}^{(m+2)}(p_{1},\ldots,p_{m+2})\;.
\label{eq:nnloreal}
\end{eqnarray}
Single real radiation 
singularities correspond to one parton becoming soft or collinear, while 
double real radiation singularities occur if two partons become soft or
collinear simultaneously. Singular terms in these limits can be identified 
by requiring a minimum number of invariants tending to zero in a given 
kinematical configuration. This number depends on the limit under 
consideration and follows from the phase space volume available to a given 
configuration. A detailed discussion of the 
kinematical definition of double unresolved 
limits can be found in~\cite{campbell,cg}. 

We must distinguish the following 
configurations:
\begin{itemize}
\item[(a)] One unresolved parton but the experimental observable selects only
$m$ jets.
\item[(b)] Two colour-connected unresolved partons (colour-connected).
\item[(c)] Two unresolved partons that are not colour connected but share a common
radiator (almost colour-unconnected).
\item[(d)] Two unresolved partons that are well separated from each other 
in the colour 
chain (colour-unconnected).
\end{itemize}

The first configuration  
was treated already in the context of antenna subtraction at NLO in 
Section~\ref{sec:nloant} above. In the context of the construction of 
 ${\rm d}\sigma^{S}_{NNLO}$, the same single-particle subtraction terms 
can be used. These do however not yet guarantee a 
finite $(m+2)$-parton contribution in all 
single unresolved regions for two reasons: (1) while the jet function in 
${\rm d}\sigma^{S}_{NLO}$ ensured that the subtraction term is non-zero only
in the single unresolved limit it was constructed for, this is no
longer the case for single unresolved 
radiation at NNLO; (2) the subtraction terms for the remaining three double 
unresolved configurations will in general be singular in the single unresolved
regions, where they do not match the matrix element. Both problems will be 
addressed below.

The remaining
 three configurations (b)--(d) are illustrated in Figures~\ref{fig:double}, 
\ref{fig:overlap} and \ref{fig:nonoverlap}. The singular behaviour 
of the full $(m+2)$-parton matrix 
element in these configurations is the product
of double unresolved factors 
(see Section~\ref{sec:limits} below) and 
reduced $m$-parton matrix elements. Subtraction terms for all these 
configurations can be constructed using either 
single four-parton antenna functions or products of two three-parton antenna 
functions. In all cases, attention has to be paid to the matching of 
different double and single unresolved regions. 
This problem has been addressed already in several earlier publications 
on subtraction at 
NNLO~\cite{nnlosub1,nnlosub2,nnlosub3,nnlosub4,nnlosub5,our2j}, the most 
concise discussion can be found in~\cite{nnlosub5}. 
\FIGURE[h!]{ 
\epsfig{file=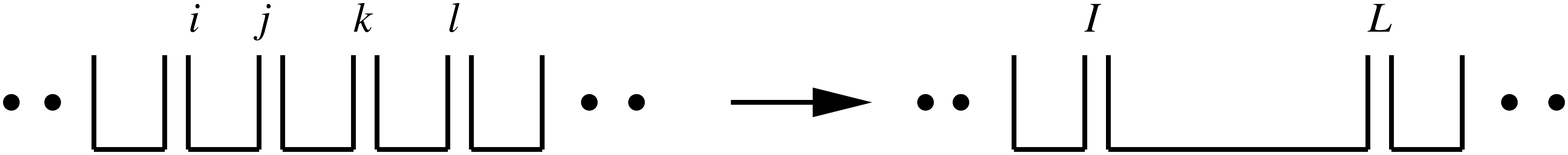,height=1.2cm}
\label{fig:double}
\caption{Colour connection of the partons showing the parent and daughter
partons for the double unresolved antenna.}
}
\FIGURE[h!]{ 
\epsfig{file=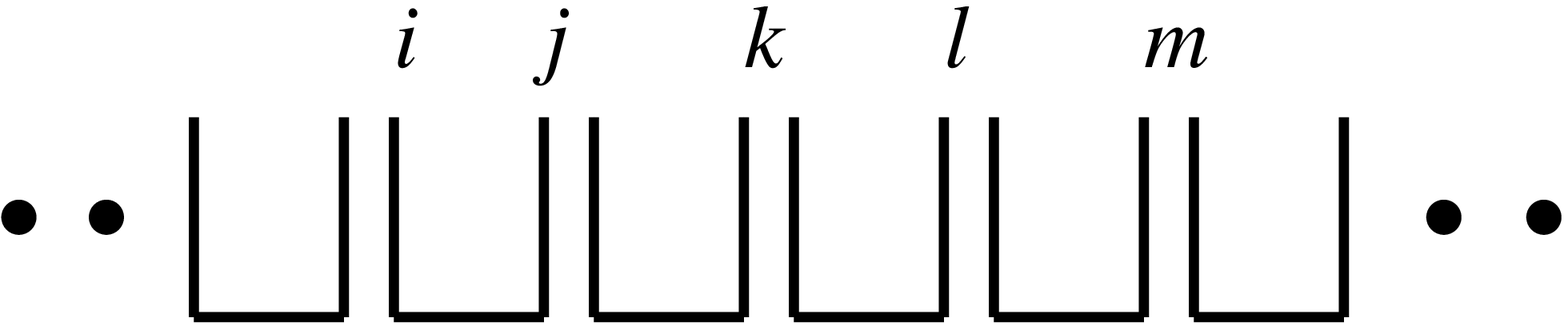,height=1.2cm}
\label{fig:overlap}
\caption{Colour connection of the partons showing the parent and daughter
partons for two adjacent single unresolved antennae.}
}
\FIGURE[h!]{ 
\epsfig{file=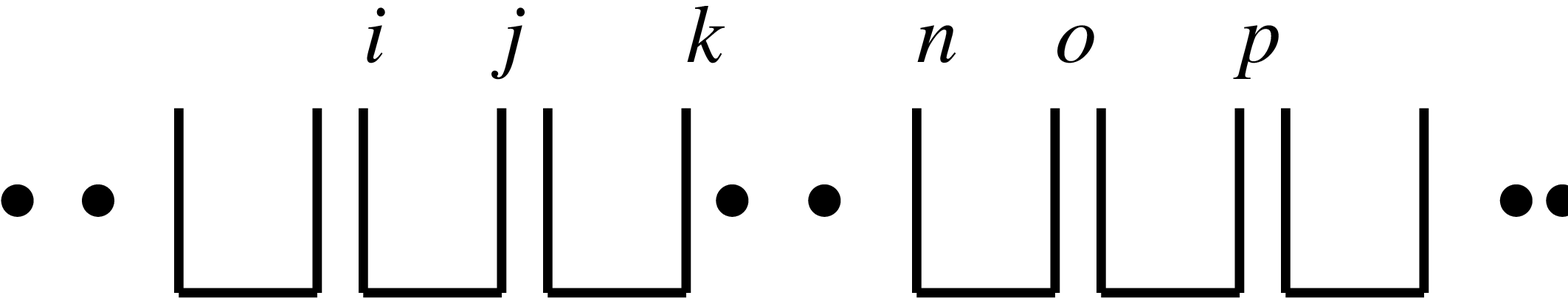,height=1.2cm}
\label{fig:nonoverlap}
\caption{Colour connection of the partons showing the parent and daughter
partons for two disconnected single unresolved antennae.}
}

In the following, we construct the subtraction
terms for all four configurations.

\subsubsection{Subtraction terms for single
unresolved partons}
\label{sec:sub2a}

Starting point for the subtraction terms for single unresolved partons 
are the NLO single unresolved antenna subtraction terms (\ref{eq:sigmasNLO}), 
\begin{eqnarray}
\lefteqn{{\rm d}\sigma_{NNLO}^{S,a}
=  {\cal N}\,\sum_{m+2}{\rm d}\Phi_{m+2}(p_{1},\ldots,p_{m+2};q)
\frac{1}{S_{{m+2}}} }\nonumber \\
&\times& \,\Bigg [ \sum_{j}\;X^0_{ijk}\,
|{\cal M}_{m+1}(p_{1},\ldots,\tilde{p}_{I},\tilde{p}_{K},\ldots,p_{m+2})|^2\,
\JET_{m}^{(m+1)}(p_{1},\ldots,\tilde{p}_{I},\tilde{p}_{K},\ldots,p_{m+2})\;
\Bigg
],\nonumber \\
\label{eq:sub2a}
\end{eqnarray}
where the NLO jet function $\JET_{m}^{(m)}$ is now replaced by 
$\JET_{m}^{(m+1)}$. In contrast to the NLO case, 
subtracting these terms from the full $(m+2)$-parton 
matrix element does not ensure a finite contribution in all single unresolved 
regions. This behaviour is related to the fact that
at NNLO the jet function 
$\JET_{m}^{(m+1)}$ allows one of the $(m+1)$ momenta 
to become unresolved, while at NLO 
$\JET_{m}^{(m)}$ required all $m$ momenta to be hard.
 As a consequence,
all momenta in the antenna function, including 
$p_j$, could be resolved while one of the momenta present in the 
reduced $(m+1)$-parton matrix element  becomes unresolved.
We distinguish between
two cases:  (1) where $\tilde{p}_I$ or $\tilde{p}_K$ become unresolved and (2) 
where any other 
momentum, $p_o$, in the matrix element becomes unresolved. Case (1)
is necessarily a double unresolved limit since $\tilde{p}_I$ and $\tilde{p}_K$ 
are
linear combinations of two momenta; this configuration is treated below in 
the discussion of the double unresolved limits of (\ref{eq:sub2a}). 

Case (2) is a single unresolved limit, since $p_j$ is resolved, while 
$p_o$ becomes unresolved. In this limit, 
${\rm d}\sigma_{NNLO}^{S,a}$ becomes singular. Its singular structure 
in this limit does not coincide with the limit of the 
full $(m+2)$-parton matrix 
element (which is already subtracted by the appropriate antenna term in 
${\rm d}\sigma_{NNLO}^{S,a}$ when $j=o$). 
The unresolved momentum $p_o$ 
in this situation is 
not colour-connected to the $p_j$ of $X^0_{ijk}$. 
The form of the spurious singular terms in 
${\rm d}\sigma_{NNLO}^{S,a}$ is therefore equal to the product of the antenna 
function with a singular limit of the 
reduced $(m+1)$-parton matrix element. It 
is equivalent to the product of two 
almost colour-unconnected or colour-unconnected  antenna 
functions with the reduced $m$-parton matrix element:
\begin{eqnarray*}
\lefteqn{X^0_{ijk}\,
|{\cal M}_{m+1}(p_{1},\ldots,\tilde{p}_{I},\tilde{p}_{K},\ldots,p_n,
p_o,p_p,\ldots,p_{m+2})|^2}\\
&-&X^0_{ijk}\,X^0_{nop}\,
|{\cal M}_{m}(p_{1},\ldots,\tilde{p}_{I},\tilde{p}_{K},\ldots,\tilde{p}_N,
\tilde{p}_P,\ldots,p_{m+2})|^2\stackrel{p_o~ 
\mbox{{\scriptsize  unresolved}}}{\longrightarrow} 0
\;,
\end{eqnarray*}
where $\tilde{p}_N=\tilde{p}_K$ is allowed.
Such structures 
appear as double unresolved subtraction terms for the configurations 
(c) and (d) defined above, and it will be shown in the appropriate subsections 
below, that the simple collinear limits of these double subtraction terms 
yield the correct behaviour to cancel the spurious single unresolved
poles of ${\rm d}\sigma_{NNLO}^{S,a}$.

To construct the double unresolved subtraction 
terms, we need to 
investigate the behaviour of the 
subtraction term (\ref{eq:sub2a}) in all its double unresolved limits. 
We distinguish the cases of colour-connected and colour-unconnected 
double unresolved limits. In the colour-connected, or type (b), 
double unresolved 
limits, (\ref{eq:sub2a}) can be non-vanishing only for one particular 
momentum configuration: two neighbouring pairs of colour-connected 
momenta becoming independently collinear, where one of the pairs lies 
inside the antenna, while the other pair consists of the remaining 
antenna momentum and its colour-connected neighbour. Each configuration of 
this type is contained precisely twice in (\ref{eq:sub2a}), since 
there are two possibilities of attributing the inside/outside pair.
Where appropriate, we shall 
call these limits colour-neighbouring in the following, and decompose 
colour-connected limits into genuinely colour-connected and 
colour-neighbouring limits.
In the colour-unconnected, type (c,d), double unresolved limits, 
(\ref{eq:sub2a}) is always non-vanishing and yields twice the double 
unresolved limit of the $(m+2)$-parton matrix element. This factor of two 
comes from the fact that each double unresolved limit requires both the 
unresolved antenna momentum $p_j$ and one other momentum $p_o$
appearing in the
$(m+1)$--parton matrix element to become unresolved; the role of the 
$p_j$ and $p_o$  can be interchanged, resulting in two identical terms 
contributing to the same limit. 

To summarise the above discussion, we conclude that (\ref{eq:sub2a}) 
yields twice the $(m+2)$-parton matrix element in all 
colour-unconnected,
almost colour-unconnected and colour-neighbouring
double unresolved limits, while vanishing in all 
genuinely colour-connected double unresolved limits.

\subsubsection{Subtraction terms for two colour-connected unresolved partons}
\label{sec:sub2b}

When the two unresolved partons $j$ and $k$ are adjacent, we construct the 
subtraction term starting from the four-particle tree-level antenna 
$X^0_{ijkl}$, which is an  appropriately normalised four-particle 
matrix element~\cite{our2j,chi,h}. 
By construction, it contains all colour-connected 
double unresolved limits of the $(m+2)$-parton matrix element
associated with partons $j$ and $k$, but it can 
also be singular in {\em single} unresolved limits associated with $j$ or $k$, where 
it does not coincide with limits of the matrix element. To ensure a finite 
subtraction term in all these single unresolved limits, we therefore
subtract the 
appropriate limits of the four-particle tree antennae, which are products of 
two tree-level three-particle antennae, such that the colour-connected double 
subtraction term reads:
\begin{eqnarray}
\lefteqn{{\rm d}\sigma_{NNLO}^{S,b}
=  {\cal N}\,\sum_{m+2}{\rm d}\Phi_{m+2}(p_{1},\ldots,p_{m+2};q)
\frac{1}{S_{{m+2}}}} \nonumber \\
&\times& \,\Bigg [ \sum_{jk}\;\left( X^0_{ijkl}
- X^0_{ijk} X^0_{IKl} - X^0_{jkl} X^0_{iJL} \right)\nonumber \\
&\times&
|{\cal M}_{m}(p_{1},\ldots,\tilde{p}_{I},\tilde{p}_{L},\ldots,p_{m+2})|^2\,
\JET_{m}^{(m)}(p_{1},\ldots,\tilde{p}_{I},\tilde{p}_{L},\ldots,p_{m+2})\;
\Bigg]\;,
\label{eq:sub2b}
 \end{eqnarray}
where the sum runs over all colour-adjacent pairs $j,k$ and implies the 
appropriate selection of hard momenta $i,l$.

As before, the subtraction term involves the $m$-parton amplitude 
evaluated with on-shell momenta $p_{1},\ldots,\tilde{p}_{I},\tilde{p}_{L},\ldots,p_{m+2}$
where now $\tilde{p}_{I}$ and $\tilde{p}_{L}$ are a linear combination of
$p_i$, $p_j$, $p_k$ and $p_l$.  As for the NLO antenna of the previous section, 
the tree antenna function $X^0_{ijkl}$
depends only on $p_i,p_j,p_k,p_l$.  Particles $i$ and $l$ play the
 role of the
radiators while $j$ and $k$ are the radiated partons.  
Antenna factorisation of squared matrix element and 
phase space in this configuration is illustrated pictorially in 
Figure~\ref{fig:sub2a}.
\FIGURE[h!]{ 
\label{fig:sub2a}
\epsfig{file=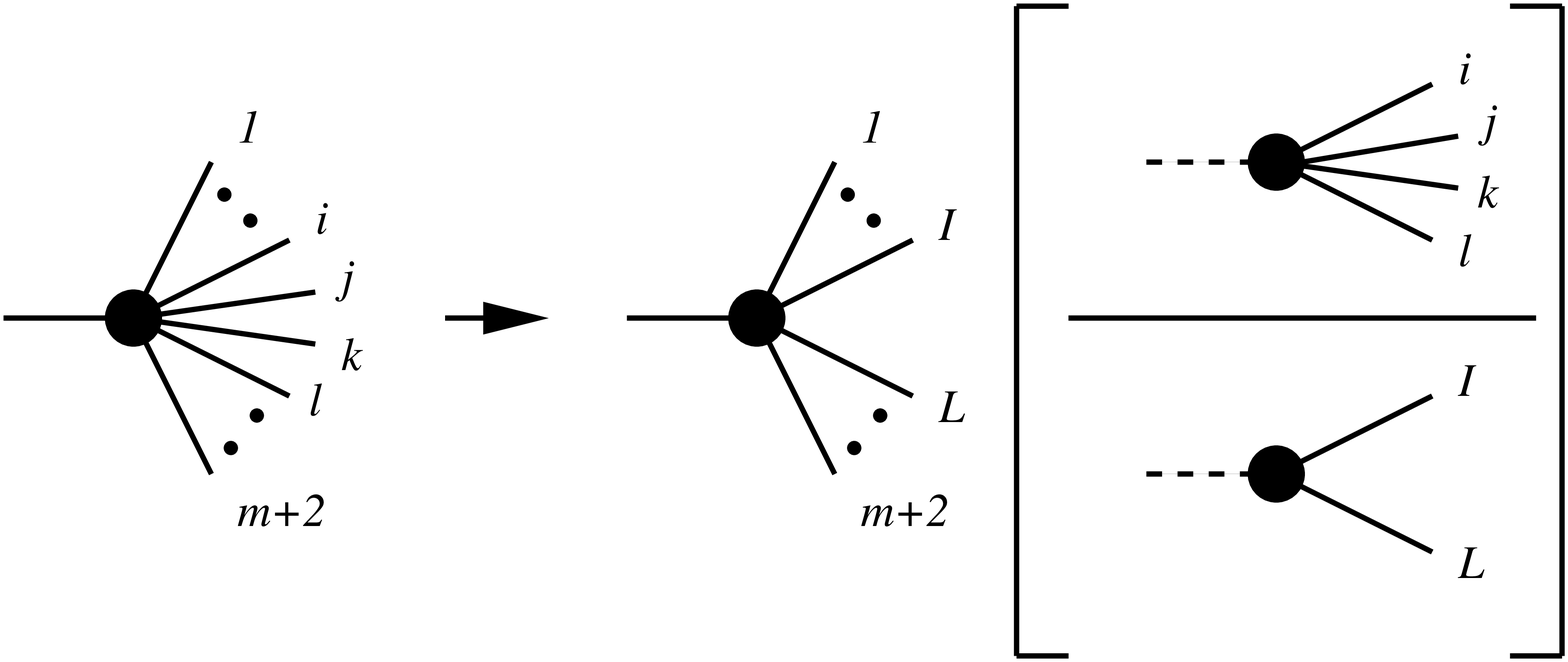,,height=4cm}
\caption{Illustration of NNLO antenna factorisation representing the
factorisation of both the squared matrix elements and the $(m+2)$-particle 
phase
space when the unresolved particles are colour connected. 
The term in square brackets
represents both the antenna function $X^0_{ijkl}$ and the antenna phase space
${\rm d}\Phi_{X_{ijkl}}$.}
}

Once again, the jet function $\JET^{(m)}_m$ in (\ref{eq:sub2b}) 
depends only on the parent momenta 
$\tilde{p}_{I},\tilde{p}_{L}$ and not $p_i,\ldots,p_l$. 
One can therefore carry out the integration
over the unresolved antenna phase space (or part thereof)
analytically, exploiting the 
factorisation of the phase space,
\begin{equation}
\label{eq:psx4}
\d \Phi_{m+2}(p_{1},\ldots,p_{m+2};q)  = 
\d \Phi_{m}(p_{1},\ldots,\tilde{p}_{I},\tilde{p}_{L},\ldots,p_{m+1};q)\cdot 
\d \Phi_{X_{ijkl}} (p_i,p_j,p_k,p_l;\tilde{p}_{I}+\tilde{p}_{L})\;.
\end{equation}
The factorisation~\cite{nnlosub1,ggh,uwer} is obtained by 
redefining a set of four massless on-shell momenta (radiator, two unresolved
partons, radiator) into two on-shell massless momenta. 
The NNLO antenna phase space $\d \Phi_{X_{ijkl}}$ 
is proportional to the four-particle phase space,
as can be seen by using $m=2$ in the above formula 
such that
\begin{equation}
{\rm d}\Phi_{4}= P_2\; {\rm d}\Phi_{X_{ijkl}}\;.
\end{equation}
One suitable representation of momenta for analytic integration purposes 
is the tripole phase space mapping of~\cite{ggh} 
where parton $i$ is the emitter and parton $l$ the spectator,
\begin{equation}
\tilde{p}_I^\mu = p_i^\mu+p_j^\mu+p_k^\mu - \frac{y_{ijk,l}}{1-y_{ijk,l}}p_l^\mu,
\qquad\qquad
\tilde{p}_L^\mu = \frac{1}{1-y_{ijk,l}}p_l^\mu,
\end{equation}
with
\begin{equation}
y_{ijk,l} =
\frac{p_l.p_i+p_l.p_j+p_l.p_k}{p_i.p_j+p_i.p_k+p_i.p_l+p_j.p_k+p_j.p_l+p_k.p_l}.
\end{equation}
For numerical implementation the mapping of~\cite{nnlosub1} is suitable.

For the analytic integration, 
we can use (\ref{eq:psx4}) to rewrite
each of the genuine four-particle subtraction terms  in the form, 
\begin{equation}
|{\cal M}_{m}|^2\,
\JET_{m}^{(m)}\; 
{\rm d}\Phi_{m}
\int {\rm d} \Phi_{X_{ijkl}}\;X^0_{ijkl},
\end{equation}
as before.
The analytic integral of the four-particle antenna subtraction term is 
therefore the antenna function
integrated over the fully inclusive 
antenna phase space, again including a normalisation factor to account for 
powers of the QCD coupling constant,
\begin{equation}
\label{eq:x4int}
{\cal X}^0_{ijkl}(s_{ijkl}) = 
\left(8\pi^2 \,(4\pi)^{-\e}\, e^{\e\gamma}\right)^2 
\int {\rm d} \Phi_{X_{ijkl}}\;X^0_{ijkl}\;.
\end{equation}
This integration is performed analytically in $d$ dimensions using the 
techniques explained in~\cite{ggh}
to make the infrared singularities explicit.

The double unresolved subtraction term (\ref{eq:sub2b}) correctly approximates 
the $(m+2)$-parton matrix element (\ref{eq:nnloreal}) in all 
double unresolved limits where colour-connected partons 
$j$ and $k$ become unresolved, except in the configuration where parton $j$
becomes collinear with parton $i$, while parton $k$ becomes collinear with 
parton $l$. In this (so-called colour-neighbouring)
configuration, the tree-level four-parton 
antenna function $X_{ijkl}$
correctly approximates the $(m+2)$-parton matrix element, while
the products of tree-level three-parton 
antenna functions $X^0_{ijk} X^0_{IKl}$ and $X^0_{jkl} X^0_{iJL}$
in (\ref{eq:sub2b}) are non-vanishing, and are each equal to 
the double unresolved limit of the matrix element. This 
colour-neighbouring momentum configuration 
was already discussed in the context of the 
double unresolved limits of ${\rm d}\sigma_{NNLO}^{S,a}$ in the previous 
subsection.
It can be seen easily that the two products of tree-level three-parton 
antenna functions 
$X^0_{ijk} X^0_{IKl}$ and $X^0_{jkl} X^0_{iJL}$
in (\ref{eq:sub2b}) exactly cancel the spurious 
double unresolved singular terms in present in (\ref{eq:sub2a}) in these 
colour-neighbouring double unresolved
limits. 

In all  genuinely colour-connected limits,  the four-parton 
tree-level antenna function  $X_{ijkl}$ 
in (\ref{eq:sub2b}) correctly matches 
the singular 
structure of the $(m+2)$-parton matrix element, while the 
products of tree-level three-parton 
antenna functions $X^0_{ijk} X^0_{IKl}$ and $X^0_{jkl} X^0_{iJL}$
vanish (although these terms are singular at first 
sight, they are not sufficiently singular to overcome the small 
phase space volume associated with the double unresolved limits under 
consideration). 

By construction, (\ref{eq:sub2b}) vanishes in all single unresolved 
limits: the antenna functions cancel each other in these limits, and 
singularities in the $m$-parton matrix element are forbidden by the 
jet function.

\subsubsection{Subtraction terms for two almost colour-unconnected unresolved partons}
\label{sec:sub2c}

There are double unresolved configurations where the unresolved partons are
separated by a hard radiator parton, for example, $i,j,k,l,m$ where $j$
and $l$ are unresolved as illustrated in Fig.~\ref{fig:sub2c}.
These configurations are called almost colour-unconnected.
In this case, we take the strongly
ordered approach~\cite{ant} where
$i,j,k$ form an antenna with hard partons $I$ and $K$ yielding an ordered
amplitude involving $I,K,l,m$.   As usual, the momenta of the hard
radiator partons $I$ and $K$ are labelled $\tilde{p}_I$ and 
$\tilde{p}_K$ and are constructed from $p_i,p_j,p_k$.
The cases where $l$ is unresolved are then 
treated using an antenna $K,l,m$ with hard partons $K$ and $L$.
Here the momenta of the hard
radiator partons $K$ and $L$ are labelled $\doubletilde{p}_K$ and
$\doubletilde{p}_M$ which are made from $\tilde{p}_K,p_l,p_m$.
The other case where first $k,l,m$ form an 
antenna followed by $i,j,K$ is also
included.
\FIGURE[!t]{ 
\label{fig:sub2c}
\epsfig{file=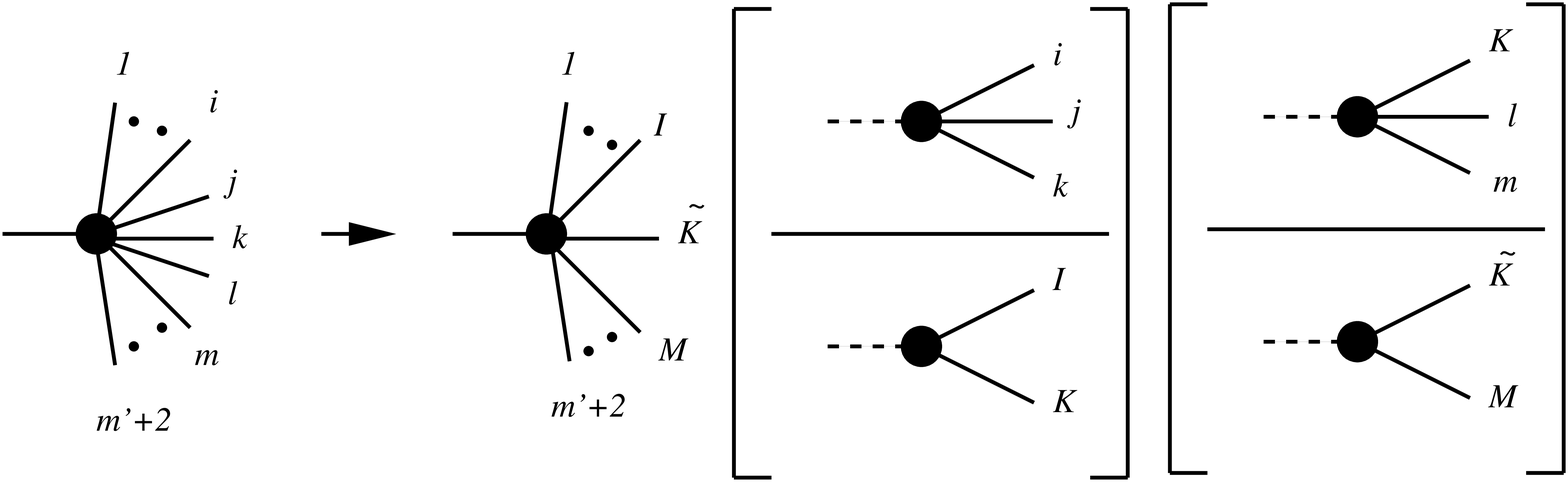,height=4cm}
\caption{Illustration of NNLO antenna factorisation representing the
factorisation of both the squared matrix elements and the 
$\protect{(m^{\prime}+2)}$-particle
 phase
space when the unresolved particles are almost colour-unconnected. 
}}

In constructing the subtraction term, we have to account for the fact that 
the previously defined subtraction term ${\rm d}\sigma_{NNLO}^{S,a}$ 
(\ref{eq:sub2a}) already contributes in the almost 
 colour-unconnected configurations by subtracting twice the limit 
of the ($m^\prime+2$)-parton matrix element in these limits (to avoid confusion 
between the indivial unresolved momenta and the total number of particles, 
we denote the latter $(m^\prime+2)$ instead of $(m+2)$ in this subsection).
Therefore,
the almost colour-unconnected subtraction term reads,
\begin{eqnarray}
{\rm d}\sigma_{NNLO}^{S,c}
&= & - {\cal N}\,\sum_{m^\prime+2}{\rm d}\Phi_{m^\prime+2}(p_{1},\ldots,p_{m^\prime+2};q)
\frac{1}{S_{{m^\prime+2}}} \nonumber \\
&\times& \,\Bigg [  \sum_{j,l}\;X^0_{ijk}\;x^0_{mlK}\,
|{\cal M}_{m^\prime}(p_{1},\ldots,\tilde{p}_{I},{\doubletilde{p}}_{K},{\doubletilde{p}}_{M},\ldots,p_{m^\prime+2})|^2\,\nonumber \\
&&\hspace{3cm}\times
\JET_{m^\prime}^{(m^\prime)}(p_{1},\ldots,\tilde{p}_{I},{\doubletilde{p}}_{K},{\doubletilde{p}}_{M},\ldots,p_{m^\prime+2})\;
\phantom{\Bigg]}
\nonumber \\
&& \,+ \sum_{j,l}\;X^0_{klm}\;x^0_{ijK}\,
|{\cal M}_{m^\prime}(p_{1},\ldots,{\doubletilde{p}}_{I},{\doubletilde{p}}_{K},\tilde{p}_{M},\ldots,p_{m^\prime+2})|^2\,\nonumber \\
&&\hspace{3cm}\times
\JET_{m^\prime}^{(m^\prime)}(p_{1},\ldots,{\doubletilde{p}}_{I},{\doubletilde{p}}_{K},\tilde{p}_{M},\ldots,p_{m^\prime+2})\;
\Bigg
]\;,
\label{eq:sub2c}
 \end{eqnarray}
where $x^0_{mlK}$ denotes a sub-antenna, containing only the collinear limit 
of $m$ with $l$, but not the collinear limit of $l$ with $K$; 
in the soft limit of $l$, this sub-antenna yields 
half the soft eikonal factor. 
The jet function $\JET_{m^\prime}^{(m^\prime)}$ and the matrix element 
depend only on the momenta associated 
with the $m^\prime$-particle final state. Therefore, the integral over the 
phase space associated with the 
leftmost antenna  function can be carried out analytically, like in the 
single unresolved case discussed above.  After this 
integration, the sub-antennae can be recombined to full antenna functions 
again. 

If required, it is also possible to carry out the integrals over both 
phase spaces associated with the two antenna functions.  In this case,
one can 
exploit the factorisation of the phase space to
analytically perform the integrals,
\begin{eqnarray}
\label{eq:ps33}
\lefteqn{
\d \Phi_{m^\prime+2}(p_{1},\ldots,p_{m^\prime+2};q)  =}\nonumber \\
&&
\d \Phi_{m^\prime+1}(p_{1},\ldots,\tilde{p}_{I},\tilde{p}_{K},\ldots,
p_{m^\prime+2};q)\cdot 
\d \Phi_{X_{ijk}} (p_i,p_j,p_k)\;\nonumber \\
&=&
\d \Phi_{m^\prime}(p_{1},\ldots,\tilde{p}_{I},{\doubletilde{p}}_{K},{\doubletilde{p}}_{M},\ldots,p_{m^\prime+2};q)\cdot 
\d \Phi_{X_{Klm}} (\tilde{p}_K,p_l,p_m;
{\doubletilde{p}}_{K}+{\doubletilde{p}}_{M})\nonumber \\
&& \hspace{2cm}
\cdot
\d \Phi_{X_{ijk}} (p_i,p_j,p_k;\tilde{p}_{I}+\tilde{p}_{K})\;.
\end{eqnarray}
In carrying out the analytic integration, some care has to be taken in 
the second dipole integral $\d \Phi_{X_{Klm}}$, which will pick up 
$\e$-dependent factors from the first integral (both integrals are
fully independent only in four dimensions) related to 
the correlation of unit volumes involved in the two integrals.
As a consequence, the analytic integration of the above formula will 
not yield 
the product of two independent integrated NLO antenna functions.  

Concerning the single unresolved limits of ${\rm d}\sigma_{NNLO}^{S,c}$,
(\ref{eq:sub2c}) becomes singular only 
if one of the antenna partons $j$ or $l$ is unresolved. In these cases, 
it exactly cancels the spurious single unresolved singularities encountered 
in  ${\rm d}\sigma_{NNLO}^{S,a}$ for the 
configuration of an unresolved momentum $p_o$ in the ($m^\prime+1$)-parton matrix 
element, which is almost colour-unconnected to the momenta in the antenna 
function. In the double unresolved limits of ${\rm d}\sigma_{NNLO}^{S,c}$,
momenta $j$ and $l$ are both unresolved, yielding minus the double unresolved
limit of the $(m^\prime+2)$-parton matrix element, as required to cancel the 
oversubtraction incurred by ${\rm d}\sigma_{NNLO}^{S,a}$ in these limits. 

\subsubsection{Subtraction terms for two colour-unconnected
unresolved partons}
\label{sec:sub2d}
\FIGURE[h!]{ 
\label{fig:sub2d}
\epsfig{file=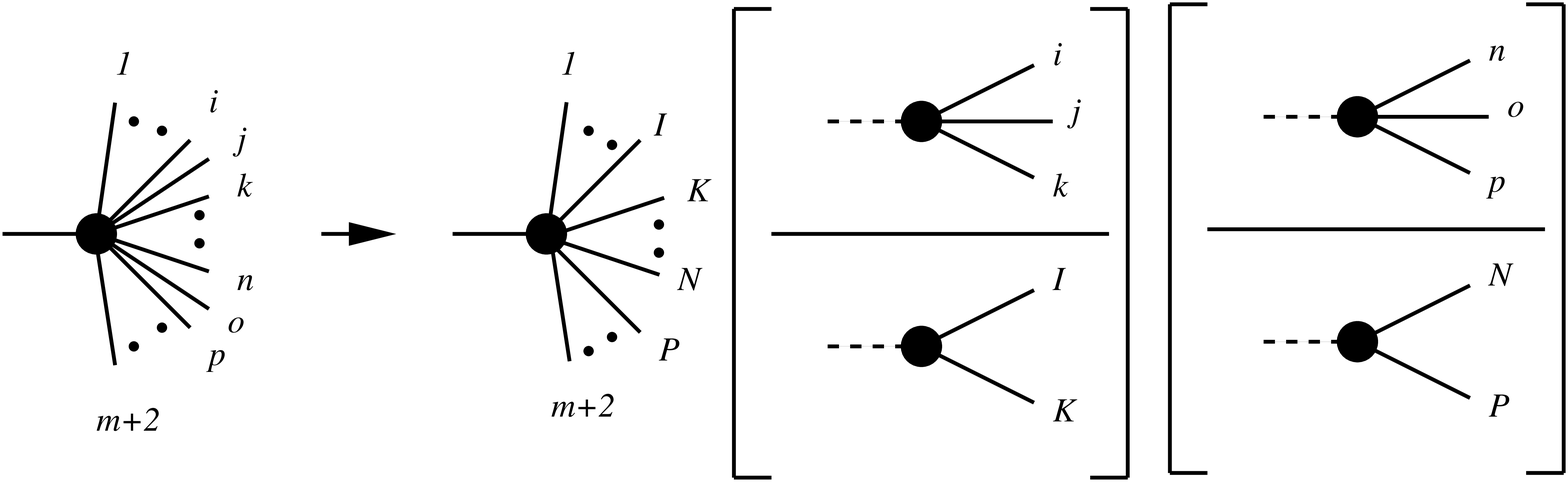,height=4cm}
\caption{Illustration of NNLO antenna factorisation representing the
factorisation of both the squared matrix elements and the $(m+2)$-particle phase
space when the unresolved particles are colour-unconnected. 
}}

When two unresolved partons $j$ and $o$ are completely disconnected, 
the $(m+2)$-parton matrix element factorises into the product of two 
uncorrelated single unresolved factors with a reduced $m$-parton matrix 
element. In this limit, 
the single unresolved subtraction term ${\rm d}\sigma_{NNLO}^{S,a}$ 
(\ref{eq:sub2a}) subtracts a contribution which is twice the limit of
the full $(m+2)$-parton matrix element. The dedicated 
colour-unconnected double unresolved subtraction term has to 
compensate for this oversubtraction. 
This subtraction term is a sum over independent three-particle tree-level 
 antennae 
as illustrated in Figure~\ref{fig:sub2d},
\begin{eqnarray}
{\rm d}\sigma_{NNLO}^{S,d}
&= & - {\cal N}\,\sum_{m+2}{\rm d}\Phi_{m+2}(p_{1},\ldots,p_{m+2};q)
\frac{1}{S_{{m+2}}} \nonumber \\
&\times& \,\Bigg [ \sum_{j,o}\;X^0_{ijk}\;X^0_{nop}\,
|{\cal M}_{m}(p_{1},\ldots,\tilde{p}_{I},\tilde{p}_{K},\ldots,\tilde{p}_{N},\tilde{p}_P,\ldots,p_{m+2})|^2\,\nonumber \\
&\times&
\JET_{m}^{(m)}(p_{1},\ldots,\tilde{p}_{I},\tilde{p}_{K},\ldots,\tilde{p}_{N},\tilde{p}_P,\ldots,p_{m+2})\;\Bigg
]\;,
\label{eq:sub2d}
 \end{eqnarray}
where the summation over $o$ is such that it only includes 
colour-connected antenna 
configurations $X^0_{nop}$ which have no 
common momentum with $X^0_{ijk}$.

The subtraction term involves the $m$-parton amplitude 
evaluated with on-shell momenta $p_{1},\ldots,\tilde{p}_{I},\tilde{p}_{K},\ldots,\tilde{p}_{N},\tilde{p}_P,\ldots,p_{m+2}$
where now $\tilde{p}_{I}$ and $\tilde{p}_{K}$ are a linear combination of
$p_i$, $p_j$, $p_k$ and $\tilde{p}_N$ and $\tilde{p}_P$ are made from $p_n$,
$p_o$ and $p_p$.  The sum in the above equation is such that no product of 
two antenna configurations appears twice. 
Each antenna is fully independent, and the phase space factorises as,
\begin{eqnarray}
\label{eq:psx33}
\d \Phi_{m+2}(p_{1},\ldots,p_{m+2};q)  &=& 
\d \Phi_{m}(p_{1},\ldots,\tilde{p}_{I},\tilde{p}_{K},\ldots,\tilde{p}_{N},\tilde{p}_P,\ldots,p_{m+2};q)\nonumber \\ 
&&\times \d \Phi_{X_{ijk}} (p_i,p_j,p_k;\tilde{p}_{I}+\tilde{p}_{K})
\cdot\d \Phi_{X_{nop}} (p_n,p_o,p_p;\tilde{p}_{N}+\tilde{p}_P)\;.\nonumber \\
\end{eqnarray}
This contribution is either integrated only once, and then cancelled 
with subtraction terms present in the $(m+1)$-parton contribution, or
integrated twice to make its full infrared pole structure explicit. 
Integration can be obtained by repeated use of 
(\ref{eq:x3int}),
so that each term in the sum is in the form, 
\begin{displaymath}
|{\cal M}_{m}|^2\,
\JET_{m}^{(m)}\; 
{\cal X}^0_{ijk}(s_{ijk})\;{\cal X}^0_{nop}(s_{nop}).
\end{displaymath}

The subtraction term (\ref{eq:sub2d}) contributes to single unresolved 
limits if either $p_j$ or $p_o$ are unresolved. In these limits, it 
cancels precisely the spurious poles associated with the momentum $p_o$ 
becoming unresolved in the ($m+1$)-parton matrix element of (\ref{eq:sub2a}).
In the double unresolved limits ($p_j$ and $p_o$) it is designed for, it 
cancels the oversubtraction incurred by (\ref{eq:sub2a}), thus ensuring 
a proper subtraction of the singular terms of the $(m+2)$-parton matrix 
element in these.

\subsubsection{Correction terms in the $m$-jet region}

The full double radiation subtraction term is given as sum of all 
subtraction terms constructed above:
\begin{equation}
{\rm d}\sigma_{NNLO}^{S} = {\rm d}\sigma_{NNLO}^{S,a} 
+{\rm d}\sigma_{NNLO}^{S,b}+{\rm d}\sigma_{NNLO}^{S,c}
+{\rm d}\sigma_{NNLO}^{S,d} \;.
\label{eq:sub2}
\end{equation}
As outlined in the previous subsections, this subtraction term 
correctly approximates the $(m+2)$-parton matrix element contribution to 
$m$-jet final states as defined in (\ref{eq:nnloreal}) in all double and 
single unresolved regions. Although individual terms in (\ref{eq:sub2}) 
contain spurious singularities in these limits, they cancel among each
other in the sum. 

The integrated form of (a) corresponds to an ($m+1$)-parton configuration, 
while the integrated forms of (b), (c) and (d) are either ($m+1$)-parton 
or $m$-parton configurations (for all but the four-parton antenna terms 
in (b), we can actually choose which type of configuration we want to
integrate). 
 They are added with the two-loop 
$m$-parton and the one-loop $(m+1)$-parton contributions to $m$-jet final 
states to yield integrands free of explicit infrared poles.

\subsection{Single unresolved loop subtraction terms}

The $(m+1)$-parton one-loop contribution to $m$-jet final states at NNLO is
\begin{eqnarray}
{\rm d}\sigma^{V,1}_{NNLO}
&=& {\cal N}\,\sum_{m+2}{\rm d}\Phi_{m+1}(p_{1},\ldots,p_{m+1};q)
\frac{1}{S_{{m+1}}} \nonumber \\ &&\times
|{\cal M}^1_{m+1}(p_{1},\ldots,p_{m+1})|^2\;
\JET_{m}^{(m+1)}(p_{1},\ldots,p_{m+1})\;,
\label{eq:nnloonel}
\end{eqnarray}
where we introduced a shorthand notation for the one-loop 
corrected contribution to the $(m+1)$-parton squared matrix element,
\begin{equation}
|{\cal M}^1_{m+1}(p_{1},\ldots,p_{m+1})|^2 
= 2 \,{\rm Re}\, \left({\cal M}^{{\rm loop}}_{m+1}(p_{1},\ldots,p_{m+1})\,
{\cal M}^{{\rm tree},*}_{m+1}(p_{1},\ldots,p_{m+1})\right)\;,
\end{equation}
where the subleading contributions in colour are again implicit. 
This expression 
contains two types of infrared singularities. The 
renormalised one-loop virtual correction 
${\cal M}^{{\rm loop}}_{m+1}$ 
to the $(m+1)$-parton matrix contains explicit infrared 
poles, which can be expressed using the infrared singularity operators defined 
in Section~\ref{sec:operators} below. On the other hand, one of the 
$(m+1)$-partons can be unresolved, leading to infrared singularities which
become explicit only after integration over the unresolved patch of the 
final state $(m+1)$-parton phase space. The single unresolved limits of 
one- and two-loop amplitudes are investigated in detail in~\cite{onelstr,
oneloopsoft,onelstr1,onelstr2,twolstr}.

To carry out the numerical integration over the $(m+1)$-parton phase,
weighted by the appropriate jet function, we have to construct an 
infrared subtraction term ${\rm d}\sigma^{VS,1}_{NNLO}$, which purpose 
is twofold: on the one hand, it should fully account for the explicit 
infrared singularities arising from the one-loop correction, while on the 
other hand it should approximate (\ref{eq:nnloonel}) in all single unresolved 
limits. After subtraction of  ${\rm d}\sigma^{VS,1}_{NNLO}$ from 
${\rm d}\sigma^{V,1}_{NNLO}$, the integrand itself should be free from explicit
infrared poles, and all singular limits due to single unresolved phase 
space configurations must be properly accounted for. 
We require three types of subtraction terms:
\begin{itemize}
\item[(a)] The explicit infrared poles of the $(m+1)$-parton one-loop matrix 
element are removed. 
\item[(b)] The single unresolved limits of the 
$(m+1)$-parton one-loop matrix 
element are subtracted. 
\item[(c)] Oversubtracted explicit and implicit pole terms are removed.
\end{itemize}
All three types of  subtraction terms are constructed below.

\subsubsection{Subtraction of explicit infrared poles}
It is a well known fact from NLO calculations, that the explicit 
infrared poles of one-loop matrix elements cancel with the 
corresponding infrared poles obtained by integrating out all single 
unresolved configurations from the real radiation matrix elements 
contributing to the same (infrared safe) observable~\cite{kln}. 
We can therefore subtract all explicit poles present in 
${\rm d}\sigma^{V,1}_{NNLO}$ with the subtraction term
\begin{eqnarray}
\lefteqn{{\rm d}\sigma_{NNLO}^{VS,1,a}
=   {\cal N}\,\sum_{m+1}{\rm d}\Phi_{m+1}(p_{1},\ldots,p_{m+1};q)
\frac{1}{S_{{m+1}}} }\nonumber \\
&\times& \,\Bigg [ \sum_{ik}\;  - {\cal X}^0_{ijk}(s_{ik}) \,
|{\cal M}_{m+1}(p_{1},\ldots,{p}_{i},{p}_{k},\ldots,p_{m+1})|^2\,
\JET_{m}^{(m+1)}(p_{1},\ldots,{p}_{i},{p}_{k},\ldots,p_{m+1})\;
\Bigg
],\nonumber \\
\label{eq:subv2a}
\end{eqnarray}
where the sum runs over all colour-connected pairs of momenta $(p_i,p_k)$.
The symbol ${\cal X}^0_{ijk}(s_{ik})$ denotes the normalised 
integral of the three-parton antenna function, with a parton of type 
$j$ emitted between partons $i$ and $k$,
with the total invariant mass 
of the antenna particles given by the invariant mass of $p_i$ and 
$p_k$. Since in the single unresolved mapping of the antenna phase space
(\ref{eq:psx3}), $p_i+p_j+p_k = \tilde{p}_I + \tilde{p}_K$, we 
can identify the momenta $p_i$ and $p_k$ of the $(m+1)$-parton phase space 
with the momenta $\tilde{p}_I$ and $\tilde{p}_K$ obtained in the 
factorisation of the $(m+2)$-parton phase space in 
Section~\ref{sec:sub2a}. Consequently, using the antenna factorisation of the
$(m+2)$-parton phase space in (\ref{eq:sub2a}) and carrying out the 
phase space integral for each antenna $X_{ijk}$, we find 
\begin{equation}
{\rm d}\sigma_{NNLO}^{VS,1,a} = - {\rm d}\sigma_{NNLO}^{S,a} \;,
\label{eq:nlocancel}
\end{equation}
which is of course merely a consequence of the cancellation of infrared 
poles at NLO. 

It is evident that subtraction of ${\rm d}\sigma^{VS,1,a}_{NNLO}$
from ${\rm d}\sigma^{V,1}_{NNLO}$ ensures an integrand free from 
explicit infrared poles over the whole region of integration. In the 
single unresolved regions, both these functions develop further 
infrared singularities, which do not coincide. Therefore, we have to 
introduce further subtraction terms to ensure a finite integrand 
in all single unresolved regions. 

\subsubsection{Subtraction terms for one-loop single-unresolved 
contributions}
\FIGURE[h!]{ 
\label{fig:subv}
\epsfig{file=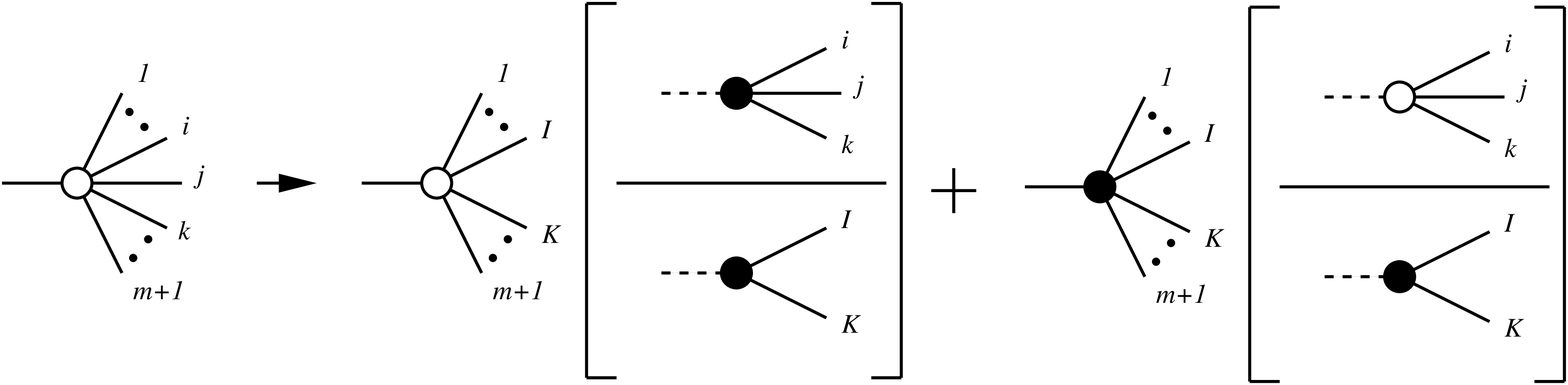,height=3.5cm}
\caption{Illustration of NNLO antenna factorisation representing the
factorisation of both the one-loop
``squared" matrix elements (represented by the white blob)
and the $(m+1)$-particle phase
space when the unresolved particles are colour connected. 
The terms in square brackets
represent both the three-particle tree-level antenna function $X^0_{ijk}$ 
and the three-particle one-loop antenna function $X^1_{ijk}$
and the antenna phase space.}}

In simple unresolved limits, the behaviour of $(m+1)$-parton 
one-loop amplitudes is described 
by the sum of two different contributions~\cite{onelstr,oneloopsoft,
onelstr1,onelstr2,twolstr}: a simple unresolved tree level factor times a
$m$-parton one-loop amplitude and a simple unresolved one-loop factor 
times a $m$-parton tree-level amplitude, as 
illustrated in Figure~\ref{fig:subv}. Accordingly, we construct 
the one-loop single unresolved subtraction term as
\begin{eqnarray}
{\rm d}\sigma_{NNLO}^{VS,1,b}
&= & {\cal N}\,\sum_{m+1}{\rm d}\Phi_{m+1}(p_{1},\ldots,p_{m+1};q)
\frac{1}{S_{{m+1}}} \nonumber \\
&\times& \,\sum_{j} \Bigg [X^0_{ijk}\,
|{\cal M}^1_{m}(p_{1},\ldots,\tilde{p}_{I},\tilde{p}_{K},\ldots,p_{m+1})|^2\,
\JET_{m}^{(m)}(p_{1},\ldots,\tilde{p}_{I},\tilde{p}_{K},\ldots,p_{m+1})\;
\nonumber \\
&&\phantom{\sum_{j} }+\;X^1_{ijk}\,
|{\cal M}_{m}(p_{1},\ldots,\tilde{p}_{I},\tilde{p}_{K},\ldots,p_{m+1})|^2\,
\JET_{m}^{(m)}(p_{1},\ldots,\tilde{p}_{I},\tilde{p}_{K},\ldots,p_{m+1})\;\Bigg
].\nonumber \\
\label{eq:subv2b}
\end{eqnarray}
In this expression, we have introduced the one-loop three-parton antenna
function $X^1_{ijk}$, which depends only on the antenna momenta
$p_i,p_j,p_k$. It correctly describes all simple unresolved limits of the
difference between an $(m+1)$-parton one-loop corrected 
squared matrix element and the product of a tree-level antenna function 
with the  $m$-parton one-loop corrected squared matrix element.
It can therefore be constructed out of one-loop three-parton and two-parton 
matrix elements, as outlined in Section~\ref{sec:notation} below. 
It should be noted that  $X^1_{ijk}$ is renormalised at a scale corresponding 
to the invariant mass of the antenna partons, $s_{ijk}$, while the one-loop
$(m+1)$-parton matrix element is renormalised at a scale $\mu^2$ (which is 
often chosen to be $q^2$ in jet production in $e^+e^-$-collisions). To 
ensure correct subtraction of terms arising from renormalisation, 
we have to substitute 
\begin{equation}
X^1_{ijk} \to X^1_{ijk} + \frac{\beta_0}{\e}\,
\frac{\left(4\pi\right)^{\e} e^{-\e\gamma}}{8\pi^2} X^0_{ijk} \left(
\left(s_{ijk}\right)^{-\e} - \left(\mu^2\right)^{-\e} \right)  
\end{equation}
in (\ref{eq:subv2b}).
The terms arising from this substitution
 will in general be kept apart in the construction of the 
colour-ordered subtraction terms, since they all share a  common 
colour structure $\beta_0$.

In contrast to all other expressions appearing in the construction of the 
one-loop single unresolved subtraction terms, $X^1_{ijk}$ can never
be related to integrals of tree-level subtraction terms. Therefore, this 
component of the subtraction term must cancel with parts of the two-loop 
$m$-parton contribution and we must 
integrate it over the three-parton antenna phase space. This can be 
accomplished using the techniques described in~\cite{ggh} and yields
\begin{equation}
\label{eq:x31int}
{\cal X}^1_{ijk}(s_{ijk}) = \left(8\pi^2 \,(4\pi)^{-\e}\, e^{\e\gamma}\right) 
\int {\rm d} \Phi_{X_{ijk}}\;X^1_{ijk}.
\end{equation}

\subsubsection{Compensation terms for oversubtracted poles}
By construction, (\ref{eq:subv2b}) correctly approximates the one-loop
$(m+1)$-parton contribution to $m$-jet final states in all single unresolved 
limits. However, outside these limits, where ($m+1$)-parton
configurations form $m$-jet final states because of the experimental jet
resolution criteria, this subtraction term no longer coincides with the 
squared matrix element, and induces spurious explicit infrared poles. 

To compensate for these explicit infrared poles outside the 
singular regions, we introduce a further subtraction term,
\begin{eqnarray}
\lefteqn{{\rm d}\sigma_{NNLO}^{VS,1,c}
=   {\cal N}\,\sum_{m+1}{\rm d}\Phi_{m+1}(p_{1},\ldots,p_{m+1};q)
\frac{1}{S_{{m+1}}} }\nonumber \\
&\times& \,\Bigg [ \sum_{ik}\; {\cal X}^0_{ijk}(s_{ik}) \,
\sum_o X^0_{nop} \,
|{\cal M}_{m}(p_{1},\ldots,p_i,p_k,\ldots,\tilde{p}_{N},\tilde{p}_{P},\ldots,p_{m+1})|^2\,
\nonumber \\ 
&& \hspace{3cm} \times 
\JET_{m}^{(m)}(p_{1},\ldots,p_i,p_k,\ldots,\tilde{p}_{N},\tilde{p}_{P},\ldots,p_{m+1})\;
\Bigg
],
\label{eq:subv2c}
\end{eqnarray}
where any of the two momenta 
$p_i$ and $p_k$ can be equal to $p_n$, $p_p$, $\tilde{p}_N$ or 
$\tilde{p}_P$, but not to the unresolved 
momentum $p_o$. Individual terms in ${\rm d}\sigma_{NNLO}^{VS,1,c}$ 
cancel with individual terms in  ${\rm d}\sigma_{NNLO}^{S,b}$ (\ref{eq:sub2b}),
${\rm d}\sigma_{NNLO}^{S,c}$ (\ref{eq:sub2c}) and
  ${\rm d}\sigma_{NNLO}^{S,d}$ (\ref{eq:sub2d}), after 
one of the two three-parton antenna 
phase space integrals is carried out. Any remaining terms of 
${\rm d}\sigma_{NNLO}^{VS,1,c}$ can be integrated over the antenna phase space
$\d \Phi_{X_{nop}}$. In the case of $p_i$ or $p_k$ coinciding with one of
the momenta of the antenna phase space, some care has to be taken in carrying 
out the integrals, which differ from the standard tree-level three-parton 
antenna integrals by normalisation factors coming from 
${\cal X}^0_{ijk}(s_{ik})$.

The subtraction term (\ref{eq:subv2c}) cancels the explicit infrared poles 
of (\ref{eq:subv2b}) in the region where all $(m+1)$ partons are theoretically 
resolved, thus ensuring a finite integrand. In the single unresolved 
regions, the sum of (\ref{eq:subv2c}) and (\ref{eq:subv2a}) vanishes, as 
can be seen rather easily from the nature of NLO antenna subtraction. 
Consequently, only ${\rm d}\sigma_{NNLO}^{VS,1,b}$ contributes in this 
region, as required to cancel the singularities
of the $(m+1)$-parton squared matrix 
element.

\subsubsection{Correction terms in the $m$-jet region}

The full one-loop single real radiation
 subtraction term is the sum of all 
subtraction terms constructed above:
\begin{equation}
{\rm d}\sigma_{NNLO}^{VS,1} = {\rm d}\sigma_{NNLO}^{VS,1,a} 
+{\rm d}\sigma_{NNLO}^{VS,1,b}+{\rm d}\sigma_{NNLO}^{VS,1,c}\;.
\label{eq:sub2vall}
\end{equation}
As outlined in the previous subsections, this subtraction term 
correctly approximates the one-loop $(m+1)$-parton squared
matrix element contribution to 
$m$-jet final states as defined in (\ref{eq:nnloonel}) in all 
single unresolved regions and removes all explicit infrared poles.

While ${\rm d}\sigma_{NNLO}^{VS,1,a}$ cancels fully with 
${\rm d}\sigma_{NNLO}^{S,a}$, parts of the remaining two terms 
have to be integrated to yield $m$-parton configurations, which are 
then added  
with the two-loop 
$m$-parton contributions
${\rm d}\sigma_{NNLO}^{V,2}$. The precise nature of cancellations between 
the terms 
appearing in the  integrated forms of ${\rm d}\sigma_{NNLO}^{S,(bcd)}$ 
and terms in the unintegrated 
${\rm d}\sigma_{NNLO}^{VS,1,(bc)}$ differs considerably among the 
different colour structures. Therefore,  no generic formula for these 
cancellations can be stated. In Section~\ref{sec:3j}, we will illustrate 
these cancellations on the example of the subleading colour contribution  
to $e^+e^- \to 3j$ at NNLO.

\subsection{Comparison with other approaches}

The NNLO antenna subtraction approach which we derive in detail in this section
was previously sketched in our earlier publications~\cite{our2j,our3j}, 
where specific applications to $e^+e^- \to 2j$ and $e^+e^-\to 3j$ were 
considered. The case two-jet production is special in several aspects, 
particularly since the double unresolved tree-level and single unresolved 
one-loop subtraction terms exactly coincide with the full matrix elements they are 
supposed to approximate. Moreover, the jet functions for two- and three-parton 
final states fulfil special relations which are in general not fulfilled 
for $m$- and $(m+1)$-parton final states. Most notably (see (5) 
in~\cite{our3j}),
\begin{equation}
\JET_3^{(3)}(p_i,p_j,p_k) = 
\JET_2^{(2)}(\tilde{p}_I,\tilde{p}_K) -
\JET_2^{(3)}(p_i,p_j,p_k)\;, 
\end{equation}
 which can be used to express (3.17) of~\cite{our2j} in the form of 
(\ref{eq:sub2}) presented in this paper.

Several other approaches to the handling of infrared singularities 
at NNLO have been proposed in the literature. One can distinguish two 
substantially different lines of making  the infrared singularities in 
real radiation at NNLO explicit: either through a direct expansion 
of matrix element and phase space or through 
subtraction terms. In expanding the full real radiation matrix elements 
and the full multi-parton phase space in dimensional regularisation
through sector decomposition~\cite{secdec,ggh}, one 
arrives at a Laurent expansion in $\e$ with coefficients 
in terms of distributions in the Lorentz 
invariants associated with the process. These coefficients still contain the 
full phase space integration, but they are finite and the integrals 
can  be carried out numerically. 
Using this approach, NNLO results have been 
obtained for $e^+e^- \to 2j$~\cite{babis2j},
$pp \to H+X$~\cite{babishiggs} and more recently muon decay~\cite{babismu} . 
In approaches invoking subtraction terms,
such as the antenna subtraction proposed here, the real radiation singularities 
from multiparton matrix elements are removed by subtracting appropriate 
approximations, which coincide with the matrix elements in all singular 
limits, but are sufficiently simple to be integrated analytically. 

 A first formulation of 
the antenna subtraction method at NNLO was presented 
in Refs.~\cite{nnlosub1,onelstr1}. In 
this work, all antenna phase space mappings required at NNLO were derived 
and documented. These mappings are to be used in the implementation of our
formulation of NNLO antenna subtraction as well. Concerning the construction 
of subtraction terms, Ref.~\cite{nnlosub1} addresses processes involving gluons
only, and restricts its results to the leading colour contributions (which 
are sufficient in purely gluonic processes at NNLO). The 
antenna subtraction terms 
in~\cite{nnlosub1} are presented only in their unintegrated form. To implement 
them in an actual calculation, they still need to be integrated over the 
appropriate antenna phase spaces to make their infrared singularities explicit.

A generalisation of the NLO dipole subtraction formalism to NNLO was 
presented in~\cite{nnlosub2,onelstr2}, where subtraction terms for all 
partonic configurations in all colour structures were derived. Again, these 
were presented only in their unintegrated form, while their analytical 
integration over the NNLO dipole phase spaces is a necessary prerequisite 
for their implementation into a numerical programme. 

Finally, a third approach to the construction of subtraction 
terms at NNLO was presented in~\cite{nnlosub4}, where an iterative procedure 
of subtraction and subsequent cancellation of oversubtracted terms is 
invoked. Using this procedure, NNLO subtraction terms for the $C_FT_R$ 
colour factor in $e^+e^-\to 2j$ are constructed in~\cite{nnlosub4},
where they are also integrated analytically and implemented into a numerical 
programme. The agreement of~\cite{nnlosub4} with 
earlier results~\cite{babis2j,our2j} illustrates the potential of this method.
However, the full set of subtraction terms for all relevant partonic 
configurations in all colour structures is at present not yet derived 
(and thus not available in integrated form either) for this method. 

More specific issues about subtraction at NNLO were addressed in two further 
works: \cite{nnlosub3} deals with overlapping collinear divergences 
in the initial state, and proposes a method for their systematic separation. 
More recently, a detailed study of the matching of single and double unresolved
regions in subtraction at NNLO was presented in~\cite{nnlosub5}.

To implement a subtraction method at NNLO, one requires explicit 
expressions for all subtraction terms (double real radiation at tree level and 
single radiation at one-loop) in their unintegrated form, as well as the
integrals of these expressions over the appropriate (double unresolved or 
single unresolved) phase spaces. Up to now, none of the approaches available
in the literature provided these to an extent 
which would permit implementation for a realistic observable.

In the following sections, we will present the NLO and NNLO antenna functions
for all partonic configurations and all colour factors. These are 
the building blocks of NLO and NNLO 
antenna subtraction terms 
required to implement our formulation of the colour-ordered antenna 
subtraction method. We provide the integrals of these subtraction terms 
for the kinematical situation of final state radiation. The method itself 
can in principle be extended to processes with hadrons in the initial state.
In this case, the antenna subtraction terms 
could be constructed from the same building blocks, but 
 different phase space factorisations and consequently different integrals 
are required.

\section{Notation and structure of antenna functions}
\label{sec:notation}
In this paper we derive antennae for all possible pairs of hard partons, 
quark-antiquark (Section~\ref{sec:qq}), quark-gluon (Section~\ref{sec:qg}) 
and gluon-gluon (Section~\ref{sec:gg}). Underlying the antenna is a colour
connected pair of hard partons, that emit radiation between them.   The
three-particle antennae involve one unresolved parton, while the four-particle
antennae involve two unresolved partons. The antenna may be at the tree or
one-loop level.
\begin{table}[t!]
\begin{center}
\begin{tabular}{cll}\hline
\rule[0mm]{0mm}{5mm}
 & tree level & one loop \\[2mm] \hline 
\rule[0mm]{0mm}{5mm}
\underline{quark-antiquark} && \\[2mm]
\rule[0mm]{0mm}{5mm}
$q g\bar q$ & $A_3^0(q,g,\bar q)$ & $A_3^1(q,g,\bar q)$, 
$\tilde{A}_3^1(q,g,\bar q)$, $\hat{A}_3^1(q,g,\bar q)$ \\[2mm]
$q g g\bar q$ & $A_4^0(q,g,g,\bar q)$, $\tilde{A}_4^0(q,g,g,\bar q)$ &\\[2mm]
$q q' \bar q'\bar q$ & $B_4^0(q,q',\bar q',\bar q)$ & \\[2mm]
$q q \bar q \bar q$ & $C_4^0(q,q,\bar q,\bar q)$ & 
 \\[2mm] \hline 
\rule[0mm]{0mm}{5mm}
\underline{quark-gluon} && \\[2mm]
\rule[0mm]{0mm}{5mm}
$q g g$ & $D_3^0(q,g,g)$ & $D_3^1(q,g,g)$, 
$\hat{D}_3^1(q,g,g)$ \\[2mm]
$q g g g$ & $D_4^0(q,g,g,g)$ &\\[2mm]
$q q' \bar q' $ & $E_3^0(q,q',\bar q')$
&$E_3^1(q,q',\bar q')$, 
$\tilde{E}_3^1(q,q',\bar q')$, $\hat{E}_3^1(q,q',\bar q')$\\[2mm]
$q q' \bar q' g$ & $E_4^0(q,q',\bar q',g)$,
$\tilde{E}_4^0(q,q',\bar q',g)$& \\[2mm] \hline 
\rule[0mm]{0mm}{5mm}
\underline{gluon-gluon} && \\[2mm]
\rule[0mm]{0mm}{5mm}
$g g g$ & $F_3^0(g,g,g)$ & $F_3^1(g,g,g)$, 
$\hat{F}_3^1(g,g,g)$ \\[2mm]
$g g g g$ & $F_4^0(g,g,g,g)$ &\\[2mm]
$g q \bar q $ & $G_3^0(g,q,\bar q)$
&$G_3^1(g,q,\bar q)$, 
$\tilde{G}_3^1(g,q,\bar q)$, $\hat{G}_3^1(g,q,\bar q)$\\[2mm]
$g q \bar qg $ & $G_4^0(g,q,\bar q,g)$,
$\tilde{G}_4^0(g,q,\bar q,g)$&\\[2mm]
$q\bar q q' \bar q'$ & $H_4^0(q,\bar q,q',\bar q')$
& \\[2mm] \hline 
\end{tabular}
\end{center}
\caption{List of tree-level and one-loop colour-ordered antenna functions.
The tilde denotes subleading colour contributions, and the hat flavour-number 
dependent corrections.}
\label{tab:nota}
\end{table}

Each antenna is determined by both the external state and the pair of hard
partons it collapses to.   In general we denote the antenna function as $X$.
For antennae that collapse onto a hard quark-antiquark pair, 
$X = A$ for $qg \bar q$ and $qgg\bar q$,  $X=B$ for $qq^\prime
\bar q^\prime\bar q$ and $X=C$ for $q
q\bar q\bar q$ final states.   Similarly, for quark-gluon antenna, we have 
$X = D$ for $qg g$ and $qggg$ and $X=E$ for $qq^\prime
\bar q^\prime$ and $qq^\prime
\bar q^\prime g$ final states. Finally, we characterise the gluon-gluon antennae as
$X=F$ for $ggg$ and $gggg$,  $X=G$ for $gq\bar q$ and 
$gq\bar q g$ and $X=H$ for $q\bar q q' \bar q'$ final states.
Some of these antenna functions decompose into a leading colour and 
a subleading colour contribution. Where appropriate,
the  subleading colour contribution to $X$ is denoted as $\tilde{X}$. Finally,
the one loop antenna functions also contain contributions from 
closed quark loops, which we denote as $\hat{X}$. The notation for the 
different tree-level and one-loop antenna functions is summarised in 
Table~\ref{tab:nota}.

Each antenna is a function of the invariants formed by the momenta of the 
final state particles.  

It will prove useful to introduce the operators $\Poles$ and $\Finite$ that
select either the singular or finite contribution from a particular antenna.
For example,
$$\Poles({\cal X})$$ extracts the singular contribution from the antenna $X$
after integration over the unresolved phase space in terms of the infrared
singularity operators of Section~\ref{sec:operators}.
It should be noted that 
$\Poles({\cal X})$ also contains finite terms
arising from the expansion of 
the infrared singularity operators. These terms are on the one hand 
of the type  
$(\e \ln s)^n$ (where $s$ is the invariant mass of any pair of particles 
contained in the antenna, or the total invariant mass of all antenna 
particles), on the other hand they contain transcendental constants resulting 
from the expansion of the normalisation factors appearing in the different
infrared singularity operators.
 To extract the {\it remaining} finite 
contribution, we introduce 
$$\Finite({\cal X}) \equiv 
{\cal X} - \Poles({\cal X})\;.$$
Generally,
\begin{equation}
{\cal X} = \Poles({\cal X}) + \Finite({\cal X}) + {\cal O}(\epsilon).
\end{equation}

The one-loop antenna functions contain explicit poles from the loop integration.
Therefore,  the operators $\Poles$ and $\Finite$ can also be applied to 
their unintegrated forms $X$. The action of these 
operators is again to decompose the unintegrated 
antenna in terms of infrared singularity operators describing the pole terms 
and a finite remainder. 

All antenna functions are derived from physical matrix elements: 
the quark-antiquark antenna functions from 
$\gamma^* \to q\bar q~+$~(partons)~\cite{our2j}, the quark-gluon antenna 
functions from $\tilde\chi \to \tilde g~+$~(partons)~\cite{chi} and 
the gluon-gluon antenna functions from $H\to$~(partons)~\cite{h}. The
tree-level antenna functions are obtained by normalising the 
colour-ordered three- and four-parton tree-level 
squared matrix elements to the squared matrix element 
for the basic two-parton process,
\begin{eqnarray}
X_{ijk}^0 = S_{ijk,IK}\, \frac{|{\cal M}^0_{ijk}|^2}{|{\cal M}^0_{IK}|^2}\;,\nonumber\\
X_{ijl}^0 = S_{ijkl,IL}\, \frac{|{\cal M}^0_{ijkl}|^2}{|{\cal M}^0_{IL}|^2}\;,
\end{eqnarray}
where $S$ denotes the symmetry factor associated to the antenna, which accounts
both for potential identical particle symmetries and for the presence 
of more than one antenna in the basic two-parton process. 
The one-loop antenna functions are obtained from the colour-ordered 
renormalised one-loop three-parton matrix elements as
\begin{equation}
X_{ijk}^1 = S_{ijk,IK}\, \frac{|{\cal M}^1_{ijk}|^2}{|{\cal M}^0_{IK}|^2} - 
X_{ijk}^0\, \frac{|{\cal M}^1_{IK}|^2}{|{\cal M}^0_{IK}|^2} \;.
\end{equation}

The numerical implementation of the three- and four-parton antenna 
phase space~\cite{nnlosub1} requires the partonic emissions to be ordered.
Ordering of emissions means that the two hard radiator partons 
defining the antenna are identified, and that each unresolved parton can 
become singular only with the
two particles which are adjacent to it, i.e.\
with the two radiators for three-parton antenna functions and with one 
radiator and with the other unresolved parton for the four-parton antenna 
functions. For the sake of numerical implementation, this implies two 
requirements: (1) the separation of multiple antenna configurations present 
in a single antenna function for three- and four-parton antenna functions and
(2) the separation of non-ordered emissions (present only at subleading 
colour in the four-parton antenna functions) into terms that can be 
identified with a particular ordering of the momenta. 

In the colour-ordered quark-gluon and gluon-gluon antenna functions derived 
from physical matrix elements for neutralino decay~\cite{chi} and 
Higgs boson decay~\cite{h}, it is in general not possible to identify the 
hard radiators and the unresolved partons in a unique manner. The reason for 
this ambiguity is in the cyclic nature of the colour orderings, which 
becomes evident already in the three-parton antenna functions: each pair of 
two partons can in principle act as hard radiators, resulting in more than 
one antenna configuration present in a single antenna function. For the 
three-parton antenna functions $D_3^0$ and $F_3^0$, we illustrate in 
(\ref{eq:smalld}) and 
(\ref{eq:smallf}) below how these different antenna configurations can be 
disentangled, resulting in new sub-antenna functions where the hard radiators 
can be uniquely identified. Such a decomposition is also possible for 
four-parton antenna functions which display the same ambiguity. 
However, in the four-parton case repeated partial fractioning is required 
to extract individual sub-antenna configurations. Since this procedure 
introduces new denominators in the antenna terms, we will use it 
only for the numerical implementation. 
At this point, it should be pointed out that the 
cyclic ambiguity is inevitable if physical matrix elements involving 
gluons as radiators are used to 
derive the antenna functions, since the cyclic symmetry in colour space 
is enforced by gauge invariance.

The decomposition of non-ordered emissions into different terms
is discussed in detail in Section~\ref{sec:qq}
using 
the quark-antiquark antenna function $\tilde{A}_4^0$, 
which describes the emission of two gluons at subleading colour, as an example.
In this antenna 
the gluons behave effectively like photons, coupling only to the quark and 
the antiquark. Consequently, each gluon can become collinear with both 
hard radiators. In (\ref{eq:A40t}) below, we illustrate how this 
antenna function can be decomposed into individual terms corresponding 
to different well-defined orderings by repeated partial fractioning in the 
invariants. The same procedure has to be applied for all other antenna
functions which are not sufficiently ordered. Since this repeated 
partial fractioning yields increases the number of terms in the analytic
expressions for the antenna functions quite considerably, we restrict 
ourselves to quoting only the full antenna functions, not their 
ordered decompositions. 

Repeated partial fractioning generates denominators which do not 
correspond to physical propagators of four-parton matrix elements. Therefore,
it is in general not possible to integrate the ordered terms analytically, 
at least not by merely applying the methods of~\cite{ggh}. Analytical 
phase space integration of functions involving non-propagator-type 
denominators is in principle possible~\cite{babisdy,twolstr}
and requires a 
more involved reduction procedure and a larger set of master integrals.  
To avoid this problem we use the ordered sub-antenna functions 
in the numerical implementation, and ensure that all ordered 
contributions to a given antenna function are taken together with the 
same phase space factorisation (but different phase space mappings). 
With this procedure, the ordered contributions to the sub-antenna functions 
can be recombined to form the 
full antenna functions (related to physical matrix elements) 
quoted in the following three sections, which are then integrated analytically 
to make their infrared pole structure explicit. This recombination may 
be obtained 
by considering more than one colour-ordering for the full $(m+2)$-parton 
matrix element, and by explicit subtraction of unwanted spurious  
singularities. This problem is discussed more explicitly using
 the four-parton quark-gluon antenna function $D_4^0$ of
Sections~\ref{sec:qg} and~\ref{sec:D40limit} as an example.

\section{Colour-ordered infrared singularity operators}
\label{sec:operators}

In order to express the singularity structure in a way that cancellations
between real radiation and virtual radiation can be made explicit, 
it is convenient to extract the infrared singularity structure using the 
${\bf I}^{(1)}$-operator~\cite{catani}. This  operator
describes the singularity structure of virtual one-loop amplitudes. At the 
two-loop level, the infrared singularity structure is described 
by  ${\bf I}^{(1)}$-operators, a hard radiation operator ${\bf H}^{(2)}$,
the QCD $\beta$-function and the gluon splitting constant $K$. The same 
pole structure is recovered~\cite{our2j,chi,h} 
in the sum of double real radiation and one-loop single real radiation 
contributions, such that the sum of double-virtual, virtual single-unresolved
and double-unresolved corrections is finite. Some cancellations, 
related to the one-loop correction to the soft gluon 
current ${\bf S}^{(2)}$, take place
only between virtual single-unresolved
and double-unresolved corrections. 

In the  formulation of~\cite{catani}, ${\bf I}^{(1)}$
is a tensor in colour space, and contains imaginary parts 
from the analytic continuation of loop amplitudes from the Euclidian to 
the Minkowskian region. In the following, we will only consider 
colour-ordered matrix elements, for which ${\bf I}^{(1)}$ is a scalar in
colour space. Moreover, using the ${\bf I}^{(1)}$-operator to describe 
real radiation singularities, only its real part is relevant, since 
contributions from its imaginary part cancel once the double-virtual 
two-loop-times-tree-level and one-loop-times-one-loop parts are 
added together.

The real radiation infrared singularity operators that appear 
in the integrated form of an antenna function 
are denoted by $(ij=q\bar q, qg, g \bar q, gg)$
$${\bf I}^{(1)}_{ij} (\e,s_{ij}) \qquad \mbox{and}\qquad  
{\bf I}^{(1)}_{ij,\Flavour} (\e,s_{ij})\;,
$$
where the second operator describes contributions arising from the 
splitting of a gluon into a quark-antiquark pair and which are proportional 
to the number of light quark flavours $N_F$.

The operators are
given by
\begin{eqnarray}
{\bf I}^{(1)}_{q\bar q} (\e,s_{q\bar q})
&=& - \frac{e^{\e \gamma}}{2\Gamma(1-\e)}\, \left[
\frac{1}{\e^2}+\frac{3}{2\e} \right] \, \Re(-s_{q\bar q})^{-\e} \;,
\nonumber \\ 
{\bf I}^{(1)}_{q g}(\e,s_{q g})
 &=& - \frac{e^{\e \gamma}}{2\Gamma(1-\e)}\, \left[
\frac{1}{\e^2}+\frac{5}{3\e} \right] \, \Re(-s_{qg})^{-\e}  \;,
\nonumber\\
{\bf I}^{(1)}_{gg}(\e,s_{gg})
 &=& - \frac{e^{\e \gamma}}{2\Gamma(1-\e)}\, \left[
\frac{1}{\e^2}+\frac{11}{6\e} \right]\, \Re(-s_{gg})^{-\e} \;,
\nonumber\\ 
{\bf I}^{(1)}_{q\bar q,\Flavour}(\e,s_{q\bar q}) &=& 0  \;,
\nonumber\\ 
{\bf I}^{(1)}_{q g,\Flavour}(\e,s_{q g}) &=&  \frac{e^{\e \gamma}}{2\Gamma(1-\e)}\, 
\frac{1}{6\e} \, \Re(-s_{qg})^{-\e}\;, \nonumber \\
{\bf I}^{(1)}_{g g,\Flavour}(\e,s_{gg}) &=&  \frac{e^{\e \gamma}}{2\Gamma(1-\e)}\, 
\frac{1}{3\e} \, \Re(-s_{gg})^{-\e} \;.
\label{eq:Ione}
\end{eqnarray}
The antiquark-gluon operators are obtained by charge conjugation:
$${\bf I}^{(1)}_{g\bar q} (\e,s_{g\bar q}) = 
{\bf I}^{(1)}_{qg} (\e,s_{g\bar q})\qquad \mbox{and}\qquad  
{\bf I}^{(1)}_{g\bar q,\Flavour} (\e,s_{g\bar q}) = 
{\bf I}^{(1)}_{qg,\Flavour} (\e,s_{g\bar q})\;.
$$

At NNLO, several new infrared singularity operators appear;  the 
one-loop soft 
gluon current 
 ${\bf S}^{(2)}(\e,q^2)$~\cite{oneloopsoft},
 \begin{eqnarray}
 {\bf S}^{(2)}(\e,q^2)&=& \, \Bigg[ -\frac{1}{4\e^4} - \frac{3}{4\e^3}
+ \frac{1}{\e^2}\,\left(-\frac{13}{4}  + \frac{7\pi^2}{24} \right)
+ \frac{1}{\e}\,\left(-\frac{51}{4}  + \frac{7\pi^2}{8} 
+ \frac{14}{3} \zeta_3  \right)\nonumber \\&& \hspace{1.5cm}
+\left(-\frac{205}{4}  + \frac{91\pi^2}{24} 
+ 14 \zeta_3 +\frac{7\pi^4}{480} \right) + {\cal O}(\e) \Bigg] (q^2)^{-2\e} 
\; ,
\end{eqnarray}
  and the hard radiation functions 
${\bf H}^{(2)}_{ij}(\e,q^2)$, which decompose into virtual 
contributions ${\bf H}^{(2)}_{ij,V}(\e,q^2)$ and 
real contributions ${\bf H}^{(2)}_{ij,R}(\e,q^2)$. 
For the antenna functions presented here, the virtual contributions 
are given by,
\begin{eqnarray}
{\bf H}^{(2)}_{V,A}(\e,q^2) &=& 
\frac{e^{\e \gamma}}{\Gamma(1-\e)}\, \left[
\frac{1}{\e^2}\left(\frac{43}{8}-\frac{\pi^2}{6}\right)+\frac{1}{\e}
\left(\frac{839}{24}-\frac{\pi^2}{2}-11\zeta_3\right) 
\right] \, (q^2)^{-2\e}\;,\nonumber \\ 
{\bf H}^{(2)}_{V,\tilde{A}}(\e,q^2) &=& 
\frac{e^{\e \gamma}}{\Gamma(1-\e)}\, \left[
\frac{1}{\e^2}\left(\frac{43}{8}-\frac{\pi^2}{6}\right)+\frac{1}{\e}
\left(\frac{51}{2}-\frac{\pi^2}{4}-15\zeta_3\right) 
\right] \, (q^2)^{-2\e}\;,\nonumber  \\
{\bf H}^{(2)}_{V,\hat{A}}(\e,q^2) &=& 
\frac{e^{\e \gamma}}{\Gamma(1-\e)}\, \left[ - \frac{19}{12\e}  
\right] \, (q^2)^{-2\e}\;,\nonumber   \\
{\bf H}^{(2)}_{V,D}(\e,q^2) &=& 
\frac{e^{\e \gamma}}{\Gamma(1-\e)}\, \Bigg[
\frac{1}{6\e^3}+
\frac{1}{\e^2}\left(\frac{109}{4}-\frac{2\pi^2}{3}\right)\nonumber \\
&&\hspace{2cm}+\frac{1}{\e}
\left(\frac{17791}{108}-\frac{133\pi^2}{72}-52\zeta_3\right) 
\Bigg] \, (q^2)^{-2\e}\;,\nonumber  \\ 
{\bf H}^{(2)}_{V,\hat{D}}(\e,q^2) &=& 
\frac{e^{\e \gamma}}{\Gamma(1-\e)}\, \left[
-\frac{275}{36\e}
\right] \, (q^2)^{-2\e}\;,\nonumber  \\
{\bf H}^{(2)}_{V,E}(\e,q^2) &=& 
\frac{e^{\e \gamma}}{\Gamma(1-\e)}\, \left[
-\frac{2}{\e^2}+\frac{1}{\e}
\left(-\frac{326}{27}+\frac{\pi^2}{9}\right) 
\right] \, (q^2)^{-2\e} \;,\nonumber \\
{\bf H}^{(2)}_{V,\tilde{E}}(\e,q^2) &=& 
\frac{e^{\e \gamma}}{\Gamma(1-\e)}\, \left[ 
-\frac{1}{6\e^3}-
\frac{35}{36\e^2}+\frac{1}{\e}
\left(-\frac{509}{108}+\frac{17\pi^2}{72}\right) 
\right] \, (q^2)^{-2\e} \;,\nonumber  \\
{\bf H}^{(2)}_{V,F}(\e,q^2) &=& 
\frac{e^{\e \gamma}}{\Gamma(1-\e)}\, \Bigg[
\frac{1}{3\e^3}+
\frac{1}{\e^2}\left(30-\frac{2\pi^2}{3}\right)\nonumber \\
&&\hspace{2cm}+\frac{1}{\e}
\left(\frac{20009}{108}-\frac{79\pi^2}{36}-52\zeta_3\right) 
\Bigg] \, (q^2)^{-2\e} \;,\nonumber \\ 
{\bf H}^{(2)}_{V,\hat{F}}(\e,q^2) &=& 
\frac{e^{\e \gamma}}{\Gamma(1-\e)}\, \left[
-\frac{37}{3\e}
\right] \, (q^2)^{-2\e} \;,\nonumber \\
{\bf H}^{(2)}_{V,G}(\e,q^2) &=& 
\frac{e^{\e \gamma}}{\Gamma(1-\e)}\, \left[
-\frac{14}{3\e^2}+\frac{1}{\e}
\left(-\frac{805}{27}+\frac{2\pi^2}{9}\right) 
\right] \, (q^2)^{-2\e} \;,\nonumber \\
{\bf H}^{(2)}_{V,\tilde{G}}(\e,q^2) &=& 
\frac{e^{\e \gamma}}{\Gamma(1-\e)}\, \left[ 
-\frac{1}{3\e^3}-
\frac{41}{18\e^2}+\frac{1}{\e}
\left(-\frac{325}{27}+\frac{17\pi^2}{36}\right) 
\right] \, (q^2)^{-2\e} \;,\nonumber \\
{\bf H}^{(2)}_{V,\hat{G}}(\e,q^2) &=& 
\frac{e^{\e \gamma}}{\Gamma(1-\e)}\, \left[
-\frac{2}{9\e^2}+\frac{7}{9\e}
\right] \, (q^2)^{-2\e} \;,
\end{eqnarray}
while the real radiation contributions are,
\begin{eqnarray}
{\bf H}^{(2)}_{R,A}(\e,q^2) &=& 
\frac{e^{\e \gamma}}{\Gamma(1-\e)}\, \left[
\frac{1}{\e^2}\left(-\frac{43}{8}+\frac{\pi^2}{6}\right)+\frac{1}{\e}
\left(-\frac{29795}{864} + \frac{37\pi^2}{96} + \frac{51}{4}\zeta_3
\right) 
\right] \, (q^2)^{-2\e}\;,\nonumber  \\ 
{\bf H}^{(2)}_{R,\tilde{A}}(\e,q^2) &=& 
\frac{e^{\e \gamma}}{\Gamma(1-\e)}\, \left[
\frac{1}{\e^2}\left(-\frac{43}{8}+\frac{\pi^2}{6}\right)+\frac{1}{\e}
\left(-\frac{845}{32} + \frac{\pi^2}{2} + 13\zeta_3
\right) 
\right] \, (q^2)^{-2\e}\;,\nonumber  \\
{\bf H}^{(2)}_{R,B}(\e,q^2) &=& 
\frac{e^{\e \gamma}}{\Gamma(1-\e)}\, \left[ 
 \frac{1}{\e} \left( \frac{317}{216} +\frac{\pi^2}{48} \right)
\right] \, (q^2)^{-2\e} \;,\nonumber  \\
{\bf H}^{(2)}_{R,C}(\e,q^2) &=& 
\frac{e^{\e \gamma}}{\Gamma(1-\e)}\, \left[ 
 \frac{1}{\e} \left( \frac{13}{16} -\frac{\pi^2}{8} 
+\frac{1}{2}\zeta_3\right)
\right] \, (q^2)^{-2\e}  \;,\nonumber \\
{\bf H}^{(2)}_{R,D}(\e,q^2) &=& 
\frac{e^{\e \gamma}}{\Gamma(1-\e)}\, \Bigg[
-\frac{1}{6\e^3}+
\frac{1}{\e^2}\left(-\frac{109}{4}+\frac{2\pi^2}{3}\right)\nonumber \\
&& \hspace{2cm}
+\frac{1}{\e}
\left(-\frac{71261}{432}+\frac{97\pi^2}{48}+52\zeta_3\right) 
\Bigg] \, (q^2)^{-2\e} \;,\nonumber \\
{\bf H}^{(2)}_{R,E}(\e,q^2) &=& 
\frac{e^{\e \gamma}}{\Gamma(1-\e)}\, \left[
\frac{2}{\e^2}+\frac{1}{\e}
\left(\frac{518}{27}-\frac{7\pi^2}{72}\right) 
\right] \, (q^2)^{-2\e} \;,\nonumber \\
{\bf H}^{(2)}_{R,\tilde{E}}(\e,q^2) &=& 
\frac{e^{\e \gamma}}{\Gamma(1-\e)}\, \left[ 
\frac{1}{6\e^3}+
\frac{35}{36\e^2}+\frac{1}{\e}
\left(\frac{1045}{216}-\frac{17\pi^2}{72}\right) 
\right] \, (q^2)^{-2\e} \;,\nonumber \\
{\bf H}^{(2)}_{R,F}(\e,q^2) &=& 
\frac{e^{\e \gamma}}{\Gamma(1-\e)}\, \Bigg[
-\frac{1}{3\e^3}+
\frac{1}{\e^2}\left(-30+\frac{2\pi^2}{3}\right)\nonumber \\
&& \hspace{2cm}
+\frac{1}{\e}
\left(-\frac{4991}{27}+\frac{109\pi^2}{48}+\frac{105}{2}\zeta_3\right) 
\Bigg] \, (q^2)^{-2\e} \;,\nonumber \\
{\bf H}^{(2)}_{R,G}(\e,q^2) &=& 
\frac{e^{\e \gamma}}{\Gamma(1-\e)}\, \left[
\frac{14}{3\e^2}+\frac{1}{\e}
\left(\frac{4463}{108}-\frac{17\pi^2}{72}\right) 
\right] \, (q^2)^{-2\e} \;,\nonumber \\
{\bf H}^{(2)}_{R,\tilde{G}}(\e,q^2) &=& 
\frac{e^{\e \gamma}}{\Gamma(1-\e)}\, \left[ 
\frac{1}{3\e^3}+
\frac{41}{18\e^2}+\frac{1}{\e}
\left(\frac{1327}{108}-\frac{17\pi^2}{36}\right) 
\right] \, (q^2)^{-2\e} \;,\nonumber \\
{\bf H}^{(2)}_{R,H}(\e,q^2) &=& 
\frac{e^{\e \gamma}}{\Gamma(1-\e)}\, \left[
\frac{2}{9\e^2}-\frac{16}{27\e} \right] \, (q^2)^{-2\e} \;.
 \end{eqnarray}
Combining the real and virtual hard radiation functions yields 
the hard radiation terms from the two-loop virtual 
corrections~\cite{twol,3jme,our2j,chi,h},
\begin{eqnarray}
{\bf H}^{(2)}_{q\bar q}(\e,q^2) &=& 
N^2\,{\bf H}^{(2)}_{q\bar q,N^2}(\e,q^2) 
+ {\bf H}^{(2)}_{q\bar q,1}(\e,q^2) 
+ \frac{1}{N^2}\, {\bf H}^{(2)}_{q\bar q,1/N^2}(\e,q^2) \nonumber \\&&
+ N N_F\, {\bf H}^{(2)}_{q\bar q,N N_F}(\e,q^2)
+ \frac{N_F}{N} {\bf H}^{(2)}_{q\bar q,N_F/N}(\e,q^2)\;, \\
{\bf H}^{(2)}_{qg}(\e,q^2) &=& 
N^2\,{\bf H}^{(2)}_{qg,N^2}(\e,q^2) 
+ N N_F\, {\bf H}^{(2)}_{qg,N N_F}(\e,q^2)
+ \frac{N_F}{N} {\bf H}^{(2)}_{qg,N_F/N}(\e,q^2) \nonumber \\&&
+ N_F^2\, {\bf H}^{(2)}_{qg,N_F^2}(\e,q^2)\;, \\
{\bf H}^{(2)}_{gg}(\e,q^2) &=&
N^2\,{\bf H}^{(2)}_{gg,N^2}(\e,q^2) 
+ N N_F\, {\bf H}^{(2)}_{gg,N N_F}(\e,q^2)
+ \frac{N_F}{N} {\bf H}^{(2)}_{gg,N_F/N}(\e,q^2) \nonumber \\&&
+ N_F^2\, {\bf H}^{(2)}_{gg,N_F^2}(\e,q^2)\;,
\end{eqnarray}
with
\begin{eqnarray}
{\bf H}^{(2)}_{q\bar q,N^2}(\e,q^2) 
&=&
\frac{1}{2} \left( {\bf H}^{(2)}_{V,A}(\e,q^2)
+ {\bf H}^{(2)}_{R,A}(\e,q^2)\right) \nonumber \\
&=&
\frac{e^{\e \gamma}}{4\e\Gamma(1-\e)} \, \left( 
\frac{409}{432} - \frac{11\pi^2}{48} + \frac{7}{2}\zeta_3
\right) (q^2)^{-2\e} , \\
{\bf H}^{(2)}_{q\bar q,1/N^2}(\e,q^2) 
&=&
\frac{1}{2} \left( {\bf H}^{(2)}_{V,\hat{A}}(\e,q^2) +
 {\bf H}^{(2)}_{R,\hat{A}}(\e,q^2)  + {\bf H}^{(2)}_{R,C}(\e,q^2)
 \right)\nonumber  \\
&=&
\frac{e^{\e \gamma}}{4\e\Gamma(1-\e)} \, \left( 
- \frac{3}{16} + \frac{\pi^2}{4} - 3 \zeta_3
\right) (q^2)^{-2\e} ,
\label{eq:H2qqb1overN} \\
{\bf H}^{(2)}_{q\bar q,1}(\e,q^2) 
&=& - {\bf H}^{(2)}_{qq,N^2}(\e,q^2) - 
{\bf H}^{(2)}_{q\bar q,1/N^2}(\e,q^2)\,,
\\{\bf H}^{(2)}_{q\bar q,N N_F}(\e,q^2)
&=&  \frac{1}{2}\left({\bf H}^{(2)}_{V,\hat{A}}(\e,q^2) 
+ {\bf H}^{(2)}_{R,B}(\e,q^2) \right) \nonumber \\
&=&
\frac{e^{\e \gamma}}{4\e\Gamma(1-\e)} \, \left( 
-\frac{25}{108} + \frac{\pi^2}{24}
\right) (q^2)^{-2\e} , \\
{\bf H}^{(2)}_{q\bar q,N_F/N}(\e,q^2) 
&=&
 - {\bf H}^{(2)}_{q\bar q,N N_F}(\e,q^2)\,,  \\
{\bf H}^{(2)}_{qg,N^2}(\e,q^2) 
&=&
\frac{1}{2} \left( {\bf H}^{(2)}_{V,D}(\e,q^2)
+ {\bf H}^{(2)}_{R,D}(\e,q^2)\right)\nonumber  \\
&=&
\frac{e^{\e \gamma}}{4\e\Gamma(1-\e)} \, \left( 
-\frac{97}{216} + \frac{25\pi^2}{72}
\right) (q^2)^{-2\e} , \\
{\bf H}^{(2)}_{qg,N N_F}(\e,q^2)
&=&  \frac{1}{2}\left({\bf H}^{(2)}_{V,\hat{D}}(\e,q^2) 
+ {\bf H}^{(2)}_{V,E}(\e,q^2)
+ {\bf H}^{(2)}_{R,E}(\e,q^2)\right)\nonumber  \\
&=&
\frac{e^{\e \gamma}}{4\e\Gamma(1-\e)} \, \left( 
-\frac{19}{18} + \frac{\pi^2}{36}
\right) (q^2)^{-2\e} , \\
{\bf H}^{(2)}_{qg,N_F/N}(\e,q^2) 
&=&
 - \frac{1}{2} \left( {\bf H}^{(2)}_{V,\tilde{E}}(\e,q^2)
+ {\bf H}^{(2)}_{R,\tilde{E}}(\e,q^2)\right)\nonumber  \\
&=&
\frac{e^{\e \gamma}}{4\e\Gamma(1-\e)} \, \left( 
-\frac{1}{4}
\right) (q^2)^{-2\e} , \\
{\bf H}^{(2)}_{qg,N_F^2}(\e,q^2)
&=&  \frac{e^{\e \gamma}}{4\e\Gamma(1-\e)} \, \frac{5}{27} (q^2)^{-2\e}\;,
 \\
{\bf H}^{(2)}_{gg,N^2}(\e,q^2) 
&=&
\frac{1}{2} \left( {\bf H}^{(2)}_{V,F}(\e,q^2)
+ {\bf H}^{(2)}_{R,F}(\e,q^2)\right) \nonumber \\
&=&
\frac{e^{\e \gamma}}{4\e\Gamma(1-\e)} \, \left( 
\frac{5}{6} + \frac{11\pi^2}{72} +\zeta_3
\right) (q^2)^{-2\e} , \\
{\bf H}^{(2)}_{gg,N N_F}(\e,q^2)
&=&  \frac{1}{2}\left({\bf H}^{(2)}_{V,\hat{F}}(\e,q^2) 
+ {\bf H}^{(2)}_{V,G}(\e,q^2)
+ {\bf H}^{(2)}_{R,G}(\e,q^2)\right)\nonumber  \\
&=&
\frac{e^{\e \gamma}}{4\e\Gamma(1-\e)} \, \left( 
-\frac{89}{54} - \frac{\pi^2}{36}
\right) (q^2)^{-2\e} , \\
{\bf H}^{(2)}_{gg,N_F/N}(\e,q^2) 
&=&
 - \frac{1}{2} \left( {\bf H}^{(2)}_{V,\tilde{G}}(\e,q^2)
+ {\bf H}^{(2)}_{R,\tilde{G}}(\e,q^2)\right) \nonumber \\
&=&
\frac{e^{\e \gamma}}{4\e\Gamma(1-\e)} \, \left( 
-\frac{1}{2}
\right) (q^2)^{-2\e} , \\
{\bf H}^{(2)}_{gg,N_F^2}(\e,q^2)
&=&  - \frac{1}{2} \left( {\bf H}^{(2)}_{V,\hat{G}}(\e,q^2)
+ {\bf H}^{(2)}_{R,H}(\e,q^2)\right)\nonumber  \\
&=&  \frac{e^{\e \gamma}}{4\e\Gamma(1-\e)} \, \frac{10}{27} (q^2)^{-2\e}\;.
\end{eqnarray}
The relation of these operators to the singularity structure of 
physical multi-parton
two-loop matrix elements was discussed in~\cite{twol,3jme,our2j,chi,h}

Finally, the  NNLO singularity structures also contain the 
QCD $\beta$-function (\ref{eq:qcdbeta}) 
and the collinear coefficient $K$. In a colour-ordered 
decomposition, these are 
\begin{equation}
\beta_0 = b_0 N + b_{0,\Flavour} N_F \qquad \mbox{with} \qquad
b_0 = \frac{11}{6}\;, b_{0,\Flavour}= -\frac{1}{3}
\end{equation}
and
\begin{equation}
K = k_0 N + k_{0,\Flavour} N_F \qquad \mbox{with} \qquad
k_0 = \frac{67}{18}-\frac{\pi^2}{6} \;, k_{0,\Flavour}= -\frac{5}{9}.
\end{equation}

\section{Quark-antiquark antennae}
\label{sec:qq}
The quark-antiquark antenna functions are derived by appropriately 
normalising the colour-ordered QCD real radiation corrections 
to $\gamma^* \to q\bar q$, described to NNLO accuracy in~\cite{our2j}.

The overall normalisation is given by defining the tree-level two-parton
quark-antiquark 
antenna function
\begin{equation}
{\cal A}_2^0(s_{12}) \equiv 1\;.
\end{equation}

The one-loop two-parton quark-antiquark antenna is then:
\begin{eqnarray}
{\cal A}_2^1(s_{12}) &=& 
 (s_{12})^{-\e} \Bigg[ -\frac{1}{\e^2} - \frac{3}{2\e} - 4 + 
\frac{7\pi^2}{12} 
+ \left( -8 + \frac{7\pi^2}{8} + \frac{7}{3}\zeta_3 \right)
\e \nonumber \\
&& \hspace{2cm}
+ \left( - 16
          + \frac{7\pi^2}{3}
          + \frac{7}{2}\zeta_3
          - \frac{73\pi^4}{1440} \right) \e^2  + {\cal O}(\e^3) \Bigg] \;, 
\end{eqnarray}
with 
\begin{eqnarray}
\Poles\left({\cal A}_2^1(s_{12}) \right) &=& 
2 {\bf I}^{(1)}_{q\bar q} \left(\e,s_{12} \right),
\label{eq:a21poles}
\\
\Finite\left({\cal A}_2^1(s_{12}) \right) &=& - 4.
\end{eqnarray}

\subsection{Three-parton tree-level antenna functions}

The tree-level three-parton quark-antiquark antenna is: 
\begin{equation}
\label{eq:A30}
A_3^0(1_q,3_g,2_{\bar q}) = 
\frac{1}{s_{123}} \, \left(
 \frac{s_{13}}{s_{23}} +  \frac{s_{23}}{s_{13}}
+ 2 \frac{s_{12} s_{123}}{s_{13}s_{23}}
\right) + {\cal O} (\e) \;,
\end{equation}
yielding the integrated antenna function according to (\ref{eq:x3int}):
\begin{eqnarray}
{\cal A}_3^0(s_{123}) &=&  
   \left(s_{123} \right)^{-\e}\Bigg[
\frac{1}{\e^2} + \frac{3}{2\e} + \frac{19}{4} - 
\frac{7\pi^2}{12} 
+ \left( \frac{109}{8} - \frac{7\pi^2}{8} - \frac{25}{3}\zeta_3 \right)
\e \nonumber \\
&& \hspace{2cm}
+ \left( \frac{639}{16}
          - \frac{133\pi^2}{48}
          - \frac{25}{2}\zeta_3
          - \frac{71\pi^4}{1440} \right) \e^2  + {\cal O}(\e^3) \Bigg]\;, 
\end{eqnarray}
with 
\begin{eqnarray}
\Poles\left({\cal A}_3^0(s_{123}) \right) &=& 
- 2 {\bf I}^{(1)}_{q\bar q} \left(\e,s_{123} \right)\;,\\
\Finite\left({\cal A}_3^0(s_{123}) \right) &=& \frac{19}{4}\;.
\end{eqnarray}

This antenna 
 can be split symmetrically into two sub-antennae,
which coincide with the $q\to qg$ dipole functions:
\begin{equation}
A_3^0(1,3,2)= a_3^0(1,3,2) + a_3^0(2,3,1)\; ,
\end{equation}
with
\begin{equation}
a_3^0(1,3,2) = 
\frac{1}{s_{123}} \, \left(
  \frac{s_{23}}{s_{13}}
+ 2 \frac{s_{12} s_{123}}{(s_{13}+s_{23})s_{13}}
\right) + {\cal O} (\e)\;.
\end{equation}

\subsection{Three-parton one-loop antenna functions}

At one loop, one finds three different three-particle antenna functions, 
corresponding to the leading and subleading colour structures 
$A_3^1(1_q,3_g,2_{\bar q})$, 
$\tilde{A}_3^1(1_q,3_g,2_{\bar q})$  and to 
the contribution from a closed quark loop 
$\hat{A}_3^1(1_q,3_g,2_{\bar q})$.

Introducing
\begin{equation}
R(y,z)  = \log y \log z - \log y \log (1-y) - \log z \log (1-z) 
+ \frac{\pi^2}{6} - \Li_2(y) - \Li_2(z)
\end{equation}
and $y_{ij}  = s_{ij}/s_{123}$,
the one-loop antenna functions are given by:
\begin{eqnarray}
\Poles\left(A_3^1(1_q,3_g,2_{\bar q})\right) &=& 
2 \left( {\bf I}^{(1)}_{qg} (\e,s_{13}) 
+ {\bf I}^{(1)}_{qg} (\e,s_{23})
- {\bf I}^{(1)}_{q\bar q} (\e,s_{123}) \right) A_3^0(1,3,2) \;,\\
\Finite\left(A_3^1(1_q,3_g,2_{\bar q})\right) &=& -
\left(  R(y_{13},y_{23}) 
+ \frac{5}{3} \log y_{13} +  \frac{5}{3} \log y_{23} \right)
A_3^0(1,3,2) \nonumber \\
&&        + \frac{1}{s_{123}}
          + \frac{s_{12}+s_{23}}{2s_{123}s_{13}}
          + \frac{s_{12}+s_{13}}{2s_{123}s_{23}}
          - \frac{s_{13}}{2s_{123}(s_{12}+s_{13})}
\nonumber \\ &&
          - \frac{s_{23}}{2s_{123}(s_{12}+s_{23})}
          + \frac{\log y_{13}}{s_{123}} \left( 2 - \frac{1}{2}\,
 \frac{s_{13}s_{23}}{(s_{12}
                        +s_{23})^2} + 2
\frac{s_{13}-s_{23}}{s_{12}+s_{23}} \right)\nonumber \\
&&            + \frac{\log y_{23}}{s_{123}} \left( 2 
         - \frac{1}{2}\, \frac{s_{13}s_{23}}{(s_{12}+s_{13})^2} + 2
\frac{s_{23}-s_{13}}{s_{12}+s_{13}} \right)\;,
\label{eq:A31} \\
\Poles\left(\tilde{A}_3^1(1_q,3_g,2_{\bar q})\right) &=& 
2 \left( {\bf I}^{(1)}_{q\bar q} (\e,s_{12}) 
- {\bf I}^{(1)}_{q\bar q} (\e,s_{123}) \right) A_3^0(1,3,2)\;, \\
\Finite\left(\tilde{A}_3^1(q,g,\bar q)\right) &=& 
-\left(  R(y_{12},y_{13}) + 
  R(y_{12},y_{23})  
+ \frac{3}{2} \log y_{12}  \right)
A_3^0(1,3,2) \nonumber \\
&& -\frac{s_{12}+s_{23}}{2s_{123}s_{13}} 
-  \frac{s_{12}+s_{13}}{2s_{123}s_{23}}  
          + \frac{s_{12}}{2s_{123}(s_{12}+s_{13})}
          + \frac{s_{12}}{2s_{123}(s_{12}+s_{23})}
\nonumber \\ &&
          + \frac{2s_{12}}{s_{123}(s_{13}+s_{23})}
          + \frac{2\log y_{12}}{s_{123}}
 \left( 2\frac{s_{12}}{s_{13}+s_{23}} 
+ \frac{s_{12}^2}{(s_{13}+s_{23})^2}  \right)
\nonumber \\ &&
          + \frac{\log y_{13}}{2s_{123}}\,  \left( 
           \frac{s_{12} s_{13}}{(s_{12}+s_{23})^2}
          + 4 \frac{s_{12}}{s_{12}+s_{23}}
          + \frac{s_{13}}{s_{12}+s_{23}}\right)
\nonumber \\ &&
          + \frac{\log y_{23} }{2s_{123}}\, \left( 
           \frac{s_{12} s_{23}}{(s_{12}+s_{13})^2}
          + 4 \frac{s_{12}}{s_{12}+s_{13}}
          + \frac{s_{23}}{s_{12}+s_{13}}\right)
\nonumber \\ &&
          +  R(y_{12},y_{13}) \frac{2 s_{12}+s_{13}}{s_{123}
s_{23}}
          +  R(y_{12},y_{23}) 
             \frac{2 s_{12}+s_{23}}{s_{123}s_{13}}\;, \label{eq:A31t}\\
\Poles\left(\hat{A}_3^1(1_q,3_g,2_{\bar q})\right) &= &
2 \left( {\bf I}^{(1)}_{qg,\Flavour} (\e,s_{13}) + {\bf I}^{(1)}_{qg,\Flavour} 
(\e,s_{23})
 \right) A_3^0(1,3,2)\;,
\\
\Finite\left(\hat{A}_3^1(1_q,3_g,2_{\bar q})\right) &=& \frac{1}{6} \left(
\log y_{13} + \log y_{23} \right)  A_3^0(1,3,2)\;.\label{eq:A31h}
\end{eqnarray}
Note that application of the $\Finite$-operator in the above expression 
yields only the ${\cal O}(\e^0)$-terms of the antenna functions. 
These antenna functions contain higher powers in 
$\e$ as well, and these are relevant to the integrated antennae
 listed below.

The integrated antennae are defined in (\ref{eq:x31int}). They read:
\begin{eqnarray}
{\cal A}_3^1(s_{123}) &= &
 (s_{123})^{-2\e} \Bigg[ -\frac{1}{4\e^4} - \frac{31}{12\e^3} 
+ \frac{1}{\e^2} \left(-\frac{53}{8}+\frac{11\pi^2}{24}
\right) + \frac{1}{\e} \left(-\frac{647}{24} + \frac{22\pi^2}{9}
+\frac{23}{3} \,\zeta_3 \right)
\nonumber \\&&
+\left(-\frac{5231}{48}
          + \frac{17\pi^2}{2}
          + \frac{689}{18}\zeta_3   - \frac{41\pi^4}{480}
   \right) + {\cal O}(\e) \Bigg]\;,\\
\tilde{{\cal A}}_3^1(s_{123}) &= &
 (s_{123})^{-2\e} \Bigg[ \frac{1}{\e^2} \left( -\frac{5}{8} +\frac{\pi^2}{6} \right)
+ \frac{1}{\e} \left( -\frac{19}{4} +\frac{\pi^2}{4} 
+7\zeta_3 \right)
\nonumber \\&&
+\left(-\frac{105}{4}
          + \frac{27\pi^2}{16}
          + \frac{27}{2}\zeta_3   + \frac{7\pi^4}{90}
   \right) + {\cal O}(\e) \Bigg]\;,\\
\hat{{\cal A}}_3^1(s_{123}) &= &
 (s_{123})^{-2\e} \Bigg[ \frac{1}{3\e^3} + \frac{1}{2\e^2} 
+ \frac{1}{\e} \left(\frac{19}{12}-\frac{7\pi^2}{36}
\right) \nonumber \\&&
+ \left(\frac{109}{24} - \frac{7\pi^2}{24}
-\frac{25}{9} \,\zeta_3 \right)
+ {\cal O}(\e) \Bigg]\;,\end{eqnarray}
with
\begin{eqnarray}
 \Poles\left({\cal A}_3^1(s_{123})\right)  &=& 
- {\cal A}_2^1(s_{123}) \left( 
 2 {\bf I}^{(1)}_{q\bar q} (\e,s_{123}) + 
{\cal A}_3^0(s_{123}) \right) 
+ \frac{2b_0}{\e}\, (s_{123})^{-\e}\,
  {\bf I}^{(1)}_{q\bar q} (\e,s_{123}) \nonumber \\ 
&& - {\bf H}^{(2)}_{V,A}(\e,s_{123})
+ {\bf S}^{(2)}_{V}(\e,s_{123})\;,
\\
 \Finite\left({\cal A}_3^1(s_{123})\right)  &=&
         - \frac{6581}{48}
          + \frac{787\pi^2}{96}
          + \frac{17\pi^4}{360}
          + \frac{143}{3}\zeta_3
 \;,\\
 \Poles\left(\tilde{{\cal A}}_3^1(s_{123})\right)  &=& 
- {\cal A}_2^1(s_{123}) \left( 
 2 {\bf I}^{(1)}_{q\bar q} (\e,s_{123}) + 
{\cal A}_3^0(s_{123}) \right) 
- {\bf H}^{(2)}_{V,\tilde{A}}(\e,s_{123})
\label{eq:A31tpoles}
\;,\\
 \Finite\left(\tilde{{\cal A}}_3^1(s_{123})\right)  &=&
         - \frac{845}{8}
          + \frac{217\pi^2}{32}
          + \frac{9\pi^4}{40}
          + \frac{75}{2}\zeta_3\;, \\
 \Poles\left(\hat{{\cal A}}_3^1(s_{123})\right)  &=& 
 \frac{2b_{0,\Flavour}}{\e}\, (s_{123})^{-\e} \,
{\bf I}^{(1)}_{q\bar q} (\e,s_{123})
- {\bf H}^{(2)}_{V,\hat{A}}(s_{123})
\;,\\
 \Finite\left(\hat{{\cal A}}_3^1(\e,s_{123})\right)  &=&
          \frac{109}{24}
          - \frac{8}{3}\zeta_3\;.
\end{eqnarray}

\subsection{Four-parton tree-level antenna functions}

The tree-level four-parton quark-antiquark antenna contains three final 
states: 
quark-gluon-gluon-antiquark at leading and 
subleading colour, $A_4^0$ and $\tilde{A}_4^0$ 
and quark-antiquark-quark-antiquark for non-identical quark
flavours $B_4^0$ as well as the identical-flavour-only contribution
$C_4^0$. The quark-antiquark-quark-antiquark final state with 
identical quark flavours is thus described by the sum of antennae 
for non-identical flavour and identical-flavour-only.
The 
antennae for the $qgg\bar q$ final state are:
\begin{eqnarray}
\label{eq:A40}
A_4^0(1_q,3_g,4_g,2_{\bar q}) &=& 
a_4^0(1,3,4,2) + a_4^0(2,4,3,1)\;,  \\
\tilde{A}_4^0(1_q,3_g,4_g,2_{\bar q}) &=&
\tilde{a}_4^0(1,3,4,2) + \tilde{a}_4^0(2,4,3,1)  +
\tilde{a}_4^0(1,4,3,2) + \tilde{a}_4^0(2,3,4,1)\;,  
\label{eq:A40t}
\end{eqnarray}
where the sub-antennae are given by
\begin{eqnarray}
a_4^0(1,3,4,2) &=& \frac{1}{s_{1234}} \Bigg\{
       \frac{1}{2\Sac\Sbd\Scd}   \left[
            2 \Sab \Sad
          + 2 \Sab \Sbc
          + 2 \Sab^2
          + \Sad^2
          + \Sbc^2
          \right]\nonumber \\&&
       + \frac{1}{2\Sac \Sbd \Sacd \Sbcd}   \left[
            3 \Sab \Scd^2
          - 4 \Sab^2 \Scd
          + 2 \Sab^3
          -  \Scd^3
          \right]\nonumber \\&&
       + \frac{1}{\Sac \Sbd \Sacd}   \left[
            3 \Sab \Sbc
          - 3 \Sab \Scd
          + 4 \Sab^2
          - \Sbc \Scd
          + \Sbc^2
          + \Scd^2
          \right]\nonumber \\&&
       + \frac{3}{2\Sac \Sbd}   \left[
            2 \Sab
          + \Sad
          + \Sbc
          \right]
       + \frac{1}{\Sac \Scd}   \left[
            4 \Sab
          + 3 \Sbc
          + 2 \Sbd
          \right]\nonumber \\&&
       + \frac{1}{\Sac \Sacd^2}   \left[
            \Sab \Scd
          + \Sbc \Scd
          + \Sbd \Scd
          \right]\nonumber \\&&
       + \frac{1}{\Sac \Sacd \Sbcd}   \left[
            3 \Sab \Sbd
          + 6 \Sab \Scd
          - 4 \Sab^2
          - 3 \Sbd \Scd
          - \Sbd^2
          - 3 \Scd^2
          \right]\nonumber \\&&
       + \frac{1}{\Sac \Sacd}   \left[
          - 6 \Sab
          - 3 \Sbc
          - \Sbd
          + 2 \Scd
          \right]\nonumber \\&&
       + \frac{1}{\Sbd \Scd \Sacd}   \left[
            2 \Sab \Sad
          + 2 \Sab \Sbc
          + 2 \Sab^2
          + 2 \Sad \Sbc
          + \Sad^2
          + \Sbc^2
          \right]\nonumber \\&&
       + \frac{1}{\Sbd \Sacd}   \left[
          - 4 \Sab
          - \Sad
          - \Sbc
          + \Scd
          \right]
       + \frac{1}{\Scd^2}   \left[
            \Sab
          + 2 \Sac 
          - 2 \Sad
          - \Scd
          \right]\nonumber \\&&
       + \frac{1}{\Scd^2 \Sacd^2}   \left[
            2 \Sab \Sad^2
          + 2 \Sad^2 \Sbc
          + 2 \Sad^2 \Sbd
          \right]
       - \frac{2 \Sab \Sad \Sbd}{\Scd^2 \Sacd \Sbcd} \nonumber \\&&
       + \frac{1}{\Scd^2 \Sacd}   \left[
          - 2 \Sab \Sad
          - 4 \Sad \Sbd
          + 2 \Sad^2
          \right]\nonumber \\&&
       + \frac{1}{\Scd \Sacd \Sbcd}   \left[
          - 2 \Sab \Sad
          - 4 \Sab^2
          + 2 \Sad \Sbd
          - \Sad^2
          - \Sbd^2
          \right]\nonumber \\&&
       + \frac{1}{\Scd \Sacd}   \left[
          - 8 \Sab
          - 2 \Sbc
          - 2 \Sbd
          \right]
       + \frac{1}{\Sacd^2}   \left[
            \Sab
          + \Sbc  
          + \Sbd
          \right]\nonumber \\&&
       + \frac{3}{2\Sacd \Sbcd}   \left[
            2 \Sab
          +  \Sad
          -  \Sbd
          -  \Scd
          \right]
       + \frac{1}{2 \Sacd}
+ {\cal O} (\e)
\Bigg\}\; ,\\
\tilde{a}_4^0(1,3,4,2) &=& \frac{1}{s_{1234}} \Bigg\{
         \frac{1}{\Sac\Sbd\Sacd\Sbcd}\left[ 
            \frac{3}{2} \Sab \Scd^2
          - 2 \Sab^2 \Scd
          + \Sab^3
          - \frac{1}{2} \Scd^3
          \right] \nonumber \\ &&
       + \frac{1}{\Sac\Sbd \Sacd}   \left[
            3 \Sab \Sbc
          - 3 \Sab \Scd
          + 4 \Sab^2
          - \Sbc \Scd
          + \Sbc^2
          + \Scd^2
          \right]\nonumber \\ &&
       + \frac{\Sab^3}{\Sac \Sbd (\Sac+\Sbc) (\Sad+\Sbd)}
       + \frac{1}{\Sac \Sbd (\Sac+\Sbc)}   \left[
            \frac{1}{2} \Sab \Sad
          + \Sab^2
          \right]\nonumber \\ &&
       + \frac{1}{\Sac \Sbd (\Sad+\Sbd)}   \left[
            \frac{1}{2} \Sab \Sbc
          + \Sab^2
          \right]
       + \frac{1}{\Sac \Sbd}   \left[
            3 \Sab
          + \frac{3}{2} \Sad
          + \frac{3}{2} \Sbc
          \right]\nonumber \\ &&
       + \frac{1}{\Sac \Sacd^2}   \left[
            \Sab \Scd
          + \Sbc \Scd
          + \Sbd \Scd
          \right]
       + \frac{ 2 \Sab^3}{\Sac \Sacd \Sbcd (\Sac+\Sbc)} \nonumber \\ &&
       + \frac{1}{\Sac \Sacd \Sbcd}   \left[
            3 \Sab \Scd
          - \Sbd \Scd
          - 2 \Scd^2
          \right]\nonumber \\ &&
       + \frac{1}{\Sac \Sacd (\Sac+\Sbc)}   \left[
            \Sab \Sbd
          + \Sab \Scd
          + 2 \Sab^2
          \right]\nonumber \\ &&
       + \frac{1}{\Sac \Sacd}   \left[
          - \Sbc
          - \Sbd
          + 2 \Scd
          \right]
       + \frac{1}{\Sac \Sbcd (\Sac+\Sbc) }  \left[
            \Sab \Sad
          + \Sab \Scd
          + 2 \Sab^2
          \right]\nonumber \\ &&
       + \frac{1}{\Sac \Sbcd }  \left[
          - 2 \Sab
          - 2 \Sad
          + \Sbd
          + 2 \Scd
          \right]\nonumber \\ &&
       + \frac{2\Sab^3}{\Sac (\Sac+\Sbc) (\Sad+\Sbd) (\Sac+\Sad)}\nonumber \\ &&
        + \frac{1}{\Sac (\Sac+\Sbc) (\Sac+\Sad)}   \left[
            \Sab \Sbd
          + 2 \Sab^2
          \right]\nonumber \\ &&
       + \frac{1}{\Sac (\Sad+\Sbd)(\Sac+\Sad)}   \left[
            \Sab \Sbc
          + 2 \Sab^2
          \right]\nonumber \\ &&
       + \frac{2\Sab}{\Sac (\Sac+\Sad)}
       - \frac{2}{\Sac}
       + \frac{1}{\Sacd^2}   \left[
            \Sab
          + \Sbc
          + \Sbd
          \right]\nonumber \\ &&
       + \frac{1}{\Sacd \Sbcd}   \left[
            \Sab
          - \Scd
          \right]
       + \frac{1}{\Sacd} + {\cal O} (\e)\Bigg\}\;.
\end{eqnarray}
In $A_4^0$ the gluonic emissions are colour-ordered, while 
in $\tilde{A}_4^0$ the gluons are photon-like, implying no ordering.
Because of colour-ordering,  $A_4^0$ can be used with a single 
ordered phase space mapping. In contrast,  $\tilde{A}_4^0$ can not be 
used with a unique ordered phase space mapping. The above decomposition
into $\tilde{a}_4^0$ yields however ordered terms, since the combination 
$\tilde{a}_4^0(1,3,4,2) + \tilde{a}_4^0(2,4,3,1)$ contains only 
single emission singularities 
in $1/s_{13}$ and $1/s_{24}$, corresponding to the ordered $(1,3,4,2)$
phase space mapping. 
On the other hand $\tilde{a}_4^0(1,4,3,2) + \tilde{a}_4^0(2,3,4,1)$
contains only single emission singularities in  $1/s_{14}$ and $1/s_{23}$,
corresponding to the ordered $(1,4,3,2)$
phase space mapping. Since the decomposition of $\tilde{A}_4^0$ 
is symmetric, all four $\tilde{a}_4^0$ yield identical integrals if 
integrated over the tripole phase space. It should be noted that it is 
not possible to analytically 
integrate an individual  $\tilde{a}_4^0$  over the tripole 
phase space using the reduction and integration techniques described 
in~\cite{ggh}, since the extra polynomial denominators present there enlarge 
the set of basis integrals considerably. When the four
$\tilde{a}_4^0$ are added together
these  polynomial denominators cancel, and the tripole integrals can be 
carried out.

The integrals of these antenna functions are according to (\ref{eq:x4int}):
\begin{eqnarray}
{\cal A}_4^0(s_{1234}) &=& 
\left(s_{1234}\right)^{-2\e} 
\Bigg[ \frac{3}{4\e^4} + \frac{65}{24\e^3}
+ \frac{1}{\e^2} \left( \frac{217}{18} -\frac{13\pi^2}{12} \right)
\nonumber \\&&
\hspace{1cm}
+ \frac{1}{\e} \left( \frac{43223}{864} -\frac{589\pi^2}{144} 
-\frac{71}{4}\zeta_3 \right)
\nonumber \\&&
\hspace{1cm}
+\left(\frac{1076717}{5184}
          - \frac{7955\pi^2}{432}
          - \frac{1327}{18}\zeta_3   + \frac{373\pi^4}{1440}
   \right)+ {\cal O}(\e)  \Bigg] \;, \\
\tilde{{\cal A}}_4^0(s_{1234}) &=&
2\;\left(s_{1234}\right)^{-2\e} \Bigg[ \frac{1}{2\e^4} + \frac{3}{2\e^3}
+ \frac{1}{\e^2} \left( \frac{13}{2} -\frac{3\pi^2}{4} \right)
\nonumber \\&&
+ \frac{1}{\e} \left( \frac{845}{32} -\frac{9\pi^2}{4} 
-\frac{40}{3}\zeta_3 \right)
+\left(\frac{6921}{64}
          - \frac{473\pi^2}{48}
          - 40\zeta_3 + \frac{17\pi^4}{144}
   \right) + {\cal O}(\e) \Bigg]\;, \nonumber \\
\end{eqnarray}
with
\begin{eqnarray}
\Poles\left( {\cal A}_4^0(s_{1234})\right) &=& 
2\, \left[{\bf I}^{(1)}_{q \bar q} \left(\e,s_{1234} \right) 
\right]^2  -2\, e^{-\e \gamma} \frac{\Gamma (1-2\e)}{\Gamma (1-\e)}
\left(\frac{b_{0}}{\e} + k_{0} \right) \,
 {\bf I}^{(1)}_{q\bar q}(2\e,s_{1234}) \nonumber \\ &&
-  {\bf H}^{(2)}_{R,A}(\e,s_{1234})
- {\bf S}^{(2)}_{V}(\e,s_{1234})\;, \\
\Finite\left( {\cal A}_4^0(s_{1234})\right) &=& 
                 \frac{811037}{5184}
          - \frac{2321\pi^2}{288}
          - \frac{13\pi^4}{160}
          - \frac{4217}{72}\zeta_3 \;,
\\
\Poles\left( \tilde{{\cal A}}_4^0(s_{1234})\right) &=&
4 \left[{\bf I}^{(1)}_{q \bar q} \left(\e,s_{1234} \right) 
\right]^2 -  2\, {\bf H}^{(2)}_{R,\tilde{A}}(\e,s_{1234})\;,\\
\Finite\left( \tilde{{\cal A}}_4^0(s_{1234})\right) &=&
            \frac{6921}{32}
          - \frac{259\pi^2}{16}
          - \frac{3\pi^4}{10}
          - 78\zeta_3
\;.
\end{eqnarray}

The non-identical quark antenna is:
\begin{eqnarray}
\label{eq:B40}
B_4^0(1_q,3_{q'},4_{\bar q'},2_{\bar q}) &=& 
b_4^0(1,3,4,2) + 
b_4^0(2,3,4,1) + 
b_4^0(1,4,3,2) + 
b_4^0(2,4,3,1) \; ,
\end{eqnarray}
with a sub-antenna function given by
\begin{eqnarray}
b_4^0(1,3,4,2) &=& \frac{1}{s_{1234}} \Bigg\{
         \frac{1}{\Scd^2 \Sacd^2}   \left[
            \Sab \Sac \Sad
          + \Sac \Sad \Sbc
          - \Sac^2 \Sbd
          \right]\nonumber \\ &&
       + \frac{1}{\Scd^2 \Sacd \Sbcd}   \left[
          - \Sab \Sac \Sbd
          + \Sac \Sad \Sbc
          - \Sac \Sbd^2
          \right]
       + \frac{1}{\Scd \Sacd^2}   \left[
            \Sab \Sac
          + \Sac \Sbc
          \right]\nonumber \\ &&
       + \frac{1}{2\Scd \Sacd \Sbcd}   \left[
            2\Sab \Sac
          + \Sab^2
          \right]
       + \frac{\Sab}{2 \Sacd \Sbcd}
       + {\cal O} (\e)\Bigg\}\;.
\end{eqnarray}

In $B_4^0$, the secondary quark emission is ordered, such that a single 
ordered phase space mapping can be used. 

This subtraction term yields the integral
\begin{eqnarray}
{\cal B}_4^0(s_{1234}) &=& 
\left(s_{1234}\right)^{-2\e} 
  \Bigg[ 
- \frac{1}{12\e^3} - \frac{7}{18\e^2}
+ \frac{1}{\e} \left( -\frac{407}{216} +\frac{11\pi^2}{72} \right)
\nonumber \\ &&
\hspace{1cm}
+ \left( - \frac{11753}{1296} +\frac{77\pi^2}{108} 
+\frac{67}{18}\zeta_3 \right)  
+ {\cal O}(\e) \Bigg]\;,
\end{eqnarray}
with 
\begin{eqnarray}
 \Poles\left( {\cal B}_4^0(s_{1234})\right)
&=&
- 2 e^{-\e \gamma} \frac{\Gamma (1-2\e)}{\Gamma (1-\e)}
\left(\frac{b_{0,\Flavour}}{\e} + k_{0,\Flavour} \right) \,
 {\bf I}^{(1)}_{q\bar q}(2\e,s_{1234}) \nonumber \\ &&
- {\bf H}^{(2)}_{R,B}(\e,s_{1234})\;, \\
 \Finite\left( {\cal B}_4^0(s_{1234})\right)
&=&          - \frac{11753}{1296}
          - \frac{7\pi^2}{72}
          + \frac{133}{36}\zeta_3\;.
\end{eqnarray}

The identical-flavour-only quark-antiquark-quark-antiquark antenna is:
\begin{eqnarray}
\label{eq:C40}
C_4^0(1_q,3_q,4_{\bar q},2_{\bar q}) &=& c_4^0(1,2,3,4) + c_4^0(1,4,3,2)\; ,
\end{eqnarray}
with
\begin{eqnarray}
c_4^0(1,2,3,4) &=& \frac{1}{s_{1234}} \Bigg\{
       - \frac{ \Sab \Sac \Sad}{2\Sbc \Scd \Sabc \Sacd}
       + \frac{1}{2 \Sbc \Scd \Sacd \Sbcd}  \left[
          -  \Sab \Sac \Sbd
          +  \Sac \Sad \Sbd
          \right]\nonumber \\ &&
       - \frac{\Sac\Sbd^2}{2 \Sbc \Scd \Sbcd^2}
       - \frac{\Sab \Sac} {\Sbc \Sabc \Sacd}\nonumber \\ &&
       + \frac{1}{2 \Sbc \Sabc \Sbcd}   \left[
          - \Sab \Sad
          - \Sab \Scd
          - \Sab^2
          + \Sac \Sbd
          \right]\nonumber \\ &&
       + \frac{1}{2 \Sbc \Sacd \Sbcd}   \left[
             \Sab \Sad
          +  \Sab \Scd
          +  \Sab^2
          +  \Sac \Sbd
          \right]
       - \frac{\Sac}{2 \Sabc \Sacd} \nonumber \\ &&
       + \frac{1}{ \Sbc \Sbcd^2}   \left[
            \Sab \Sbd
          + \Sad \Sbd
          \right]
       + \frac{1}{2 \Sabc \Sbcd}   \left[
          - \Sab
          + \Sad
          \right]
 + {\cal O} (\e) \Bigg\}\;.
\end{eqnarray}

It integrates to 
\begin{eqnarray}
{\cal C}_4^0(s_{1234}) &=& 
\frac{1}{2} \,\left(s_{1234}\right)^{-2\e} 
\Bigg[ 
 \frac{1}{\e} \left( -\frac{13}{16} +\frac{\pi^2}{8} - 
\frac{1}{2}\zeta_3\right)
\nonumber \\
&& \hspace{1cm}
+ \left(  -\frac{339}{32} +\frac{17\pi^2}{24} 
+\frac{21}{4}\zeta_3 -\frac{2\pi^4}{45} \right)  
+ {\cal O}(\e) 
\Bigg]\;,
\end{eqnarray}
with
\begin{eqnarray}
 \Poles\left( {\cal C}_4^0(s_{1234})\right) &=& 
- \frac{1}{2} {\bf H}^{(2)}_{R,C}(\e,s_{1234})\;,\\
\Finite\left( {\cal C}_4^0(s_{1234})\right) &=& 
- \frac{339}{64} +\frac{17\pi^2}{48} 
+\frac{21}{8}\zeta_3 -\frac{\pi^4}{45}\;.
\end{eqnarray}

All antenna functions listed in this section 
agree with the four-parton matrix elements
in the Appendix of ~\cite{ERT}, taking account of the different 
normalisation used here. They also all agree with~\cite{weinzierlnew}.

\section{Quark-gluon antennae}
\label{sec:qg}

The quark-gluon antenna functions are obtained from the QCD real radiation
corrections to the decay of a heavy neutralino into a massless gluino and a 
gluon, $\tilde \chi \to \tilde g g $, which is described 
in detail in~\cite{chi}. 

The overall normalisation is given by defining the tree-level two-parton
quark-gluon 
antenna function
\begin{equation}
{\cal D}_2^0(s_{13}) \equiv 1\;.
\end{equation}
In this equation, and in all subsequent equations in the section, we 
label the primary quark momentum as $(1)_q$ and the momenta of gluons or 
of a secondary quark-antiquark pair as $(3)_i$, $(4)_j$ and
$(5)_k$. This 
non-consecutive labelling of momenta is introduced in view of applying the 
 quark-gluon antenna functions in an actual calculation, where they will always
appear in a pair: quark-gluon antenna and antiquark-gluon antenna, with 
 $(1)_q$ and $(2)_{\bar q}$ denoting the primary quark and antiquark momenta.

The one-loop two-parton quark-gluon antenna contains two 
contributions, corresponding to the different colour and flavour structures:
\begin{eqnarray}
{\cal D}_2^1(s_{13}) &=& 
 2\, (s_{13})^{-\e} \,\Bigg[ -\frac{1}{\e^2} - \frac{5}{3\e}  + 
\frac{7\pi^2}{12} 
+ \left( -1 + \frac{7}{3}\zeta_3 \right)
\e 
\nonumber \\
&& \hspace{2cm}
+ \left( - 3
- \frac{73\pi^4}{1440} \right) \e^2  + {\cal O}(\e^3) \Bigg] \;, \\
\hat{{\cal D}}_2^1(s_{13}) &=& 
  2\, (s_{13})^{-\e} \frac{1}{6\e} \;,
\end{eqnarray}
with 
\begin{eqnarray}
\Poles\left({\cal D}_2^1(s_{13}) \right) &=& 
4 {\bf I}^{(1)}_{q g} \left(\e,s_{13} \right) \label{eq:d21poles}\;,\\
\Finite\left({\cal D}_2^1(s_{13}) \right) &=& 0\;,\\
\Poles\left(\hat{{\cal D}}_2^1(s_{13 }) \right) &=& 
4 {\bf I}^{(1)}_{q g,\Flavour} \left(\e,s_{13} \right)\label{eq:d21hpoles}\;,\\
\Finite\left(\hat{{\cal D}}_2^1(s_{13}) \right) &=& 0\;.
\end{eqnarray}

The pole terms in the above expression have to be compared to the pole 
terms of the one-loop correction to the quark-antiquark antenna function
${\cal A}_2^1$ in (\ref{eq:a21poles}), containing 
$2 {\bf I}^{(1)}_{q \bar q}$. The factor $4$ in 
(\ref{eq:d21poles}),(\ref{eq:d21hpoles})
 appears since the tree level quark-gluon antenna function 
${\cal D}_2^0(s_{13})$  contains two distinct 
quark-gluon antennae, 
in contrast to the single quark-antiquark antenna contained in 
${\cal A}_2^0(s_{12})$, as will be seen below in constructing the 
three-parton tree-level antenna functions. 

\subsection{Three-parton tree-level antenna functions}

The tree-level three-parton quark-gluon antenna contains two final states: 
quark-gluon-gluon and quark-quark-antiquark. The 
antenna corresponding to the first final state is: 
\begin{eqnarray}
D_3^0(1_q, 3_g, 4_g) &=& 
\frac{1}{s_{134}^2} \, \Bigg(
\frac{2 s_{134}^2 s_{14}}{ s_{13}  s_{34}}
+\frac{2 s_{134}^2 s_{13}}{ s_{14}  s_{34}}
+ \frac{s_{14} s_{34}
+   s_{34}^2}{ s_{13}} 
\nonumber \\ && \hspace{1cm}
+ \frac{s_{13} s_{34}  
+   s_{34}^2}{ s_{14}}
+ \frac{2s_{13}s_{14}}{ s_{34}} + 5  s_{134}
+  s_{34}\Bigg) + {\cal O}(\e)
\;.
\end{eqnarray}
Its integrated form is
\begin{eqnarray}
{\cal D}_3^0(s_{134}) &=&  
2\,   \left(s_{134} \right)^{-\e}\Bigg[
\frac{1}{\e^2} + \frac{5}{3\e} + \frac{17}{3} - 
\frac{7\pi^2}{12} 
+ \left( \frac{209}{12} - \frac{35\pi^2}{36} - \frac{25}{3}\zeta_3 \right)
\e \nonumber \\
&& \hspace{2cm}
+ \left( \frac{421}{8}
          - \frac{119\pi^2}{36}
          - \frac{125}{9}\zeta_3
          - \frac{71\pi^4}{1440} \right) \e^2  + {\cal O}(\e^3) \Bigg] 
\;,\end{eqnarray}
with 
\begin{eqnarray}
\Poles\left({\cal D}_3^0(s_{134}) \right) &=& 
- 4 {\bf I}^{(1)}_{qg} \left(\e,s_{134} \right)\;,\\
\Finite\left({\cal D}_3^0(s_{134}) \right) &=& \frac{34}{3}\;.
\end{eqnarray}
This tree-level antenna function contains two antennae, corresponding 
to the 
configurations:  (gluon ($3_g$)  radiated between quark and gluon ($4_g$)) and 
 (gluon ($4_g$) radiated between quark and gluon ($3_g$)). 
The separation between these 
is not free from an ambiguity, since the collinear 
limit of the two gluons has
be split between the two configurations. 
We decompose
\begin{equation}
D_3^0(1, 3, 4) = d_3^0(1, 3, 4) + d_3^0(1, 4, 3) \; ,
\end{equation}
where the sub-antenna is given by
\begin{equation}
d_3^0(1, 3, 4) = 
\frac{1}{s_{134}^2} \, \Bigg(
\frac{2 s_{134}^2 s_{14}}{ s_{13}  s_{34}}
+ \frac{s_{14} s_{34}
+   s_{34}^2}{ s_{13}} 
+ \frac{s_{13}s_{14}}{ s_{34}} + \frac{5}{2}  s_{134}
+  \frac{1}{2}s_{34}\Bigg)
+ {\cal O}(\e).
\label{eq:smalld}
\end{equation}
The function could be further decomposed into two dipoles if needed 
for configurations discussed in Section~\ref{sec:sub2b}.

The tree-level three-parton quark-gluon
antenna corresponding to the quark-quark-antiquark final state is:
\begin{equation}
E_3^0(1_{q},3_{q'},4_{\bar q'}) = 
\frac{1}{s_{134}^2} \, \left(
 \frac{s_{13}^2+s_{14}^2}{s_{34}} + s_{13} + s_{14}
\right)+  {\cal O}(\e)\;.
\end{equation}
Phase space integration yields:
\begin{eqnarray}
{\cal E}_3^0(s_{134}) &=&  
 2\,  \left(s_{134} \right)^{-\e}\Bigg[
- \frac{1}{6\e} -\frac{1}{2}  + \left( -\frac{3}{2} + \frac{7\pi^2}{72} \right)
\e 
\nonumber \\ && \hspace{1cm}
+ \left( -\frac{9}{2} 
          + \frac{7\pi^2}{24}
          - \frac{25}{18}\zeta_3\right) \e^2  + {\cal O}(\e^3) \Bigg] 
\;,
\end{eqnarray}
with 
\begin{eqnarray}
\Poles\left({\cal E}_3^0(s_{134}) \right) &=& 
- 4 {\bf I}^{(1)}_{q g,\Flavour} \left(\e,s_{134} \right)\;,\\
\Finite\left({\cal E}_3^0(s_{134}) \right) &=& -1\;.
\end{eqnarray}

\subsection{Three-parton one-loop antenna functions}

At one loop, the correction to the quark-gluon-gluon antenna contains 
a leading colour term $D_3^1(1_q,3_g,4_g)$ and a quark loop term
$\hat{D}_3^1(1_q,3_g,4_g)$. These read: 
\begin{eqnarray}
\Poles\left(D_3^1(1_q,3_g,4_g)\right) &=& 
2 \bigg( {\bf I}^{(1)}_{qg} (\e,s_{13}) + {\bf I}^{(1)}_{qg} (\e,s_{14})
+ {\bf I}^{(1)}_{gg} (\e,s_{34}) 
\nonumber \\ && \hspace{3mm} - 2 {\bf I}^{(1)}_{qg} (\e,s_{134}) \bigg) 
D_3^0(1,3,4)\;,  \\
\Finite\left(D_3^1(1_q,3_g,4_g)\right) &=& -
\bigg(  R(y_{13},y_{34}) 
+  R(y_{14},y_{34}) +  R(y_{13},y_{14})
+ \frac{5}{3} \log y_{13} \nonumber \\ && 
+  \frac{5}{3} \log y_{14} 
+ \frac{11}{6} \log y_{34} \bigg)
D_3^0(1,3,4)    +  \frac{1}{3s_{34}}\;, \\
\Poles\left(\hat{D}_3^1(1_q,3_g,4_g)\right) &=& 
2 \Bigg( {\bf I}^{(1)}_{qg,\Flavour} (\e,s_{13}) 
+ {\bf I}^{(1)}_{qg,\Flavour} (\e,s_{14})
\nonumber \\ && \hspace{3mm}
+ {\bf I}^{(1)}_{gg,\Flavour} (\e,s_{34}) 
- 2 {\bf I}^{(1)}_{qg,\Flavour} (\e,s_{134}) \Bigg) D_3^0(1,3,4) \;,\\
\Finite\left(\hat{D}_3^1(1_q,3_g,4_g)\right) &=& \frac{1}{6}\left(
\log y_{13} +  \log y_{14} 
+ 2\log y_{34} \right)
D_3^0(1,3,4)    -  \frac{1}{3s_{34}}\;.
\end{eqnarray}

Integration of these antenna functions yields
\begin{eqnarray}
{\cal D}_3^1(s_{134}) &= &
 (s_{134})^{-2\e} \Bigg[ -\frac{1}{2\e^4} - \frac{16}{3\e^3} 
+ \frac{1}{\e^2} \left(-\frac{619}{36}+\frac{5\pi^2}{4}
\right) + \frac{1}{\e} \left(-\frac{8941}{108} + \frac{23\pi^2}{4}
+\frac{88}{3} \,\zeta_3 \right)
\nonumber \\&&
+\left(-\frac{20353}{54}
          + \frac{5473\pi^2}{216}
          + 105 \zeta_3   - \frac{11\pi^4}{720}
   \right) + {\cal O}(\e) \Bigg]\;, \\
\hat{{\cal D}}_3^1(s_{134}) &= &
 (s_{134})^{-2\e} \Bigg[   \frac{2}{3\e^3} 
+ \frac{10}{9\e^2} +
 \frac{1}{\e} \left(\frac{139}{36} - \frac{7\pi^2}{18}
 \right)
\nonumber \\ && \hspace{1cm}
+\left(\frac{443}{36}
          - \frac{35\pi^2}{54}
          - \frac{50}{9} \zeta_3   
   \right) + {\cal O}(\e) \Bigg]\;,
\end{eqnarray}
with
\begin{eqnarray}
 \Poles\left({\cal D}_3^1(s_{134})\right)  &=& 
- {\cal D}_2^1(s_{134}) \left( 
 4 {\bf I}^{(1)}_{qg} (\e,s_{134}) + 
{\cal D}_3^0(s_{134}) \right) 
+ \frac{2b_0}{\e}\, (s_{134})^{-\e}\,
 \left[ 2\,{\bf I}^{(1)}_{qg} (\e,s_{134})\right] \nonumber \\ 
&& - {\bf H}^{(2)}_{V,D}(\e,s_{134})
+ 2 {\bf S}^{(2)}(\e,s_{134})\;,
\\
 \Finite\left({\cal D}_3^1(s_{134})\right)  &=&
         - \frac{32455}{54}
          + \frac{16573\pi^2}{432}
          + \frac{49\pi^4}{90}
          + \frac{3283}{18}\zeta_3
 \;,\\
 \Poles\left(\hat{{\cal D}}_3^1(s_{134})\right)  &=& 
- \hat{{\cal D}}_2^1(s_{134}) \left( 
 4 {\bf I}^{(1)}_{qg} (\e,s_{134}) + 
{\cal D}_3^0(s_{134}) \right) 
\nonumber \\ 
&& + \frac{2b_{0,\Flavour}}{\e}\, (s_{134})^{-\e}\,
 \left[ 2\,{\bf I}^{(1)}_{qg} (\e,s_{134})\right] 
- {\bf H}^{(2)}_{V,\hat{D}}(\e,s_{134})\;,
\\
 \Finite\left(\hat{{\cal D}}_3^1(s_{134})\right)  &=&
            \frac{287}{12}
          - \frac{32}{3}\zeta_3\;.
\end{eqnarray}
Again, the factors of 2 in front of the infrared singularity operators and 
of the soft gluon current arise from the fact that the basic two-parton
process contains two quark-gluon antennae.

At one loop, the correction to the quark-quark-antiquark antenna contains 
a leading colour and 
a subleading colour term 
$E_3^1(1_q,3_{q'},4_{\bar q'})$, $\tilde{E}_3^1(1_q,3_{q'},4_{\bar q'})$ 
as well as
 a quark loop term
$\hat{E}_3^1(1_q,3_{q'},4_{\bar q'})$. These read: 
\begin{eqnarray}
\Poles\left(E_3^1(1_q,3_{q'},4_{\bar q'})\right) &=& 
2 \left( {\bf I}^{(1)}_{q\bar q} (\e,s_{13}) + {\bf I}^{(1)}_{q\bar q} 
(\e,s_{14}) -2{\bf I}^{(1)}_{q g} 
(\e,s_{134}) 
\right) E_3^0(1,3,4)\;,\nonumber \\ \\
\Finite\left(E_3^1(1_q,3_{q'},4_{\bar q'})\right) &=& 
-
\bigg( 
 R(y_{13},y_{34}) 
+  R(y_{14},y_{34}) 
+ \frac{3}{2} \log y_{13} +  \frac{3}{2} \log y_{14} 
\nonumber \\ && \hspace{1cm}
+ \frac{13}{6} \log y_{34}  - \frac{40}{9}\bigg)
E_3^0(1,3,4)    
\nonumber \\ && 
+  R(y_{13},y_{34}) \frac{s_{13}}{s_{134}^2}
+  R(y_{14},y_{34}) \frac{s_{14}}{s_{134}^2}\;,
\\
\Poles\left(\tilde{E}_3^1(1_q,3_{q'},4_{\bar q'})\right) &=& 
2 \left( {\bf I}^{(1)}_{q\bar q} (\e,s_{34})  
\right) E_3^0(1,3,4) \;,\\
\Finite\left(\tilde{E}_3^1(1_q,3_{q'},4_{\bar q'})\right) 
&=& -4 E_3^0(1,3,4) \;,  \\ 
\Poles\left(\hat{E}_3^1(1_q,3_{q'},4_{\bar q'})\right) &=& 
- 4  {\bf I}^{(1)}_{q g,\Flavour} (\e,s_{134})  E_3^0(1,3,4)\;,\\
\Finite\left(\hat{E}_3^1(1_q,3_{q'},4_{\bar q'})\right) &=& 
\left( -\frac{10}{9} + \frac{2}{3} \log y_{34}
\right) E_3^0(1,3,4)   \;. 
\end{eqnarray}

Integration of these antenna functions yields
\begin{eqnarray}
{\cal E}_3^1(s_{134}) &= &
 (s_{134})^{-2\e} \Bigg[ 
 \frac{11}{18\e^2}
+ \frac{1}{\e} \left(\frac{74}{27} - \frac{\pi^2}{9}
 \right) \nonumber \\ && \hspace{1cm} +\left(\frac{3023}{216}
          - \frac{181\pi^2}{216}
          - \frac{130}{9} \zeta_3 
   \right) + {\cal O}(\e) \Bigg]\;,\\
\tilde{{\cal E}}_3^1(s_{134}) &= &
 (s_{134})^{-2\e} \Bigg[ 
 \frac{1}{6\e^3} + \frac{35}{36\e^2}
+\frac{1}{\e} \left( \frac{509}{108}
          - \frac{\pi^2}{4} \right) \nonumber \\ && \hspace{1cm}
+ \left(\frac{1670}{81}
          - \frac{35\pi^2}{24} - \frac{31}{9}\zeta_3 \right) 
+ {\cal O}(\e) \Bigg]\;,\\
\hat{{\cal E}}_3^1(s_{134}) &= &
 (s_{134})^{-2\e} \Bigg[ 
 \frac{1}{3\e} 
+\left(\frac{172}{81}
          - \frac{11\pi^2}{108} \right) + {\cal O}(\e) \Bigg]\;,
\end{eqnarray}
with
\begin{eqnarray}
 \Poles\left({\cal E}_3^1(s_{134})\right)  &=& 
- {\cal D}_2^1(s_{134}) \left( 
 4 {\bf I}^{(1)}_{qg,\Flavour} (\e,s_{134}) + 
{\cal E}_3^0(s_{134}) \right) \nonumber \\ 
&&
+ \frac{2b_0}{\e}\, (s_{134})^{-\e}\,
  2\,{\bf I}^{(1)}_{qg,\Flavour} (\e,s_{134}) 
 - {\bf H}^{(2)}_{V,E}(\e,s_{134})\;,
\\
 \Finite\left({\cal E}_3^1(s_{134})\right)  &=&
          \frac{9071}{216}
          - \frac{143\pi^2}{54}
          - \frac{26}{3}\zeta_3\;,
\\
 \Poles\left(\tilde{{\cal E}}_3^1(s_{134})\right)  &=& 
- {\bf H}^{(2)}_{V,\tilde{E}}(\e,s_{134})\;,
\\
 \Finite\left(\tilde{{\cal E}}_3^1(s_{134})\right)  &=&
            \frac{1670}{81}
          - \frac{595\pi^2}{432}
          - \frac{61}{18}\, \zeta_3\;, \\
 \Poles\left(\hat{{\cal E}}_3^1(s_{134})\right)  &=& 
- \hat{{\cal D}}_2^1(s_{134}) \left( 
 4 {\bf I}^{(1)}_{qg,\Flavour} (\e,s_{134}) + 
{\cal E}_3^0(s_{134}) \right) \nonumber \\ 
&& 
+ \frac{2b_{0,\Flavour}}{\e}\, (s_{134})^{-\e}\,
 \left[ 2\,{\bf I}^{(1)}_{qg,\Flavour} (\e,s_{134})\right] 
+ 2 \left[2\, {\bf I}^{(1)}_{qg,\Flavour}(\e,s_{134}) \right]^2
\nonumber \\ 
&& - 2 e^{-\e \gamma} \frac{\Gamma (1-2\e)}{\Gamma (1-\e)}
\left(\frac{b_{0,\Flavour}}{\e} + k_{0,\Flavour} \right) 
\left[2\, {\bf I}^{(1)}_{qg,\Flavour}(2\e,s_{134}) \right]\nonumber \\ 
&& 
- 2\, {\bf H}^{(2)}_{qg,N_F^2}(\e,s_{134})\;,
\\
 \Finite\left(\hat{{\cal E}}_3^1(s_{134})\right)  &=&
            \frac{91}{81}
          + \frac{\pi^2}{72}\;.
\end{eqnarray}
The last contribution corresponds to the quark loop correction 
to the quark-quark-antiquark antenna, and enters into physical 
cross sections 
multiplied with a factor $N_F^2$. In the case of quark-gluon antenna
functions, there is no contribution with this factor coming from 
four-parton tree-level antennae. Therefore, $\Poles(\hat{{\cal E}}_3^1)$
contains the full infrared singularity structure of the two-loop two-parton 
quark-gluon antenna at $N_F^2$. Note that the full hard radiation factor 
${\bf H}^{(2)}_{qg,N_F^2}$ (and not only the virtual hard radiation) 
is present here. 

\subsection{Four-parton tree-level antenna functions}

The tree-level four-parton quark-gluon antenna contains two final 
states: 
quark-gluon-gluon-gluon, $D_4^0$ 
and quark-quark-antiquark-gluon at leading and subleading colour,
$E_4^0$ and $\tilde{E}_4^0$. 
The 
antenna for the $qggg$ final state is:
\begin{eqnarray}
D_4^0(1_q,3_g,4_g,5_g) &=& d_4^0(1,3,4,5) + d_4^0(1,5,4,3)\;, 
\end{eqnarray}
with 
\begin{eqnarray}
d_4^0(1,3,4,5) &=& \frac{1}{s_{1345}^2} \Bigg\{
         \frac{\Sad }{2 \Sac  \Sae \Scd \Sde } \left[
            3 \Sad \Sce^2
          + 3 \Sad^2 \Sce
          + 2 \Sad^3
          + \Sce^3
         \right]\nonumber \\ &&
       +  \frac{\Sad }{\Sac \Sae \Scd}  \left[
            6 \Sad \Sce
          + 3 \Sad \Sde
          + 4 \Sad^2
          + 3 \Sce \Sde
          + 3 \Sce^2
          + \Sde^2
         \right]\nonumber \\ &&
       +  \frac{\Sad }{\Sac \Sae}  \left[
            3 \Sad
          + 3 \Sce
          + 3 \Sde
         \right]
       + \frac{\Sad  }{\Sac \Scd \Sace \Scde}  \Big[
            3 \Sad \Sde^2
          + 3 \Sad^2 \Sde \nonumber \\ &&
          + 2 \Sad^3
          + \Sde^3
         \Big]
       + \frac{\Sad }{\Sac  \Sace \Scde}  \left[
            3 \Sad \Sde
          + 3 \Sad^2
          - 2 \Sce \Sde
          - \Sce^2
          + \Sde^2
         \right]\nonumber \\ &&
       + \frac{\Sad^3 }{\Sac  \Scd \Sace} 
       + \frac{\Sce^3 \Sde }{2\Sac \Sae  \Sacd \Sade}
       - \frac{\Sce^3 }{2\Sac \Sae  \Sacd}\nonumber \\ &&
       + \frac{\Sae  }{\Sac \Sde \Sacd \Sade}  \left[
            3 \Sae \Sce^2
          + 3 \Sae^2 \Sce
          + \Sae^3
          + \Sce^3
         \right]\nonumber \\ &&
       + \frac{1 }{\Sac \Scd \Sde}  \Big[
            9 \Sad \Sae \Sce
          + 4 \Sad \Sae^2
          + 6 \Sad \Sce^2
          + 6 \Sad^2 \Sae
          + 9 \Sad^2 \Sce
          + 4 \Sad^3\nonumber \\ &&
          + 3 \Sae \Sce^2
          + 4 \Sae^2 \Sce
          + 2 \Sae^3
          + \Sce^3
         \Big]
       + \frac{ 1}{\Sac \Scd \Scde}  \Big[
            \Sad \Sae \Sde
          + 4 \Sad \Sae^2
          + \Sad \Sde^2\nonumber \\ &&
          + 6 \Sad^2 \Sae
          + 3 \Sad^2 \Sde
          + 4 \Sad^3
          + \Sae \Sde^2
          - 2 \Sae^2 \Sde
          + 2 \Sae^3
         \Big]\nonumber \\ &&
       + \frac{1}{ \Sac \Scd } \Big[
            17 \Sad \Sae
          + 16 \Sad \Sce
          + 11 \Sad \Sde
          + 15 \Sad^2
          + 12 \Sae \Sce
          + 5 \Sae \Sde\nonumber \\ &&
          + 10 \Sae^2
          + 5 \Sce \Sde
          + 5 \Sce^2
          + 2 \Sde^2
         \Big]
       + \frac{\Scd}{ \Sac  \Sacd^2}  \Big[
            2 \Sae \Sce
          + 2 \Sae \Sde
          + \Sae^2\nonumber \\ &&
          + 2 \Sce \Sde
          + \Sce^2
          + \Sde^2
         \Big] 
       + \frac{\Sce }{\Sac  \Sde \Sade \Sace}  \left[
            3 \Scd \Sce^2
          + 3 \Scd^2 \Sce
          + \Scd^3
          + \Sce^3
         \right]\nonumber \\ &&
       + \frac{\Sce  }{\Sac  \Sade \Sace}  \left[
          - 6 \Scd \Sce
          + 3 \Scd \Sde
          - 3 \Scd^2
          + 3 \Sce \Sde
          - 3 \Sce^2
          - \Sde^2
         \right]\nonumber \\ &&
       + \frac{\Sce  }{\Sac \Sace^2 } \left[
            2 \Sad \Scd
          + 2 \Sad \Sde
          + \Sad^2
          + 2 \Scd \Sde
          + \Scd^2
          + \Sde^2
         \right]\nonumber \\ &&
       +  \frac{1}{\Sac \Sde \Sacd \Scde}  \left[
            5 \Sae \Sce^3
          + 9 \Sae^2 \Sce^2
          + 7 \Sae^3 \Sce
          + 2 \Sae^4
          + \Sce^4
         \right]\nonumber \\ &&
       + \frac{1}{ \Sac \Sde \Sacd}  \left[
            2 \Sae \Scd \Sce
          - 4 \Sae \Sce^2
          + \Sae^2 \Scd
          - 5 \Sae^2 \Sce
          - 2 \Sae^3
          + \Scd \Sce^2
          - \Sce^3
         \right]\nonumber \\ &&
       + \frac{ 1}{\Sac \Sde \Sade}  \Big[
            3 \Sae \Scd \Sce
          + 2 \Sae \Scd^2
          + 2 \Sae \Sce^2
          - \Sae^2 \Scd
          + \Sae^2 \Sce
          + \Sae^3\nonumber \\ &&
          - 3 \Scd \Sce^2
          - 3 \Scd^2 \Sce
          - \Scd^3
          - \Sce^3
         \Big]
       + \frac{1}{ \Sac \Sde \Sace \Scde } \Big[
          - 5 \Sad \Sce^3
          + 9 \Sad^2 \Sce^2\nonumber \\ && 
          - 7 \Sad^3 \Sce
          + 2 \Sad^4
          + \Sce^4
         \Big]
       + \frac{1}{ \Sac \Sde \Sace}  \Big[
            4 \Sad \Scd^2
          + 6 \Sad \Sce^2
          + 7 \Sad^2 \Scd
          - 7 \Sad^2 \Sce\nonumber \\ &&
          + 6 \Sad^3
          + 4 \Scd \Sce^2
          + 3 \Scd^2 \Sce
          + \Scd^3
         \Big]
       + \frac{ 1}{\Sac \Sde \Scde}  \Big[
            10 \Sad \Sae^2
          + 6 \Sad \Sce^2\nonumber \\ &&
          + 10 \Sad^2 \Sae
          - 7 \Sad^2 \Sce
          + 6 \Sad^3
          + 6 \Sae \Sce^2
          + 7 \Sae^2 \Sce
          + 6 \Sae^3
         \Big]\nonumber \\ &&
       +  \frac{1}{\Sac \Sde}  \Big[
            17 \Sad \Sae
          + 8 \Sad \Scd
          + 6 \Sad \Sce
          + 18 \Sad^2
          + 5 \Sae \Scd
          + 4 \Sae \Sce
          + 5 \Sae^2\nonumber \\ &&
          - \Scd \Sce
          - \Scd^2
          + 2 \Sce^2
         \Big]
       + \frac{ 1}{\Sac \Sacd \Sade}  \bigg[
            3 \Sae \Sce \Sde
          + 6 \Sae \Sce^2
          + 6 \Sae^2 \Sce\nonumber \\ &&
          + \Sae^2 \Sde
          + 2 \Sae^3
          + 3 \Sce^2 \Sde
          + \frac{3}{2} \Sce^3
         \bigg]
       +  \frac{1}{\Sac \Sacd \Scde}  \Big[
            6 \Sae \Sce \Sde
          + 9 \Sae \Sce^2\nonumber \\ &&
          + \Sae \Sde^2
          + 9 \Sae^2 \Sce
          + 3 \Sae^2 \Sde
          + 3 \Sae^3
          + \Sce \Sde^2
          + 3 \Sce^2 \Sde
          + 3 \Sce^3
         \Big]\nonumber \\ &&
       + \frac{1}{ \Sac \Sacd}  \Big[
            4 \Sae \Scd
          - 13 \Sae \Sce
          - 6 \Sae \Sde
          - 7 \Sae^2
          + 4 \Scd \Sce
          + 3 \Scd \Sde\nonumber \\ &&
          - 5 \Sce \Sde
          - 7 \Sce^2
          - 2 \Sde^2
         \Big]
       +  \frac{1}{\Sac \Sade}  \Big[
          - 2 \Sae \Scd
          - \Sae \Sce\nonumber \\ &&
          + \Sae^2
          + 6 \Scd \Sce
          - \Scd \Sde
          + 2 \Scd^2
          - 3 \Sce \Sde
          + 4 \Sce^2
         \Big]\nonumber \\ &&
       + \frac{1}{ \Sac \Sace}  \left[
            2 \Sad \Scd
          - \Sad \Sce
          + 6 \Sad^2
          - \Scd \Sce
          - 2 \Scd \Sde
          + 3 \Sce \Sde
          - \Sce^2
          - \Sde^2
         \right]\nonumber \\ &&
       + \frac{1}{ \Sac \Scde } \Big[
            2 \Sad \Sae
          + 2 \Sad \Sce
          + 2 \Sad \Sde
          + \Sad^2
          + 4 \Sae \Sce
          + 2 \Sae \Sde
          + \Sae^2\nonumber \\ &&
          + \Sce \Sde
          + 2 \Sce^2
         \Big]
       + \frac{1}{ \Sac}  \left[
            14 \Sad
          + 2 \Sae
          + 2 \Scd
          + 3 \Sce
          - 2 \Sde
         \right]\nonumber \\ &&
       - \frac{4\Sac \Sae^2 }{ \Sde \Sade \Scde} 
       + \frac{ \Sad \Sae}{ \Scd \Sacd^2}  \left[
          - 4 \Sce
          - 4 \Sde
         \right]
       - \frac{ 4\Sad \Sae^2 \Sde }{\Scd^2 \Sacd \Scde } \nonumber \\ &&
       + \frac{ \Sad }{\Scd^2 \Sacd}  \left[
            2 \Sad \Sae
          + 2 \Sad \Sce
          + 2 \Sad \Sde
          + 4 \Sae \Sce
          - 4 \Sae \Sde
          - 2 \Sce \Sde
          - 2 \Sde^2
         \right]\nonumber \\ &&
       + \frac{ \Sad^2}{ \Scd^2 \Sacd^2 } \left[
            2 \Sae \Sce
          + 2 \Sae \Sde
          + 2 \Sae^2
          + 2 \Sce \Sde
          + \Sce^2
          + \Sde^2
         \right]\nonumber \\ &&
       +  \frac{\Sae}{2 \Scd \Sde \Sacd \Sade } \left[
            3 \Sae \Sce^2
          + 3 \Sae^2 \Sce
          + 2\Sae^3
          + \Sce^3
         \right]
       +  \frac{\Sae}{ \Scd \Sde \Sade}  \bigg[
            2 \Sac \Sae\nonumber \\ &&
          + 2 \Sae \Sce
          + \frac{3}{2} \Sce^2
         \bigg]
       + \frac{ \Sae }{\Scd \Sacd \Scde } \left[
          - 4 \Sad \Sae
          - 2 \Sad \Sde
          - 4 \Sad^2
          - 8 \Sae^2
          - 2 \Sde^2
         \right]\nonumber \\ &&
       + \frac{ \Sae \Sce }{\Sde^2 \Scde}  \left[
          - 4 \Sad
          - 4 \Sae
          + 4 \Sce
         \right]
       + \frac{ \Sae \Sce}{ \Sde \Sace \Scde}  \left[
            2 \Sad
          + \Sae
          + \frac{1}{2} \Sce
         \right]\nonumber \\ &&
       + \frac{ \Sae \Sce^2 }{\Sde^2 \Scde^2 } \left[
            4 \Sad
          + 2 \Sae
         \right]
       + \frac{ \Sae}{ \Sde^2 \Sade}  \Big[
            2 \Sac \Sae
          + 2 \Sae \Scd
          + 2 \Sae \Sce\nonumber \\ &&
          - 2 \Scd \Sce
          - 2 \Sce^2
         \Big]
       +  \frac{\Sae}{ \Sde^2}  \left[
            2 \Sae
          + 2 \Scd
          - 6 \Sce
         \right]\nonumber \\ &&
       + \frac{ \Sae}{ \Sde \Sacd \Scde}  \left[
          - 6 \Sad \Sae
          - 2 \Sad \Sce
          - 2 \Sad^2
          - 6 \Sae \Sce
          - 4 \Sae^2
          - 8 \Sce^2
         \right]\nonumber \\ &&
       + \frac{ \Sae }{\Sde \Sacd } \left[
          - \Sad
          + \frac{5}{2} \Sae
          + \frac{1}{2} \Scd
          + \frac{15}{2} \Sce
         \right]
       + \frac{ \Sae}{ \Sde \Sade^2 } \left[
            4 \Scd \Sce
          + 2 \Scd^2
          + 2 \Sce^2
         \right]\nonumber \\ &&
       + \frac{ \Sae}{ \Sde \Sade \Sace}  \left[
            \frac{3}{2} \Sae \Scd
          - \frac{1}{2} \Sae \Sce
          - \Sae^2
          - \frac{3}{2} \Scd \Sce
          - \frac{3}{2} \Scd^2
          - \Sce^2
         \right]\nonumber \\ &&
       + \frac{ \Sae}{ \Sde \Sade}  \left[
          - 2 \Sae
          + \frac{11}{2} \Scd
          + 4 \Sce
         \right]
       + \frac{ \Sae }{\Sde \Sace } \left[
          - 2 \Sad
          - 2 \Scd
         \right]\nonumber \\ &&
       + \frac{ \Sae}{ \Sde \Scde}  \left[
          - 6 \Sad
          - 6 \Sae
          - 6 \Sce
         \right]
       + \frac{ 15\Sae }{2\Sde }
       +  \frac{\Sae^2}{ \Sde^2 \Sade^2}  \Big[
            2 \Sac \Scd
          + 2 \Sac \Sce\nonumber \\ &&
          + 2 \Scd \Sce
          + \Scd^2
          + \Sce^2
         \Big]
       + \frac{\Sde }{\Scd^2  \Scde}  \left[
          - 8 \Sad \Sae
          + 4 \Sad \Sde
          - 4 \Sad^2
          + 4 \Sae \Sde
         \right]\nonumber \\ &&
       + \frac{\Sde^2 }{\Scd^2  \Scde^2}  \left[
            4 \Sac \Sae
          + 4 \Sad \Sae
          + 2 \Sad^2
          + 2 \Sae^2
         \right]
       + \frac{ 1}{\Scd^2}  \Big[
          - 2 \Sac \Sae
          + 2 \Sad \Sae\nonumber \\ &&
          + 2 \Sad \Sce
          - 6 \Sad \Sde
          + 2 \Sad^2
          - 2 \Sae \Sce
          - 2 \Sae \Sde
          + \Sce^2
          + \Sde^2
         \Big]\nonumber \\ &&
       + \frac{ 1}{\Scd \Sde \Sacd } \bigg[
            \frac{1}{2} \Sad \Sae \Sce
          + 3 \Sad \Sae^2
          + \Sad^2 \Sae
          + \frac{1}{2} \Sad^2 \Sce
          + \Sad^3
          + \frac{3}{2} \Sae \Sce^2\nonumber \\ &&
          + \frac{5}{2} \Sae^2 \Sce
          + \Sae^3
          + \frac{1}{2} \Sce^3
         \bigg]
       + \frac{ 1}{\Scd \Sde}  \Big[
            4 \Sac \Sae
          + 16 \Sad \Sae
          + 11 \Sad \Sce
          + 7 \Sad^2\nonumber \\ &&
          + 16 \Sae \Sce
          + 8 \Sae^2
          + \frac{9}{2} \Sce^2
         \Big]
       + \frac{ \Sde}{\Scd  \Sace \Scde}  \bigg[
            2 \Sad \Sae
          - 6 \Sad \Sde
          - 6 \Sad^2\nonumber \\ &&
          + \frac{3}{2} \Sae \Sde
          - \Sae^2
          - \frac{5}{2} \Sde^2
         \bigg]
       + \frac{\Sde }{\Scd  \Scde^2}  \left[
            8 \Sac \Sae
          + 8 \Sad \Sae
          + 4 \Sad^2
          + 4 \Sae^2
         \right]\nonumber \\ &&
       + \frac{ 1}{\Scd \Sacd \Sade}  \bigg[
            9 \Sae \Sce \Sde
          + \frac{9}{2} \Sae \Sce^2
          + 5 \Sae \Sde^2
          + 6 \Sae^2 \Sce
          + \frac{11}{2} \Sae^2 \Sde
          + \frac{7}{2} \Sae^3\nonumber \\ &&
          + 6 \Sce \Sde^2
          + \frac{9}{2} \Sce^2 \Sde
          + \Sce^3
          + \frac{5}{2} \Sde^3
         \bigg]
       + \frac{ 1}{\Scd \Sacd \Sace}  \bigg[
            \frac{9}{2} \Sae \Sce \Sde
          + 3 \Sae \Sce^2\nonumber \\ &&
          + \frac{3}{2} \Sae \Sde^2
          + \frac{5}{2} \Sae^2 \Sce
          + \frac{3}{2} \Sae^2 \Sde
          + \Sae^3
          + \frac{9}{2} \Sce \Sde^2
          + 6 \Sce^2 \Sde
          + \frac{5}{2} \Sce^3
          + \Sde^3
         \bigg]\nonumber \\ &&
       + \frac{ 1}{2 \Scd \Sacd } \Big[
          - 5 \Sad \Sae
          + 5 \Sad \Sce
          - 3 \Sad \Sde
          - 7 \Sad^2
          - 22 \Sae \Sce
          - 19 \Sae \Sde\nonumber \\ &&
          - 35 \Sae^2
          - 27 \Sce \Sde
          - 17 \Sce^2
          - 13 \Sde^2
         \Big]\nonumber \\ &&
       + \frac{ 1}{\Scd \Sade \Scde}  \left[
          - 6 \Sac \Sae \Sde
          - 2 \Sac \Sae^2
          + 6 \Sae \Sde^2
          + 5 \Sae^2 \Sde
          + 2 \Sae^3
          + 4 \Sde^3
         \right]\nonumber \\ &&
       + \frac{ 1}{\Scd \Sade}  \left[
            \frac{13}{2} \Sac \Sae
          + \frac{13}{2} \Sae \Sce
          + \frac{1}{2} \Sae^2
          + 7 \Sce \Sde
          + 3 \Sce^2
          - \Sde^2
         \right]\nonumber \\ &&
       + \frac{ 1}{\Scd \Sace}  \bigg[
          + \Sad \Sce
          + 7 \Sad \Sde
          + 4 \Sad^2
          + \frac{3}{2} \Sae \Sce
          + \frac{1}{2} \Sae \Sde
          + \Sae^2
          + \frac{7}{2} \Sce \Sde\nonumber \\ &&
          + \frac{5}{2} \Sce^2
          + \frac{11}{2} \Sde^2
         \bigg]
       + \frac{ 1}{\Scd \Scde } \Big[
          - 8 \Sac \Sae
          - 14 \Sad \Sae
          + 7 \Sad \Sde
          - 8 \Sad^2\nonumber \\ &&
          + 7 \Sae \Sde
          - 16 \Sae^2
          - \frac{9}{2} \Sde^2
         \Big]
       + \frac{ 1}{2\Scd}  \left[
            23 \Sad
          - 13 \Sae
          + 5 \Sce
          + 4 \Sde
         \right]\nonumber \\ &&
       + \frac{ 1}{2\Sde \Sacd \Sade } \Big[
          - 3 \Sae \Scd \Sce
          -  \Sae \Scd^2
          - 3 \Sae \Sce^2
          - 3 \Sae^2 \Sce
          - 3 \Sae^3
          - 3 \Scd \Sce^2\nonumber \\ &&
          - 3 \Scd^2 \Sce
          -  \Scd^3
          - \Sce^3
         \Big]
       +  \frac{1}{\Sacd^2}  \left[
            \Sae^2
          + 4 \Sce \Sde
          + 2 \Sce^2
          + 2 \Sde^2
         \right]\nonumber \\ &&
       + \frac{1}{ \Sacd \Sade}  \Big[
            \frac{3}{2} \Sae \Scd
          - 3 \Sae \Sce
          - 6 \Sae \Sde
          - \frac{9}{2} \Sae^2
          + \frac{3}{2} \Scd \Sde
          + \frac{1}{2} \Scd^2
          - 3 \Sce \Sde\nonumber \\ &&
          - \frac{3}{2} \Sce^2
          - 4 \Sde^2
         \Big]
       + \frac{1}{ \Sacd \Sace}  \left[
          + \frac{3}{2} \Sae \Scd
          - \frac{9}{2} \Sae \Sce
          - \frac{3}{2} \Sae^2
          - 9 \Sce \Sde
          - 6 \Sce^2
         \right]\nonumber \\ &&
       + \frac{ 1}{2\Sacd \Scde}  \Big[
          - 5 \Sad \Sce
          - 3 \Sad \Sde
          + 3 \Sad^2
          - 22 \Sae \Sce
          - 12 \Sae \Sde
          + 12 \Sae^2\nonumber \\ &&
          + \Sce \Sde
          - \Sce^2
          + \Sde^2
         \Big]
       + \frac{1}{2 \Sacd } \left[
          - 3 \Sad
          + 29 \Sae
          + 18 \Sce
          + 23 \Sde
         \right]\nonumber \\ &&
       + \frac{1}{ 2\Sade \Sace } \left[
            3 \Sae \Sce
          - 3 \Sae \Sde
          - 3 \Sae^2
          + 9 \Sce \Sde
          - 2\Sde^2
         \right]\nonumber \\ &&
       + \frac{ 1}{2\Sade \Scde } \left[
          - 2 \Sae \Sce
          + 3 \Sae \Sde
          +   \Sae^2
          - 3 \Sce \Sde
          +   \Sce^2
          + 5 \Sde^2
         \right]\nonumber \\ &&
       + \frac{ 1}{\Sade } \left[
          - 4 \Sae
          - 4 \Sde
         \right]
       + \frac{1}{ \Sace^2}  \left[
            4 \Sad \Sde
          + \Sad^2
          + 2 \Scd \Sde
          + 2 \Sde^2
         \right]\nonumber \\ &&
       +  \frac{1}{\Sace \Scde}  \Big[
            2 \Sad \Sae
          + 7 \Sad \Sce
          - 6 \Sad \Sde
          - 4 \Sad^2
          - \frac{3}{2} \Sae \Sce
          + \frac{3}{2} \Sae \Sde
          - \Sae^2\nonumber \\ &&
          + \frac{5}{2} \Sce \Sde
          - \frac{5}{2} \Sce^2
          - \frac{5}{2} \Sde^2
         \Big]
       + \frac{ 1}{\Sace}  \left[
            2 \Sad
          - 2 \Sce
          + 7 \Sde
         \right]\nonumber \\ &&
       + \frac{1}{ \Scde^2 } \left[
            6 \Sac \Sae
          + 8 \Sad \Sae
          + 3 \Sad^2
          + 4 \Sae^2
         \right]\nonumber \\ &&
       +  \frac{1}{\Scde}  \left[
            \frac{7}{2} \Sad
          + \frac{15}{2} \Sae
          + 3 \Sce
          - \frac{5}{2}\Sde
         \right]       + 8 + {\cal O}(\e)\Bigg\}\;.
\end{eqnarray}

In $D_4^0(1,3,4,5)$, the gluonic emissions are colour-ordered. However,
as explained in~\cite{chi}, both possible colour-orderings (normal and 
reverse) are contained in the same colour-ordered squared matrix element. 
Moreover, since this squared matrix element originates  from an amplitude 
which is cyclic in the partonic colour indices, it also contains singular 
limits in the momentum configuration $(5_g,1_q,3_g)$,
as outlined in detail in Section~\ref{sec:D40limit} below. Using this antenna 
function as NNLO subtraction term, this limit has to be accounted for 
properly. 

For the above reasons, it is not possible to use a single ordered 
phase space parametrisation in the 
 numerical implementation of  $D_4^0(1,3,4,5)$. Instead, 
$D_4^0(1,3,4,5)$ has to be split into different ordered pieces by repeated 
partial fractioning on pairs of invariants. Since this decomposition is 
not symmetric, the individual pieces can no longer be integrated analytically. 
Therefore the full $D_4^0(1,3,4,5)$ and not the sub-antennae 
must be used as a subtraction term.

The integral of this antenna function is 
\begin{eqnarray}
{\cal D}_4^0(s_{1345}) &=& 
\left(s_{1345}\right)^{-2\e} 
\Bigg[ \frac{5}{2\e^4} + \frac{37}{4\e^3}
+ \frac{1}{\e^2} \left( \frac{398}{9} -\frac{11\pi^2}{3} \right)
\nonumber \\&&
+ \frac{1}{\e} \left( \frac{28319}{144} -\frac{55\pi^2}{4} 
-\frac{188}{3}\zeta_3 \right)
\nonumber \\&&
+\left(\frac{2201527}{2592}
          - \frac{529\pi^2}{8}
          + \frac{511\pi^4}{720}
          - \frac{722}{3}\zeta_3
   \right)+ {\cal O}(\e)  \Bigg] \;,
\end{eqnarray}
with
\begin{eqnarray}
\Poles\left( {\cal D}_4^0(s_{1345})\right) &=& 
2\, \left[2 {\bf I}^{(1)}_{qg} \left(\e,s_{1345} \right) 
\right]^2  - 2
e^{-\e \gamma} \frac{\Gamma (1-2\e)}{\Gamma (1-\e)}
\left(\frac{b_{0}}{\e} + k_{0} \right) \,
 \left[2 {\bf I}^{(1)}_{qg}(2\e,s_{1345})\right] \nonumber \\ &&
-   {\bf H}^{(2)}_{R,D}(\e,s_{1345})
- 2 \,{\bf S}^{(2)}_{V}(\e,s_{1345})\;, \\
\Finite\left( {\cal D}_4^0(s_{1345})\right) &=&
            \frac{1935847}{2592}
          - \frac{4271\pi^2}{108}
          - \frac{73\pi^4}{144}
          - \frac{7483}{36}\zeta_3 \;.
\end{eqnarray}

The leading colour and subleading colour
 quark-quark-antiquark-gluon antennae are:
\begin{eqnarray}
E_4^0(1_q,3_{q'},4_{\bar q'},5_g) &=& \frac{1}{s_{1345}^2} \Bigg\{
 - \frac{\Sac \Sad} {\Sae \Scde}   
      + \frac{\Sad}{ \Sae \Scd \Sde}  \left[
            \Sac \Sce
          + \Sac^2
          + \Sad^2
         \right]\nonumber \\ &&
      + \frac{\Sad}{ \Sae \Scd \Sacd}  \left[
            2 \Sad \Sce
          + 2 \Sad \Sde
          - \Sce^2
          + \Sde^2
         \right]\nonumber \\ &&
      + \frac{\Sad}{ \Sae \Sde \Scde}  \left[
            \Sac \Sce
          + \Sac^2
          + \Sad^2
         \right]
      + \frac{\Sad}{ \Sae \Sde}  \left[
            2 \Sac
          + 2 \Sad
          + \Sce
         \right]\nonumber \\ &&
      - \frac{4 \Sad \Sae^2\Sce}{ \Scd^2  \Sacd \Scde } 
      + \frac{\Sad}{ \Scd^2 \Sacd}  \Big[
          - 4 \Sad \Sae
          - 4 \Sad \Sce
          - 4 \Sad \Sde\nonumber \\ &&
          + 8 \Sae \Sde
          + 4 \Sae^2
          + 4 \Sce \Sde
          + 4 \Sde^2
         \Big]\nonumber \\ &&
      + \frac{\Sad}{ \Scd \Sde \Sacd}  \left[
          - 2 \Sad \Sae
          - \Sad \Sce
          - 2 \Sad^2
          - \Sae \Sce
          - \Sae^2
         \right]\nonumber \\ &&
      + \frac{\Sad}{ \Scd \Sacd^2 } \left[
          - 4 \Sae \Sce
          - 4 \Sae \Sde
          - 2 \Sae^2
          - 4 \Sce \Sde
          - 2 \Sce^2
          - 2 \Sde^2
         \right]\nonumber \\ &&
      - \frac{\Sad}{ \Scde} 
      + \frac{\Sad^2}{ \Scd^2 \Sacd^2}  \left[
          - 4 \Sae \Sce
          - 4 \Sae \Sde
          - 2 \Sae^2
          - 4 \Sce \Sde
          - 2 \Sce^2
          - 2 \Sde^2
         \right]\nonumber \\ &&
      + \frac{1}{\Sae \Scd \Scde}  \left[
            \Sac \Sad^2
          + \Sac^2 \Sad
          + \Sac^2 \Sce
          + \Sac^3
          - \Sad^2 \Sce
          + \Sad^3
         \right]\nonumber \\ &&
      + \frac{1}{\Sae \Scd}  \Big[
            \Sac \Sad
          + 3 \Sac \Sce
          + \Sac \Sde
          + 2 \Sac^2
          - \Sad \Sce
          + \Sad \Sde\nonumber \\ &&
          + 4 \Sad^2
          + \Sce \Sde
          + \Sce^2
         \Big]
      + \frac{1}{\Sae \Sacd}  \Big[
          - \Sac \Sde
          + \Sad \Sce\nonumber \\ &&
          + 2 \Sad \Sde
          - \Sce \Sde
          + \Sde^2
         \Big]
      + \frac{1}{\Sae}  \left[
            \Sac
          + 2 \Sad
          + \Sce
          + \Sde
         \right]\nonumber \\ &&
      + \frac{\Sae}{ \Scd \Sacd \Scde}  \left[
          - 2 \Sad \Sce
          + 8 \Sad^2
          - 2 \Sae \Sce
          + 2 \Sae^2
          + 2 \Sce^2
         \right]\nonumber \\ &&
      + \frac{\Sce}{\Scd^2  \Scde } \Big[
            4 \Sac \Sad
          + 8 \Sac \Sae
          - 4 \Sac \Sce
          + 4 \Sac^2
          - 4 \Sad \Sce
          - 4 \Sae \Sce\nonumber \\ &&
          + 4 \Sae^2
         \Big]
      + \frac{\Sce^2 }{\Scd^2 \Scde^2}  \left[
          - 4 \Sac \Sad
          - 4 \Sac \Sae
          - 2 \Sac^2
          - 4 \Sad \Sae
          - 2 \Sad^2
          - 2 \Sae^2
         \right]\nonumber \\ &&
      + \frac{1}{\Scd^2}  \Big[
          - 4 \Sac \Sae
          + 4 \Sac \Sce
          - 4 \Sac \Sde
          - 2 \Sac^2
          + 4 \Sad \Sae
          + 4 \Sad \Sce\nonumber \\ &&
          + 4 \Sad \Sde
          + 4 \Sae \Sce
          - 4 \Sae \Sde
          - 2 \Sae^2
          - 2 \Sde^2
         \Big]\nonumber \\ &&
      + \frac{1}{\Scd \Sacd}  \Big[
          - \Sad \Sae
          - 7 \Sad \Sce
          + \Sad \Sde
          + 7 \Sad^2
          + \Sae \Sce
          + 7 \Sae \Sde
\nonumber \\ &&          + 6 \Sae^2
          + 2 \Sce \Sde
          + \Sce^2
          + 3 \Sde^2
         \Big]
      + \frac{1}{\Scd \Scde}  \Big[
            6 \Sac \Sae
          + 2 \Sac \Sce
          + 4 \Sac^2\nonumber \\ &&  
          - 4 \Sad \Sce
          + 6 \Sad^2
          + 5 \Sae^2
         \Big]
      + \frac{1}{\Scd}  \left[
            3 \Sac
          + 2 \Sad
          + 6 \Sae
          + 6 \Sce
          + \Sde
         \right]\nonumber \\ &&
      + \frac{\Sce}{ \Sde \Scde^2 } \left[
            2 \Sac \Sad
          + 2 \Sac \Sae
          + \Sac^2
          + 2 \Sad \Sae
          + \Sad^2
          + \Sae^2
         \right]\nonumber \\ &&
      + \frac{1}{\Sde \Sacd \Scde}  \Big[
          - \Sac \Sad \Sae
          + \Sac \Sad \Sce
          - \Sac \Sad^2
          + \Sac \Sae \Sce
          + 3 \Sad \Sae^2\nonumber \\ &&
          + 3 \Sad^2 \Sae
          + \Sad^3
          + \Sae^3
         \Big]
      + \frac{1}{\Sde \Sacd}  \left[
          - \Sac \Sae
          + \Sad \Sae
          - \Sad^2
          + \Sae^2
         \right]\nonumber \\ &&
      + \frac{1}{\Sde \Scde } \Big[
          - 2 \Sac \Sae
          + 2 \Sac \Sce
          - \Sac^2
          + 3 \Sad \Sae
          + \Sad \Sce
          + 4 \Sad^2\nonumber \\ &&
          + \Sae \Sce
         \Big]
      + \frac{1}{\Sde } \left[
          - 2 \Sac
          + 2 \Sad
          - \Scd 
         \right]\nonumber \\ &&
      + \frac{1}{\Sacd^2  }\left[
          - 2 \Sae \Sce
          - 2 \Sae \Sde
          - \Sae^2
          - 2 \Sce \Sde
          - \Sce^2
          - \Sde^2
         \right]\nonumber \\ &&
      + \frac{1}{\Sacd \Scde}  \left[
            \Sac \Sad
          - 3 \Sac \Sae
          - \Sac^2
          + 4 \Sad \Sae
          - \Sad^2
          - 3 \Sae^2
         \right]\nonumber \\ &&
      + \frac{1}{\Sacd } \left[
          - 3 \Sac
          + 6 \Sad 
          - 3 \Sce
          + 2 \Sde
         \right]
       + 2 + {\cal O}(\e)\Bigg\} \;,\\
\tilde{E}_4^0(1_q,3_{q'},4_{\bar q'},5_g) &=& 
\tilde{e}_4^0(1,3,4,5) + 
\tilde{e}_4^0(1,4,3,5)\;,
\end{eqnarray}
with
\begin{eqnarray}
\tilde{e}_4^0(1,3,4,5) &=& \frac{1}{s_{1345}^2} \Bigg\{
         \frac{1}{\Sce \Sde}  \left[
           2 \Sac \Scd
          + 2 \Sac \Sae
          + 2 \Sac^2
          + \Scd \Sae
          + \Sae^2
         \right]\nonumber \\ &&
       +  \frac{\Sce}{\Sde \Scde^2}  \left[
            2 \Sac \Sad 
          + 2 \Sac \Sae 
          + \Sac^2 
          + 2 \Sad \Sae 
          + \Sad^2 
          + \Sae^2 
         \right]\nonumber \\ &&
       +  \frac{1}{\Sde \Scde}  \Big[
          - 2 \Sac \Sad
          - 4 \Sac \Sae
          + \Sac \Sce
          - 2 \Sac^2
          - 2 \Sad \Sae\nonumber \\ &&
          + \Sad \Sce
          + \Sae \Sce
          - 2 \Sae^2
         \Big] + {\cal O}(\e)
\Bigg\}\;.
\end{eqnarray}
Concerning the phase space mapping, $E_4^0$ can be decomposed into ordered 
contributions, which can use an ordered parametrisation of the phase space, 
by repeated partial fractioning of the invariants (like in $\tilde{A}_4^0$).
Due to the lack of symmetry, the result is considerably 
larger than the above expression and not very instructive. It is therefore not 
quoted here.  $\tilde{E}_4^0$ is already ordered (since quark (1) decouples 
from all singular limits). 

In the leading colour term,
the gluonic emission is colour-ordered (either in the  
$q,\bar q'$-antenna denoted here, or in the $q,q'$-antenna). The 
second colour-ordering is obtained by interchanging $q'\leftrightarrow 
\bar q'$. 
Note, in particular, that gluon emission inside the $q',\bar q'$-antenna 
is only possible at subleading colour (and constitutes in fact the full 
subleading colour contribution). 

Integration yields 
\begin{eqnarray}
{\cal E}_4^0(s_{1345}) &=& 
\frac{1}{2}\left(s_{1345}\right)^{-2\e} 
  \Bigg[ 
- \frac{5}{6\e^3} - \frac{17}{4\e^2}
+ \frac{1}{\e} \left( -\frac{2239}{108} +\frac{5\pi^2}{4} \right)
\nonumber \\ &&\hspace{2cm}
+ \left( - \frac{20521}{216} +\frac{51\pi^2}{8} 
+\frac{200}{9}\zeta_3 \right)  
+ {\cal O}(\e) \Bigg] \;,\\
\tilde{{\cal E}}_4^0(s_{1345}) &=& 
\left(s_{1345}\right)^{-2\e} 
  \Bigg[ 
- \frac{1}{6\e^3} - \frac{35}{36\e^2}
+ \frac{1}{\e} \left( -\frac{1045}{216} +\frac{\pi^2}{4} \right)
\nonumber \\ &&\hspace{2cm}
+ \left( - \frac{28637}{1296} +\frac{35\pi^2}{24} 
+\frac{40}{9}\zeta_3 \right)  
+ {\cal O}(\e) \Bigg]\;,
\end{eqnarray}
with 
\begin{eqnarray}
 \Poles\left( {\cal E}_4^0(s_{1345})\right)
&=&
 2\, \left[2 {\bf I}^{(1)}_{qg} \left(\e,s_{1345} \right) \right] 
\, \left[ 2
{\bf I}^{(1)}_{qg,\Flavour} \left(\e,s_{1345} \right) 
\right]  \nonumber \\ &&
- e^{-\e \gamma} \frac{\Gamma (1-2\e)}{\Gamma (1-\e)} \Bigg[
\left(\frac{b_{0,\Flavour}}{\e} + k_{0,\Flavour} \right) \,
\left[2 {\bf I}^{(1)}_{qg}(2\e,s_{1345})\right]\nonumber \\ && 
+ \left(\frac{b_{0}}{\e} + k_{0} \right) \,
\left[2 {\bf I}^{(1)}_{qg,\Flavour}(2\e,s_{1345})\right] 
\Bigg]
-  \frac{1}{2} {\bf H}^{(2)}_{R,E}(\e,s_{1345}) \;, \\
 \Finite\left( {\cal E}_4^0(s_{1345})\right)
&=&       - \frac{20521}{432}
          + \frac{1097\pi^2}{864}
          + \frac{391}{36}\zeta_3\;,\\
 \Poles\left( \tilde{{\cal E}}_4^0(s_{1345})\right)
&=&
-   {\bf H}^{(2)}_{R,\tilde{E}}(\e,s_{1345}) \;, \\
 \Finite\left( \tilde{{\cal E}}_4^0(s_{1345})\right)
&=&       - \frac{28637}{1296}
          + \frac{595\pi^2}{432}
          + \frac{79}{18}\zeta_3\;.
\end{eqnarray}

\section{Gluon-gluon antennae}
\label{sec:gg}

The gluon-gluon antenna functions are obtained from the QCD real radiation
corrections to the decay of a massive Higgs boson into 
two gluons, $H \to gg$, which is described 
in detail in~\cite{h}.

The overall normalisation is given by defining the tree-level two-parton
gluon-gluon
antenna function
\begin{equation}
{\cal F}_2^0(s_{12}) \equiv 1\;.
\end{equation}

The one-loop two-parton gluon-gluon antenna contains two 
contributions, corresponding to the different colour and flavour structures:
\begin{eqnarray}
{\cal F}_2^1(s_{12}) &=& 
2\, (s_{12})^{-\e} \Bigg[ -\frac{1}{\e^2} - \frac{11}{6\e}  + 
\frac{7\pi^2}{12} 
+ \left( -1 + \frac{7}{3}\zeta_3 \right)
\e 
\nonumber \\
&& \hspace{2cm}
+ \left( - 3
- \frac{73\pi^4}{1440} \right) \e^2  + {\cal O}(\e^3) \Bigg] \;, \\
\hat{{\cal F}}_2^1(s_{12}) &=& 
2\, (s_{12})^{-\e} \frac{1}{3\e} \;, 
\end{eqnarray}
with 
\begin{eqnarray}
\Poles\left({\cal F}_2^1(s_{12}) \right) &=& 
4\, {\bf I}^{(1)}_{g g} \left(\e,s_{12} \right)\;, \label{eq:f21poles}\\
\Finite\left({\cal F}_2^1(s_{12}) \right) &=& 0\;,\\
\Poles\left(\hat{{\cal F}}_2^1(s_{12 }) \right) &=& 
4\, {\bf I}^{(1)}_{g g,\Flavour} \left(\e,s_{12} \right)
\;,\label{eq:f21hpoles}\\
\Finite\left(\hat{{\cal F}}_2^1(s_{12}) \right) &=& 0\;.
\end{eqnarray}

The pole terms in the above expression have to be compared to the pole 
terms of the one-loop correction to the quark-antiquark antenna function
${\cal A}_2^1$ in (\ref{eq:a21poles}), and the quark-gluon antenna functions
${\cal D}_2^1$ in (\ref{eq:d21poles}) and $\hat{{\cal D}}_2^1$ in
(\ref{eq:d21hpoles}). 
The factor $4$ in 
(\ref{eq:f21poles}),(\ref{eq:f21hpoles})
 appears since the tree-level gluon-gluon antenna function 
${\cal F}_2^0$  contains two distinct 
gluon-gluon antennae. This situation is 
like for  ${\cal D}_2^0$, which contains two quark-gluon antennae, but
in contrast to the single quark-antiquark antenna contained in 
${\cal A}_2^0$.

\subsection{Three-parton tree-level antenna functions}

The tree-level three-parton gluon-gluon antenna contains two final states: 
gluon-gluon-gluon and gluon-quark-antiquark. The 
antenna corresponding to the first final state is: 
\begin{eqnarray}
F_3^0(g_1, g_2, g_3) &=& 
\frac{2}{s_{123}^2} \, \Bigg(
\frac{s_{123}^2 s_{12}}{ s_{13}  s_{23}}
+\frac{s_{123}^2 s_{13}}{ s_{12}  s_{23}}
+\frac{s_{123}^2 s_{23}}{ s_{12}  s_{13}}
+ \frac{s_{12} s_{13}}{s_{23}}
+ \frac{s_{12} s_{23}}{s_{13}}
+ \frac{s_{13} s_{23}}{s_{12}}
\nonumber \\ && \hspace{1cm}
+ 4 s_{123} + {\cal O}(\e)\Bigg)\;.
\end{eqnarray}
It yields the integral:
\begin{eqnarray}
{\cal F}_3^0(s_{123}) &=&  
  3\, \left(s_{123} \right)^{-\e}\Bigg[
\frac{1}{\e^2} + \frac{11}{6\e} + \frac{73}{12} - 
\frac{7\pi^2}{12} 
+ \left( \frac{451}{24} - \frac{77\pi^2}{72} - \frac{25}{3}\zeta_3 \right)
\e \nonumber \\
&& \hspace{2cm}
+ \left( \frac{2729}{48}
          - \frac{511\pi^2}{144}
          - \frac{275}{18}\zeta_3
          - \frac{71\pi^4}{1440} \right) \e^2  + {\cal O}(\e^3) \Bigg]\;, 
\end{eqnarray}
with 
\begin{eqnarray}
\Poles\left({\cal F}_3^0(s_{123}) \right) &=& 
- 6 {\bf I}^{(1)}_{gg} \left(\e,s_{123} \right)\;,\\
\Finite\left({\cal F}_3^0(s_{123}) \right) &=& \frac{73}{4}\;.
\end{eqnarray}
As can be seen from the pole structure, 
this tree-level antenna function contains three antenna 
configurations, corresponding 
to the three possible  
configurations of emitting a gluon in between a gluon pair.
The separation between these 
is not free from an ambiguity, but is in fact fixed by the decomposition 
used in the case of the quark-gluon-gluon antenna in (\ref{eq:smalld}) 
above. We decompose
\begin{equation}
F_3^0(1, 2, 3) = f_3^0(1, 3, 2) + f_3^0(3, 2, 1) + f_3^0(2, 1, 3)  \; ,
\end{equation}
where
\begin{equation}
f_3^0(1, 3, 2) = 
\frac{1}{s_{123}^2} \, \Bigg(
2\frac{s_{123}^2 s_{12}}{ s_{13}  s_{23}}
+ \frac{s_{12} s_{13}}{s_{23}}
+ \frac{s_{12} s_{23}}{s_{13}}
+ \frac{8}{3} s_{123} + {\cal O}(\e)\Bigg)\;.
\label{eq:smallf}
\end{equation}

The tree-level three-parton gluon-gluon
antenna corresponding to the gluon-quark-antiquark final state is:
\begin{equation}
G_3^0(1_g,3_q,4_{\bar q}) = 
\frac{1}{s_{134}^2} \, \left(
 \frac{s_{13}^2+s_{14}^2}{s_{34}} 
+ {\cal O}(\e) \right)\;.
\end{equation}
Its integrated form reads:
\begin{eqnarray}
{\cal G}_3^0(s_{134}) &=&  
  \left(s_{134} \right)^{-\e}\Bigg[
- \frac{1}{3\e} -\frac{7}{6}  + \left( -\frac{15}{4}+\frac{7\pi^2}{36}\right)
\e 
\nonumber \\ && \hspace{1cm}+ \left( -\frac{93}{8} 
          + \frac{49\pi^2}{72}
          - \frac{25}{9}\zeta_3\right) \e^2  + {\cal O}(\e^3) \Bigg]\;, 
\end{eqnarray}
with 
\begin{eqnarray}
\Poles\left({\cal G}_3^0(s_{134}) \right) &=& 
- 2 {\bf I}^{(1)}_{g g,\Flavour} \left(\e,s_{134} \right)\;,\\
\Finite\left({\cal G}_3^0(s_{134}) \right) &=& -\frac{7}{6}\;.
\end{eqnarray}

\subsection{Three-parton one-loop antenna functions}

At one loop, the correction to the gluon-gluon-gluon antenna contains 
a leading colour term $F_3^1(g_1,g_2,g_3)$ and a quark loop term
$\hat{F}_3^1(g_1,g_2,g_3)$. These read: 
\begin{eqnarray}
\Poles\left(F_3^1(g_1,g_2,g_3)\right) &=& 
2 \bigg( {\bf I}^{(1)}_{gg} (\e,s_{12}) + {\bf I}^{(1)}_{gg} (\e,s_{13})
+ {\bf I}^{(1)}_{gg} (\e,s_{23}) \nonumber \\
&& \hspace{2cm}- 2\,{\bf I}^{(1)}_{gg} (\e,s_{1 2 3}) \bigg) F_3^0(1,2,3)
\;,\\
\Finite\left(F_3^1(1,2,3)\right) &=& -
\bigg(  R(y_{12},y_{13}) 
+  R(y_{13},y_{23}) +  R(y_{12},y_{23})
 + \frac{11}{6} \log y_{12} 
\nonumber \\ && \hspace{1cm} 
+  \frac{11}{6} \log y_{13} 
+ \frac{11}{6} \log y_{23} \bigg)
F_3^0(1,2,3)  \nonumber \\ && \hspace{1cm} 
  +  \frac{1}{3s_{12}}+  \frac{1}{3s_{13}}
+  \frac{1}{3s_{23}}+  \frac{1}{3s_{123}}\;, \\
\Poles\left(\hat{F}_3^1(g_1,g_2,g_3)\right) &= &
2 \bigg( {\bf I}^{(1)}_{gg,\Flavour} (\e,s_{12}) 
+ {\bf I}^{(1)}_{gg,\Flavour} (\e,s_{13})
+ {\bf I}^{(1)}_{gg,\Flavour} (\e,s_{23}) \nonumber \\ &&\hspace{2cm}
- 2\, {\bf I}^{(1)}_{gg,\Flavour} (\e,s_{1 2 3}) \bigg) F_3^0(1,2,3)\;, \\
\Finite\left(\hat{F}_3^1(g_1,g_2,g_3)\right) &=& \frac{1}{3}\Bigg[\left(
\log y_{12} +  \log y_{13} 
 + \log y_{23} \right)
F_3^0(1,2,3)   \nonumber \\ && \hspace{2cm} -  \frac{1}{s_{12}} 
-  \frac{1}{s_{13}} -  \frac{1}{s_{23}} -  \frac{1}{s_{123}}\Bigg]\;. 
\end{eqnarray}

Integration of these antenna functions yields
\begin{eqnarray}
{\cal F}_3^1(s_{123}) &= &
 (s_{123})^{-2\e} \Bigg[ -\frac{3}{4\e^4} - \frac{33}{4\e^3} 
+ \frac{1}{\e^2} \left(-\frac{85}{3}+\frac{15\pi^2}{8}
\right) + \frac{1}{\e} \left(-\frac{9827}{72} + \frac{55\pi^2}{6}
+44 \,\zeta_3 \right)
\nonumber \\&&
+\left(-\frac{90185}{144}
          + \frac{6005\pi^2}{144}
          + \frac{506}{3} \zeta_3   - \frac{11\pi^4}{480}
   \right) + {\cal O}(\e) \Bigg]\;,\\
\hat{{\cal F}}_3^1(s_{123}) &= &
 (s_{123})^{-2\e} \Bigg[   \frac{1}{\e^3} 
+ \frac{11}{6\e^2} +
 \frac{1}{\e} \left(\frac{19}{3} - \frac{7\pi^2}{12}
 \right)
\nonumber \\ && \hspace{1cm}
+\left(\frac{499}{24}
          - \frac{77\pi^2}{72}
          - \frac{25}{3} \zeta_3   
   \right) + {\cal O}(\e) \Bigg]\;,
\end{eqnarray}
with
\begin{eqnarray}
 \Poles\left({\cal F}_3^1(s_{123})\right)  &=& 
- {\cal F}_2^1(s_{123}) \left( 
 6 {\bf I}^{(1)}_{gg} (\e,s_{123}) + 
{\cal F}_3^0(s_{123}) \right) 
+ \frac{2b_0}{\e}\, (s_{123})^{-\e}\,
 \left[ 3\,{\bf I}^{(1)}_{gg} (\e,s_{123})\right] \nonumber \\ 
&& - \frac{3}{2}\, {\bf H}^{(2)}_{V,F}(\e,s_{123})
+ 3 {\bf S}^{(2)}(\e,s_{123})\;,
\\
 \Finite\left({\cal F}_3^1(s_{123})\right)  &=&
          - \frac{146933}{144}
          + \frac{1139\pi^2}{18}
          + \frac{49\pi^4}{60}
          + \frac{902}{3}\zeta_3\;,
\\
 \Poles\left(\hat{{\cal F}}_3^1(s_{123})\right)  &=& 
- \hat{{\cal F}}_2^1(s_{123}) \left( 
 6 {\bf I}^{(1)}_{gg} (\e,s_{123}) + 
{\cal F}_3^0(s_{123}) \right) 
\nonumber \\ 
&& + \frac{2b_{0,\Flavour}}{\e}\, (s_{123})^{-\e}\,
 \left[ 3\,{\bf I}^{(1)}_{gg} (\e,s_{123})\right] 
- \frac{3}{2}\,{\bf H}^{(2)}_{V,\hat{F}}(\e,s_{123})\;,
\\
 \Finite\left(\hat{{\cal F}}_3^1(s_{123})\right)  &=&
            \frac{467}{8}
          - 24\zeta_3\;.
\end{eqnarray}
The factors of 3 in front of the infrared singularity operators and 
of the soft gluon current arise from the fact that the tree-level three-parton 
antenna function contains three antenna configurations. 

At one loop, the correction to the quark-quark-antiquark antenna contains 
a leading colour and 
a subleading colour term 
$G_3^1(1_g,3_q,4_{\bar q})$, $\tilde{G}_3^1(1_g,3_q,4_{\bar q})$ as well as
 a quark loop term
$\hat{G}_3^1(1_g,3_q,4_{\bar q})$. These read: 
\begin{eqnarray}
\Poles\left(G_3^1(1_g,3_q,4_{\bar q})\right) &=& 
2 \left( {\bf I}^{(1)}_{qg} (\e,s_{13}) + {\bf I}^{(1)}_{qg} 
(\e,s_{14}) -2\,{\bf I}^{(1)}_{g g} 
(\e,s_{134}) 
\right) G_3^0(1,3,4),
\\
\Finite\left(G_3^1(1_g,3_q,4_{\bar q})\right) &=& -
\bigg( 
 R(y_{13},y_{34}) 
+  R(y_{14},y_{34}) 
+ \frac{5}{3} \log y_{13} +  \frac{5}{3} \log y_{14} 
\nonumber \\ && \hspace{1cm}
+ \frac{13}{6} \log y_{34}  - \frac{40}{9}\bigg)
G_3^0(1,3,4)  -  \frac{s_{13} + s_{14}}{2\,s_{134}^2}\;,\\
\Poles\left(\tilde{G}_3^1(1_g,3_q,4_{\bar q})\right) &=& 
2 \left( {\bf I}^{(1)}_{q\bar q} (\e,s_{34}) \right) G_3^0(1,3,4)\;,\\
\Finite\left(\tilde{G}_3^1(1_g,3_q,4_{\bar q})\right) &=& - \left(4 + 
 R(y_{13},y_{14}) \right)
G_3^0(1,3,4)    + \frac{s_{13} + s_{14}}{2\,s_{134}^2}\;,
\\
\Poles\left(\hat{G}_3^1(1_g,3_q,4_{\bar q})\right) &=& 
- 4  {\bf I}^{(1)}_{g g,\Flavour} (\e,s_{134})  G_3^0(1,3,4)\;,\\
\Finite\left(\hat{G}_3^1(1_g,3_q,4_{\bar q})\right) &=& 
\left( -\frac{10}{9} + \frac{2}{3} \log y_{34} 
+ \frac{1}{6} \log y_{13}  + \frac{1}{6} \log y_{14} 
\right) G_3^0(1,3,4).\hspace{3mm}   
\end{eqnarray}

Integration of these antenna functions yields
\begin{eqnarray}
{\cal G}_3^1(s_{134}) &= &
 (s_{134})^{-2\e} \Bigg[ 
 \frac{11}{18\e^2}
+ \frac{1}{\e} \left(\frac{169}{54} - \frac{\pi^2}{9}
 \right) \nonumber \\ && \hspace{1cm} +\left(\frac{446}{27}
          - \frac{205\pi^2}{216}
          - \frac{130}{9} \zeta_3 
   \right) + {\cal O}(\e) \Bigg]\;,\\
\tilde{{\cal G}}_3^1(s_{134}) &= &
 (s_{134})^{-2\e} \Bigg[ 
 \frac{1}{6\e^3} + \frac{41}{36\e^2}
+\frac{1}{\e} \left( \frac{325}{54}
          - \frac{\pi^2}{4} \right) \nonumber \\ && \hspace{1cm}
+ \left(\frac{18457}{648}
          - \frac{41\pi^2}{24} - \frac{37}{9}\zeta_3 \right) 
+ {\cal O}(\e) \Bigg]\;,\\
\hat{{\cal G}}_3^1(s_{134}) &= &
 (s_{134})^{-2\e} \Bigg[ 
\frac{7}{18\e} 
+\left(\frac{895}{324}
          - \frac{11\pi^2}{108} \right) + {\cal O}(\e) \Bigg]\;,
\end{eqnarray}
with
\begin{eqnarray}
 \Poles\left({\cal G}_3^1(s_{134})\right)  &=& 
- {\cal F}_2^1(s_{134}) \left( 
 2 {\bf I}^{(1)}_{gg,\Flavour} (\e,s_{134}) + 
{\cal G}_3^0(s_{134}) \right) \nonumber \\ 
&&
+ \frac{2b_0}{\e}\, (s_{134})^{-\e}\,
  {\bf I}^{(1)}_{gg,\Flavour} (\e,s_{134}) 
 - \frac{1}{2}\,{\bf H}^{(2)}_{V,G}(\e,s_{134})\;,
\\
 \Finite\left({\cal G}_3^1(s_{134})\right)  &=&
          \frac{1445}{27}
          - \frac{337\pi^2}{108}
          - \frac{26}{3}\zeta_3\;,
\\
 \Poles\left(\tilde{{\cal G}}_3^1(s_{134})\right)  &=& 
- \frac{1}{2}\,{\bf H}^{(2)}_{V,\tilde{G}}(\e,s_{134})\;,
\\
 \Finite\left(\tilde{{\cal G}}_3^1(s_{134})\right)  &=&
            \frac{18457}{648}
          - \frac{697\pi^2}{432}
          - \frac{73}{18}\, \zeta_3\;,
\\
 \Poles\left(\hat{{\cal G}}_3^1(s_{134})\right)  &=& 
- \hat{{\cal F}}_2^1(s_{134}) \left( 
 2 {\bf I}^{(1)}_{gg,\Flavour} (\e,s_{134}) + 
{\cal G}_3^0(s_{134}) \right) \nonumber \\ 
&& 
+ \frac{2b_{0,\Flavour}}{\e}\, (s_{134})^{-\e}\,
 {\bf I}^{(1)}_{gg,\Flavour} (\e,s_{134}) 
- \frac{1}{2}\, {\bf H}^{(2)}_{V,\hat{G}}(\e,s_{134})\;,
\\
 \Finite\left(\hat{{\cal G}}_3^1(s_{134})\right)  &=&
            \frac{85}{324}
          - \frac{17\pi^2}{108}\;.
\end{eqnarray}
In contrast to the $N_F^2$ one-loop three-parton 
quark-gluon antenna function  
$\hat{{\cal E}}_3^1$, which had no corresponding 
tree-level four-parton antenna function with the same colour and flavour
structure, 
$\hat{{\cal G}}_3^1$ is complemented by an  antenna function 
containing two quark-antiquark pairs, ${\cal H}_4^0$, which is derived 
below. Correspondingly, $\hat{{\cal G}}_3^1$ does not contain 
the full $N_F^2$-terms for gluon-gluon final states.

\subsection{Four-parton tree-level antenna functions}

The tree-level four-parton quark-gluon antenna contains three final 
states: 
gluon-gluon-gluon-gluon, $F_4^0$, 
and gluon-quark-antiquark-gluon at leading and subleading colour,
$G_4^0$ and $\tilde{G}_4^0$ and quark-antiquark-quark-antiquark, $H_4^0$. 
The 
antenna for the $gggg$ final state is:
\begin{eqnarray}
F_4^0(g_1,g_2,g_3,g_4) &=&  \hspace{3mm} f_4^0(1,2,3,4) + f_4^0(4,3,2,1) +  
f_4^0(2,3,4,1) + f_4^0(1,4,3,2)  \nonumber \\ && 
+ f_4^0(3,4,1,2) + f_4^0(2,1,4,3) +  
f_4^0(4,1,2,3) + f_4^0(3,2,1,4) \; ,
\end{eqnarray}
with 
\begin{eqnarray}
f_4^0(1,2,3,4) &=& \frac{1}{s_{1234}^2} \Bigg\{
       - \frac{2\Scd \Sac \Sad^2 }{\Sbc^2 \Sabc \Sbcd  } 
       + \frac{1}{\Sbc^2  } \Big[
            2 \Sab \Sad
          - 2 \Sab \Sbd
          + 2 \Sab \Scd\nonumber \\ &&
          + \Sab^2
          - 2 \Sac \Sad
          - 2 \Sac \Sbd
          - 2 \Sac \Scd
          - 2 \Sad \Sbd
          + 2 \Sad \Scd
          + \Sad^2
          + \Scd^2
          \Big]\nonumber \\ &&
       + \frac{\Sac }{\Sbc^2 \Sabc  } \left[
            4 \Sac \Sad
          + 4 \Sac \Sbd
          + 4 \Sac \Scd
          - 8 \Sad \Scd
          - 2 \Sad^2
          - 4 \Sbd \Scd
          - 4 \Scd^2
          \right]\nonumber \\ &&
       + \frac{\Sac^2}{\Sbc^2  \Sabc^2  } \left[
            4 \Sad \Sbd
          + 4 \Sad \Scd
          + 2 \Sad^2
          + 4 \Sbd \Scd
          + 2 \Sbd^2
          + 2 \Scd^2
          \right] \nonumber \\ &&
       + \frac{1}{4 \Sbc \Sab \Scd \Sad  } \left[
            2 \Sac \Sbd^3
          + 3 \Sac^2 \Sbd^2
          + 2 \Sac^3 \Sbd
          + \Sac^4
          + \Sbd^4
          \right]\nonumber \\ &&
       + \frac{1}{\Sbc \Sab \Scd  } \Big[
            6 \Sac \Sad \Sbd
          + 2 \Sac \Sad^2
          + 6 \Sac \Sbd^2
          + 3 \Sac^2 \Sad
          + 6 \Sac^2 \Sbd
          + 2 \Sac^3\nonumber \\ &&
          + 3 \Sad \Sbd^2
          + 2 \Sad^2 \Sbd
          + \Sad^3
          + 2 \Sbd^3
          \Big]
       + \frac{\Sbd }{\Sbc \Sab \Sabd  } \left[
            \Sbd \Scd
          + \Sbd^2
          + 2 \Scd^2
          \right]\nonumber \\ &&
       + \frac{1}{\Sbc \Sab \Sbcd \Sabd  } \left[
            2 \Sac \Scd^3
          + 3 \Sac^2 \Scd^2
          + 2 \Sac^3 \Scd
          + \Sac^4
          + \Scd^4
          \right]\nonumber \\ &&
       + \frac{1}{\Sbc \Sab \Sbcd  } \Big[
            2 \Sac \Sad \Scd
          + 2 \Sac \Sad^2
          + 2 \Sac \Scd^2
          + 3 \Sac^2 \Sad
          + 3 \Sac^2 \Scd
          + 2 \Sac^3\nonumber \\ &&
          + 2 \Sad \Scd^2
          - \Sad^2 \Scd
          + \Sad^3
          \Big]
       + \frac{1}{\Sbc \Sab  } \Big[
            20 \Sac \Sad
          + 14 \Sac \Sbd
          + 9 \Sac^2
          + 16 \Sad \Sbd\nonumber \\ &&
          + 4 \Sad \Scd
          + 19 \Sad^2
          + 7 \Sbd^2
          - 10 \Scd^2
          \Big]
       + \frac{\Sad }{\Sbc \Sabc \Sbcd  } \Big[
          - \Sac \Sad
          - 4 \Sac \Scd
          - 4 \Sac^2\nonumber \\ &&
          + \Sad \Scd
          - 4 \Sad^2
          - 4 \Scd^2
          \Big]
       + \frac{1}{\Sbc \Sabc \Sacd  } \Big[
            6 \Sad \Sbd \Scd
          + 6 \Sad \Sbd^2
          + 6 \Sad^2 \Sbd
          + \Sad^3\nonumber \\ &&
          + 6 \Sbd \Scd^2
          + 9 \Sbd^2 \Scd
          + 6 \Sbd^3
          + \Scd^3
          \Big]
       + \frac{1}{\Sbc \Sabc \Sabd  } \bigg[
          - 3 \Sad \Sbd \Scd
          - \frac{3}{2} \Sad \Sbd^2\nonumber \\ &&
          - \frac{3}{2} \Sad \Scd^2
          - \Sad^2 \Sbd
          - \Sad^2 \Scd
          + \frac{3}{4} \Sad^3
          - 3 \Sbd \Scd^2
          - 3 \Sbd^2 \Scd
          - \Sbd^3
          - \Scd^3
          \bigg]\nonumber \\ &&
       + \frac{1}{4\Sbc \Sabc  } \Big[
          - 7 \Sac \Sad
          + 18 \Sac \Sbd
          - 16 \Sac \Scd
          - 11 \Sac^2
          - 41 \Sad \Sbd
          - 36 \Sad \Scd\nonumber \\ &&
          - 63 \Sad^2
          - 16 \Sbd \Scd
          - 21 \Sbd^2
          - 18 \Scd^2
          \Big]
       + \frac{1}{\Sbc \Sacd  } \Big[
            7 \Sab \Sad
          + 2 \Sab \Sbd
          + 8 \Sab \Scd\nonumber \\ &&
          - 4 \Sab^2
          + 4 \Sad \Sbd
          - 3 \Sad \Scd
          - \Sad^2
          + 2 \Sbd \Scd
          + 3 \Sbd^2
          - 3 \Scd^2
          \Big]\nonumber \\ &&
       + \frac{1}{8\Sbc  } \Big[
           21 \Sab
          + 69 \Sac
          + 14 \Sad
          + 69 \Sbd
          + 21 \Scd
          \Big]\nonumber \\ &&
       + \frac{1}{2\Sab^2 \Scd^2  } \left[
          - 2 \Sac \Sad \Sbc \Sbd
          + \Sac^2 \Sbd^2
          + \Sad^2 \Sbc^2
          \right]\nonumber \\ &&
       + \frac{1}{\Sab \Scd \Sabc \Sbcd  } \left[
            4 \Sad \Sbd^3
          + 6 \Sad^2 \Sbd^2
          + 4 \Sad^3 \Sbd
          + \Sad^4
          + \Sbd^4
          \right]\nonumber \\ &&
       + \frac{1}{\Sab \Scd \Sabc \Sacd  } \left[
            4 \Sad \Sbd^3
          + 6 \Sad^2 \Sbd^2
          + 4 \Sad^3 \Sbd
          + \Sad^4
          + \Sbd^4
          \right]\nonumber \\ &&
       + \frac{1}{8 \Sab \Scd \Sabc  } \Big[
            12 \Sad \Sbc \Sbd
          + 12 \Sad \Sbc^2
          - 12 \Sad \Sbd^2
          - 6 \Sad^2 \Sbc
          - 18 \Sad^2 \Sbd\nonumber \\ &&
          - 4 \Sad^3
          + 27 \Sbc \Sbd^2
          + 21 \Sbc^2 \Sbd
          + 3 \Sbc^3
          + 5 \Sbd^3
          \Big]
       + \frac{1}{8 \Sab \Scd \Sbcd  } \Big[
            12 \Sac \Sad \Sbd\nonumber \\ &&
          + 18 \Sac \Sad^2
          + 3 \Sac \Sbd^2
          + 12 \Sac^2 \Sad
          + 3 \Sac^2 \Sbd
          + 3 \Sac^3
          + 12 \Sad \Sbd^2
          + 18 \Sad^2 \Sbd\nonumber \\ &&
          + 12 \Sad^3
          + 3 \Sbd^3
          \Big]
       + \frac{1}{8 \Sab \Scd  } \Big[
            16 \Sac \Sad
          + 31 \Sac \Sbc
          + 45 \Sac \Sbd
          + 25 \Sac^2\nonumber \\ &&
          - 8 \Sad \Sbc
          + 16 \Sad \Sbd
          + 6 \Sad^2
          + 31 \Sbc \Sbd
          + 21 \Sbc^2
          + 25 \Sbd^2
          \Big] \nonumber \\ &&
       + \frac{5}{8 \Sab \Sabc \Sbcd  } \Big[
            12 \Sad \Sbd \Scd
          + 12 \Sad \Sbd^2
          + 4 \Sad \Scd^2
          + 12 \Sad^2 \Sbd
          + 6 \Sad^2 \Scd\nonumber \\ &&
          + 4 \Sad^3
          + 4 \Sbd \Scd^2
          + 6 \Sbd^2 \Scd
          + 4 \Sbd^3
          +   \Scd^3
          \Big]
       + \frac{5}{8\Sab \Sabc  } \Big[
            4 \Sad \Sbc
          - 8 \Sad \Sbd\nonumber \\ &&
          - 4 \Sad \Scd
          - 6 \Sad^2
          + 2 \Sbc \Sbd
          + \Sbc \Scd
          - \Sbc^2
          - 3 \Sbd \Scd
          - 3 \Sbd^2
          - \Scd^2
          \Big]\nonumber \\ &&
       + \frac{3}{8 \Sab \Sbcd  } \Big[
            4 \Sac \Sad
          + 2 \Sac \Sbd
          + \Sac \Scd
          + \Sac^2
          + 8 \Sad \Sbd
          + 4 \Sad \Scd
          + 6 \Sad^2\nonumber \\ &&
          + 3 \Sbd \Scd
          + 3 \Sbd^2
          + \Scd^2
          \Big]
       + \frac{3}{8\Sab  } \left[
          - \Sac
          - 4 \Sad
          + \Sbc
          - 2 \Sbd
          - \Scd
          \right]\nonumber \\ &&
       + \frac{3}{8 \Scd \Sabc \Sbcd  } \Big[
          - 4 \Sac \Sad \Sbd
          - 6 \Sac \Sad^2
          -  \Sac \Sbd^2
          - 4 \Sac^2 \Sad
          -  \Sac^2 \Sbd
          -  \Sac^3\nonumber \\ &&
          - 4 \Sad \Sbd^2
          - 6 \Sad^2 \Sbd
          - 4 \Sad^3
          -  \Sbd^3
          \Big]
       + \frac{3}{8 \Scd \Sabc  } \Big[
            4 \Sac \Sad
          -  \Sac \Sbc
          +  \Sac \Sbd\nonumber \\ &&
          +  \Sac^2
          - 4 \Sad \Sbc
          + 4 \Sad \Sbd
          + 6 \Sad^2
          -  \Sbc \Sbd
          +  \Sbc^2
          +  \Sbd^2
          \Big]\nonumber \\ &&
       + \frac{1}{\Sbd \Sabc \Sacd  } \left[
          - 2 \Sad
          + \Sbd
          - 2 \Scd
          \right]
       + \frac{1}{\Sabc^2  } \Big[
            2 \Sad \Sbd
          + 2 \Sad \Scd
          + \Sad^2\nonumber \\ &&
          + 2 \Sbd \Scd
          + \Sbd^2
          + \Scd^2
          \Big]
       + \frac{1}{8\Sabc \Sbcd  } \Big[
          - 12 \Sac \Sad
          - 6 \Sac \Sbd
          - 3 \Sac \Scd\nonumber \\ &&
          - 3 \Sac^2
          - 24 \Sad \Sbd
          - 12 \Sad \Scd
          + 38 \Sad^2
          - 9 \Sbd \Scd
          - 9 \Sbd^2
          - 3 \Scd^2
          \Big]\nonumber \\ &&
       + \frac{1}{8\Sabc   } \left[
          - 6 \Sac
          + 45 \Sad
          - 3 \Sbc
          + 58 \Sbd
          + 36 \Scd
          \right]
       + \frac{35}{8}
+{\cal O}(\e) \Bigg\}\; .
\end{eqnarray}
In $F_4^0$, the gluonic emissions are colour-ordered. Since the 
original colour structure is a trace over the gluon colour indices~\cite{h}, 
$F_4^0$ is symmetric under cyclic interchanges of the momenta.
Therefore,
each pair of adjacent momenta can act as hard emitter pair for the antenna 
function. $F_4^0$ thus contains four different colour-ordered antennae.
For numerical implementation, these have to be separated from each other 
by repeated partial fractioning of the associated invariants. This 
fractioning is not made explicit in the above expression $f_4^0$. 

The integral of this antenna function is 
\begin{eqnarray}
{\cal F}_4^0(s_{1234}) &=& 
2 \,
\left(s_{1234}\right)^{-2\e} 
\Bigg[ 
\frac{5}{2\e^4}  + \frac{121}{12\e^3} + \frac{1}{\e^2} 
\left(\frac{436}{9} - \frac{11\pi^2}{3} \right) \nonumber \\
&&+ \frac{1}{\e} \left(
            \frac{23455}{108}
          - \frac{1067\pi^2}{72}
          - \frac{379}{6}\zeta_3 \right) 
\nonumber \\
&&
+ \left(    \frac{304951}{324}
          - \frac{7781\pi^2}{108}
          - \frac{2288}{9}\zeta_3
          + \frac{479\pi^4}{720} 
\right)
+ {\cal O}(\e)  \Bigg]\;, 
\end{eqnarray}
with
\begin{eqnarray}
\Poles\left( {\cal F}_4^0(s_{1234})\right) &=& 
4\, \left[2 {\bf I}^{(1)}_{gg} \left(\e,s_{1234} \right) 
\right]^2  - 4\,
e^{-\e \gamma} \frac{\Gamma (1-2\e)}{\Gamma (1-\e)}
\left(\frac{b_{0}}{\e} + k_{0} \right) \,
 \left[2 {\bf I}^{(1)}_{gg}(2\e,s_{1234})\right] \nonumber \\ &&
- 2\, {\bf H}^{(2)}_{R,F}(\e,s_{1234})
- 4\, \,{\bf S}^{(2)}_{V}(\e,s_{1234})\;, \\
\Finite\left( {\cal F}_4^0(s_{1234}\right) &=& 
           \frac{271741}{162}
          - \frac{2335\pi^2}{27}
          - \frac{397\pi^4}{360}
          - \frac{2651}{6}\zeta_3 \;.
\end{eqnarray}

The leading colour and subleading colour
 gluon-quark-antiquark-gluon antennae are:
\begin{eqnarray}
G_4^0(1_g,3_q,4_{\bar q},2_g) &=& g_4^0(1,3,4,2) +
g_4^0(2,4,3,1)\;, \\
\tilde{G}_4^0(1_g,3_q,4_{\bar q},2_g) &=&
\tilde{g}_4^0(1,3,4,2) + \tilde{g}_4^0(1,4,3,2) +
\tilde{g}_4^0(2,3,4,1) + \tilde{g}_4^0(2,4,3,1)\;, 
\end{eqnarray}
with
\begin{eqnarray}
g_4^0(1,3,4,2) &=& \frac{1}{s_{1234}^2} \Bigg\{
        \frac{1}{\Sab^2 \Scd^2   } \left[
            2 \Sac \Sad \Sbc \Sbd
          -  \Sac^2 \Sbd^2
          -  \Sad^2 \Sbc^2
          \right]
       + \frac{\Sbc }{\Sab \Sac \Scd   } \left[
             \Sbc^2
          +  \Sbd^2
          \right]\nonumber \\ &&
       + \frac{\Sbc }{\Sab \Sac \Sacd   } \left[
             \Sbc^2
          - 2 \Sbd \Scd
          +  \Sbd^2
          +  \Scd^2
          \right]
       + \frac{\Sbc }{\Sab \Sac   } \left[
             \Sad
          + 2 \Sbd
          -  \Scd
          \right]\nonumber \\ &&
       - \frac{\Sac}{\Sab    }
       + \frac{1}{\Sab \Scd \Sacd   } \Big[
          - 2 \Sad \Sbc^2
          + 2 \Sad \Sbd^2
          + 2 \Sad^2 \Sbc
          + 2 \Sad^2 \Sbd
          +  \Sbc \Sbd^2\nonumber \\ &&
          +  \Sbc^2 \Sbd
          +  \Sbc^3
          +  \Sbd^3
          \Big]
       + \frac{1}{\Sab \Scd   } \Big[
            2 \Sac \Sad
          + 4 \Sac \Sbc
          + 3 \Sac \Sbd
          + 2 \Sac^2\nonumber \\ &&
          -  \Sad \Sbc
          + 4 \Sad^2
          \Big]
       + \frac{1}{\Sab \Sacd   } \left[
            2 \Sad \Sbd
          - 4 \Sbc \Sbd
          +  \Sbc \Scd
          +  \Sbd \Scd
          \right]\nonumber \\ &&
       + \frac{\Sab }{\Scd \Sacd \Sbcd   } \left[
            2 \Sab \Sad
          +  \Sab^2
          + 4 \Sad \Sbd
          + 4 \Sad^2
          + 4 \Sbd^2
          \right]\nonumber \\ &&
       + \frac{\Sab}{ \Sacd \Sbcd   } \left[
            6 \Sad
          + 6 \Sbd
          + 3 \Scd
          \right]
       + \frac{2\Sab^2 \Sad \Sbd}{ \Scd^2 \Sacd \Sbcd   }
       + \frac{1}{2\Sac \Sbd   } \left[
            \Sab \Scd
          - \Sad \Sbc
          \right]\nonumber \\ &&
       + \frac{\Sbd }{\Sac \Scd \Sbcd   } \left[
          - 2 \Sab \Sbd
          +  \Sab^2
          + 2 \Sbd^2
          \right]
       + \frac{1}{\Sac \Scd   } \Big[
          - 2 \Sab \Sbc
          + 2 \Sab \Sbd\nonumber \\ &&
          -  \Sab^2
          + 2 \Sbc \Sbd
          - 2 \Sbc^2
          - 2 \Sbd^2
          \Big]
       + \frac{\Scd }{\Sac \Sacd^2   } \Big[
          - 2 \Sab \Sbc
          - 2 \Sab \Sbd
          -  \Sab^2\nonumber \\ &&
          - 2 \Sbc \Sbd
          -  \Sbc^2
          -  \Sbd^2
          \Big]
       + \frac{1}{\Sac \Sacd \Sbcd   } \Big[
            6 \Sab \Sbd \Scd
          + 4 \Sab \Sbd^2
          + 3 \Sab \Scd^2\nonumber \\ &&
          - 3 \Sab^2 \Sbd
          - 3 \Sab^2 \Scd
          +  \Sab^3
          - 3 \Sbd \Scd^2
          - 4 \Sbd^2 \Scd
          - 2 \Sbd^3
          -  \Scd^3
          \Big]\nonumber \\ &&
       + \frac{1}{\Sac \Sacd   } \Big[
            8 \Sab \Sbc
          - 4 \Sab \Sbd
          - 4 \Sab \Scd
          + 4 \Sab^2
          - 2 \Sbc \Scd
          + 6 \Sbc^2
          + 2 \Sbd^2\nonumber \\ &&
          + 2 \Scd^2
          \Big]
       + \frac{1}{\Sac \Sbcd   } \left[
          - 4 \Sab \Sbd
          - 2 \Sab \Scd
          +  \Sab^2
          + 3 \Sbd \Scd
          + 4 \Sbd^2
          +  \Scd^2
          \right]\nonumber \\ &&
       + \frac{1}{\Sac   } \left[
            2 \Sab
          + 2 \Sbc
          - 2 \Sbd
          - 2 \Scd
          \right]
       + \frac{\Sad}{ \Scd^2 \Sacd   } \Big[
          - 4 \Sab \Sad
          + 8 \Sab \Sbd
          + 2 \Sab^2\nonumber \\ &&
          - 4 \Sad \Sbc
          - 4 \Sad \Sbd
          + 4 \Sbc \Sbd
          + 4 \Sbd^2
          \Big]
       + \frac{\Sad^2 }{\Scd^2 \Sacd^2   } \Big[
          - 4 \Sab \Sbc
          - 4 \Sab \Sbd\nonumber \\ &&
          - 2 \Sab^2
          - 4 \Sbc \Sbd
          - 2 \Sbc^2
          - 2 \Sbd^2
          \Big]
       + \frac{1}{\Scd^2   } \Big[
          - 4 \Sab \Sac
          + 4 \Sab \Sad
          -  \Sab^2\nonumber \\ &&
          + 4 \Sac \Sbc
          - 2 \Sac \Sbd
          - 2 \Sac^2
          + 2 \Sad \Sbc
          \Big]
       + \frac{1}{\Scd \Sacd   } \Big[
            4 \Sab \Sbc
          + 2 \Sab \Sbd
          + 5 \Sab^2\nonumber \\ &&
          - 8 \Sad \Sbc
          + 6 \Sad \Sbd
          + 6 \Sad^2
          + 6 \Sbc^2
          + 4 \Sbd^2
          \Big]
       + \frac{1}{\Scd   } \left[
            2 \Sab
          + 2 \Sac 
          + 6 \Sad
          \right]\nonumber \\ &&
       + \frac{1}{\Sacd^2   } \left[
          - 2 \Sab \Sbc
          - 2 \Sab \Sbd
          -  \Sab^2
          - 2 \Sbc \Sbd
          -  \Sbc^2
          -  \Sbd^2
          \right]\nonumber \\ &&
       + \frac{1}{\Sacd   } \left[
          - 4 \Sab
          + 4 \Sad
          - 4 \Sbc 
          - 2 \Sbd
          + 3 \Scd
          \right]
       - \frac{7}{2}         
+{\cal O}(\e)\Bigg\}\;,
\\
\tilde{g}_4^0(1,3,4,2) &=& \frac{1}{s_{1234}^2} \Bigg\{
         \frac{ \Sab^2 + 2\Sbc^2+2 \Sab\Sbc}{\Sac \Sad  } 
       + \frac{\Sab \Sbd}{2\Sac \Sbc \Sacd \Sbcd  } \left[
            \Sbd^2
          + \Sab^2
          \right] \nonumber \\ &&
       + \frac{\Sab}{2\Sac \Sbc \Sacd  } \left[
            \Sbd \Scd
          - \Sbd^2
          - \Scd^2
          - \Sab^2
          \right]
       + \frac{1}{2\Sac \Sbd  } \left[
            \Sab \Scd
          - \Sad \Sbc
          \right]\nonumber \\ &&
       + \frac{\Scd}{\Sac \Sacd^2  } \left[
          - 2 \Sab \Sbc 
          - 2 \Sab \Sbd 
          -  \Sab^2 
          - 2 \Sbc \Sbd 
          -  \Sbc^2 
          -  \Sbd^2 
          \right]\nonumber \\ &&
       + \frac{1}{\Sac \Sacd \Sbcd  } \Big[
            \frac{7}{2} \Sab \Sbd \Scd
          + \frac{1}{2}\Sab \Sbd^2
          + \frac{5}{2} \Sab \Scd^2
          - 3 \Sab^2 \Scd
          + \frac{1}{2} \Sab^3
          - 2 \Sbd \Scd^2\nonumber \\ &&
          - 2 \Sbd^2 \Scd
          -   \Scd^3
          \Big]
       + \frac{1}{\Sac \Sacd  } \Big[
          - 2 \Sab \Sbd
          - 4 \Sab \Scd
          -  \Sab^2
          - 2 \Sbc \Scd
          +  \Sbc^2\nonumber \\ &&
          + 2 \Sbd \Scd
          -  \Sbd^2
          \Big]
       + \frac{1}{\Sac \Sbcd  } \Big[
          - 2 \Sab \Sbd
          - 2 \Sab \Scd
          +  \Sab^2
          + 2 \Sbd \Scd\nonumber \\ &&
          + 2 \Sbd^2
          +  \Scd^2
          \Big]
       + \frac{1}{\Sac  } \left[
            2 \Sab
          + 2 \Sbc
          - 2 \Sbd
          \right]
       + \frac{1}{\Sacd^2  } \Big[
          - 2 \Sab \Sbc
          - 2 \Sab \Sbd\nonumber \\ &&
          -  \Sab^2
          - 2 \Sbc \Sbd
          -  \Sbc^2
          -  \Sbd^2
          \Big]
       + \frac{\Sab}{\Sacd \Sbcd  } \left[
          - 2 \Scd
          -  \Sab
          \right]\nonumber \\ &&
       + \frac{1}{\Sacd   }\left[
          - 2 \Sab
          -  \Sbc
          -  \Sbd
          -  \Scd
          \right]
       - \frac{1}{2}  +{\cal O}(\e)\Bigg\}\;.
\end{eqnarray}
In the leading colour contribution $G_4^0$, the gluonic emissions are 
colour-ordered in between the quark-antiquark pair. However, like all
gluon-gluon antenna functions at leading colour, $G_4^0$ is cyclic 
in the colour indices, such that it also contains a configuration where the 
two gluons form the hard emitter pair, emitting the quark-antiquark pair
inside the antenna. As before, the different antenna configurations 
can be separated from each other for the numerical integration by
repeated partial fractioning of the invariants. 

Integration yields 
\begin{eqnarray}
{\cal G}_4^0(s_{1234}) &=& 
\frac{1}{2}\, \left(s_{1234}\right)^{-2\e} 
  \Bigg[ 
-\frac{3}{2\e^3}  - \frac{155}{18\e^2} + \frac{1}{\e} 
\left(-\frac{523}{12} + \frac{79\pi^2}{36} \right)
\nonumber \\ && \hspace{3.8cm}
+  \left(
  -\frac{16579}{81}
          + \frac{1385\pi^2}{108}
          + 37\zeta_3 \right) 
+ {\cal O}(\e) \Bigg]\;,\\
\tilde{{\cal G}}_4^0(s_{1234}) &=& 
\left(s_{1234}\right)^{-2\e} 
  \Bigg[ - 
 \frac{1}{3\e^3} - \frac{41}{18\e^2} 
+ \frac{1}{\e} \left( -\frac{1327}{108}
          + \frac{\pi^2}{2}
\right) \nonumber \\ && \hspace{3.8cm}
+ \left(- \frac{4864}{81}
          + \frac{41\pi^2}{12}
          + \frac{86}{9}\zeta_3
\right)
+ {\cal O}(\e) \Bigg]\;,
\end{eqnarray}
with 
\begin{eqnarray}
\Poles\left( {\cal G}_4^0(s_{1234})\right)
&=&
  2\, \left[2 {\bf I}^{(1)}_{gg} \left(\e,s_{1234} \right) \right] 
\, \left[ 2
{\bf I}^{(1)}_{gg,\Flavour} \left(\e,s_{1234} \right) 
\right]  \nonumber \\ &&
- e^{-\e \gamma} \frac{\Gamma (1-2\e)}{\Gamma (1-\e)} \Bigg[
\left(\frac{b_{0,\Flavour}}{\e} + k_{0,\Flavour} \right) \,
\left[2 {\bf I}^{(1)}_{gg}(2\e,s_{1234})\right]\nonumber \\ && 
+ \left(\frac{b_{0}}{\e} + k_{0} \right) \,
\left[2 {\bf I}^{(1)}_{gg,\Flavour}(2\e,s_{1234})\right] 
\Bigg]
-  \frac{1}{2} {\bf H}^{(2)}_{R,G}(\e,s_{1234})\;, \\
 \Finite\left( {\cal G}_4^0(s_{1234})\right)
&=&  
 - \frac{16579}{162}
          + \frac{155\pi^2}{48}
          + \frac{649}{36}\zeta_3
\;,\\
 \Poles\left( \tilde{{\cal G}}_4^0(s_{1234})\right)
&=&
-   {\bf H}^{(2)}_{R,\tilde{G}}(\e,s_{1234}) \;,\\
 \Finite\left( \tilde{{\cal G}}_4^0(s_{1234})\right)
&=&      - \frac{4864}{81}
          + \frac{697\pi^2}{216}
          + \frac{85}{9}\zeta_3\;.
\end{eqnarray}

The quark-antiquark-quark-antiquark antenna is
\begin{eqnarray}
H_4^0(1_q,2_{\bar q},3_{q'},4_{\bar q'}) &=& \frac{1}{s_{1234}^2} \Bigg\{ 
         \frac{2}{\Sab^2 \Scd^2} \left[
           \Sac \Sbd - \Sad \Sbc
          \right]^2 
       + \frac{1}{\Sab \Scd}   \Big[
          - 2 \Sac \Sbd
          +  \Sac^2 \nonumber \\ &&
          - 2 \Sad \Sbc
          +  \Sad^2
          +  \Sbc^2
          +  \Sbd^2
          \Big] + 2 + {\cal O}(\e) \Bigg\}\;,
\end{eqnarray}
where only different quark flavours need to be considered, since the identical 
flavour contribution to this final state is finite. 

The integral of this antenna term is
\begin{eqnarray}
{\cal H}_4^0 (s_{1234}) &=&  \left(s_{1234}
\right)^{-2\e} \Bigg[ \frac{1}{9\e^2}  + \frac{7}{9\e} + \left(
  \frac{677}{162}
          - \frac{\pi^2}{6}\right) \Bigg]\;,
\end{eqnarray}
with
\begin{eqnarray}
\Poles\left( {\cal H}_4^0(s_{1234})\right)
&=&
 2 \left[2 {\bf I}^{(1)}_{gg,\Flavour} \left(\e,s_{1234} 
\right) \right]^2 
 \nonumber \\ &&
- 2 e^{-\e \gamma} \frac{\Gamma (1-2\e)}{\Gamma (1-\e)} 
\left(\frac{b_{0,\Flavour}}{\e} + k_{0,\Flavour} \right) \,
\left[2 {\bf I}^{(1)}_{gg,\Flavour}(2\e,s_{1234})\right]
 \nonumber \\ &&
-   {\bf H}^{(2)}_{R,H}(\e,s_{1234}) \;,\\
 \Finite\left( {\cal H}_4^0(s_{1234}) \right) &=& 
\frac{677}{162}
          + \frac{11\pi^2}{36}\;.
\end{eqnarray}
The contribution from identical quarks only to this final state is 
finite, and thus no gluon-gluon antenna function is defined 
for two identical quark-antiquark pairs.

\section{Infrared limits of the antenna subtraction terms}
\label{sec:limits}
The antenna subtraction terms defined in the three previous sections 
encapsulate all single and double unresolved limits of tree-level 
QCD matrix elements and all single unresolved limits of one-loop 
QCD matrix elements. In this section, we list the behaviour of 
all tree-level three-parton and four-parton antenna functions. 
Using this information, it is then possible to employ these antenna functions 
in the construction of complete infrared subtraction functions for 
QCD matrix elements at NNLO.

\subsection{Generalised collinear and soft factors}
The factorisation properties of tree-level QCD squared matrix elements 
at NLO and NNLO have been investigated in detail 
in~\cite{singleun,cs,audenigel,campbell,cg,campbellandother}. At NLO, only 
a single particle can become unresolved, either soft or collinear.
In these limits, the  $(m+1)$-parton matrix element 
factorises into a reduced  $m$-parton matrix element times a
soft eikonal factor or a collinear splitting function. At NNLO, two 
particles can become unresolved in several possible configurations:
double soft, soft/collinear, double single collinear, triple collinear. 
In each of these limits, the  $(m+2)$-parton matrix element 
factorises into a reduced  $m$-parton matrix element times a 
generalised double unresolved factor (double soft factor, soft/collinear 
splitting function, double single collinear splitting function,
triple collinear splitting function). In the following, we list all 
generalised single and double unresolved factors. 

\subsubsection{Single unresolved factors}

When deriving limits of the three-parton antenna functions
in single unresolved configurations, we encounter 
well-known soft eikonal factors when a gluon is soft 
and three different Altarelli-Parisi splitting functions when two 
partons are collinear. These are listed below.

When a soft gluon ($b$)  
is emitted between two hard partons ($a$ and $c$), the 
 eikonal factor $S_{abc}$ factorises off the squared matrix element:
\begin{equation}
S_{abc}\equiv\frac{2s_{ac}}{s_{ab}s_{bc}}.
\label{eq:eikonal}
\end{equation} 
When two partons become collinear,
we have different splitting functions  
corresponding to various final state configurations:  
a quark splits into a quark and a gluon ($P_{qg\to Q}$), 
a gluon splits into a quark-antiquark pair ($P_{q\bar q\to G}$)  
or a gluon splits into two gluons ($P_{gg\to G}$).
These are given by:
\begin{eqnarray}
P_{qg\to Q}(z)&=&\left(\frac{1+(1-z)^2-\e z^2}{z}\right) ,\nonumber\\
P_{q\bar{q}\to G}(z)&=&\left(\frac{z^2+(1-z)^2 -\e}{1-\e}\right),\nonumber \\
P_{gg\to G}(z)&=& 2\left(\frac{z}{1-z}+ \frac{1-z}{z}+z(1-z)\right)\;.
\label{eq:apkernels}
\end{eqnarray}
In these equations, $z$ is the momentum fraction of one of the collinear 
partons and the label $q$ appearing in these splitting function 
can stand for a quark or antiquark:  
$P_{qg\to Q}= P_{\bar{q}g\to \bar{Q}}$ by charge conjugation.
$Q$ or $G$ appearing in the collinear splitting functions denotes the 
parent particle of the two collinear partons $i$ and $j$. In the discussion
below, the momentum associated with this parent particle will be denoted 
by $(ij)$.

\subsubsection{Double unresolved factors}

To describe
the limiting behaviours of the four-parton antenna functions 
in double unresolved configurations, 
we require generalised soft and 
collinear factors. These were first derived in~\cite{audenigel,campbell,
cg,campbellandother}, 
and are listed below. We follow largely the notation of~\cite{campbell}.
\begin{enumerate}

\item Double soft factors

When  two colour connected gluons $(b)$ and $(c)$ are simultaneously soft,
the double soft gluon function for  four partons in the final state 
is given by
\begin{eqnarray}
\lefteqn{
S_{abcd}(s_{ad},s_{ab},s_{cd},s_{bc},s_{abc},s_{bcd})=
\frac{2s_{ad}^2}{s_{ab}s_{bcd}s_{abc}s_{cd}}}\nonumber \\
&&+ \frac{2s_{ad}}{s_{bc}} \left( \frac{1}{s_{ab}s_{cd}}
 + \frac{1}{s_{ab}s_{bcd}} + \frac{1}{s_{cd}s_{abc}}
 - \frac{4}{s_{abc}s_{bcd}} \right)
+ \frac{2(1-\e)}{s_{bc}^2} \left( \frac{s_{ab}}{s_{abc}}
 + \frac{s_{cd}}{s_{bcd}} -1 \right)^2.\nonumber \\
\end{eqnarray}
Here $a$ and $d$ are the hard partons surrounding the soft pair.
\\
In the case when two unconnected gluons become simultaneously soft 
the corresponding double soft gluon factor is the product of 
two eikonal factors $S_{abc}$, given in 
(\ref{eq:eikonal}).   
\\
For the emission of a soft quark-antiquark pair resulting from 
the splitting of an intermediate gluon emitted itself from a primary 
pair of hard partons,
the soft factor takes the form,
\begin{eqnarray}
S_{ab}(c_q,d_{\bar q})=
\frac{2}{s^2_{cd}\,(s_{ac}+s_{ad})(s_{bc}+s_{bd})} 
\left(s_{ab}s_{cd}-s_{ac}s_{bd}-s_{bc}s_{ad}\right) \nonumber \\
+\frac{2}{s^2_{cd}}
\left(\frac{s_{ac}s_{ad}}{(s_{ac}+s_{ad})^2}
+ \frac{s_{bc}s_{bd}}{(s_{bc}+s_{bd})^2}\right)\;,
\end{eqnarray}
with $a$ and $b$ the adjacent hard partons.
Note that this formula is different from the one that can be found in 
\cite{cg} for this limit, where 
 the second line of the above formula is absent. 

\item Soft-collinear factors 

When a gluon ($a$) is soft, partons ($d$,$a$,$b$,$c$) 
are colour connected 
and partons $b$ and $c$ are collinear then 
the limiting behaviour of the squared matrix element is described by
the soft-collinear factor, which is subsequently multiplied with the 
appropriate simple collinear splitting function
\begin{equation}
\label{eq:softfactor}
S_{d;abc}(z,s_{ab},s_{bc},s_{abc},s_{ad},s_{bd},s_{cd}) =
\frac{(s_{bd}+s_{cd})}{s_{ab}s_{ad}}
\left( z + \frac{s_{ab}+z s_{bc}}{s_{abc}} \right)\;.
\end{equation}
In here, $z$ is always the collinear momentum fraction of  
parton $b$ in the collinear pair ($bc$), with $b$ being colour-connected 
to the soft parton $a$.

\item Triple collinear splitting functions 

There are five different triple collinear splitting functions
depending on the nature of the partons which become collinear. 
\\
The colour-ordered triple collinear splitting
function $P_{ggg \rightarrow G}$ is given by
\begin{eqnarray}
\lefteqn{P_{abc \rightarrow G}(w,x,y,
 s_{ab},s_{bc},s_{abc}) =
 2 \times \Biggl\{ } \nonumber \\
 &+& \frac{(1-\e)}{s_{ab}^2s_{abc}^2}\frac{(xs_{abc}-(1-y)s_{bc})^2}{(1-y)^2}
    +\frac{2(1-\e)s_{bc}}{s_{ab}s_{abc}^2} +\frac{3(1-\e)}{2s_{abc}^2}
\nonumber\\
 &+& \frac{1}{s_{ab}s_{abc}} \left(
     \frac{(1-y(1-y))^2}{yw(1-w)}-2\frac{x^2+xy+y^2}{1-y}
    +\frac{xw-x^2y-2}{y(1-y)} +2\e \frac{x}{(1-y)} \right)\nonumber \\
 &+& \frac{1}{2s_{ab}s_{bc}} \left(
    3x^2 - \frac{2(2-w+w^2)(x^2+w(1-w))}{y(1-y)}
    + \frac{1}{yw} + \frac{1}{(1-y)(1-w)} \right) \Biggr\} \nonumber \\
 &+& ( s_{ab} \leftrightarrow s_{bc}, w \leftrightarrow y)\;,
\end{eqnarray}
with $w$, $x$ and $y$ 
being the momentum fractions carried by the collinear gluons $a,b,c$, 
and with $w=(1-x-y)$.  
This splitting function is symmetric under the exchange of the outer gluons
($a$ and $c$),
and contains poles only in $s_{ab}$ and $s_{bc}$.
\\
$\tilde{P}_{q g_{1}g_{2} \to Q}$ is the triple collinear splitting function 
obtained when quark $q$ is collinear to two gluons, $g_1$ and $g_2$, 
which are not colour connected, i.e.\ behave like photons~\cite{audenigel}. 
It reads,
\begin{eqnarray}
\lefteqn{\tilde{P}_{q g_1g_2 \rightarrow Q}
 (w,x,y,s_{qg_1},s_{qg_2},s_{qg_1g_2}) =}\nonumber\\
&+& \frac{1}{2s_{qg_1}s_{qg_2}} \frac{w}{xy}
\left( 1+w^2-\e(x^2+xy+y^2)-\e^2 xy \right) \nonumber \\
 &+&\frac{1}{s_{qg_1}s_{qg_1g_2}} \frac{1}{xy} \left( w(1-x+\e^2
xy)+(1-y)^3-\e(1-y)
  (x^2+xy+y^2)+\e^2 xy \right) \nonumber \\
 &-&\frac{(1-\e)}{s_{qg_1g_2}^2} \left( (1-\e)
  \frac{s_{qg_1}}{s_{qg_2}} - \e \right)
  + ( s_{qg_1} \leftrightarrow s_{qg_2}, x \leftrightarrow y) ,
\end{eqnarray}
with $x$,$y$ and $w$ the momentum fractions carried by the collinear particles
and with $w=(1-x-y)$.  
For the case when the two gluons are instead colour connected,
the colour-ordered splitting function reads,
\begin{eqnarray}
\lefteqn{P_{qg_1g_2 \rightarrow Q}
 (w,x,y,s_{qg_1},s_{qg_2},s_{g_1g_2},s_{qg_1g_2}) =}\nonumber\\
&+& \frac{1}{s_{qg_1}s_{g_1g_2}} \left( (1-\e) \left(
  \frac{1+w^2}{y}+\frac{1+(1-y)^2}{(1-w)} \right)
  +2\e \left( \frac{w}{y}+\frac{1-y}{1-w} \right) \right) \nonumber \\
 &+&\frac{1}{s_{qg_1}s_{qg_1g_2}} \Bigg( (1-\e) \left( \frac{
  (1-y)^3+w(1-x)-2y}{y(1-w)} \right) \nonumber \\ && \hspace{3cm}
  - \e \left( \frac{2(1-y)(y-w)}{y(1-w)} -x \right) -\e^2 x \Bigg) 
\nonumber
\\
 &+&\frac{1}{s_{g_1g_2}s_{qg_1g_2}} \Bigg( (1-\e) \left( \frac{
  (1-y)^2 (2-y)+x^3+2xw-2-y}{y(1-w)} \right)
\nonumber \\ && \hspace{3cm} +2\e \frac{(xw-y-2yw)}{y(1-w)}
\Bigg) \nonumber \\
 &+&(1-\e) \Bigg( \frac{2\left( x{s_{qg_1g_2}}-(1-w)s_{qg_1} \right)^2}
  {s_{g_1g_2}^2s_{qg_1g_2}^2(1-w)^2} \nonumber \\ && \hspace{3cm}
  +\frac{1}{s_{qg_1g_2}^2} \left( 4\frac{s_{qg_1}}{s_{g_1g_2}}
   +(1-\e) \frac{s_{g_1g_2}}{s_{qg_1}} + (3-\e) \right) \Bigg).
\end{eqnarray} 

Similarly the clustering of a gluon with a quark-antiquark pair into
a parent gluon again has two distinct functions depending on whether 
the final state 
is colour-ordered or not, In the colour connected case, the gluon 
is emitted outside the quark-antiquark pair, and one obtains the 
colour-ordered splitting function,
\begin{eqnarray}
\lefteqn{P_{g{\bar q}q \rightarrow G}
 (w,x,y,s_{g\bar q},s_{\bar q q},s_{g{\bar q}q}) =}\nonumber\\
&-& \frac{1}{s_{g{\bar q}q}^2} \left( 4\frac{s_{g{\bar q}}}{s_{\bar q q}}
 +(1-\e)\frac{s_{\bar q q}}{s_{g{\bar q}}} + (3-\e) \right)
 - \frac{2 \left( xs_{g{\bar q}q}-(1-w)s_{g{\bar q}} \right)^2}{s_{\bar q q}^2
  s_{g{\bar q}q}^2(1-w)^2} \nonumber \\
&+& \frac{1}{s_{g{\bar q}}s_{g{\bar q}q}} \left( \frac{(1-y)}{w(1-w)}-y-2w-\e
 -\frac{2x(1-y)(y-w)}{(1-\e)w(1-w)} \right) \nonumber \\
&+&  \frac{1}{s_{g{\bar q}}s_{\bar q q}} \left( \frac{x\left( (1-w)^3-w^3
\right)}
 {w(1-w)} -\frac{2x^2 \left( 1-yw-(1-y)(1-w) \right)}{(1-\e)w(1-w)} \right)
\nonumber \\
&+& \frac{1}{s_{{\bar q}q}s_{g{\bar q}q}} \left( \frac{(1+w^3+4xw)}{w(1-w)}
 +\frac{2x \left( w(x-y)-y(1+w) \right)}{(1-\e)w(1-w)} \right),
\end{eqnarray} 
while in the case where the gluon is emitted between the quark-antiquark pair, 
which is subleading in colour, one obtains a QED-like splitting function,
\begin{eqnarray}
\lefteqn{\tilde{P}_{q g {\bar q} \rightarrow G}
 (w,x,y,s_{qg},s_{g\bar q},s_{\bar q q},s_{qg{\bar q}}) =}\nonumber\\
&-& \frac{1}{s_{qg{\bar q}}^2} \left( (1-\e)
 \frac{s_{q{\bar q}}}{s_{qg}} +1 \right)
 + \frac{1}{s_{g\bar q}s_{qg}} \left( (1+x^2)-\frac{x+2wy}{1-\e} \right)
\nonumber \\
&-& \frac{1}{s_{qg}s_{qg{\bar q}}} \left( 1+2x+\e-\frac{2(1-y)}{(1-\e)} \right)
+ ( s_{qg} \leftrightarrow s_{g\bar q}, w \leftrightarrow y).
\end{eqnarray}
Lastly, the clustering of a quark-antiquark
pair~($q'\bar{q}'$) and a quark~($q$) to form a parent 
quark $Q$ with the same flavour as $q$.
The splitting function depends upon
whether or not the quarks are of identical flavour,
\begin{eqnarray}
\label{qQQQ}
P_{q{\overline{q}'}q' \rightarrow Q} =
 P_{q{\overline{q}}'q' \rightarrow Q}^{{\rm non-ident.}}
-\frac{\delta_{qq'}}{N} P_{q{\overline{q}}'q' \rightarrow Q}^{{\rm
ident.}},
\end{eqnarray}
where $\delta_{qq'}=1$ for identical flavour quarks.
If quarks $q_1$, ${q'}_3$ and ${\overline{q}'}_4$ are clustered to form $Q$,
\begin{eqnarray}
\lefteqn{P_{q\overline{q}'q' \rightarrow Q}^{{\rm non-ident.}}(w,x,y,
 s_{q\overline{q}'},s_{q'\overline{q}'},s_{qq'\overline{q}'}) =}\nonumber\\
 &-& \frac{1}{s_{qq'\overline{q}'}^2} \left( (1-\e)
 +\frac{2s_{q\overline{q}'}}{s_{q'\overline{q}'}} \right)  
-\frac{2\left( x s_{qq'\overline{q}'}-(1-w) s_{q\overline{q}'} \right)^2}
  {s_{q'\overline{q}'}^2s_{qq'\overline{q}'}^2(1-w)^2}\nonumber\\
  &+&\frac{1}{s_{q'\overline{q}'}s_{qq'\overline{q}'}} \left(
  \frac{1+x^2+(x+w)^2}{(1-w)} - \e(1-w) \right).
\end{eqnarray}
When the flavours of the clustering quarks are the same, there is an
additional contribution coming from the interference
terms of the four-quark matrix elements, which reads
\begin{eqnarray}
\lefteqn{P_{q\overline{q}'q' \rightarrow Q}^{{\rm ident.}}(w,x,y,
 s_{q\overline{q}'},s_{q\overline{q}'},s_{qq'\overline{q}'}) =} \nonumber\\
 &-& \frac{(1-\e)}{s_{qq'\overline{q}'}^2} \left(
  \frac{2s_{q\overline{q}'}}{s_{q'\overline{q}'}}+2+\e \right)
  -\frac{1}{2s_{q\overline{q}'}s_{q'\overline{q}'}} \left(
  \frac{x(1+x^2)}{(1-y)(1-w)}-\e x \left(
  \frac{2(1-y)}{(1-w)}+1+\e \right) \right)
\nonumber \\
 &+& \frac{1}{s_{q'\overline{q}'}s_{qq'\overline{q}'}} \left(
  \frac{1+x^2}{(1-y)}+\frac{2x}{(1-w)} - \e \left(
  \frac{(1-w)^2}{(1-y)}+(1+x)+\frac{2x}{(1-w)}+\e(1-w) \right) \right)
\nonumber \\
 &+& ( s_{q\overline{q}'} \leftrightarrow s_{q'\overline{q}'}, 
y \leftrightarrow w).
\end{eqnarray}
This identical flavour splitting function contains poles
 when ${\overline{q}'}$
clusters with both ${q'}$ and $q$. It is symmetric under $q_1 \leftrightarrow 
{q'}_3$.
\item Double collinear splitting functions

If two distinct pairs of  
partons become simultaneously 
collinear, the $(m+2)$-parton 
squared matrix element factorises into the product of 
two simple collinear splitting functions (\ref{eq:apkernels}) with
the $m$-parton squared matrix element.
\end{enumerate}

\subsection{Quark-antiquark antennae}
The three-parton and four-parton quark-antiquark antenna functions 
were derived in Section~\ref{sec:qq}
from the real radiation corrections to $\gamma^*\to q\bar q$. Their
behaviour in all limits where one or two partons become unresolved are
described below.

\subsubsection{Three-parton antenna functions}
We have only one three-parton quark-antiquark antenna function:
$A_{3}^{0}(1_q,3_g,2_{\bar q})$, which 
has the following limits:
\begin{enumerate}
\item Soft limit:
\begin{eqnarray}
A_{3}^0(1,3,2)
& \stackrel{3_g \to 
0}{\longrightarrow}& 
S_{132}\;. 
\end{eqnarray}
\item Collinear limit:
\begin{eqnarray}       
A_{3}^0(1,3,2)
& \stackrel{1_q \parallel
3_g}{\longrightarrow}&  \frac{1}{s_{13}}\,P_{qg\to Q}(z).
\end{eqnarray}
where the momenta of the quark $(1_q)$ and the antiquark $(2_{\bar q})$ 
can be interchanged.
\end{enumerate}

\subsubsection{Four-parton antenna functions}

The NNLO real radiation corrections to $\gamma^*\to q\bar q$ yield
four different four-parton quark-antiquark antenna functions:
the leading and subleading colour quark-antiquark-gluon-gluon 
antennae $A_4^0(1_q,3_g,4_g,2_{\bar q})$, 
$\tilde{A}_4^0(1_q,3_g,4_g,2_{\bar q})$, as well as the 
antennae with two quark-antiquark pairs of different and identical flavour
$B_4^0(1_q,3_{q'},4_{\bar q'},2_{\bar q})$ and 
$C_4^0(1_q,3_{q},4_{\bar q},2_{\bar q})$. Their single and double 
unresolved limits are described in the following. 
We shall always 
restrict ourselves to the non-vanishing limits only.

In $A_4^0(1,3,4,2)$, the colour-ordering of the gluonic emissions ensures 
that only neighbouring partons can form singular configurations. 
While singularities are present in $1/s_{13}$ and $1/s_{24}$, no singularities 
in $1/s_{14}$ and $1/s_{23}$ appear.

For this colour-ordered 
antenna function, we find the following non-vanishing double unresolved limits:
\begin{enumerate}
\item Double soft and soft-collinear limits: 
\begin{eqnarray}
A_4^0(1,3,4,2)&\stackrel{3_g \to
  0,4_g \to 0}{\longrightarrow}& S_{1342}\;, \nonumber\\ 
A_4^0(1,3,4,2)&\stackrel{3_g \to 0,4_g\parallel 2_{\bar q}}
{\longrightarrow}& S_{1;342}(z)\;\frac{1}{s_{24}}\; P_{qg\to Q}(z)\;,
\nonumber\\
A_4^0(1,3,4,2)&\stackrel{4_g \to 0,3_g\parallel 1_q}
{\longrightarrow}&S_{2;431}(z)\;\frac{1}{s_{13}}\; P_{qg\to Q}(z)\;.
\end{eqnarray}

\item Triple collinear limits
\begin{eqnarray}
A_4^0(1,3,4,2)&\stackrel{1_q \parallel 3_g \parallel 4_g}
{\longrightarrow}& P_{134 \to Q}(w,z,y)\;,
\end{eqnarray}
with quark ($1_q$) and antiquark ($2_{\bar q}$) being interchangeable.   
\item Double collinear limits:
\begin{eqnarray} 
A_4^0(1,3,4,2)&\stackrel{1_q\parallel 3_g,2_{\bar q}\parallel g_4}
{\longrightarrow}& \frac{1}{s_{13}}\; P_{qg\to Q}(z)\;
\frac{1}{s_{24}}\; P_{\bar qg\to \bar Q}(y)\;.
\end{eqnarray}
\end{enumerate}
We find the following non-vanishing single unresolved limits:
\begin{enumerate}
\item Soft limits:
\begin{eqnarray}
A_4^0(1,i,j,2)&\stackrel{i_g \to 0}
{\longrightarrow}& S_{1ij}\;A_3^0(1,j,2)\; , \nonumber\\ 
A_4^0(1,i,j,2)&\stackrel{j_g \to 0}
{\longrightarrow}& S_{ij2}\;A_3^0(1,i,2)\; . 
\end{eqnarray}
\item Collinear limits (quark-gluon):
\begin{eqnarray}
A_4^0(1,3,4,2)&\stackrel{1_q \parallel 3_g}
{\longrightarrow}& \frac{1}{s_{13}}\; P_{qg\to Q}(z)\;A_3^0((13),4,2)\;,
\nonumber\\ 
A_4^0(1,3,4,2)&\stackrel{2_{\bar q} \parallel 4_g}
{\longrightarrow}& \frac{1}{s_{24}}\; P_{qg\to Q}(z)\;A_3^0(1,3,(24))\;,
\end{eqnarray}
where the parent parton of the collinear partons ($1$) and
($3$) (or ($2$) and ($4$))  is denoted $(13)$ (or $(24)$) 
in the three-parton antenna function $A_3^0$.
\item Collinear limit (gluon-gluon):
\begin{eqnarray}
A_4^0(1,3,4,2)&\stackrel{3_g \parallel 4_g}
{\longrightarrow}& 
\frac{1}{s_{34}}\; P_{gg\to G}(z)\;A_3^0(1,(34),2) \,+\,{\rm ang.}\;,
\end{eqnarray}
where (ang.) means that angular terms are also present here. These 
do however cancel after integration over the unresolved phase space. 
A detailed discussion of these angular terms can be found 
in Section~\ref{sec:ang} below.
\end{enumerate}

The limits of the other colour-ordered subtraction term, 
$A_4^0(1,4,3,2)$,  can be inferred from the limits 
of $A_4^0(1,3,4,2)$ listed above  
 by interchanging 
the momenta of the two final state gluons $(3_g)$ and $(4_g)$.

The subleading colour contribution $\tilde{A}_4^0(1,3,4,2)$ is 
not colour-ordered, and therefore symmetric under the interchange of the 
two gluon momenta $3$ and $4$.
We find the following non-vanishing double unresolved limits:
\begin{enumerate}
\item Double soft and soft-collinear limits:
\begin{eqnarray}
\tilde{A}_4^0(1,3,4,2)&\stackrel{3_g \to
  0,4_g \to 0}{\longrightarrow}& S_{132}S_{142}\;, 
\nonumber\\ 
 \tilde{A}_4^0(1,i,j,2)&\stackrel{g_i \to 0,g_j\parallel q_2}
{\longrightarrow}&
\tilde{S}_{1;ij2}(z)\;\frac{1}{s_{j2}}\; P_{qg\to Q}(z)\;,
\end{eqnarray}
with $i$,$j$ standing each for one of the two intermediate 
gluons. Quark ($1_q$) and antiquark ($2_{\bar q}$)
are interchangeable.   
\item Triple collinear limits:
\begin{eqnarray}
\tilde{A}_4^0(1,3,4,2)&\stackrel{1_q \parallel 3_g \parallel 4_g}
{\longrightarrow}& 
\tilde{P}_{134 \to Q}(w,x,y)\;,
\end{eqnarray}
with quark ($1_q$) and antiquark ($2_{\bar q}$) being interchangeable. 
\item Double collinear limits:
\begin{eqnarray} 
\tilde{A}_4^0(1,i,j,2)&\stackrel{1_q\parallel i_g,2_{\bar q}\parallel j_g}
{\longrightarrow}& \frac{1}{s_{1i}}\; P_{qg\to Q}(z)\;
\frac{1}{s_{2j}}\; P_{\bar qg\to \bar Q}(y)\;,
\end{eqnarray}
with $i$ and $j$ each standing for one of the  two gluons 
$(3_g)$ or $(4_g)$; quark $(1_q)$ and antiquark 
$(2_{\bar q})$ can also be exchanged.
\end{enumerate}

The non-vanishing single unresolved limits are:
\begin{enumerate}
\item Soft limits:
\begin{eqnarray}
\tilde{A}_4^0(1,i,j,2)&\stackrel{i_g \to 0}
{\longrightarrow}& S_{1i2}\;A_3^0(1,j,2)\;, 
\end{eqnarray}
where $i$ and $j$ represent the  two gluons $(3_g)$ and $(4_g)$.   
\item Collinear limits:
\begin{eqnarray}
\tilde{A}_4^0(1,i,j,2)&\stackrel{1_q \parallel i_g}
{\longrightarrow}& \frac{1}{s_{1i}}\; P_{qg\to Q}(z)\;A_3^0((1i),j,2)
\;.\end{eqnarray}
\end{enumerate}
Again, $(1_q)$ and $(2_{\bar q})$ can be interchanged and $i$ can stand 
for one of the two gluons $(3_g)$ or $(4_g)$.

For the antenna function containing  two quark-antiquark pairs of 
non-identical flavours, 
$B_4^0(1_q,3_{q'},4_{\bar q'},2_{\bar q})$, 
we find the following double unresolved limits:
\begin{enumerate}
\item Double soft limit: 
\begin{eqnarray}
B_4^0(1,3,4,2)&\stackrel{3_{q'} \to
  0, 4_{\bar q'} \to 0}{\longrightarrow}& 
S_{12}(3,4)\;.
\end{eqnarray}
\item Triple collinear limit:
\begin{eqnarray}
B_4^0(1,3,4,2)&\stackrel{1_q \parallel 3_{q'} \parallel 4_{\bar q'}}
{\longrightarrow}&
P^{{\rm non-ident.}}_{1  
4 3\to Q}(x,y)\;.
\end{eqnarray}
The quark ($1_q$) and antiquark ($2_{\bar q}$) momenta can be interchanged.
\end{enumerate}

$B_4^0(1,3,4,2)$ contains only one single unresolved limit
($3_{q}\parallel 4_{\bar{q}}$):
\begin{eqnarray}
B_4^0(1,3,4,2)&\stackrel{3_{q'} \parallel 4_{\bar q'}}
{\longrightarrow}& \frac{1}{s_{34}} \;P_{q\bar{q}\to G}(z) \; 
A_3^0(1,(34),2) + {\rm ang.}\;,
\end{eqnarray}
where (ang.) means that angular terms are also obtained here.

Finally we consider the identical-flavour-only antenna function 
denoted by $C_4^0(1_q,3_q,4_{\bar{q}},2_{\bar{q}})$.
This function is proportional has only one non-vanishing double 
unresolved limit: triple collinear  ($2_{\bar q}\parallel 
4_{\bar q}\parallel 3_q$),
\begin{eqnarray}
C_4^0(1,3,4,2)&\stackrel{2_{\bar q} \parallel 3_q\parallel  4_{\bar{q}}}
 {\longrightarrow}& \frac{1}{2}\;
P^{{\rm ident.}}_{234\to \bar{Q}}(w,x,y) .
\label{eq:c40lim}
\end{eqnarray}   
This antenna function has no further non-vanishing 
(double or single unresolved) limits, although terms in 
$1/s_{24}$ and $1/s_{34}$ are present in the function.
In the respective limits, the coefficients of these terms 
vanish. 

\subsection{Quark-gluon antennae}

The quark-gluon antenna functions were derived in Section~\ref{sec:qg}
above. We describe their behaviour in all single and double unresolved 
limits in the following. 
Again, we will only list the non-vanishing contributions.
In all these functions, the quark $(1_q)$ can 
also represent an antiquark. 
 
\subsubsection{Three-parton antenna functions} 
The tree-level three-parton quark-gluon antenna functions 
 represent  two final states: quark-gluon-gluon ($D_{3}^0(1_q,3_g,4_g)$) and 
quark-quark-antiquark  ($E_{3}^0(1_q,3_{q'},4_{\bar{q}'})$).

The tree-level antenna function $D_3^0(1_q,3_g,4_g)$
contains two colour orderings corresponding to 
the following configurations: gluon $(3_g)$ radiated between quark $(1_q)$ 
and gluon $(4_g)$ 
denoted by $d_3^0(1,3,4)$ and 
gluon $(3_g)$ radiated between quark $(1_q)$ 
and gluon $(4_g)$ 
denoted by $d_3^0(1,4,3)$, such that 
\begin{displaymath}
D_3^0(1,3,4) = d_3^0(1,3,4) + d_3^0(1,4,3) \; .
\end{displaymath}
The separation between these 
is not free from an ambiguity. It is constructed in such a way that 
the collinear limit of the two gluons has to be split between 
the two configurations. The decomposition used here is  stated in 
(\ref{eq:smalld}). 

The simple unresolved limits of $D_3^0(1,3,4)$ and $d_3^0(1,3,4)$
are stated below. For the sake of clarity, we also list some non-trivial
vanishing limits:
\begin{enumerate}
\item Soft limits:
\begin{eqnarray}
D_3^0(1,i,j)&\stackrel{i_g \to
  0}{\longrightarrow}& S_{1ij}\;, \nonumber\\
D_3^0(1,i,j)&\stackrel{j_g \to
  0}{\longrightarrow}& S_{1ji}\;, \nonumber\\
d_3^0(1,i,j)&\stackrel{i_g \to
  0}{\longrightarrow}& S_{1ij} 
\;, \nonumber\\
d_3^0(1,i,j)&\stackrel{j_g \to
  0}{\longrightarrow}& 0 \;.
\end{eqnarray}
\item Collinear limits:
\begin{eqnarray}
D_3^0(1,3,4)&\stackrel{ 3_g \parallel
4_g}{\longrightarrow}& \frac{1}{s_{34}}\,P_{gg\to G}(z)\;, \nonumber \\
d_3^0(1,3,4)&\stackrel{ 3_g \parallel
4_g}{\longrightarrow}& \frac{1}{s_{34}}\,
\left(P_{gg\to G}(z)-\frac{2 z}{1-z} - z(1-z)\right)\;, \nonumber\\
d_3^0(1,4,3)&\stackrel{ 3_g \parallel
4_g}{\longrightarrow}& \frac{1}{s_{34}}\,
\left(P_{gg\to G}(z)-\frac{2(1-z)}{z}- z(1-z) \right)\;, \nonumber\\
D_3^0(1,i,j)&\stackrel{ 1_q \parallel
i_g}{\longrightarrow}& \frac{1}{s_{1i}}\,P_{qg \to  Q}(z)\;, \nonumber \\
D_3^0(1,i,j)&\stackrel{ 1_q \parallel
j_g}{\longrightarrow}& \frac{1}{s_{1j}}\,P_{qg \to  Q}(z)\;, \nonumber \\
d_3^0(1,i,j)&\stackrel{ 1_q \parallel
i_g}{\longrightarrow}& \frac{1}{s_{1i}}\,P_{qg \to Q}(z)\;, \nonumber \\
d_3^0(1,i,j)&\stackrel{ 1_q \parallel
j_g}{\longrightarrow}& 0 \;.
\end{eqnarray}
\end{enumerate}

The antenna function 
$E_{3}^0(1_q,3_{q'},4_{\bar{q}'})$ has only one singular behaviour, 
when the quark-antiquark pair  becomes collinear:
\begin{eqnarray}
E_3^0(1,3,4)&\stackrel{ 3_{q'}\parallel
4_{\bar{q}'}}{\longrightarrow}& \frac{1}{s_{34}}\,P_{q\bar{q}\to G}(z)\;.
\end{eqnarray}

\subsubsection{Four-parton antenna functions}
\label{sec:D40limit}
The NNLO real radiation corrections to $\tilde{\chi}\to \tilde{g}g$ yield
three different four-parton quark-gluon antenna functions:
the quark-gluon-gluon-gluon 
antenna contains only a leading colour (colour-ordered)
term $D_4^0(1_q,3_g,4_g,5_g)$, while the  
quark-quark-antiquark-gluon antennae have 
leading colour and subleading colour contributions
$E_4^0(1_q,3_{q'},4_{\bar q'},5_g)$ and 
$\tilde{E}_4^0(1_q,3_{q'},4_{\bar q'},5_g)$. By construction, no 
identical-flavour contribution appears. 
The non-vanishing single and double 
unresolved limits of these antenna functions 
are described in the following. As before, some of the single unresolved 
limits also contain angular terms, which will be discussed in 
more detail in Section~\ref{sec:ang} below.

The colour-ordered  quark-gluon-gluon-gluon antenna function 
 $D_4^0(1_q,3_g,4_g,5_g)$ contains several different antenna configurations: 
due to the cyclic nature of the colour indices of the $\tilde{\chi} \to 
\tilde{g} ggg$ matrix element~\cite{chi}, each pair of two neighbouring
partons can represent the two hard partons forming the antenna. This
behaviour is in contrast to the quark-antiquark antennae described in the 
previous subsection, where the colour-ordering ensures that 
the primary quark-antiquark pair always forms the hard partons. 
Consequently, $D_4^0$ contains considerably more 
double and single unresolved limits than the four-parton quark-antiquark 
antenna functions studied above. 

For the double unresolved limits we find:
\begin{enumerate}
\item Double soft and soft-collinear limits:
\begin{eqnarray}
D_4^0(1,3,4,5)&\stackrel{ 3_g \to 0, 4_g \to 0}
{\longrightarrow}&\; S_{1345}\;,\nonumber\\
D_4^0(1,3,4,5)&\stackrel{ 4_g \to 0, 5_g \to 0}
{\longrightarrow}&\; S_{1543}\;,\nonumber\\
D_4^0(1,3,4,5)&\stackrel{ 3_g \to 0, 5_g \to 0}
{\longrightarrow}&\; S_{134}\;S_{154}\;,
\end{eqnarray}
\begin{eqnarray} 
D_4^0(1,3,4,5)&\stackrel{ 1_q\parallel 5_g, 3_g \to 0}
{\longrightarrow}& S_{4;315}(z) \;
\frac{1}{s_{15}}\; P_{qg\to Q}(1-z)
\;,\nonumber \\
D_4^0(1,3,4,5)&\stackrel{ 4_g\parallel 5_g, 3_g \to 0}
{\longrightarrow}& S_{1;345}(z) \;
\frac{1}{s_{45}}\; P_{gg\to G}(z)
\;,\nonumber \\
D_4^0(1,3,4,5)&\stackrel{ 1_q\parallel 3_g, 4_g \to 0}
{\longrightarrow}& S_{5;431}(z) \;
\frac{1}{s_{13}}\; P_{qg\to Q}(z)
\;,\nonumber \\
D_4^0(1,3,4,5)&\stackrel{ 1_q\parallel 5_g, 4_g \to 0}
{\longrightarrow}& S_{3;451}(z) \;
\frac{1}{s_{15}}\; P_{qg\to Q}(z)
\;,\nonumber \\
D_4^0(1,3,4,5)&\stackrel{ 1_q\parallel 3_g, 5_g \to 0}
{\longrightarrow}& S_{4;513}(z) \;
\frac{1}{s_{13}}\; P_{qg\to Q}(1-z)
\;,\nonumber \\
D_4^0(1,3,4,5)&\stackrel{ 3_g\parallel 4_g, 5_g \to 0}
{\longrightarrow}& S_{1;543}(z) \;
\frac{1}{s_{34}}\; P_{gg\to G}(z)\;.
\end{eqnarray}
\item Triple collinear limits:
\begin{eqnarray}
D_4^0(1,3,4,5)&\stackrel{ 1_q\parallel 3_g  \parallel 4_g }
{\longrightarrow}& P_{134 \to Q}(w,x,y)\;,\nonumber\\
D_4^0(1,3,4,5)&\stackrel{ 1_q\parallel 5_g \parallel 4_g}
{\longrightarrow}& P_{154 \to Q}(w,x,y)\;,\nonumber\\
D_4^0(1,3,4,5)&\stackrel{ 1_q \parallel 3_g \parallel 5_g}
{\longrightarrow}& \tilde{P}_{135 \to Q}(w,x,y)\;,\nonumber\\
D_4^0(1,3,4,5)&\stackrel{ 3_g\parallel 4_g \parallel 5_g}
{\longrightarrow}& P_{345 \to G}(w,x,y)\;.
\end{eqnarray}
In these, the presence of the triple collinear  limit $(1_q\parallel 3_g
\parallel 5_g)$ is 
particularly noteworthy. This limit appears due to the cyclic nature of the 
$D_4^0$ antenna function. In contrast to the other two triple collinear limits 
involving $(1_q)$, where the leading colour triple collinear splitting 
function $P_{1ij \to Q}$ appears, this limit is controlled by the 
subleading colour splitting function $\tilde{P}_{135 \to Q}$, since the 
gluons $(3_g)$ and $(5_g)$ are not directly colour-connected. Applying 
$D_4^0$ as antenna subtraction term to a physical multi-parton matrix element,
special care has to be taken about this particular limit, which is a priori
oversubtracted. 
\item Double collinear limits:
\begin{eqnarray}
D_4^0(1,3,4,5)&\stackrel{ 1_q \parallel 3_g,  4_g \parallel 5_g}
{\longrightarrow}& 
\frac{1}{s_{13} s_{45}}\;
P_{qg\to Q}(z)\; P_{g g\to G}(y)\;,\nonumber \\
D_4^0(1,3,4,5)&\stackrel{1_q \parallel 5_g,  3_g \parallel 4_g }
{\longrightarrow}& 
\frac{1}{s_{15} s_{34}}\;
P_{qg\to Q}(z)\; P_{ g g\to G }(y)\;.
\end{eqnarray}
\end{enumerate}
For the single unresolved limits we obtain:
\begin{enumerate}
\item Soft limits:
\begin{eqnarray}
D_4^0(1,3,4,5)&\stackrel{{3}_g\to 0}
{\longrightarrow}&S_{134}\; D_3^0(1,4,5)\;,\nonumber\\
D_4^0(1,3,4,5)&\stackrel{{4}_g\to 0}
{\longrightarrow}&S_{345}\; D_3^0(1,3,5)\;,\nonumber\\
D_4^0(1,3,4,5)&\stackrel{{5}_g\to 0}
{\longrightarrow}&S_{154}\; D_3^0(1,3,4)\;.
\end{eqnarray}
\item Collinear limits 
\begin{eqnarray}
D_4^0(1,3,4,5)&\stackrel{1_q \parallel 3_g}
{\longrightarrow}&\frac{1}{s_{13}}\;P_{qg\to Q}(z)\; 
D_3^0((13),4,5)\;,\nonumber\\
D_4^0(1,3,4,5)&\stackrel{1_q \parallel 5_g}
{\longrightarrow}&\frac{1}{s_{15}}\;P_{qg\to Q}(z)\; 
D_3^0((15),3,4)\;,\nonumber\\
D_4^0(1,3,4,5)&\stackrel{3_g \parallel 4_g}
{\longrightarrow}&
\frac{1}{s_{34}}\;P_{gg\to G}(z)\; 
D_3^0((1,(34),5) + {\rm ang.}\;,\nonumber\\
D_4^0(1,3,4,5)&\stackrel{4_g \parallel 5_g}
{\longrightarrow}&
\frac{1}{s_{45}}\;P_{gg\to G}(z)\; 
D_3^0((1,3,(45)) + {\rm ang.}\;.
\end{eqnarray}
\end{enumerate}

The leading colour quark-quark-antiquark-gluon antenna function 
$E_4^0(1_q,3_{q'},4_{\bar q'},5_g)$ is colour-ordered: the gluon 
$5_g$ is emitted only between the primary quark $(1_q)$ and the secondary 
antiquark $(4_{\bar q'})$. The second colour-ordering 
$E_4^0(1_q,4_{\bar q'},3_{q'},5_g)$ is obtained by exchanging the 
secondary quark and antiquark momenta. 

For the double unresolved limits of this antenna function we find:
\begin{enumerate}
\item Double soft and soft-collinear limits:
\begin{equation}
E_4^0(1,3,4,5)\stackrel{ 3_{q'} \to 0, 4_{\bar{q}'} \to 0}
{\longrightarrow} S_{15}(3,4)\;,
\end{equation}
\begin{equation} 
E_4^0(1,3,4,5)\stackrel{ 3_{q'}\parallel 4_{\bar q'}, 5_g \to 0}
{\longrightarrow}   S_{1;543}(z) \;
\frac{1}{s_{34}}\; P_{q\bar q\to G}(z)\;.
\end{equation}
\item Triple collinear limits:
\begin{eqnarray}
E_4^0(1,3,4,5)&\stackrel{ 1_q\parallel 3_{q'}  \parallel 4_{\bar q'}}
{\longrightarrow} &
P^{{\rm non-ident.}}_{134 \to Q}(w,x,y) 
\;,\nonumber\\
E_4^0(1,3,4,5)& \stackrel{ 3_{q'}  \parallel 4_{\bar q'}\parallel 5_g }
{\longrightarrow}& P_{543 \to G}(w,x,y) \;.
\end{eqnarray}
\item Double collinear limit:
\begin{eqnarray} 
E_4^0(1,3,4,5)&\stackrel{ 1_q\parallel 5_g , 3_{q'}  \parallel 4_{\bar q'}}
{\longrightarrow}& 
\frac{1}{s_{34} s_{15}}\;
P_{qg\to Q}(z)\; P_{q \bar{q}\to G}(y)\;.
\end{eqnarray}
\end{enumerate}
For the single unresolved limits we obtain:
\begin{enumerate}
\item Soft limit:
\begin{eqnarray}
E_4^0(1,3,4,5)&\stackrel{5_g\to 0}
{\longrightarrow}& S_{154}\; E_3^0(1,3,4)\;,
\end{eqnarray}
\item Collinear limits: 
\begin{eqnarray}
E_4^0(1,3,4,5)&\stackrel{3_{q'} \parallel 4_{\bar q'}}
{\longrightarrow}&
 \frac{1}{s_{34}}\;P_{q\bar q\to G}(z)\; 
D_3^0(1,(34),5) + {\rm ang.} \;,\nonumber\\
E_4^0(1,3,4,5)&\stackrel{ 4_{\bar q'}\parallel 5_g}
{\longrightarrow}&  
\frac{1}{s_{45}}\;P_{qg\to Q}(z)\; E_3^0(1,3,(45))\;, \nonumber \\
E_4^0(1,3,4,5)&\stackrel{ 1_q \parallel 5_g}
{\longrightarrow}&  
\frac{1}{s_{15}}\;P_{qg\to Q}(z)\; E_3^0((15),3,4)\;.
\end{eqnarray}
\end{enumerate}

In the subleading colour antenna
$\tilde{E}_4^0(1_q,3_{q'},4_{\bar q'},5_g)$ 
the gluon $(5_g)$ is radiated between 
the secondary  quark-antiquark pair($3_{q'}, 4_{\bar q'}$).
As a consequence, no singular structure  involves the primary quark $(1_q)$,
thus limiting the number of singular configurations contained in this antenna.
It contains only one double unresolved configuration, triple collinear 
$(3_{q'}\parallel 4_{{\bar q}'}\parallel 5_g)$:
\begin{eqnarray}
\tilde{E}_4^0(1,3,4,5)&\stackrel{ 3_{q'} \parallel 4_{\bar q'} \parallel
5_g}
{\longrightarrow}& 
\tilde{P}_{534\to G}(w,x,y). 
\end{eqnarray}
The single unresolved limits always involve the gluon $(5_g)$. They read:
\begin{enumerate}
\item Soft limit: 
\begin{eqnarray}
\tilde{E}_4^0(1,3,4,5)&\stackrel{ g_5 \to 0}
{\longrightarrow}&  S_{354}\; E_3^0(1,3,4)\;.
\end{eqnarray}
\item Collinear limits: 
\begin{eqnarray}
\tilde{E}_4^0(1,i,j,5)\stackrel{ i \parallel g_5}
{\longrightarrow}&  \frac{1}{s_{i5}}\;P_{qg\to Q}(z)
\;E_3^0(1,(i5),j)\; ,
\end{eqnarray}
where $i$ and $j$ can both play the role of the quark $(3_{q'})$ 
or the antiquark $(4_{{\bar q}'})$.
\end{enumerate}

\subsection{Gluon-gluon antennae}

The gluon-gluon antenna functions were derived in Section~\ref{sec:gg}
above. Their behaviour in all single and double unresolved limits is 
summarised in the following, where only the non-vanishing contributions 
are given. 

\subsubsection{Three-parton antenna functions}

There are two tree-level three-parton gluon-gluon antenna functions:
gluon-gluon-gluon $F_3^0(1_g,2_g,3_g)$ and 
gluon-quark-antiquark $G_3^0(1_g,3_q,4_{\bar{q}})$.

The tree-level antenna function $F_3^0(1_g,2_g,3_g)$  contains three antenna
configurations since each pair of gluons can represent the hard 
partons, emitting the remaining third gluon. Each of these configurations 
is denoted by $f_3^0(i,j,k)$:
\begin{displaymath}
F_3^0(1,2,3) = f_3^0(1,2,3) + f_3^0(1,3,2) + f_3^0(2,1,3)  \; .
\end{displaymath}
 As with the quark-gluon-gluon antenna function,
this decomposition into antenna configurations is not unambiguous, the
decomposition used here is stated in (\ref{eq:smallf}).

The simple unresolved limits of $F_3^0(1,2,3)$ and $f_3^0(1,2,3)$ are:
\begin{enumerate}
\item Soft limits:
\begin{eqnarray}
F_3^0(1,2,3)&\stackrel{1_g \to
  0}{\longrightarrow}& S_{213} \;,\nonumber\\
F_3^0(1,2,3)&\stackrel{2_g \to
  0}{\longrightarrow}& S_{123} \;,\nonumber\\
F_3^0(1,2,3)&\stackrel{3_g \to
  0}{\longrightarrow}& S_{132} \;,\nonumber\\
f_3^0(1,2,3)&\stackrel{2_g \to
  0}{\longrightarrow}& S_{123}  \;,\nonumber\\
f_3^0(1,2,3)&\stackrel{1_g \to
  0}{\longrightarrow}& 0 \;, \nonumber\\
f_3^0(1,2,3)&\stackrel{3_g \to
  0}{\longrightarrow}& 0 \;,
\end{eqnarray}
where we also include the non-trivial vanishing limits for clarity.
\item Collinear limits:
\begin{eqnarray}
F_3^0(1,2,3)&\stackrel{ 1_g \parallel
1_g}{\longrightarrow}& \frac{1}{s_{12}}\,P_{gg\to G}(z) \;,\nonumber \\
F_3^0(1,2,3)&\stackrel{ 1_g \parallel
2_g}{\longrightarrow}& \frac{1}{s_{13}}\,P_{gg\to G}(z) \;,\nonumber \\
F_3^0(1,2,3)&\stackrel{ 2_g \parallel
3_g}{\longrightarrow}& \frac{1}{s_{23}}\,P_{gg\to G}(z) \;,\nonumber\\
f_3^0(1,2,3)&\stackrel{ 1_g \parallel
2_g}{\longrightarrow}& \frac{1}{s_{12}}\,
\left(P_{gg\to G}(z)-\frac{2 z}{1-z} - z(1-z)\right)\;,\nonumber \\
f_3^0(1,2,3)&\stackrel{ 2_g \parallel
3_g}{\longrightarrow}& \frac{1}{s_{23}}\,
\left(P_{gg\to G}(z)-\frac{2 (1-z)}{z} - z(1-z)\right)\;, \nonumber \\
f_3^0(1,2,3)&\stackrel{ 1_g \parallel
3_g}{\longrightarrow}& 0  \;.
\end{eqnarray}
\end{enumerate}

The gluon-quark-antiquark
antenna function
 has only one unresolved configuration: if  
quark and antiquark are collinear. In this case,
\begin{eqnarray}
G_3^{0}(1,3,4)&\stackrel{ 3_q \parallel
4_{\bar{q}}}{\longrightarrow}& \frac{1}{s_{34}}\,P_{ q\bar{q}\to G}(z). 
\end{eqnarray}

\subsubsection{Four-parton antenna functions}
The NNLO real radiation corrections to $H\to gg$ yield
four different four-parton gluon-gluon antenna functions:
the gluon-gluon-gluon-gluon 
antenna contains only a leading colour (colour-ordered)
term $F_4^0(1_g,2_g,3_g,4_g)$, while the  
gluon-quark-antiquark-gluon antennae contain both 
leading colour and subleading colour contributions
$G_4^0(1_g,3_{q},4_{\bar q},2_g)$ and 
$\tilde{G}_4^0(1_g,3_{q},4_{\bar q},2_g)$. 
Finally, there is also a quark-antiquark-quark-antiquark antenna function
$H_4^0(1_q,2_{\bar q},3_{q'},4_{\bar q'})$, 
where the quarks are of different flavour. Angular  terms are
indicated where appropriate and will be discussed in detail in 
Section~\ref{sec:ang} below.

Like already observed in the quark-gluon case, 
the colour-ordered  gluon-gluon-gluon-gluon antenna function 
 $F_4^0(1_g,2_g,3_g,4_g)$ contains several different antenna configurations: 
due to the cyclic nature of the colour indices of the $H \to 
g ggg$ matrix element~\cite{h}, each pair of two neighbouring
partons can represent the two hard partons forming the antenna. 
In the following, the set $(ijkl)$ represents any of the ordered permutations
$(1234,2341,3412,4123)$.

The double unresolved limits are: 
\begin{enumerate}
\item Double soft and soft-collinear limits: 
\begin{eqnarray}
F_4^0(1,2,3,4)&\stackrel{ j_g \to 0, k_g \to 0}
{\longrightarrow}&\; S_{ijkl}\;, \nonumber\\
F_4^0(1,2,3,4)&\stackrel{ j_g \to 0, l_g \to 0}
{\longrightarrow}&\; S_{ijk}\; S_{kli}\;,
\end{eqnarray}
\begin{eqnarray} 
F_4^0(1,2,3,4)&\stackrel{ i_g\parallel j_g, k_g \to 0}
{\longrightarrow}& S_{l;kji}(z) \;
\frac{1}{s_{ij}}\; P_{gg\to G}(z)\;,
\nonumber \\
F_4^0(1,2,3,4)&\stackrel{ i_g\parallel j_g, l_g \to 0}
{\longrightarrow}& S_{k;lij}(z) \;
\frac{1}{s_{ij}}\; P_{gg\to G}(z)\;.
\end{eqnarray}
\item Triple collinear limits:
\begin{eqnarray}
F_4^0(1,2,3,4)&\stackrel{ i_g \parallel j_g  \parallel k_g }
{\longrightarrow}& P_{ijk \to G}(w,x,y)\;.
\end{eqnarray}
\item Double collinear limits:
\begin{eqnarray}
F_4^0(1,2,3,4)&\stackrel{ i_g \parallel j_g,  k_g \parallel l_g}
{\longrightarrow}& 
\frac{1}{s_{ij} s_{kl}}\;
P_{gg\to G}(z)\; P_{g g\to G}(y)\;.
\end{eqnarray}
\end{enumerate}

In the single unresolved limits we have:
\begin{enumerate}
\item Soft limits:
\begin{eqnarray}
F_4^0(1,2,3,4)&\stackrel{{j}_g\to 0}
{\longrightarrow}&S_{ijk}\; F_3^0(i,k,l)\;.
\end{eqnarray}
\item Collinear limits:
\begin{eqnarray}
F_4^0(1,2,3,4)&\stackrel{i_g \parallel j_g}
{\longrightarrow}&\frac{1}{s_{ij}}\;P_{gg\to G}(z)\; 
F_3^0((ij),k,l) + {\rm ang.}\;.
\end{eqnarray}
\end{enumerate}

The leading colour gluon-gluon-quark-antiquark antenna function 
$G_4^0(1_g,3_{q},4_{\bar q},2_g)$ is colour-ordered: the gluon 
$(1_g)$ is emitted between gluon $(2_g)$ and the 
quark $(3_{q})$, while 
(again due to the cyclic nature of the colour indices)
gluon $(2_g)$ is emitted between gluon $(1_g)$ and the 
antiquark $(4_{\bar q})$.
 The second colour-ordering 
$G_4^0(1_g,4_{\bar q},3_{q},2_g)$ is obtained by exchanging the 
quark and antiquark momenta. 

The double unresolved limits of this antenna function are:
\begin{enumerate}
\item Double soft and soft-collinear limits:
\begin{equation}
G_4^0(1,3,4,2)\stackrel{ 3_{q} \to 0, 4_{\bar{q}} \to 0}
{\longrightarrow} S_{12}(3,4)\;,
\end{equation}
\begin{eqnarray} 
G_4^0(1,3,4,2)&\stackrel{ 3_{q}\parallel 4_{\bar q}, 2_g \to 0}
{\longrightarrow}&   S_{1;243}(z) \;
\frac{1}{s_{34}}\; P_{q\bar q\to G}(z)\;, \nonumber \\
G_4^0(1,2,3,4)&\stackrel{ 3_{q}\parallel 4_{\bar q}, 1_g \to 0}
{\longrightarrow}&   S_{2;134}(z) \;
\frac{1}{s_{34}}\; P_{q\bar q\to G}(z)\;.
\end{eqnarray}
\item Triple collinear limits:
\begin{eqnarray}
G_4^0(1,3,4,2)&\stackrel{ 1_g \parallel 3_{q}  \parallel 4_{\bar q}}
{\longrightarrow} &
P_{134 \to G}(w,x,y)\;, 
\nonumber\\
G_4^0(1,3,4,2)&\stackrel{ 2_g \parallel 3_{q}  \parallel 4_{\bar q}}
{\longrightarrow} &
P_{243 \to G}(w,x,y) \;.
\end{eqnarray}
\item Double collinear limit:
\begin{eqnarray} 
G_4^0(1,3,4,2)&\stackrel{ 1_g\parallel 2_g , 3_{q}  \parallel 4_{\bar q}}
{\longrightarrow}& 
\frac{1}{s_{12}s_{34}}\;
P_{gg\to G}(z)\; P_{q \bar{q}\to G}(y)\;.
\end{eqnarray}
\end{enumerate}
For the single unresolved limits we have:
\begin{enumerate}
\item Soft limits:
\begin{eqnarray}
G_4^0(1,3,4,2)&\stackrel{1_g\to 0}
{\longrightarrow}& S_{213}\; G_3^0(2,3,4) \nonumber\;, \\
G_4^0(1,3,4,2)&\stackrel{2_g\to 0}
{\longrightarrow}& S_{124}\; G_3^0(1,3,4)\;.
\end{eqnarray}
\item Collinear limits: 
\begin{eqnarray}
G_4^0(1,3,4,2)&\stackrel{1_{g} \parallel 2_{g}}
{\longrightarrow}&
 \frac{1}{s_{12}}\;P_{gg\to G}(z)\; 
G_3^0((12),3,4) + {\rm ang.} \;, \nonumber \\
G_4^0(1,3,4,2)&\stackrel{3_{q} \parallel 4_{\bar q}}
{\longrightarrow}&
 \frac{1}{s_{34}}\;P_{q\bar q\to G}(z)\; 
F_3^0(1,(34),2) + {\rm ang.} \;, \nonumber \\
G_4^0(1,3,4,2)&\stackrel{ 1_g \parallel 3_q}
{\longrightarrow}&  
\frac{1}{s_{13}}\;P_{qg\to Q}(z)\; G_3^0(2,(13),4)\;,\nonumber\\
G_4^0(1,3,4,2)&\stackrel{ 4_{\bar q}\parallel 2_g}
{\longrightarrow}&  
\frac{1}{s_{24}}\;P_{qg\to Q}(z)\; G_3^0(1,3,(24))\;.
\end{eqnarray}
\end{enumerate}

In the subleading colour antenna
$\tilde{G}_4^0(1_q,3_{q},4_{\bar q},2_g)$ one of the 
gluons 
is radiated between 
the quark-antiquark pair($3_{q}, 4_{\bar q}$), while the other 
is outside the quark-antiquark system. 
As a consequence, no singular structure  involves both gluons at once.

This antenna function 
contains only one type of double unresolved configuration, 
the triple collinear limit: 
\begin{eqnarray}
\tilde{G}_4^0(1,3,4,2)&\stackrel{1_g \parallel 3_{q} \parallel 4_{\bar q} }
{\longrightarrow}& 
\tilde{P}_{314\to G}(w,x,y)\;, \nonumber \\ 
\tilde{G}_4^0(1,3,4,2)&\stackrel{2_g \parallel 3_{q} \parallel 4_{\bar q} }
{\longrightarrow}& 
\tilde{P}_{324\to G}(w,x,y). 
\end{eqnarray}
The single unresolved limits always involve one of the gluons. They read:
\begin{enumerate}
\item Soft limits: 
\begin{eqnarray}
\tilde{G}_4^0(1,3,4,2)&\stackrel{ g_1 \to 0}
{\longrightarrow}&  S_{314}\; G_3^0(2,3,4) \;,\nonumber \\
\tilde{G}_4^0(1,3,4,2)&\stackrel{ g_2 \to 0}
{\longrightarrow}&  S_{324}\; G_3^0(1,3,4)\;.
\end{eqnarray}
\item Collinear limits: 
\begin{eqnarray}
\tilde{G}_4^0(1,i,j,2)\stackrel{ i \parallel g_1}
{\longrightarrow}&  \frac{1}{s_{1i}}\;P_{qg\to Q}(z)
\;G_3^0(2,(i1),j)\; ,\nonumber \\
\tilde{G}_4^0(1,i,j,2)\stackrel{ i \parallel g_2}
{\longrightarrow}&  \frac{1}{s_{2i}}\;P_{qg\to Q}(z)
\;G_3^0(1,(i2),j)\;,
 \end{eqnarray}
where $i$ and $j$ can both play the role of the quark or the antiquark.
\end{enumerate}

The antenna function containing  two quark-antiquark pairs of 
non-identical flavours, 
$H_4^0(1_q,2_{\bar q},3_{q'},4_{\bar q'})$ contains only one
double unresolved limit, the double single collinear configuration:
\begin{eqnarray}
H_4^0(1,2,3,4)&\stackrel{1_{q} \parallel 2_{\bar q}
,3_{q'} \parallel 4_{\bar q'}
}{\longrightarrow}& 
\frac{1}{s_{12}s_{34}} \;P_{q\bar q \to G}(z)  \;P_{q\bar q \to G} (y) \;.
\end{eqnarray}
The two  single collinear limits are:
\begin{eqnarray}
H_4^0(1,2,3,4)&\stackrel{1_{q} \parallel 2_{\bar q}}
{\longrightarrow}& 
\frac{1}{s_{12}}\; P_{q\bar q \to G}(z) \; G_3^0((12),3,4)  + {\rm ang.}
\;,\nonumber\\
H_4^0(1,2,3,4)&\stackrel{3_{q'} \parallel 4_{\bar q'}
}{\longrightarrow}& 
\frac{1}{s_{34}}\; P_{q\bar q \to G}(z)\;  G_3^0(1,2,(34)) + {\rm ang.}\;. 
\end{eqnarray}

\subsection{Angular terms}
\label{sec:ang}

Angular terms manifest themselves in collinear limits 
of antenna functions and matrix elements when a final state gluon splits into 
a quark-antiquark pair or into two
 gluons. Several examples were listed in 
Section~\ref{sec:limits}. To obtain local subtraction terms (i.e.\ 
subtraction terms which approach the full multi-parton matrix element 
in its unresolved limits before any integrations are carried out), it 
is necessary to take proper account of these angular terms. In this section,
we illustrate an algorithmic procedure to reconstruct the angular terms 
appearing in the simple collinear limits of four-parton antenna functions.
The same procedure can be generalised in principle to reconstruct 
angular terms appearing in the single and double unresolved limits of 
multi-parton matrix elements. First steps in this direction were 
performed in~\cite{nnlosub5}. 

In the collinear limits 
arising from gluon splitting, the four-parton antenna functions do not 
yield the unpolarised splitting functions (\ref{eq:apkernels})
multiplied by a spin-averaged three-parton antenna function. 
Instead, one finds that the four-parton antenna functions factorise
into the corresponding spin-dependent tensorial splitting functions 
and tensorial three-parton antenna functions~\cite{cs,cg}.

In constructing the subtraction terms for two 
colour-connected unresolved partons,
Section~\ref{sec:sub2b}, we used the 
difference between four-parton antenna function and products of 
three-parton antenna functions (\ref{eq:sub2b}). 
The former are intended to subtract all
singularities in the double unresolved region, while the latter ensure that 
the whole subtraction term is free of singularities in all single unresolved 
regions. If the four-parton antenna appearing in (\ref{eq:sub2b})
has simple unresolved  limits where angular terms are present, 
the product of two 
three-parton antenna functions in the same equation will no 
longer subtract its simple collinear behaviour locally. The left-over
terms are however vanishing after integration over the antenna phase space,
thus not affecting the cancellation of infrared poles.

To access these angular terms correctly, one has to keep track of the 
transverse momentum components of the collinear partons.
The collinear limit of partons $p_{1}$ and $p_{2}$  
is defined~\cite{cs} as the limit $k_\perp \to 0$ of 
\begin{eqnarray}
p_{1}^{\mu}&=&z\;P^{\mu} + k_{\perp}^{\mu}-\frac{k_{\perp}^{2}}{z}\;
\frac{n^{\mu}}{2\,P\cdot n}\; , \nonumber \\
p_{2}^{\mu}&=&(1-z)\;P^{\mu} - k_{\perp}^{\mu}-\frac{k_{\perp}^{2}}{(1-z)}\;
\frac{n^{\mu}}{2\,P \cdot n}\; ,
\end{eqnarray}
with
\begin{equation}
s_{12}\equiv 2\,p_{1}
\cdot p_{2}=-\frac{k_{\perp}^{2}}{z\,(1-z)}\;.
\end{equation}
In these equations the vector $P^{\mu}$ ($P^{2}=0$) denotes the collinear 
direction, while $n^{\mu}$ is an auxiliary light-like vector, 
which is necessary to specify the transverse component $k_{\perp}$ 
($k_{\perp}\cdot P=k_{\perp} \cdot n=0$). In the small $k_{\perp}$ 
limit (i.e.\  neglecting terms that are less singular than $1/k_{\perp}^2$),   
we find the following factorisation formulae of the four-parton antenna 
functions:
\begin{eqnarray}
A_4^0(1,3,4,2)&\stackrel{3_g \parallel 4_g}
{\longrightarrow}& 
\frac{1}{s_{34}}\; P^{\mu \nu}_{gg\to G}(z)\;(A_{3}^{0})_{\mu \nu}(1,(34),2) \,
\;, \nonumber \\
B_4^0(1,3,4,2)&\stackrel{3_{q} \parallel 4_{\bar q}}
{\longrightarrow}& \frac{1}{s_{34}} \;P^{\mu \nu}_{q\bar{q}\to Q}(z) \; 
(A_3^0)_{\mu \nu}(1,(34),2) 
\;, \nonumber \\
D_4^0(1,3,4,5)&\stackrel{3_g \parallel 4_g}
{\longrightarrow}& 
 \frac{1}{s_{34}}\;P^{\mu \nu}_{gg\to G}(z)\; 
(D_3^0)_{\mu \nu}(1,(34),5) \;, \nonumber\\
D_4^0(1,3,4,5)&\stackrel{4_g \parallel 5_g}
{\longrightarrow}& 
 \frac{1}{s_{45}}\;P^{\mu \nu}_{gg\to G}(z)\; 
(D_3^0)_{\mu \nu}(1,3,(45)) \;, \nonumber\\
E_4^0(1,3,4,5)&\stackrel{3_q \parallel 4_{\bar{q}}}
{\longrightarrow}& 
 \frac{1}{s_{34}}\;P^{\mu \nu}_{q\bar q\to G}(z)\; 
(D_3^0)_{\mu \nu}(1,(34),5) \;, \nonumber\\
 F_4^0(1,2,3,4)&\stackrel{i_g \parallel j_g}
{\longrightarrow}& 
 \frac{1}{s_{ij}}\;P^{\mu \nu}_{gg\to G}(z)\; 
(F_3^0)_{\mu \nu}((ij),k,l)  \;, \nonumber\\
 G_4^0(1,3,4,2)&\stackrel{1_g \parallel 2_g}
{\longrightarrow}& 
 \frac{1}{s_{12}}\;P^{\mu \nu}_{gg\to G}(z)\; 
(G_3^0)_{\mu \nu}((12),3,4) \;, \nonumber\\
 G_4^0(1,3,4,2)&\stackrel{3_q \parallel 4_{\bar q}}
{\longrightarrow}& 
 \frac{1}{s_{34}}\;P^{\mu \nu}_{q\bar q\to G}(z)\; 
(F_3^0)_{\mu \nu}(1,(34),2) 
\;, \nonumber\\
 H_4^0(1,2,3,4)&\stackrel{1_q \parallel 2_{\bar q}}
{\longrightarrow}& 
 \frac{1}{s_{12}}\;P^{\mu \nu}_{q\bar q\to G}(z)\; 
(G_3^0)_{\mu \nu}((12),3,4)  \;, \nonumber\\
 H_4^0(1,2,3,4)&\stackrel{3_q \parallel 4_{\bar q}}
{\longrightarrow}& 
 \frac{1}{s_{34}}\;P^{\mu \nu}_{q\bar q\to G}(z)\; 
(G_3^0)_{\mu \nu}((34),1,2) \;.
\label{eq:spin}
\end{eqnarray}
The spin-dependent splitting functions $P^{\mu\nu}$ appearing in these 
equations were given in \cite{cg}.
The tensorial three-parton antenna functions $(X_3^0)_{\mu\nu}$
can be derived by analogy with 
the scalar three-parton antenna functions of 
Sections~\ref{sec:qq}--\ref{sec:gg} from physical squared matrix elements. 
Their tensorial structure is obtained 
by 
leaving the polarisation index of the gluon associated with the momentum
$P^{\mu}$ 
uncontracted. On contraction of the tensorial three-parton antenna functions 
with a physical gluon polarisation average, we recover 
their scalar counterparts.

It should be noted that the 
collinear factorisation (\ref{eq:spin}) of 
four-parton antenna functions in spin-dependent splitting functions and 
tensorial three-parton antenna functions prevents us 
a priori from using the 
scalar (spin-independent) antenna functions derived in the previous 
sections to construct local subtraction terms. 
It seems that tensorial antenna functions are required. 
However this problem can be circumenvented by explicitly isolating the 
angular terms as follows. 

For each four-parton antenna function $X_{4}^0$ 
yielding angular  terms   
in a given simple collinear limit
(gluon splitting into two partons $i$ and $j$ in the final state),
one considers an angular function denoted by 
$\Theta_{{X_3^0}}$.
This function must fulfil two properties: 
(1) it yields the correct local behaviour
 in this particular limit and (2) it  
integrates to zero over the corresponding 
unresolved phase space.
The second requirement is particularly important, since 
it 
ensures that the analytic integration will not 
be modified by the presence of these angular functions, and that the 
integrated scalar antennae are sufficient to fully describe the 
pole structure of the antenna subtraction.
In principle, a local counterterm is not even required for the numerical 
implementation; it does however allow a point-by-point check of the correct
numerical behaviour of the subtraction terms, and guarantees considerable 
improvement of the numerical stability. 

We find that the local counterterms are obtained by the replacement
\begin{eqnarray}
X_4^0(1,i,j,2) &\to& 
X_4^0(1,i,j,2) - \Theta_{{X}_3^0}(i,j,z,k_{\perp}) \;,
\label{eq:angreplace}
\end{eqnarray} 
where the angular function 
$\Theta_{{X_3^0}}(i,j,z,k_{\perp})$ is given by,
\begin{eqnarray}
\Theta_{{X_3^0}}(i,j,z,k_{\perp})=
\left [\frac{1}{s_{ij}} \;P^{\mu \nu}_{ij\to (ij)}(z,k_{\perp})
  ({X}_3^0)_{\mu \nu}(1,(ij),2)\; 
     -\frac{1}{s_{ij}} \;P_{i j\to (ij)}(z)\; {X}_3^0(1,(ij),2) \right]
\,. \nonumber \\
\end{eqnarray} 
In this equation, ${X}_3^0$ is the appropriate three-parton 
antenna function
obtained from $X_4^0$ when partons $i$ and $j$ build the parent 
parton $(ij)$. 
$P^{\mu \nu}_{i j\to (ij)}$ 
stands for the spin-dependent splitting function while  
$P_{i j\to (ij)}$ 
stands for the spin averaged splitting function appropriate to the
limit under consideration. The collinear momentum fraction $z$ can be 
expressed in terms of invariants formed by the momenta appearing 
in the four-parton antenna phase space. Its precise definition is 
irrelevant, as long as it yields the correct expression in the collinear 
limit.   
After the replacement (\ref{eq:angreplace}), the resulting 
four-parton antenna function is locally free from singular 
terms in the single unresolved regions.

On the other hand, the term $\Theta_{X_3^0}(i,j,z,k_{\perp})$ integrates 
to zero if integrated over the unresolved phase space, since 
\begin{equation}
\int {\rm d}\phi \;k_{\perp}^{\mu} k_{\perp}^{\nu}\;f(k_{\perp}^2)
=  -\frac{d^{\mu \nu}}{d-2} \;k_{\perp}^2\;
f(k_{\perp}^2)\;, 
\label{eq:polsum}
\end{equation}
with 
\begin{displaymath}
d^{\mu \nu}= g^{\mu \nu}-\frac{P^{\mu}n^{\nu} +n^{\mu}P^{\nu}}{n\cdot P}
\end{displaymath}
being the gluon polarisation sum in the axial gauge.
Applied to the spin-dependent splitting function 
$P^{\mu \nu}_{ij\to (ij)}$, relation (\ref{eq:polsum}) 
yields the spin-averaged splitting function $d^{\mu\nu} P_{ij\to (ij)}$.
Contraction with the tensorial antenna function 
$(X^0_3)_{\mu\nu}$ then reproduces the product of the spin-averaged 
splitting function with the 
scalar antenna function $P_{ij\to (ij)}\,(X^0_3)$.
Integration of $\Theta_{{X}_3^0}$ over the three-parton antenna 
phase space made from momenta $(i)$, $(j)$ and either (1) or (2) yields 
zero by construction.

To illustrate the angular replacement  (\ref{eq:angreplace}) 
on a specific example, we consider the four-parton antenna function 
$B_4^0(1_q,3_{q'},4_{\bar q'},2_{\bar{q}})$ in the 
$(3 \parallel 4)$ limit:
\begin{equation}  
B_4^0(1,3,4,2)\to B_4^0(1,3,4,2)  - \Theta_{A_3^0}(i,j,z,k_{\perp})\; ,
\label{eq:b40ang}
\end{equation}
with
\begin{eqnarray}
\Theta_{A_3^0}(i,j,z,k_{\perp})
&=&\frac{1}{s_{34}} \;P^{\mu \nu}_{q\bar{q}\to Q}(z,k_{\perp}) \; 
(A_3^0)_{\mu \nu}(1,(34),2)
-\frac{1}{s_{34}} \;P_{q\bar{q}\to Q}(z)
A_3^0(1,(34),2)\;, \nonumber \\    
\end{eqnarray}
where $z$ is the momentum fraction carried by one of the collinear partons
in the angular dependent function $\Theta_{A_3^0}(i,j,z,k_{\perp})$.

In the colour-connected double unresolved subtraction term (\ref{eq:sub2b}),
$B_4^0$ appears in the combination 
\begin{displaymath}
B_4^0(1,3,4,2)-E_3^0(1,3,4)\, A_3^0(1,(34),2) \;.
\end{displaymath}
In the collinear $(3 \parallel 4)$ limit, this expression is not vanishing,
but yields some residual angular terms. After substituting (\ref{eq:b40ang}),
this becomes
\begin{eqnarray}
B_4^0(1,3,4,2)&-&\left[\frac{1}{s_{34}} \;
P^{\mu \nu}_{q\bar{q}\to Q}(z,k_{\perp})
 \; (A_3^0)_{\mu \nu}(1,(34),2)-\frac{1}{s_{34}} \;P_{q\bar{q}\to Q}(z)
A_3^0(1,(34),2)\right] 
\nonumber\\
&-&E_3^0(1,3,4)\, A_3^0(1,(34),2)\, ,
\end{eqnarray}
which is free of singularities in the $3 \parallel 4$ limit, since 
(\ref{eq:spin}) holds and,
\begin{eqnarray}
E_3^0(1,3,4)&\stackrel{ 3_q\parallel
4_{\bar{q}}}{\longrightarrow}& \frac{1}{s_{34}}\,P_{g \to q\bar{q}}(z)\;.
\end{eqnarray}
The term $\Theta_{A_3^0}(i,j,z,k_{\perp})$ vanishes 
when integrated over the unresolved phase space, because of 
(\ref{eq:polsum}).

\section{The $1/N^2$ contribution to $e^+e^- \to 3$~jets at NNLO}
\label{sec:3j}

To illustrate the application of antenna factorisation  on a 
non-trivial example, in this section we derive the $1/N^2$-contribution to 
the NNLO corrections to  $e^+e^- \to 3$~jets. For completeness, and also 
for future reference, we will first discuss all NNLO contributions to 
$e^+e^- \to 3$~jets in Section~\ref{sec:allpartons}. 
We construct the $1/N^2$ double real radiation subtraction term in
Section~\ref{sec:dsub} and the  $1/N^2$ virtual single real radiation 
subtraction term in Section~\ref{sec:vsub}. In Section~\ref{sec:cancel}, we 
then show how the integrated subtraction terms cancel the $1/N^2$-poles 
of the two-loop virtual corrections. Some details of the numerical
implementation are discussed in Section.~\ref{sec:numer}.  Finally, to illustrate
the power of our approach, in Section~\ref{sec:3jir}
we show that the 
infrared poles of the two-loop (including one-loop times one-loop) correction
to $\gamma^*\to q\bar qg$ are cancelled in all colour factors by a 
combination of integrated three-parton and four-parton
antenna functions.

The ${\cal O}(1/N^2)$ colour or QED-like
contribution to three jet production in electron positron collisions
receives contributions from several different partonic channels. There are
contributions from $\gamma^*\to q\bar q ggg$ and $\gamma^*\to q\bar q q\bar qg$
at tree-level,  $\gamma^*\to q\bar q gg$ and $\gamma^*\to q\bar q q\bar q$  at
one-loop and $\gamma^*\to q\bar q g$ at two-loops. The four-parton and 
five-parton final states contain infrared singularities, which need to
be extracted using the antenna subtraction formalism. 

\subsection{The matrix elements}
\label{sec:allpartons}
First we list the tree, one-loop and two-loop
amplitudes for $\gamma^* \to n$~partons where $n \leq 5$.

\subsubsection{Tree-level matrix elements for up to five partons}

The tree-level amplitude $M^0_{q\bar q(n-2)g}$ 
for a virtual photon to produce a quark-antiquark pair
and $(n-2)$-gluons, 
$$\gamma^*(q) \to q(p_1) \bar q(p_2)  g(p_3)\ldots
g(p_n) $$
can be expressed as sum over the permutations of the colour-ordered amplitude
$\MA{0}{n}$ of the possible 
orderings for the gluon colour indices
\begin{equation}
\label{eq:qqnm2g}
M^0_{q\bar q(n-2)g} 
= i e (\sqrt{2}g)^{n-2} \sum_{(i,\ldots,k) \in P(3,\ldots, n)} \left(
T^{a_i}\cdots T^{a_n}\right)_{i_1i_2} \MA{0}{n} (p_1,p_3,\ldots,p_n,p_2)\;.
\end{equation}

The 
squared matrix elements for $n=3,\ldots,5$, summed over gluon polarisations, 
but excluding symmetry factors for identical particles,
are given by,
\begin{eqnarray}
\left|M^0_{q\bar qg}\right|^2
&=& N_3 \, A_3^0(1_q,3_g,2_{\bar q}) \, ,\\
\left|M^0_{q\bar qgg}\right|^2
&=& N_4 \,  
\left[
\sum_{(i,j) \in P(3,4)}
N A_{4}^0 (1_q,i_g,j_g,2_{\bar q})  
-\frac{1}{N}
 \tilde A_{4}^0 (1_q,3_g,4_g,2_{\bar q})  \right]\, 
,\\
\left|M^0_{q\bar qggg}\right|^2
&=& N_5 \,  
\Bigg[
\sum_{(i,j,k) \in P(3,4,5)}\left(
N^2 A_{5}^0 (1_q,i_g,j_g,k_g,2_{\bar q}) 
-
\tilde A_{5}^0 (1_q,i_g,j_g,k_g,2_{\bar q}) \right)
\nonumber \\
&& \hspace{3cm}+\left(\frac{N^2+1}{N^2}\right)
\bar A_{5}^0 (1_q,3_g,4_g,5_g,2_{\bar q}) \Bigg]\, ,
\end{eqnarray}
where,
\begin{equation}
N_n = 4 \pi  \alpha\, 
\sum_q e_q^2 \left(g^2\right)^{(n-2)}
\left(N^2-1\right)\,\left|{\cal M}^0_{q\bar q}\right|^2,
\end{equation}
and
\begin{equation}
\left|{\cal M}^0_{q\bar q}\right|^2 = 4 (1-\epsilon) q^2.
\end{equation}
The squared colour-ordered matrix elements $A^0_3$, $A^0_4$ and $\tilde A^0_4$
are given in eqs.~(\ref{eq:A30}), (\ref{eq:A40}) and (\ref{eq:A40t}) respectively.
For the five parton case,
\begin{eqnarray}
 A_5^{0}(1_q,i_g,j_g,k_g,2_{\bar q}) \left|{\cal M}^0_{q\bar q}\right|^2&=&
\bigg|\MA{0}{5}(p_1,p_i,p_j,p_k,p_2)\bigg|^2\\
\tilde A_5^{0}(1_q,i_g,j_g,k_g,2_{\bar q}) \left|{\cal M}^0_{q\bar q}\right|^2
&=&
\nonumber \\
&&
\hspace{-4.4cm}\bigg| \MA{0}{5}(p_1,p_i,p_j,p_k,p_2)
+\MA{0}{5}(p_1,p_i,p_k,p_j,p_2)
+\MA{0}{5}(p_1,p_k,p_i,p_j,p_2)
\bigg|^2,
\\ 
\bar A_{5}^0(1_q,i_g,j_g,k_g,2_{\bar q}) \left|{\cal M}^0_{q\bar q}\right|^2 
&=& 
\left|\sum_{(i,j,k) \in P(3,\ldots, 5)}
\MA{0}{5}(p_1,p_i,p_j,p_k,p_2)\right|^2\, .
\end{eqnarray}
In the sub-leading colour contribution $\tilde{A}_5^0$, gluon $k$ is 
effectively photon-like, while in the sub-sub-leading colour contribution
(also called Abelian contribution),
$\bar{A}_5^0$, all three gluons are effectively photon-like.   
Photon-like gluons do not couple to 
three- and four-gluon vertices, and there are no simple collinear 
limits as any two photon-like gluons become
collinear.  As a consequence, the only colour connected pair 
in $\bar{A}_5^0$ are the quark and
antiquark.   All subtraction terms for this five-parton contribution 
are therefore based on quark-antiquark
antennae. 


The tree-level amplitude for 
$$\gamma^*(q) \to q(p_1) \bar q(p_2) q' (p_3) \bar q'(p_4)$$
is given by
\begin{eqnarray}
M_{q\bar q q'\bar q'}^0
&=& i e_1 g^2 
\delta_{q_1q_2}\delta_{q_3q_4}
\left(\delta_{i_1i_4}\delta_{i_3i_2}-\frac{1}{N}
\delta_{i_1i_2}\delta_{i_3i_4}\right) 
\MB{0}{4} (p_1,p_2,p_3,p_4) \nonumber \\
&& \hspace{1.6cm}
+ (1\leftrightarrow 3,2\leftrightarrow 4)\;,
\end{eqnarray}
where $\delta_{q_1q_2}\delta_{q_3q_4}$ indicates the quark
flavours.
The amplitude $\MB{0}{4}(p_1,p_2,p_3,p_4)$ thus denotes the contribution 
from the $q_1 \bar q_2$--pair coupling to the vector boson.  
The identical quark amplitude is obtained
\begin{equation}
M_{q\bar q q\bar q}^0 =
M_{q\bar q q'\bar q'}^0
 - (2\leftrightarrow 4).
\end{equation}

The resulting four-quark 
squared matrix elements, summed over final state quark 
flavours and including symmetry factors are given by 
\begin{eqnarray}
\left|M_{4q}^0\right|^2 
&=&\sum_{q,q'}
\left|M_{q\bar q q'\bar q'}\right|^2 
+\sum_{q}
\left|M_{q\bar q q\bar q}\right|^2 \nonumber \\
&=& N_4\, \Bigg [  
N_F B_{4}^0 (1_q,3_q,4_{\bar q},2_{\bar q}) 
- \frac{1}{N} \, \left(
C_{4}^0 (1_q,3_q,4_{\bar q},2_{\bar q})
+C_{4}^0 (2_{\bar q},4_{\bar q},3_q,1_q) \right) \nonumber \\
&&+ N_{F,\gamma} \,\hat B_{4}^0 (1_q,3_q,4_{\bar q},2_{\bar q})
\Bigg ]\, ,
\end{eqnarray}
where 
\begin{eqnarray}
B_{4}^0 (1_q,3_q,4_{\bar q},2_{\bar q})\left|{\cal M}^0_{q\bar q}\right|^2
&=& \left| 
 \MB{0}{4}(p_1,p_2,p_3,p_4)\right|^2,\nonumber \\
C_{4}^0 (1_q,3_q,4_{\bar q},2_{\bar q})\left|{\cal M}^0_{q\bar q}\right|^2
&=& - \Re \bigg(
 \MB{0}{4}(p_1,p_2,p_3,p_4)\MB{0,\dagger}{4}(p_1,p_4,p_3,p_2)
\bigg)\, , \\
\hat B_{4}^0 (1_q,3_q,4_{\bar q},2_{\bar q})\left|{\cal M}^0_{q\bar q}\right|^2
&=& \mbox{Re} \bigg(
 \MB{0}{4}(p_1,p_2,p_3,p_4)\MB{0,\dagger}{4}(p_3,p_4,p_1,p_2)
\bigg).
\end{eqnarray}
Explicit expressions for 
$B_4^0$ and $C_4^0$ are given in eqs.~(\ref{eq:B40}) and (\ref{eq:C40}) 
respectively.
The last term, $\hat B_{4}^0$, is proportional to the 
charge weighted sum of the quark flavours,
$N_{F,\gamma}$, which for electromagnetic interactions is given by,
\begin{equation}
N_{F,\gamma} = \frac{(\sum_q e_q)^2}{\sum_q e_q^2} .
\end{equation}
It is relevant only for observables where the final state quark 
charge can be determined.

There are four colour structures in the 
tree-level amplitude for $$\gamma^*(q) \to q(p_1) \bar q(p_2) 
q' (p_3) \bar q'(p_4) g(p_5)$$
which reads 
\begin{eqnarray}
M_{q\bar q q'\bar q' g}
&=& i e_1  g^3 \sqrt{2}  
\delta_{q_1q_2} \delta_{q_3q_4} 
\nonumber \\
&\times&
\bigg[
T^{a_5}_{i_1i_4} \delta_{i_3i_2}  
\MB{0,a}{5} (p_1,p_2,p_3,p_4,p_5)
- \frac{1}{N} T^{a_5}_{i_1i_2} \delta_{i_3i_4} 
\MB{0,c}{5} (p_1,p_2,p_3,p_4,p_5)
  \nonumber \\
&& 
+ T^{a_5}_{i_3i_2} \delta_{i_1i_4}  
\MB{0,b}{5} (p_1,p_2,p_3,p_4,p_5)
- \frac{1}{N} T^{a_5}_{i_3i_4} \delta_{i_1i_2} 
\MB{0,d}{5} (p_1,p_2,p_3,p_4,p_5)\bigg]
  \nonumber \\
&& \hspace{1.6cm}
+ (1\leftrightarrow 3,2\leftrightarrow 4) \;.
\end{eqnarray}
The amplitude $\MB{0,x}{5}(p_1,p_2,p_3,p_4,p_5)$ for $x=a,\ldots,d$
denotes the contribution 
from the $q_1 \bar q_2$--pair coupling to the vector boson.  
Due to the colour decomposition,
the following relation holds between the leading and subleading colour 
amplitudes:
\begin{eqnarray}
\MB{0,e}{5}(p_1,p_2,p_3,p_4,p_5)&=&
\MB{0,a}{5} (p_1,p_2,p_3,p_4,p_5) + 
\MB{0,b}{5} (p_1,p_2,p_3,p_4,p_5) \nonumber \\
&=& 
\MB{0,c}{5} (p_1,p_2,p_3,p_4,p_5) +
\MB{0,d}{5} (p_1,p_2,p_3,p_4,p_5) \;.
\end{eqnarray}
As before, the identical quark matrix element is obtained by permuting the
antiquark momenta,
\begin{equation}
M_{q\bar q q\bar qg}^0 =
M_{q\bar q q'\bar q'g}^0
 - (2\leftrightarrow 4).
\end{equation}

The squared matrix element, summed over flavours and including 
symmetry factors is given by,
\begin{eqnarray}
\lefteqn{\left|M_{4qg}^0\right|^2 
=
\sum_{q,q'}
\left|M_{q\bar q q'\bar q' g}\right|^2 
+\sum_q \left|M_{q\bar q q\bar q g}\right|^2} \nonumber \\
&=&
N_5\, \Bigg [
N\NF \left(
B_5^{0,a}(1_q,5_g,4_{\bar q'};3_{q'},2_{\bar q})
+B_5^{0,b}(1_q,4_{\bar q'};3_{q'},5_g,2_{\bar q})
\right)\nonumber\\
&& + \frac{\NF}{N}
\left(
 B_5^{0,c}(1_q,5_g,2_{\bar q};3_{q'},4_{\bar q'})
+B_5^{0,d}(1_q,2_{\bar q};3_{q'},5_g,4_{\bar q'})
-2 B_5^{0,e}(1_q,2_{\bar q};3_{q'},4_{\bar q'};5_g)
\right)\nonumber \\
&&-C_5^{0}(1_q,3_{q},4_{\bar q},5_g,2_{\bar q})
+\left(\frac{N^2+1}{N^2}\right) \left(
\tilde C_5^{0}(1_q,3_{q},4_{\bar q},5_g,2_{\bar q}) 
+ \tilde C_5^{0}(2_{\bar q},4_{\bar q},3_{q},5_g,1_q)\right)
\nonumber \\
&&- N N_{F,\gamma}\left(
 \hat B_5^{0,a}(1_q,5_g,4_{\bar q'};3_{q'},2_{\bar q})
+\hat B_5^{0,b}(1_q,4_{\bar q'};3_{q'},5_g,2_{\bar q})
-\hat B_5^{0,e}(1_q,4_{\bar q'};3_{q'},2_{\bar q},5_g)
\right)\nonumber \\
&&+\frac{N_{F,\gamma}}{N}
\left(
 \hat B_5^{0,c}(1_q,5_g,2_{\bar q};3_{q'},4_{\bar q'})
+\hat B_5^{0,d}(1_q,2_{\bar q};3_{q'},5_g,4_{\bar q'})
+\hat B_5^{0,e}(1_q,2_{\bar q};3_{q'},4_{\bar q'};5_g)
\right)\Bigg ],
\nonumber \\
\end{eqnarray}
where for $x=a,\ldots,e$
\begin{eqnarray}
B_5^{0,x}(\ldots) \left|{\cal M}^0_{q\bar q}\right|^2 &=& |\MB{0,x}{5}(p_1,p_2,p_3,p_4,p_5)|^2,\\
\hat B_5^{0,x}(\ldots) \left|{\cal M}^0_{q\bar q}\right|^2 &=& \mbox{Re} \left(\MB{0,x}{5}(p_1,p_2,p_3,p_4,p_5)
\MB{0,x,\dagger}{5}(p_3,p_4,p_1,p_2,p_5)\right),
\end{eqnarray}
and
\begin{eqnarray}
C_5^{0}(1_q,3_{q},4_{\bar q},5_g,2_{\bar q}) \left|M^0_{q\bar q}\right|^2 &=& 
-2\mbox{Re} \bigg(
 \MB{0,a}{5}(p_1,p_2,p_3,p_4,p_5)\MB{0,c,\dagger}{5}(p_1,p_4,p_3,p_2,p_5)
\nonumber \\&&
\phantom{2\mbox{Re}}+\MB{0,b}{5}(p_1,p_2,p_3,p_4,p_5)\MB{0,d,\dagger}{5}(p_1,p_4,p_3,p_2,p_5)
\nonumber \\&&
\phantom{2\mbox{Re}}+\MB{0,a}{5}(p_1,p_2,p_3,p_4,p_5)\MB{0,d,\dagger}{5}(p_3,p_2,p_1,p_4,p_5)
\nonumber \\&&
\phantom{2\mbox{Re}}+\MB{0,b}{5}(p_1,p_2,p_3,p_4,p_5)\MB{0,c,\dagger}{5}(p_3,p_2,p_1,p_4,p_5)
\bigg),\nonumber \\
&&\\
\tilde{C}_5^{0}(1_q,3_{q},4_{\bar q},5_g,2_{\bar q})
\left|M^0_{q\bar q}\right|^2
&=& - \mbox{Re} \bigg(
 \MB{0,e}{5}(p_1,p_2,p_3,p_4,p_5)\MB{0,e,\dagger}{5}(p_1,p_4,p_3,p_2,p_5)
\bigg).\nonumber \\
\end{eqnarray}

\subsubsection{One-loop matrix elements for up to four partons}

The renormalised one-loop  amplitude $M^1_{q\bar qg}$ 
for a virtual photon to produce a quark-antiquark pair
 together with a single gluon, 
$$\gamma^*(q) \to q(p_1) \bar q(p_2)  g(p_3)$$
contains a single colour structure such that
\begin{equation}
\label{eq:ggponeloop}
M^1_{q\bar qg} 
= i e \sqrt{2} g \left(\frac{g^2}{16\pi^2}\right)
T^{a_3}_{i_1i_2} \MA{1}{3} (p_1,p_3,p_2)\;.
\end{equation}
Unless stated otherwise, the renormalisation scale is set to 
$\mu^2 = q^2$. 

The interference of the one-loop amplitude with the three-parton
tree-level amplitude (\ref{eq:qqnm2g}) is given by
\begin{eqnarray}
2\Re\left(M^{0,\dagger}_{q\bar qg}M^1_{q\bar qg} \right)
&=& N_3 \left(\frac{\alpha_s}{2\pi}\right)
A_3^{(1\times 0)}(1_q,3_g,2_{\bar q})\,,
\end{eqnarray}
where
\begin{eqnarray}
\label{eq:M31}
A_3^{(1\times 0)}(1_q,3_g,2_{\bar q})
&=&
\, \bigg( N \left[
A_3^1(1_q,3_g,2_{\bar q})
+{\cal A}_2^1(s_{123}) A_3^0(1_q,3_g,2_{\bar q})\right]\nonumber \\
&&
-\frac{1}{N}  \left[
\tilde
A_3^1(1_q,3_g,2_{\bar q})
+{\cal A}_2^1(s_{123}) A_3^0(1_q,3_g,2_{\bar q})\right]
+N_F \hat A_3^1(1_q,3_g,2_{\bar q})\bigg)\, ,\nonumber \\
\end{eqnarray}
where $A^1_3$, $\tilde A^1_3$ and $\hat A^1_3$ 
are given up to ${\cal O}(\epsilon^0)$ in eqs.~(\ref{eq:A31}), 
(\ref{eq:A31t}) and (\ref{eq:A31h})
respectively.

The one-loop corrections to $\gamma^* \to 4$~partons
 have been available for
some time~\cite{onel-4}.   
The one-loop amplitude for 
$$\gamma^*(q) \to q(p_1) \bar q(p_2)  g(p_3) 
g(p_4)$$ contains two colour structures,
\begin{eqnarray}
M_{q\bar q gg}^1 
&=& i e 2 g^2 \left(\frac{g^2}{16\pi^2}\right)\nonumber \\
&\times&
\Bigg[
\sum_{(i,j) \in P(3,4)} 
\left(T^{a_i}T^{a_j}\right)_{i_1i_2} 
\bigg(N\MA{1,a}{4} (p_1,p_i,p_j,p_2)
-\frac{1}{N}\MA{1,b}{4} (p_1,p_i,p_j,p_2)\nonumber \\
&&\hspace{4cm}
+N_F\MA{1,c}{4} (p_1,p_i,p_j,p_2)\bigg)\nonumber \\
&&\hspace{4cm}
+\frac{1}{2}\delta^{a_ia_j}\delta_{i_1i_2}
\MA{1,d}{4} (p_1,p_3,p_4,p_2)
\Bigg]\;,
\end{eqnarray}
where
\begin{eqnarray}
\MA{1,d}{4} (p_1,p_3,p_4,p_2)&=&
\MA{1,d}{4} (p_1,p_4,p_3,p_2).
\end{eqnarray}

The ``squared" matrix element is the interference between the tree-level and
one-loop amplitudes,
\begin{eqnarray}
\lefteqn{2 \left | M^{0,\dagger}_{q\bar q gg} M^1_{q\bar q gg}\right|
=N_4 \left(\frac{\alpha_s}{2\pi}\right)\,} \nonumber
\\
&\times&  
\Bigg[
\sum_{(i,j) \in P(3,4)} 
\left(
N^2 
A_{4}^{1,a} (1_q,i_g,j_g,2_{\bar q})
-
A_{4}^{1,b} (1_q,i_g,j_g,2_{\bar q})
+NN_F A_{4}^{1,c} (1_q,i_g,j_g,2_{\bar q})
\right)\nonumber \\
&&-
\Bigg(
\tilde A_{4}^{1,a} (1_q,3_g,4_g,2_{\bar q})
-\tilde A_{4}^{1,d} (1_q,3_g,4_g,2_{\bar q})
-\frac{1}{N^2}
\tilde A_{4}^{1,b} (1_q,3_g,4_g,2_{\bar q}) \nonumber \\ && \hspace{2.2cm}
+\frac{N_F}{N}
\tilde A_{4}^{1,c} (1_q,3_g,4_g,2_{\bar q})
\Bigg) \Bigg]\,,
\end{eqnarray}
where for  $x=a,\ldots,d$,
\begin{eqnarray}
A_{4}^{1,x} (1_q,i_g,j_g,2_{\bar q}) \left|{\cal M}^0_{q\bar q}\right|^2&=& 
\Re \left(\MA{1,x}{4} (p_1,p_i,p_j,p_2) 
\MA{0,\dagger}{4} (p_1,p_i,p_j,p_2)\right),\\
\tilde A_{4}^{1,x} (1_q,3_g,4_g,2_{\bar q}) \left|{\cal M}^0_{q\bar q}\right|^2&=& 
\Re \left(\tilde\MA{1,x}{4} (p_1,p_3,p_4,p_2) 
\tilde\MA{0,\dagger}{4} (p_1,p_3,p_4,p_2)\right),
\end{eqnarray}
and
\begin{eqnarray}
\tilde\MA{1,x}{4} (p_1,p_3,p_4,p_2)&=&
\MA{1,x}{4} (p_1,p_3,p_4,p_2)+\MA{1,x}{4} (p_1,p_4,p_3,p_2).
\end{eqnarray}

The renormalised
singularity structure of the various contributions can be easily written in
terms of the tree-level squared matrix elements multiplied
by combinations of the infrared singularity operators of
Section~\ref{sec:operators}.
Explicitly, we find
\begin{eqnarray}
\Poles(A_{4}^{1,a}(1_q,i_g,j_g,2_{\bar q}))&=&
2\left(
 {\bf I}^{(1)}_{qg} (\e,s_{1i})
+{\bf I}^{(1)}_{gg} (\e,s_{ij})
+{\bf I}^{(1)}_{g\bar q} (\e,s_{j2})\right)A_{4}^0 (1_q,i_g,j_g,2_{\bar q})
,\nonumber  \\ \\
\Poles( A_{4}^{1,b}(1_q,i_g,j_g,2_{\bar q}))&=&
2{\bf I}^{(1)}_{q\bar q} (\e,s_{12})
A_{4}^0 (1_q,i_g,j_g,2_{\bar q}),\\
\Poles(A_{4}^{1,c}(1_q,i_g,j_g,2_{\bar q}))&=&2\left(
 {\bf I}^{(1)}_{qg,F} (\e,s_{1i})
+{\bf I}^{(1)}_{gg,F} (\e,s_{ij})
+{\bf I}^{(1)}_{g\bar q,F} (\e,s_{j2})\right)A_{4}^0 (1_q,i_g,j_g,2_{\bar q})
, \nonumber \\ \\
\Poles(\tilde A_{4}^{1,a}(1_q,3_g,4_g,2_{\bar q}))&=&\nonumber \\
&&\hspace{-4cm}
\left(
2{\bf I}^{(1)}_{gg} (\e,s_{34})
+{\bf I}^{(1)}_{qg} (\e,s_{14})
+{\bf I}^{(1)}_{g\bar q} (\e,s_{23})
+{\bf I}^{(1)}_{qg} (\e,s_{13})
+{\bf I}^{(1)}_{g\bar q} (\e,s_{24})\right)\tilde A_{4}^0 (1_q,3_g,4_g,2_{\bar q}),\nonumber \\ \\
\Poles( \tilde A_{4}^{1,b}(1_q,3_g,4_g,2_{\bar q}))&=&
2 {\bf I}^{(1)}_{q\bar q} (\e,s_{12}) 
\tilde A_{4}^0 (1_q,3_g,4_g,2_{\bar q}),
\label{eq:At41b}
\\
\Poles(\tilde A_{4}^{1,c}(1_q,3_g,4_g,2_{\bar q}))&=&\nonumber \\
&&\hspace{-4cm}\left(
2{\bf I}^{(1)}_{gg,F} (\e,s_{34})
+{\bf I}^{(1)}_{qg,F} (\e,s_{14})
+{\bf I}^{(1)}_{g\bar q,F} (\e,s_{23})
+{\bf I}^{(1)}_{qg,F} (\e,s_{13})
+{\bf I}^{(1)}_{g\bar q,F} (\e,s_{24})\right)
\nonumber \\ && \times \tilde A_{4}^0 (1_q,3_g,4_g,2_{\bar q}), \\
\Poles(\tilde A_{4}^{1,d}(1_q,3_g,4_g,2_{\bar q}))&=&\nonumber \\
&&\hspace{-4cm}
\left(
 2{\bf I}^{(1)}_{q\bar q} (\e,s_{12})
+2{\bf I}^{(1)}_{gg} (\e,s_{34})
-{\bf I}^{(1)}_{qg} (\e,s_{14})
-{\bf I}^{(1)}_{g\bar q} (\e,s_{23})
-{\bf I}^{(1)}_{qg} (\e,s_{13})
-{\bf I}^{(1)}_{g\bar q} (\e,s_{24})\right)\nonumber \\
&&\times
\tilde A_{4}^0 (1_q,3_g,4_g,2_{\bar q}).
\end{eqnarray}

As at tree-level, the one-loop amplitude for 
$$\gamma^*(q) \to q(p_1) \bar q(p_2)  q'(p_3) 
\bar q'(p_4)$$ contains two colour structures,
\begin{eqnarray}
\lefteqn{
M_{q\bar q q'\bar q'}^1
= i e_1 g^2 
\left(\frac{g^2}{16\pi^2}\right)
\delta_{q_1q_2}\delta_{q_3q_4}}\nonumber \\
&\times&
\Bigg[
\delta_{i_1i_4}\delta_{i_3i_2}
\left(
N \MB{1,a}{4}(p_1,p_2,p_3,p_4)  
- \frac{1}{N}\MB{1,b}{4}(p_1,p_2,p_3,p_4) 
+N_F \MB{1,c}{4}(p_1,p_2,p_3,p_4)  
\right)\nonumber \\
&&
-\frac{1}{N}
\delta_{i_1i_2}\delta_{i_3i_4}
\Bigg(N \MB{1,d}{4}(p_1,p_2,p_3,p_4)  
- \frac{1}{N}\MB{1,e}{4}(p_1,p_2,p_3,p_4) \nonumber \\ &&
\hspace{2.2cm} +N_F \MB{1,f}{4}(p_1,p_2,p_3,p_4)  
\Bigg) \Bigg]
+ (1\leftrightarrow 3,2\leftrightarrow 4)  \;,
\end{eqnarray}
where
\begin{equation}
\MB{1,a}{4}(p_1,p_2,p_3,p_4) + \MB{1,e}{4}(p_1,p_2,p_3,p_4) 
=
\MB{1,b}{4}(p_1,p_2,p_3,p_4) + \MB{1,d}{4}(p_1,p_2,p_3,p_4). 
\end{equation}
As before, the identical quark matrix element is obtained by permuting the antiquark momenta,
\begin{equation}
M_{q\bar q q\bar q}^1 =
M_{q\bar q q'\bar q'}^1
 - (2\leftrightarrow 4).
\end{equation}

Summing over flavours and including 
symmetry factors, we find that the ``squared" matrix element, is given by
\begin{eqnarray}
\lefteqn{2 \left | M_{4q}^{0,\dagger} M_{4q}^{1}\right|=
\sum_{q,q'}
2 \left | M_{q\bar q q'\bar q'}^{0,\dagger} M_{q\bar q q'\bar q'}^{1}\right|
+
\sum_q 2 \left | M_{q\bar q q\bar q}^{0,\dagger} M_{q\bar q q\bar q}^{1}\right|
}  \nonumber \\
&=&N_4 \left(\frac{\alpha_s}{2\pi}\right)\,
\Bigg[
N \NF B_4^{1,a}(1_q,3_q,4_{\bar q},2_{\bar q})
- \frac{\NF}{N} B_4^{1,b}(1_q,3_q,4_{\bar q},2_{\bar q})
+ N_F^2 B_4^{1,c}(1_q,3_q,4_{\bar q},2_{\bar q})
\nonumber \\
&&\hspace{0cm}-C_4^{1,d}(1_q,3_q,4_{\bar q},2_{\bar q})
+\frac{1}{N^2}C_4^{1,e}(1_q,3_q,4_{\bar q},2_{\bar q})
-\frac{N_F}{N}C_4^{1,f}(1_q,3_q,4_{\bar q},2_{\bar q})
\nonumber \\
&&\hspace{0cm}-C_4^{1,d}(2_{\bar q},4_{\bar q},3_q,1_q)
+\frac{1}{N^2}C_4^{1,e}(2_{\bar q},4_{\bar q},3_q,1_q)
-\frac{N_F}{N}C_4^{1,f}(2_{\bar q},4_{\bar q},3_q,1_q)
\nonumber \\
&&\hspace{0cm}
+N N_{F,\gamma} \hat B_4^{1,a}(1_q,3_q,4_{\bar q},2_{\bar q})
- \frac{N_{F,\gamma}}{N} \hat B_4^{1,b}(1_q,3_q,4_{\bar q},2_{\bar q})
+ N_F N_{F,\gamma}  \hat B_4^{1,c}(1_q,3_q,4_{\bar q},2_{\bar q})
\Bigg ],  \nonumber \\
\end{eqnarray}
where for $x=a,b,c$
\begin{eqnarray}
B_4^{1,x}(1_q,3_q,4_{\bar q},2_{\bar q})\left|{\cal M}^0_{q\bar q}\right|^2 &=&
\Re \left( \MB{1,x}{4}(p_1,p_2,p_3,p_4)\MB{0,\dagger}{4}(p_1,p_2,p_3,p_4)\right), \\
\hat B_4^{1,x}(1_q,3_q,4_{\bar q},2_{\bar q})\left|{\cal M}^0_{q\bar q}\right|^2 &=&
\Re \left( \MB{1,x}{4}(p_1,p_2,p_3,p_4)\MB{0,\dagger}{4}(p_3,p_4,p_1,p_2)\right),  
\end{eqnarray}
and for $x=d,e,f$
\begin{eqnarray}
C_4^{1,x}(1_q,3_q,4_{\bar q},2_{\bar q})\left|{\cal M}^0_{q\bar q}\right|^2 &=&
- \Re \left( \MB{1,x}{4}(p_1,p_2,p_3,p_4)\MB{0,\dagger}{4}(p_1,p_4,p_3,p_2)\right).  
\end{eqnarray}

Using the infrared singularity operators of
section~\ref{sec:operators}, we can extract the singular contributions of the 
renormalised one-loop contribution as,
\begin{eqnarray}
\Poles(B_{4}^{1,a}(1_q,3_q,4_{\bar q},2_{\bar q}))&=&
2\left( 
 {\bf I}^{(1)}_{q\bar q} (\e,s_{14})
+{\bf I}^{(1)}_{q\bar q} (\e,s_{23})\right)
B_{4}^0 (1_q,3_q,4_{\bar q},2_{\bar q}),\\
\Poles(B_{4}^{1,b}(1_q,3_q,4_{\bar q},2_{\bar q}))&=&\nonumber \\
&&\hspace{-4cm}
2\left( 
 2{\bf I}^{(1)}_{q\bar q} (\e,s_{14})
-2{\bf I}^{(1)}_{q\bar q} (\e,s_{13})
+2{\bf I}^{(1)}_{q\bar q} (\e,s_{23})
-2{\bf I}^{(1)}_{q\bar q} (\e,s_{24})
+{\bf I}^{(1)}_{q\bar q} (\e,s_{12})
+{\bf I}^{(1)}_{q\bar q} (\e,s_{34})
\right)\nonumber \\
&&\hspace{2cm}\times
B_{4}^0 (1_q,3_q,4_{\bar q},2_{\bar q}),\\
\Poles(C_{4}^{1,d}(1_q,3_q,4_{\bar q},2_{\bar q}))&=&
 2\left( 
 {\bf I}^{(1)}_{q\bar q} (\e,s_{13})
+{\bf I}^{(1)}_{q\bar q} (\e,s_{24})\right)
C_{4}^0 (1_q,3_q,4_{\bar q},2_{\bar q}),\\
\Poles(C_{4}^{1,e}(1_q,3_q,4_{\bar q},2_{\bar q}))&=&\nonumber \\
&&\hspace{-4cm}
2\left( 
{\bf I}^{(1)}_{q\bar q} (\e,s_{12})
+{\bf I}^{(1)}_{q\bar q} (\e,s_{14})
+{\bf I}^{(1)}_{q\bar q} (\e,s_{23})
+{\bf I}^{(1)}_{q\bar q} (\e,s_{34})
-{\bf I}^{(1)}_{q\bar q} (\e,s_{13})
-{\bf I}^{(1)}_{q\bar q} (\e,s_{24})
\right)\nonumber \\
&&\hspace{2cm}\times
C_{4}^0 (1_q,3_q,4_{\bar q},2_{\bar q}),
\label{eq:C41e} \\
\Poles(\hat B_{4}^{1,a}(1_q,3_q,4_{\bar q},2_{\bar q}))&=&
2\left( 
 {\bf I}^{(1)}_{q\bar q} (\e,s_{14})
+{\bf I}^{(1)}_{q\bar q} (\e,s_{23})\right)
\hat B_{4}^0 (1_q,3_q,4_{\bar q},2_{\bar q}),\\
\Poles(\hat B_{4}^{1,b}(1_q,3_q,4_{\bar q},2_{\bar q}))&=&\nonumber \\
&&\hspace{-4cm}
2\left( 
 2{\bf I}^{(1)}_{q\bar q} (\e,s_{14})
-2{\bf I}^{(1)}_{q\bar q} (\e,s_{13})
+2{\bf I}^{(1)}_{q\bar q} (\e,s_{23})
-2{\bf I}^{(1)}_{q\bar q} (\e,s_{24})
+{\bf I}^{(1)}_{q\bar q} (\e,s_{12})
+{\bf I}^{(1)}_{q\bar q} (\e,s_{34})
\right)\nonumber \\
&&\hspace{2cm}\times
\hat B_{4}^0 (1_q,3_q,4_{\bar q},2_{\bar q})\,.
\end{eqnarray}

\subsubsection{Two-loop matrix elements for three partons}
\label{sec:3jme}

The renormalised two-loop  amplitude $M^2_{q\bar qg}$ 
for a virtual photon to produce a quark-antiquark pair
 together with a single gluon, 
$$\gamma^*(q) \to q(p_1) \bar q(p_2)  g(p_3)$$
contains a single colour structure such that
\begin{equation}
M^2_{q\bar qg} 
= i e \sqrt{2}g \left(\frac{g^2}{16\pi^2}\right)^2
T^{a_3}_{i_1i_2} \MA{2}{3} (p_1,p_3,p_2)\;.
\end{equation}

At NNLO, there are two contributions.   One from the interference of the
two-loop and tree-level amplitudes (\ref{eq:qqnm2g}), the other from the square of the one-loop
amplitudes given in (\ref{eq:ggponeloop}).
These terms were computed in~\cite{3jme} and are given by,
\begin{eqnarray}
2\Re\left(M^{0,\dagger}_{q\bar qg}M^2_{q\bar qg} \right)
&=& N_3 \left(\frac{\alpha_s}{2\pi}\right)^2
\, A_3^{(2\times 0)}(1_q,3_g,2_{\bar q}) \, ,\\
\Re\left(M^{1,\dagger}_{q\bar qg}M^1_{q\bar qg} \right)
&=& N_3 \left(\frac{\alpha_s}{2\pi}\right)^2
\, A_3^{(1\times 1)}(1_q,3_g,2_{\bar q}) \, .
\end{eqnarray}
Following~\cite{catani}, we organise the 
infrared pole structure of the NNLO contributions renormalised in the 
\MSbar\ scheme in terms of the tree and renormalised one-loop amplitudes such
that,
\begin{eqnarray}
\label{eq:twolooppoles}
\lefteqn{\Poles\left(A_3^{(2\times 0)}(1_q,3_g,2_{\bar q})+A_3^{(1\times 1)}(1_q,3_g,2_{\bar q})\right)
}\nonumber \\
 &=&
2\bigg[-\left({\bom I}_{q\bar qg}^{(1)}(\epsilon)\right)^2 
- \frac{\beta_0}{\epsilon}{\bom I}_{q\bar qg}^{(1)}(\epsilon)\nonumber \\
&&
+e^{-\epsilon\gamma}\frac{\Gamma(1-2\epsilon)}{\Gamma(1-\epsilon)}
\left(\frac{\beta_0}{\epsilon} + K\right) {\bom I}_{q\bar qg}^{(1)}(2\epsilon)
+{\bom H}_{q\bar qg}^{(2)}\bigg] A_3^0(1_q,3_g,2_{\bar q}) \nonumber \\
&&
+2\, {\bom I}_{q\bar qg}^{(1)}(\epsilon)   \, 
A_3^{(1 \times 0)}(1_q,3_g,2_{\bar q})\;.
\end{eqnarray}
Here, 
\begin{eqnarray}
{\bom I}_{q\bar qg}^{(1)}(\epsilon) &=& N \left(
{\bom I}^{(1)}_{qg}(\epsilon,s_{13})
+{\bom I}^{(1)}_{qg}(\epsilon,s_{23})
\right)
-\frac{1}{N} {\bom I}^{(1)}_{q\bar q}(\epsilon,s_{12}) \nonumber \\
&& +N_F  
\left(
{\bom I}^{(1)}_{qg,F}(\epsilon,s_{13})
+{\bom I}^{(1)}_{qg,F}(\epsilon,s_{23})
\right),
\end{eqnarray}
with the individual ${\bom I}^{(1)}_{ij}$ defined in 
(\ref{eq:Ione}) and
\begin{eqnarray}
{\bom H}^{(2)}_{q\bar qg} &= & 
\frac{e^{\epsilon \gamma}}{4\,\epsilon\,\Gamma(1-\epsilon)} \Bigg[
\left(4\zeta_3+\frac{589}{432}- \frac{11\pi^2}{72}\right)N^2
+\left(-\frac{1}{2}\zeta_3-\frac{41}{54}-\frac{\pi^2}{48} \right)\nonumber \\
&&
+\left(-3\zeta_3 -\frac{3}{16} + \frac{\pi^2}{4}\right) \frac{1}{N^2}
+\left(-\frac{19}{18}+\frac{\pi^2}{36} \right) N N_F 
+\left(-\frac{1}{54}-\frac{\pi^2}{24}\right) \frac{N_F}{N}+ \frac{5}{27} N_F^2.
\Bigg]\;.\nonumber \\
\label{eq:Htwo}
\end{eqnarray}

We denote the finite contributions as,
\begin{eqnarray}
\label{eq:twoloopfinite}
\Finite{(A_3^{(2\times 0)}(1_q,3_g,2_{\bar q}))} &=& 
N^2 A_{3,N^2}^{(2\times 0),finite}
+A_{3,1}^{(2\times 0),finite}
+ \frac{1}{N^2}A_{3,1/N^2}^{(2\times 0),finite}\nonumber \\
&&+NN_F A_{3,NN_F}^{(2\times 0),finite}
+ \frac{N_F}{N}A_{3,N_F/N}^{(2\times 0),finite}\nonumber \\
&&
+N_F^2 A_{3,N_F^2}^{(2\times 0),finite}
+N_{F,\gamma} \left(\frac{4}{N}-N\right)A_{3,N_{F,\gamma}^2}^{(2\times 0),finite},\\
\Finite{(A_3^{(1\times 1)}(1_q,3_g,2_{\bar q}))} &=& 
N^2 A_{3,N^2}^{(1\times 1),finite}
+A_{3,1}^{(1\times 1),finite}
+ \frac{1}{N^2}A_{3,1/N^2}^{(1\times 1),finite}\nonumber \\
&&+NN_F A_{3,NN_F}^{(1\times 1),finite}
+ \frac{N_F}{N}A_{3,N_F/N}^{(1\times 1),finite}\nonumber \\
&&
+N_F^2 A_{3,N_F^2}^{(1\times 1),finite}.
\end{eqnarray}
Explicit formulae for the 
finite remainders have been given in~\cite{3jme}. These are expressed in 
terms of one-dimensional and two-dimensional harmonic polylogarithms
(HPLs and 2dHPLs)~\cite{hpl,3jmi}, which are generalisations of the 
well-known Nielsen polylogarithms~\cite{nielsen}.
A numerical implementation, which is 
required for all practical applications, is available for HPLs and 
2dHPLs~\cite{hplnum}.

\subsection{Five-parton contribution}
\label{sec:dsub}

Two different five-parton final states contribute at $1/N^2$ to three-jet final
states at NNLO:  $\gamma^*\to q\bar q ggg$ and 
 $\gamma^*\to q\bar q q\bar qg$ with identical quarks.  

The most colour suppressed contribution to the
squared matrix element for $\gamma^*\to q\bar q ggg$ is proportional 
to $1/N^2$.
It is given by,
\begin{eqnarray}
\left|M^0_{q\bar q3g}\right|^2
&=&  N_5 \, \frac{1}{N^2}  \bar A_{5}^0 (1_q,3_g,4_g,5_g,2_{\bar q}),
\label{eq:2q3g}
\end{eqnarray}
while the identical quark contribution is 
\begin{eqnarray}
\left|M^0_{4qg}\right|^2
&=&  N_5 \, \frac{1}{N^2} \, \left[\tilde C_{5}^0 (1_q,2_{\bar q},3_q,4_{\bar q},5_g)
+ \tilde C_{5}^0 (3_q,2_{\bar q},1_q,4_{\bar q},5_g)\right] . 
\label{eq:4q1g}
\end{eqnarray}

The NNLO radiation term appropriate for the three jet final state is given by
\begin{equation}
{\rm d}\sigma_{NNLO}^R = {\rm d}\sigma_{NNLO,\bar A}^R 
+ {\rm d}\sigma_{NNLO,\tilde C}^R \;,
\label{eq:deco}
\end{equation}
with
\begin{eqnarray}
{\rm d}\sigma_{NNLO,\bar A}^R&=& 
\frac{N_{{5}}}{N^2} \,  {\rm d}\Phi_{5}(p_{1},\ldots,p_{5};q)
 \, \frac{1}{3!}\,
 \bar A_{5}^0 (1_q,3_g,4_g,5_g,2_{\bar q}) \JET_{3}^{(5)}(p_{1},\ldots,p_{5})
\;,
\label{eq:a50bar} \\
{\rm d}\sigma_{NNLO,\tilde C}^R&=& 
\frac{N_{{5}}}{N^2} \,  {\rm d}\Phi_{5}(p_{1},\ldots,p_{5};q)\,
2\, \tilde C_{5}^0 (1_q,2_{\bar q},3_q,4_{\bar q},5_g) \JET_{3}^{(5)}(p_{1},\ldots,p_{5})
\;,
\label{eq:c50bar}\end{eqnarray}
where the symmetry factor in front of 
$\bar{A}_5^0$ is due to the inherent 
indistinguishability of gluons. The factor 2 in front of $\tilde C_{5}^0$ arises 
from the fact that two different momentum arrangements contribute to 
the squared matrix element (\ref{eq:4q1g}). If the quarks and antiquarks are 
not distinguished by the jet functions, these contribute equally.

The subtraction terms 
for these cross sections 
are straightforwardly obtained using the procedure
detailed in Section~\ref{sec:method}.
Because there is only a single colour connected pair at $1/N^2$,
there are no singularities
associated with almost colour-unconnected or unconnected unresolved partons 
for either of the two contributions. Therefore
\begin{eqnarray}
{\rm d} \sigma^{S,c}_{NNLO} = {\rm d} \sigma^{S,d}_{NNLO} = 0.
\end{eqnarray}

The other two types of real radiation subtraction terms are decomposed 
according to (\ref{eq:deco}) as
\begin{equation}
{\rm d}\sigma_{NNLO}^{S,i} = {\rm d}\sigma_{NNLO,\bar A}^{S,i} 
+ {\rm d}\sigma_{NNLO,\tilde C}^{S,i} \;.
\end{equation}

The subtraction terms for a single unresolved parton 
are obtained using 
(\ref{eq:sub2a}),
\begin{eqnarray}
{\rm d}\sigma_{NNLO,\bar A}^{S,a}
&= & \frac{N_{5}}{N^2}\, {\rm d}\Phi_{5}(p_{1},\ldots,p_{5};q)\,  \frac{1}{3!}
 \nonumber \\
&\times& 
\sum_{i,j,k \in P_C(3,4,5)}\;A^0_3(1_q,i_g,2_{\bar q})\,
\tilde{A}^0_{4}(\widetilde{(1i)}_q,j_g,k_g,\widetilde{(2i)}_{\bar q})\,
\,\JET_{3}^{(4)}(\widetilde{p_{1i}},p_j,p_k,\widetilde{p_{2i}})\;,
\nonumber \\
{\rm d}\sigma_{NNLO,\tilde C}^{S,a}
&= & \frac{N_{5}}{N^2}\, {\rm d}\Phi_{5}(p_{1},\ldots,p_{5};q)\, 2
 \nonumber \\
&\times & 
\bigg\{ A^0_3(1_q,5_g,2_{\bar q})\, C_4^0(\widetilde{(15)}_q,
3_q,4_{\bar q},\widetilde{(25)}_{\bar q})\, 
\JET_{3}^{(4)}(\widetilde{p_{15}},p_3,p_4,\widetilde{p_{25}}) \nonumber \\&&
+ A^0_3(1_q,5_g,4_{\bar q})\, C_4^0(\widetilde{(15)}_q,3_q,
\widetilde{(45)}_{\bar q},2_{\bar q})  \,
\JET_{3}^{(4)}(\widetilde{p_{15}},p_2,p_3,\widetilde{p_{45}}) \nonumber \\&&
+  A^0_3(3_q,5_g,2_{\bar q})\, C_4^0(1_q,
\widetilde{(35)}_q,4_{\bar q},\widetilde{(25)}_{\bar q})\, 
\JET_{3}^{(4)}(\widetilde{p_{35}},p_1,p_4,\widetilde{p_{25}}) \nonumber \\&&
+ A^0_3(3_q,5_g,4_{\bar q})\, C_4^0(1_q,\widetilde{(35)}_q,
\widetilde{(45)}_{\bar q},2_{\bar q})  \,
\JET_{3}^{(4)}(\widetilde{p_{35}},p_1,p_2,\widetilde{p_{35}}) \nonumber \\&&
-  A^0_3(1_q,5_g,3_q)\, C_4^0(\widetilde{(15)}_{q},
\widetilde{(35)}_q,4_{\bar q},2_{\bar q}) \,
\JET_{3}^{(4)}(\widetilde{p_{15}},p_2,p_4,\widetilde{p_{35}}) \nonumber \\&&
- A^0_3(2_{\bar q},5_g,4_{\bar q})\, C_4^0(1_q,3_q,
\widetilde{(45)}_{\bar q},\widetilde{(25)}_{\bar q}) \,
\JET_{3}^{(4)}(\widetilde{p_{25}},p_1,p_3,\widetilde{p_{45}}) 
\bigg\}\;.
\label{eq:sub2aslc}
\end{eqnarray}
The sum in the first contribution runs only over the three cyclic 
permutations of the gluon momenta to prevent double counting of identical
configurations obtained by interchange of $j$ and $k$.

The colour-connected double unresolved subtraction terms
are given by (see (\ref{eq:sub2b})),
\begin{eqnarray}
\label{eq:Sb}
{\rm d}\sigma_{NNLO,\bar A}^{S,b}
&= & N_{5}\, {\rm d}\Phi_{5}(p_{1},\ldots,p_{5};q)
\,\frac{1}{3!}\,
\sum_{i,j,k \in P_C(3,4,5)} 
 \biggl(
\tilde{A}^0_4(1_q,i_g,j_g,2_{\bar q})\,\nonumber \\
&&
-A^0_3(1_q,i_g,2_{\bar q})\; A^0_3(\widetilde{(1i)}_q,j_g,\widetilde{(2i)}_{\bar q})\;
-A^0_3(1_q,j_g,2_{\bar q})\; A^0_3(\widetilde{(1j)}_q,i_g,\widetilde{(2j)}_{\bar q})\;
\biggr)\;\nonumber \\
&&\hspace{2cm}\times
A^0_{3}(\widetilde{(1ij)}_q,k_g,\widetilde{(2ij)}_{\bar q})\,
\JET_{3}^{(3)}(\widetilde{p_{1ij}},p_k,\widetilde{p_{2ij}})\;, 
\label{eq:sub2bslcA} \\
{\rm d}\sigma_{NNLO,\tilde C}^{S,b}
&= & N_{5}\, {\rm d}\Phi_{5}(p_{1},\ldots,p_{5};q)\, 2
\nonumber \\
&\times& 
{C}^0_4(1_q,3_q,4_{\bar q},2_{\bar q}) \,
A^0_{3}(\widetilde{(134)}_q,5_g,\widetilde{(234)}_{\bar q})
\,\JET_{3}^{(3)}(\widetilde{p_{134}},p_5,\widetilde{p_{234}})\;.
\label{eq:sub2bslcC}
\end{eqnarray}
In the last equation, no single unresolved  terms have to be subtracted 
from 
$C_4^0$, which contains only a single triple collinear limit, but no 
single unresolved limits (\ref{eq:c40lim}).

By construction, the combination
\begin{equation}
{\rm d}\sigma_{NNLO}^{R}
-{\rm d}\sigma_{NNLO}^{S,a}
-{\rm d}\sigma_{NNLO}^{S,b}
\end{equation}
is finite in all unresolved limits. 

The momentum maps used in (\ref{eq:sub2aslc})--(\ref{eq:sub2bslcC}) are 
explicit realizations of the generic momentum maps denoted 
in Section~\ref{sec:method}. In particular, in the single unresolved 
subtraction term (\ref{eq:sub2aslc}), we map the 
three on-shell momenta $i$, $j$ and 
$k$ onto the on-shell momenta $(\widetilde{ij})$ and $(\widetilde{kj})$ 
by using the three-to-two-parton momentum map of~\cite{nnlosub1}. With 
this mapping, the single unresolved phase spaces in the collinear 
limits of the unresolved momentum $k$ with the hard momenta $i$ or $j$  
are treated symmetrically. This mapping is in contrast to the dipole phase 
space mapping~\cite{cs}, where one collinear limit is mapped out to a larger 
phase space volume than the other. In the double unresolved subtraction 
terms (\ref{eq:sub2bslcA}) and
(\ref{eq:sub2bslcC}), we map the four on-shell momenta  $i$, $j$, $k$ and 
$l$ onto the on-shell momenta $(\widetilde{ijk})$ and $(\widetilde{ljk})$ 
by using the ordered 
four-to-two-parton momentum map of~\cite{nnlosub1}. This mapping 
guarantees that both triple collinear limits of the unresolved momenta 
$j$ and $k$ with the hard momenta $i$ or $l$ are treated symmetrically. The 
single unresolved limits of $j$ with $i$ and of $k$ with $l$ are also 
mapped in a symmetric manner, ensuring the correctness of the double unresolved
limit obtained by combining both of them. However, the phase space mapping 
of~\cite{nnlosub1} used here does not yield the correct single unresolved 
behaviour of the phase space in the simple collinear limits of $k$ with $i$ 
or of $j$ with $l$ (as explained in detail in~\cite{nnlosub1}, it is not 
possible to construct a single four-to-two-parton momentum map which 
yields the correct simple unresolved behaviour in all four limits). 
We therefore decompose $\tilde{A}_4^0$ into its ordered components
(which contain only two of the four simple collinear limits), 
as explained in the context of (\ref{eq:A40t}) and use either 
the ordered mapping onto $(\widetilde{ijk})$ and $(\widetilde{ljk})$ or onto
 $(\widetilde{ikj})$ and $(\widetilde{lkj})$. For clarity, these two mappings 
are not distinguished in (\ref{eq:sub2bslcA}). Since $C_4^0$ contains no 
simple collinear limit, but only triple collinear limits, it does not need 
to be decomposed any further.

\subsection{Four-parton contribution}
 \label{sec:vsub}
At one-loop,
there are two contributions to
the colour suppressed contribution proportional to $1/N^2$, one from the four
quark final state and one from the two quark-two gluon final state.
We therefore decompose the virtual one-loop four-parton contribution as
\begin{equation}
{\rm d}\sigma_{NNLO}^{V,1} = {\rm d}\sigma_{NNLO,\bar A}^{V,1} 
+ {\rm d}\sigma_{NNLO,\tilde C}^{V,1} \;,
\end{equation}
with
\begin{eqnarray}
{\rm d}\sigma_{NNLO,\bar A}^{V,1}&=&
\frac{N_{{4}}}{N^2}\, \left(\frac{\alpha_s}{2\pi}\right)\,
{\rm d}\Phi_{4}(p_{1},\ldots,p_{4};q)\,
\frac{1}{2!}\,
\tilde A_{4}^{1,b} (1_q,3_g,4_g,2_{\bar q})\; 
\JET_{3}^{(4)}(p_{1},\ldots,p_{4}),\hspace{4mm}
\label{eq:sigvA}
\\
{\rm d}\sigma_{NNLO,\tilde C}^{V,1}&=&
\frac{N_{{4}}}{N^2}\, \left(\frac{\alpha_s}{2\pi}\right)\,
{\rm d}\Phi_{4}(p_{1},\ldots,p_{4};q)\,
2\, C_{4}^{1,e} (1_q,3_q,4_{\bar q},2_{\bar q})\; 
\JET_{3}^{(4)}(p_{1},\ldots,p_{4}),\hspace{4mm}
\label{eq:sigvC}
\end{eqnarray}
where the origin of the symmetry factors is as in the 
real radiation five-parton contributions of the previous section.

As detailed in Section~\ref{sec:method}, there are three subtraction terms; 
the subtraction of the infrared poles in the one-loop contribution,
the subtraction of the one-loop unresolved part and a compensation term
outside the singular region.
The first of these three terms is constructed  in (\ref{eq:subv2a}). 
Following 
(\ref{eq:nlocancel}), it 
is minus the integrated form of 
${\rm d}\sigma_{NNLO}^{S,a}$ given in (\ref{eq:sub2aslc}):
\begin{eqnarray}
{\rm d}\sigma_{NNLO,\bar A}^{VS,1,a}
&= &
- \frac{N_{{4}}}{N^2}\, \left(\frac{\alpha_s}{2\pi}\right)\,
 {\rm d}\Phi_{4}(p_{1},\ldots,p_{4};q)
\, \frac{1}{2!} \nonumber \\
&\times&  {\cal A}_3^0(s_{12})\,
\tilde A_{4}^0 (1_q,3_g,4_g,2_{\bar q})\,
\JET_{3}^{(4)}(p_{1},p_3,p_4,p_{2})\;,\\
{\rm d}\sigma_{NNLO,\tilde C}^{VS,1,a}
&= &
- \frac{N_{{4}}}{N^2}\, \left(\frac{\alpha_s}{2\pi}\right)\,
 {\rm d}\Phi_{4}(p_{1},\ldots,p_{4};q)
\,2 \, \nonumber \\
&\times& \Big\{ {\cal A}_3^0(s_{12}) 
+ {\cal A}_3^0(s_{14}) + {\cal A}_3^0(s_{23})+ {\cal A}_3^0(s_{34})
- {\cal A}_3^0(s_{13}) 
- {\cal A}_3^0(s_{24}) \Big\}\nonumber \\ &\times&
 C_{4}^0 (1_q,3_q,4_{\bar q},2_{\bar q})
\JET_{3}^{(4)}(p_{1},p_3,p_4,p_{2})\;.
\end{eqnarray}
Given that $\Poles({\cal A}_3^0(s_{ij})) = -2 {\bom I}^{(1)}_{q\bar q}(s_{ij})$,
we see that the pole structures of these subtraction terms 
exactly cancel
the pole structure of $\tilde A^{1,b}_4$ given in
(\ref{eq:At41b}) and of $C_4^{1,e}$ given in (\ref{eq:C41e}).

Single particle unresolved contributions have to be subtracted only from 
(\ref{eq:sigvA}), since (\ref{eq:sigvC}) vanishes in all its single unresolved 
limits, implying
\begin{equation}
{\rm d}\sigma_{NNLO,\tilde C}^{VS,1,b} = {\rm d}\sigma_{NNLO,\tilde C}^{VS,1,c}  = 0 \;.
\end{equation}
Moreover, renormalisation does not affect the colour structure considered here.

From (\ref{eq:subv2b}), we find:
\begin{eqnarray}
\label{eq:VS1b}
{\rm d}\sigma_{NNLO,\bar A}^{VS,1,b}
&= &  \frac{N_{{4}}}{N^2}\, \left(\frac{\alpha_s}{2\pi}\right)\,
{\rm d}\Phi_{4}(p_{1},\ldots,p_{4};q)
\frac{1}{2!}\, \nonumber \\
&\times& 
\sum_{i,j\in P(3,4)}
\bigg(
\tilde A_3^1(1_q,i_g,2_{\bar q})A_3^0(\widetilde{(1i)}_q,j_g,\widetilde{(2i)}_{\bar q})
\nonumber \\
&&+A_3^0(1_q,i_g,2_{\bar q})\left[
\tilde A_3^1(\widetilde{(1i)}_q,j_g,\widetilde{(2i)}_{\bar q})
+{\cal A}_2^1(s_{1234})A_3^0(\widetilde{(1i)}_q,j_g,\widetilde{(2i)}_{\bar
q})\right]
\bigg)\nonumber\\
&&\hspace{4cm}\times
\JET_{3}^{(3)}(\widetilde{p_{1i}},p_j,\widetilde{p_{2i}})\;.
\end{eqnarray}

The previous term correctly subtracts the singularities in the unresolved
limits. 
However away from the limit, where the jet algorithm combines the partons, it
introduces spurious poles.   These terms are compensated by (\ref{eq:subv2c}).
In this case there are only two hard radiators - the quark and antiquark so
that $i \equiv n = 1$ and $j \equiv p = 2$:
\begin{eqnarray}
\label{eq:VS1c}
{\rm d}\sigma_{NNLO,\bar A}^{VS,1,c}
&= & \frac{N_{{4}}}{N^2}\, \left(\frac{\alpha_s}{2\pi}\right)\,
{\rm d}\Phi_{4}(p_{1},\ldots,p_{4};q)
\frac{1}{2!}\, \nonumber \\
&\times& 
\sum_{i,j\in P(3,4)}
{\cal A}_3^0(s_{12})
A_3^0(1_q,i_g,2_{\bar q})A_3^0(\widetilde{(1i)}_q,j_g,\widetilde{(2i)}_{\bar q})
\JET_{3}^{(3)}(\widetilde{p_{1i}},p_j,\widetilde{p_{2i}})\;.
\end{eqnarray}
This term corresponds to integrating the last two terms of  ${\rm
d}\sigma_{NNLO}^{S,b}$ given in (\ref{eq:Sb}).

Taken together, the explicit poles in
(\ref{eq:VS1b}) and (\ref{eq:VS1c}) cancel.   Explicitly,  
while dropping overall factors of phase space etc., and using the
definitions given in Sections~\ref{sec:operators} and~\ref{sec:qq}, 
the pole part of (\ref{eq:VS1c})
is proportional to 
$$
-2{\bom I}^{(1)}_{q\bar q} (\epsilon,s_{12})
A_3^0(1_q,i_g,2_{\bar q})A_3^0(\widetilde{(1i)}_q,j_g,\widetilde{(2i)}_{\bar q}),
$$
while the first term in (\ref{eq:VS1b}) is
$$
\left(2{\bom I}^{(1)}_{q\bar q} (\epsilon,s_{12})
-2{\bom I}^{(1)}_{q\bar q} (\epsilon,s_{12i})\right)
A_3^0(1_q,i_g,2_{\bar q})A_3^0(\widetilde{(1i)}_q,j_g,\widetilde{(2i)}_{\bar
q}).
$$
The second term in eq.~(\ref{eq:VS1b}) is
$$
2{\bom I}^{(1)}_{q\bar q} (\epsilon,s_{12i})
A_3^0(1_q,i_g,2_{\bar q})A_3^0(\widetilde{(1i)}_q,j_g,\widetilde{(2i)}_{\bar
q}).
$$
Adding these contributions, we see that the explicit poles cancel.

By construction, the combination
\begin{equation}
{\rm d}\sigma_{NNLO}^{V,1}
-{\rm d}\sigma_{NNLO}^{VS,1,a}
-{\rm d}\sigma_{NNLO}^{VS,1,b}
-{\rm d}\sigma_{NNLO}^{VS,1,c}
\end{equation}
is free of explicit $1/\e$-poles.  
 Furthermore, it is also finite in its unresolved limits.

\subsection{Three-parton contribution}
\label{sec:cancel}
The three-parton contribution consists of the two-loop three-parton matrix
element together with the integrated forms of 
the five-parton and four-parton
subtraction terms,
$$
{\rm d}\sigma_{NNLO}^{V,2}+{\rm d}\sigma_{NNLO}^{S}+{\rm d}\sigma_{NNLO}^{VS,1}
\,.
$$
We have shown in the previous subsections that parts of the double real 
radiation subtraction term cancel already with the virtual subtraction term. 
To make these cancellations explicit, we introduce a function 
${\rm d}\sigma_{NNLO}^{T}$, which subtracts these terms from 
${\rm d}\sigma_{NNLO}^{S}$ and adds them to 
${\rm d}\sigma_{NNLO}^{VS,1}$.
After these cancellations, we observe that the three-parton 
contribution reduces to 
$$
{\rm d}\sigma_{NNLO}^{V,2}+\left({\rm d}\sigma_{NNLO}^{S}
- {\rm d}\sigma_{NNLO}^{T}\right) 
+\left({\rm d}\sigma_{NNLO}^{VS,1}  +
{\rm d}\sigma_{NNLO}^{T}\right)\,.
$$

Upon integration of the unresolved phase space, we find
\begin{eqnarray}
{\rm d}\sigma_{NNLO}^{S}- {\rm d}\sigma_{NNLO}^{T}
&= &  \frac{1}{N^2}\, \left(
\frac{1}{2}\tilde {\cal A}_4^0(s_{12})
+2 {\cal C}_4^0(s_{12})\right)
A_3^0(1_q,3_g,2_{\bar q}) \,{\rm d}\sigma_3\;,
\end{eqnarray}
where we defined the three-parton normalisation factor
\begin{equation}
{\rm d}\sigma_3 = 
N_{{3}}\, \left(\frac{\alpha_s}{2\pi}\right)^2\,
{\rm d}\Phi_{3}(p_{1},p_2,p_{3};q) \, \JET_{3}^{(3)}(p_1,p_2,p_3)\;.
\end{equation}

Using the explicit formulae for $\tilde {\cal A}_4^0$
and $C_4^0$ presented in
Section~\ref{sec:qq}, we find that the singular contribution is
given by
\begin{eqnarray}
\label{eq:Sint}
\lefteqn{\Poles({\rm d}\sigma_{NNLO}^{S}-{\rm d}\sigma_{NNLO}^{T})=}
\nonumber \\
& &  \frac{1}{N^2}\,\left[
2 \left({\bom I}_{q\bar q}^{(1)}(\epsilon,s_{12})\right)^2
- {\bom H}_{R,\tilde A}^{(2)}(\epsilon,s_{12})
- {\bom H}_{R,\tilde C}^{(2)}(\epsilon,s_{12})
\right]\,A_3^0(1_q,3_g,2_{\bar q}) \, {\rm d}\sigma_3. 
\end{eqnarray}

Similarly, integration of (\ref{eq:VS1b}) yields,
\begin{eqnarray}
\label{eq:VS1int}
{\rm d}\sigma_{NNLO}^{VS,1} + {\rm d}\sigma_{NNLO}^{T}
&= & 
 \frac{1}{N^2}\,\bigg(
\tilde {\cal A}_3^1(s_{12})A_3^0(1_q,3_g,2_{\bar q})\nonumber \\
&&
+{\cal A}_3^0(s_{12})
\left[\tilde A_3^1(1_q,3_g,2_{\bar q})+{\cal A}_2^1(s_{123})
A_3^0(1_q,3_g,2_{\bar q})\right]\bigg) 
{\rm d}\sigma_3\;, \nonumber \\
\end{eqnarray}
resulting in 
\begin{eqnarray}
\lefteqn{\Poles({\rm d}\sigma_{NNLO}^{VS,1}+{\rm d}\sigma_{NNLO}^{T})=}
\nonumber \\ &&
 \frac{1}{N^2}\,
\bigg(- {\bf H}^{(2)}_{V,\tilde{A}}(\e,s_{12})  
+2{\bom I}_{q\bar q}^{(1)}(\epsilon,s_{12})
\left[
\tilde A_3^1(1_q,3_g,2_{\bar q})+{\cal A}_2^1(s_{123})
A_3^0(1_q,3_g,2_{\bar q})\right]\bigg)
{\rm d}\sigma_3,\nonumber \\
\end{eqnarray}
where it is noteworthy that the finite part of ${\cal A}_3^0$ in 
(\ref{eq:VS1int}) is cancelled exactly by the terms proportional 
to ${\cal A}_2^1$ in $\tilde{{\cal A}}_3^1$ (\ref{eq:A31tpoles}).  

At ${\cal O}(1/N^2)$, the singular 
two-loop virtual contribution is given by
expanding (\ref{eq:twolooppoles}).
We find,
\begin{eqnarray}
\Poles({\rm d}\sigma_{NNLO}^{V,2})
&= &  \frac{1}{N^2}\,
\bigg(
\left[-2\left({\bom I}_{q\bar q}^{(1)}(\epsilon,s_{12})\right)^2
+2{\bom H}_{q\bar q,1/N^2}^{(2)}(\epsilon,s_{12})\right]
A_3^0(1_q,3_g,2_{\bar q}) \nonumber \\
&&
+2{\bom I}_{q\bar q}^{(1)}(\epsilon,s_{12})
\left[
\tilde A_3^1(1_q,3_g,2_{\bar q})+{\cal A}_2^1(s_{123})
A_3^0(1_q,3_g,2_{\bar q})\right]\bigg)
{\rm d}\sigma_3.\nonumber \\
\end{eqnarray}

Using relation between the  hard radiation operators 
${\bf H}$ 
given in
(\ref{eq:H2qqb1overN}), it is easy to see that
\begin{eqnarray}
\Poles({\rm d}\sigma_{NNLO}^{S}
+{\rm d}\sigma_{NNLO}^{VS,1}+{\rm d}\sigma_{NNLO}^{V})&=&0.
\end{eqnarray}
The three-parton contribution is thus free of explicit infrared poles.
The remaining finite contributions are fully differential in the
three-jet variables
and 
can be integrated numerically for any given infrared-safe jet definition.

\subsection{Numerical implementation}
\label{sec:numer}

Starting from the program {\tt EERAD2}~\cite{cullen}, which computes four-jet
production at NLO, we implemented the NNLO antenna subtraction method 
for the $1/N^2$ colour factor contribution to $e^+e^-\to 3j$. {\tt EERAD2}
already  contains the five-parton and four-parton 
matrix elements relevant here, as well as the NLO-type subtraction terms 
${\rm d} \sigma_{NNLO}^{S,a}$ and ${\rm d} \sigma_{NNLO}^{VS,1,a}$. 

The implementation contains three channels, classified 
by their partonic multiplicity: 
\begin{itemize}
\item
in the five-parton channel, we
integrate
\begin{equation}
{\rm d}\sigma_{NNLO}^{R} - {\rm d}\sigma_{NNLO}^{S}\;.
\end{equation}
\item in the four-parton channel, we integrate
\begin{equation}
{\rm d}\sigma_{NNLO}^{V,1} - {\rm d}\sigma_{NNLO}^{VS,1}\;.
\end{equation}
\item in the three-parton channel, we integrate
\begin{equation}
{\rm d}\sigma_{NNLO}^{V,2} +{\rm d}\sigma_{NNLO}^{S}
+ {\rm d}\sigma_{NNLO}^{VS,1}\;.
\end{equation}
\end{itemize}
The numerical integration over these channels is carried out by Monte Carlo 
methods using the {\tt VEGAS}~\cite{vegas} implementation. 

It was already demonstrated above that the integrands in the four-parton and 
three-parton channel are free of explicit infrared poles. In the 
five-parton and four-parton channel, we tested the proper implementation of 
the subtraction by generating trajectories of phase space points approaching 
a given single or double unresolved limit using the {\tt RAMBO}~\cite{rambo}
phase space generator. Along these trajectories, we observe that the 
antenna subtraction terms converge locally towards the physical matrix 
elements (it has to be pointed out that the $1/N^2$ colour structure is 
free of angular terms, since it does not involve any gluon splitting 
processes), and that the cancellations among individual 
contributions to the subtraction terms take place as expected in 
Section~\ref{sec:method}. Moreover, we checked the correctness of the 
subtraction by introducing a 
lower cut (slicing parameter) on the phase space variables, and observing 
that our results are independent of this cut (provided it is 
chosen small enough). This behaviour indicates that the 
subtraction terms ensure that the contribution of potentially singular 
regions of the final state phase space does not contribute to the numerical 
integrals, but is accounted for analytically.

\section{Infrared cancellations in $e^+e^-\to 3$~jets at NNLO}
\label{sec:3jir}
To finally illustrate the power of the antenna 
subtraction method at NNLO, we note that the 
infrared poles of the two-loop (including one-loop times one-loop) correction
to $\gamma^*\to q\bar qg$ are cancelled in all colour factors by a 
combination of integrated three-parton and four-parton
antenna functions.

The integrated five-parton
double real radiation subtraction term for three-jet production
 after cancellation of terms with the virtual single unresolved 
subtraction term reads:
\begin{equation} 
\d \sigma_{NNLO}^S - \d \sigma_{NNLO}^T=  {\cal X}_{q\bar q g,NNLO}^{S}\, 
 A_3^0(1_q, 3_g,2_{\bar q})\,{\rm d}\sigma_3
\end{equation}
with
\begin{eqnarray}
\hspace{-1cm}{\cal X}_{q\bar q g,NNLO}^{S} = && \nonumber \\
 N^2&& \bigg[ \frac{1}{2}\, {\cal D}_4^0  (s_{13}) + 
 \frac{1}{2}\, {\cal D}_4^0  (s_{23}) - \frac{1}{8}
\left({\cal D}_3^0  (s_{13}) - {\cal D}_3^0  (s_{23})\right)^2 
\nonumber \\ &&\hspace{2cm}
-\frac{1}{2} \left(\tilde{{\cal A}}_4^0(s_{12})
- {\cal A}_3^0(s_{12})\, {\cal A}_3^0(s_{12})
\right)
\bigg]\nonumber \\
+ N^0&& \bigg[ - {\cal A}_4^0  (s_{12}) 
- \frac{1}{2}\, \tilde{{\cal A}}_4^0(s_{12}) - 2\,  {\cal C}_4^0 (s_{12}) 
\nonumber \\ &&\hspace{2cm}
- \frac{1}{2} {\cal A}_3^0(s_{12}) \left(
{\cal D}_3^0  (s_{13}) + {\cal D}_3^0  (s_{23})\right)
+ {\cal A}_3^0(s_{12})\, {\cal A}_3^0(s_{12}) 
\bigg]\nonumber \\
+ \frac{1}{N^2}&& \left[ \frac{1}{2} \,\tilde{{\cal A}}_4^0(s_{12})
+ 2\,  {\cal C}_4^0 (s_{12}) \right]\nonumber \\
+ N\,N_F&& \bigg[  {\cal E}_4^0  (s_{13}) + 
 {\cal E}_4^0  (s_{23}) 
- \frac{1}{4}
\left({\cal D}_3^0  (s_{13}) \, {\cal E}_3^0  (s_{13})
+ {\cal D}_3^0  (s_{23})\,  {\cal E}_3^0  (s_{23}) \right)
\nonumber \\ &&\hspace{2cm}+ \frac{1}{4}
\left({\cal D}_3^0  (s_{13}) \, {\cal E}_3^0  (s_{23})
+ {\cal D}_3^0  (s_{23}) \, {\cal E}_3^0  (s_{13}) \right)
\bigg] \nonumber \\
+ \frac{N_F}{N}&& \bigg[ -  {\cal B}_4^0  (s_{12})
- \frac{1}{2}\, \tilde{{\cal E}}_4^0  (s_{13}) - 
 \frac{1}{2}\,\tilde{{\cal E}}_4^0  (s_{23}) - \frac{1}{2}
{\cal A}_3^0  (s_{12}) \, \left({\cal E}_3^0  (s_{13})
+ {\cal E}_3^0  (s_{23}) \right)
\bigg] .
\label{eq:dsubreal}
\end{eqnarray}

Integration of the remainder of the 
four-parton single real radiation subtraction term
produces
\begin{eqnarray}
\d \sigma_{NNLO}^{VS,1} + \d \sigma_{NNLO}^T&=& \bigg\{
{\cal X}_{q\bar q g,NNLO}^{VS}\, 
 A_3^0(1_q, 3_g,2_{\bar q}) 
+  {\cal X}_{q\bar q g,NNLO}^{VS,tree}\,
  A_3^{(1\times 0)}(1_q, 3_g,2_{\bar q})
\nonumber \\
&&\hspace{1.27cm} 
+{\cal X}_{q\bar q g,NNLO}^{VS,\beta}\, 
 \frac{\beta_0}{\e}\, A_3^0(1_q, 3_g,2_{\bar q})\bigg\}\,{\rm d}\sigma_3\;,
\end{eqnarray}
with
\begin{eqnarray}
\hspace{-1cm}{\cal X}_{q\bar q g,NNLO}^{VS} \quad = \quad
 N^2&& \bigg[ \frac{1}{2}\, {\cal D}_3^1  (s_{13}) + 
 \frac{1}{2}\, {\cal D}_3^1  (s_{23}) - \tilde{{\cal A}}_3^1(s_{12})
\bigg]\nonumber \\
+ N^0&& \bigg[ - {\cal A}_3^1  (s_{12}) 
-  \tilde{{\cal A}}_3^1(s_{12})  
\bigg]\nonumber \\
+ \frac{1}{N^2}&&  \tilde{{\cal A}}_3^1(s_{12})\nonumber \\
 + N\,N_F&& \bigg[ \frac{1}{2}\, \hat{{\cal D}}_3^1  (s_{13}) +\frac{1}{2} \,
  \hat{{\cal D}}_3^1 (s_{23})
+ \frac{1}{2}\, {\cal E}_3^1 
 (s_{13}) +\frac{1}{2} \,
 {\cal E}_3^1  (s_{23}) 
\bigg] \nonumber \\
+ \frac{N_F}{N}&& \bigg[ -  \hat{{\cal A}}_3^1  (s_{12})
-  \frac{1}{2}\, \tilde{{\cal E}}_3^1  (s_{13}) -  \frac{1}{2}\, 
 \tilde{{\cal E}}_3^1  (s_{23}) 
\bigg] \nonumber \\
+ N_F^2&&\bigg[ \frac{1}{2}\,\hat{{\cal E}}_3^1  (s_{13}) + \frac{1}{2}\, 
 \hat{{\cal E}}_3^1  (s_{23})\bigg]
\;,
\label{eq:vsubreal}
\end{eqnarray}
\begin{eqnarray}
{\cal X}_{q\bar q g,NNLO}^{VS,tree} &= &
 N \left[\frac{1}{2}\, {\cal D}_3^0  (s_{13}) + 
 \frac{1}{2}\, {\cal D}_3^0  (s_{23}) \right] 
+N_F \left[\frac{1}{2}\, {\cal E}_3^0  (s_{13}) + 
 \frac{1}{2}\, {\cal E}_3^0  (s_{23}) \right]
-\frac{1}{N}{\cal A}_3^0  (s_{12})\;,\nonumber \\
\end{eqnarray}
\begin{eqnarray}
{\cal X}_{q\bar q g,NNLO}^{VS,\beta} &= &
 N \left[\frac{1}{2}\, {\cal D}_3^0  (s_{13}) \,
\left[(s_{13})^{-\e} - (s_{123})^{-\e} \right] 
 +\frac{1}{2}\, {\cal D}_3^0  (s_{23})
\left[(s_{23})^{-\e} - (s_{123})^{-\e} \right] 
 \right] \nonumber \\
&&+N_F \left[\frac{1}{2}\, {\cal E}_3^0  (s_{13}) 
\left[(s_{13})^{-\e} - (s_{123})^{-\e} \right] 
+  \frac{1}{2}\, {\cal E}_3^0  (s_{23}) 
\left[(s_{23})^{-\e} - (s_{123})^{-\e} \right] 
\right] \nonumber \\
&&
-\frac{1}{N}\,{\cal A}_3^0  (s_{12})\,
\left[(s_{12})^{-\e} - (s_{123})^{-\e} \right] 
 \;.
\end{eqnarray}

Adding these integrated subtraction terms to the 
two-loop three-parton contribution to $\gamma^*\to q\bar q g$~\cite{3jme},
discussed in Section~\ref{sec:3jme}, 
one observes
\begin{equation}
\Poles \left( \d \sigma_{NNLO}^{S} + \d \sigma_{NNLO}^{VS,1}
 + \d \sigma_{NNLO}^{V,2}\right) =0
\end{equation}
for all colour structures (\ref{eq:twolooppoles}).
This highly non-trivial cancellation 
clearly illustrates that the antenna functions derived in this paper
correctly approximate QCD matrix elements in all infrared singular limits at 
NNLO. The formulae in this section also outline the structure of infrared 
cancellations in $e^+e^-\to 3j$ at NNLO, and indicate the structure of the 
subtraction terms in all colour factors.

The above formulae also illustrate   
several aspects of the antenna subtraction
method which were discussed in more generality in the preceeding sections.

The subtraction terms for two colour-unconnected and almost colour-unconnected 
unresolved partons $\d \sigma_{NNLO}^{S,c}$ (\ref{eq:sub2c}) and 
$\d \sigma_{NNLO}^{S,d}$ (\ref{eq:sub2d}) cancel partially with 
the virtual one-loop single unresolved subtraction term
 $\d \sigma_{NNLO}^{VS,1,c}$ (\ref{eq:subv2c}). The pattern of 
these cancellations is different for each colour factor, as can be seen 
from the different products of two tree-level three-parton antenna functions
in (\ref{eq:dsubreal}).

In Sections~\ref{sec:qg} and \ref{sec:D40limit}, we 
indicated that 
the tree-level four-parton quark-gluon-gluon-gluon antenna function
$D^0_4$ contains an unwanted double unresolved limit due to the 
cyclic nature of its colour indices. In the $N^2$--contribution to 
(\ref{eq:dsubreal}), we see how this limit can be removed 
by subtraction of the subleading colour quark-antiquark antenna function 
$\tilde{A}^0_4$ and its strongly ordered limits. We observe that a 
similar subtraction has to be carried out on the one-loop three-parton antenna 
function $D^1_3$ in (\ref{eq:vsubreal}). 

\section{Conclusions}
\label{sec:conc}

In this paper, we presented a new method for the subtraction of infrared 
singularities in the calculation of jet observables at NNLO. We introduced 
subtraction terms for double real radiation at tree level and 
single real radiation at one loop based on 
antenna functions. These antenna 
functions describe the colour-ordered radiation of unresolved 
partons between a 
pair of hard (radiator) partons. 

All antenna functions at NLO and NNLO can 
be derived from physical matrix elements: quark-antiquark antenna 
functions from the corrections to the process $\gamma^* \to q\bar 
q$~\cite{our2j}, quark-gluon antenna functions from $\tilde{\chi} \to 
\tilde{g} g$~\cite{chi} and gluon-gluon antenna functions from 
$H\to gg$~\cite{h}. We listed all three-parton tree-level and one-loop 
antenna functions as well as all four-parton tree-level antenna functions
in their explicit, unintegrated form. We provided all integrals of these 
functions over the three- and four-parton antenna phase spaces for the 
kinematical situation of a neutral massive particle decaying into massless 
partons, as is relevant to $e^+e^- \to m$~jets. Having the subtraction terms 
in both unintegrated and integrated form, they can now be directly implemented into
a parton level NNLO event generator for jet production in electron-positron 
annihilation. 

To demonstrate the application of our new method on a non-trivial example, 
we implemented the NNLO corrections 
to the subleading colour contribution to $e^+e^- \to 3$~jets. Starting from 
an existing programme for $e^+e^- \to 4$~jets at NLO~\cite{cullen}, we 
constructed the newly required subtraction terms for double unresolved 
real radiation in the five-parton channel and for single unresolved real 
radiation off one-loop matrix elements in the four-parton channel. We 
observed that after implementation of these subtraction terms, 
five-parton and four-parton channels integrate to numerically finite 
contributions to three-jet final states, and that the four-parton channel is 
free from explicit infrared singularities from the loop integrations 
and the subtraction terms. All explicit infrared singularities obtained 
from integrating the newly introduced subtraction terms cancel in the
three-parton channel after adding with the 
the genuine two-loop virtual corrections to this channel, which were 
computed previously~\cite{3jme}. A further illustration of the 
potential of our method 
is given  in Section~\ref{sec:3jir}, where we demonstrate that 
the 
infrared poles of the two-loop virtual corrections to $e^+e^- \to 3~$jets
cancel in all its seven 
colour factors against linear combinations of integrated 
NNLO antenna functions. 

An immediate application of the method presented here is 
the calculation of 
the full NNLO corrections to $e^+e^- \to 3$~jets~\cite{new3j}. 
The antenna subtraction method 
can be further generalised to NNLO corrections to jet production in 
 lepton-hadron or hadron-hadron collisions. In these kinematical situations,
the subtraction terms are constructed using the same antenna functions, but 
in different  phase space configurations: instead of the $1\to n$ decay 
kinematics considered here, $2\to n$ scattering kinematics are required, which
can also contain singular configurations due to single or double initial state 
radiation. These require new sets of integrated antenna functions.

\section*{Acknowledgements}

We would like to thank Gudrun Heinrich and Zoltan Trocsanyi for useful 
discussions on several of the issues 
presented in this paper and for very 
constructive criticism expressed on our earlier works. 

We would equally
like to thank Alejandro Daleo and Stefan Weinzierl~\cite{weinzierlnew}
 for pointing out
 a number of typographical errors in the original form 
of (5.29), (5.30), (5.38), (5.43), (6.29), (6.48), (7.13), (7.50), (7.51).
These typographical errors are corrected in the present version. 

Part of the work presented here was performed 
during the 2004 ``QCD and Collider Physics''-programme 
at the Kavli Institute for Theoretical Physics (KITP), 
Santa Barbara. The authors would like to thank the KITP for its kind 
hospitality.

This research was supported in part by the Swiss National Science Foundation
(SNF) under contracts PMPD2-106101  and 200021-101874, 
 by the UK Particle Physics and Astronomy  Research Council
and by the National 
Science Foundation under Grant No.\ PHY99-07949.

\bibliographystyle{JHEP}

\begin{thebibliography}{10}


\bibitem{dissertori}
R.K.\ Ellis, W.J.\ Stirling and B.R.\ Webber, {\it QCD and Collider Physics},
Cambridge University Press (Cambridge, 1996);\\
G.\ Dissertori, I.G.\ Knowles and M.\ Schmelling, {\it Quantum 
Chromodynamics: High Energy Experiments and Theory}, Oxford University Press
(Oxford, 2003).

\bibitem{glover}
E.W.N.~Glover,
Nucl.\ Phys.\ Proc.\ Suppl.\  {\bf 116} (2003) 3
[hep-ph/0211412].

\bibitem{mvv}
S.~Moch, J.A.M.~Vermaseren and A.~Vogt,
Nucl.\ Phys.\ B {\bf 688} (2004) 101
[hep-ph/0403192];\\
A.~Vogt, S.~Moch and J.A.M.~Vermaseren,
Nucl.\ Phys.\ B {\bf 691} (2004) 129
[hep-ph/0404111].

\bibitem{twol}
Z.~Bern, L.J.~Dixon and A.~Ghinculov, Phys.\ Rev.\ D {\bf 63} (2001)
053007 [hep-ph/0010075];\\
C.\ Anastasiou, E.W.N.~Glover, C.\ Oleari and M.E.\ Tejeda-Yeomans,
Nucl.\ Phys.\ B~{\bf 601}~(2001) 318~[hep-ph/0010212];~{\bf 601}~(2001)~347 [hep-ph/0011094];
 {\bf 605} (2001) 486 [hep-ph/0101304];\\
E.W.N.~Glover, C.~Oleari and M.E.~Tejeda-Yeomans,
Nucl.\ Phys.\ {\bf 605} (2001) 467 [hep-ph/0102201];\\
C.~Anastasiou, E.W.N.~Glover and M.E.~Tejeda-Yeomans,
Nucl.\ Phys.\ B {\bf 629} (2002) 255 [hep-ph/0201274];\\
E.W.N.~Glover and M.E.~Tejeda-Yeomans,
JHEP {\bf 0306} (2003) 033
[hep-ph/0304169];\\
E.W.N.~Glover,
JHEP {\bf 0404} (2004) 021
[hep-ph/0401119];\\
Z.~Bern, A.~De Freitas and L.J.~Dixon,
JHEP {\bf 0109} (2001) 037 [hep-ph/0109078];
JHEP {\bf 0203} (2002) 018 [hep-ph/0201161];
JHEP {\bf 0306} (2003) 028
[hep-ph/0304168];\\
A.~De Freitas and Z.~Bern,
JHEP {\bf 0409} (2004) 039
[hep-ph/0409007];\\
Z.~Bern, A.~De Freitas, L.J.~Dixon, A.~Ghinculov and H.L.~Wong,
JHEP {\bf 0111} (2001) 031 [hep-ph/0109079];\\
T.~Binoth, E.W.N.~Glover, P.~Marquard and J.J.~van der Bij,
JHEP {\bf 0205} (2002) 060
[hep-ph/0202266];\\
T.~Gehrmann and E.~Remiddi,
Nucl.\ Phys.\ B {\bf 640} (2002) 379
[hep-ph/0207020];\\
S.~Moch, P.~Uwer and S.~Weinzierl,
Phys.\ Rev.\ D {\bf 66} (2002) 114001
[hep-ph/0207043].

\bibitem{3jme}
L.W.~Garland, T.~Gehrmann, E.W.N.~Glover, A.~Koukoutsakis and E.~Remiddi,
Nucl.\ Phys.\ B {\bf 627} (2002) 107 [hep-ph/0112081] and
{\bf 642} (2002) 227 [hep-ph/0206067].


\bibitem{twolmeth}
F.V.\ Tkachov, Phys.\ Lett.\ {\bf 100B} (1981) 65;\\
K.G.\ Chetyrkin and F.V.\ Tkachov, Nucl.\ Phys.\ B {\bf 192} (1981) 159;\\
V.A.\ Smirnov, Phys.\ Lett.\ B {\bf 460} (1999) 397 [hep-ph/9905323];\\
J.B.\ Tausk, Phys.\ Lett.\ B {\bf 469} (1999) 225 [hep-ph/9909506];\\
T.~Binoth and G.~Heinrich,
Nucl.\ Phys.\ B {\bf 585} (2000) 741
[hep-ph/0004013];\\
S.~Laporta,
Int.\ J.\ Mod.\ Phys.\ A {\bf 15} (2000) 5087
[hep-ph/0102033];\\
T.\ Gehrmann and E.\ Remiddi, Nucl.\ Phys.\ B
{\bf 580} (2000) 485 [hep-ph/9912329];\\
S.~Moch, P.~Uwer and S.~Weinzierl,
J.\ Math.\ Phys.\  {\bf 43} (2002) 3363 [hep-ph/0110083];\\
S.~Weinzierl, Comput.\ Phys.\ Commun.\  {\bf 145} (2002) 357 
[math-ph/0201011];\\
C.~Anastasiou and K.~Melnikov,
Nucl.\ Phys.\ B {\bf 646} (2002) 220
[hep-ph/0207004];\\
V.A.\ Smirnov, {\it Evaluating Feynman Integrals}, Springer Tracts of 
Modern Physics (Heidelberg, 2004).


\bibitem{onel-3}
Z.~Bern, L.J.~Dixon and D.A.~Kosower,
Phys.\ Rev.\ Lett.\  {\bf 70} (1993) 2677
[hep-ph/9302280];\\
Z.~Kunszt, A.~Signer and Z.~Trocsanyi,
Phys.\ Lett.\ B {\bf 336} (1994) 529
[hep-ph/9405386];\\
Z.~Bern, L.J.~Dixon and D.A.~Kosower,
Nucl.\ Phys.\ B {\bf 437} (1995) 259
[hep-ph/9409393];\\

\bibitem{onel-4}
E.W.N.~Glover and D.J.~Miller,
Phys.\ Lett.\ B {\bf 396} (1997) 257
[hep-ph/9609474];\\
Z.~Bern, L.J.~Dixon, D.A.~Kosower and S.~Weinzierl,
Nucl.\ Phys.\ B {\bf 489} (1997) 3
[hep-ph/9610370];\\
J.M.~Campbell, E.W.N.~Glover and D.J.~Miller,
Phys.\ Lett.\ B {\bf 409} (1997) 503
[hep-ph/9706297];\\
Z.~Bern, L.J.~Dixon and D.A.~Kosower,
Nucl.\ Phys.\ B {\bf 513} (1998) 3
[hep-ph/9708239].




\bibitem{nlomult}
L.J.\ Dixon and A.\ Signer,
Phys.\ Rev.\ Lett.\  {\bf 78} (1997) 811
[hep-ph/9609460];
Phys.\ Rev.\ D {\bf 56} (1997) 4031
[hep-ph/9706285];\\
Z.\ Nagy and Z.\ Trocsanyi,
Phys.\ Rev.\ Lett.\  {\bf 79} (1997) 3604
[hep-ph/9707309];\\
S.~Weinzierl and D.A.~Kosower,
Phys.\ Rev.\ D {\bf 60} (1999) 054028
[hep-ph/9901277];\\
W.B.~Kilgore and W.T.~Giele,
Phys.\ Rev.\ D {\bf 55} (1997) 7183
[hep-ph/9610433];\\
Z.~Nagy and Z.~Trocsanyi,
Phys.\ Rev.\ Lett.\  {\bf 87} (2001) 082001
[hep-ph/0104315];\\
Z.~Nagy, Phys.\ Rev.\ Lett.\ {\bf 88}~(2002)~122003 
[hep-ph/0110315];
Phys.\ Rev.\ D {\bf 68} (2003) 094002 [hep-ph/0307268];\\
J.~Campbell and R.K.~Ellis,
Phys.\ Rev.\ D {\bf 65} (2002) 113007
[hep-ph/0202176].

\bibitem{cullen}
J.\ Campbell, M.A.\ Cullen and E.W.N.\ Glover,
Eur.\ Phys.\ J.\ C {\bf 9} (1999) 245
[hep-ph/9809429].


\bibitem{onelstr}
Z.\ Bern, L.J.\ Dixon, D.C.\ Dunbar and D.A.\ Kosower,
Nucl.\ Phys.\ B {\bf 425} (1994) 217 [hep-ph/9403226];\\
D.A.\ Kosower, Nucl.\ Phys.\ B {\bf 552} (1999) 319 [hep-ph/9901201];\\
D.A.~Kosower and P.~Uwer, Nucl.\ Phys.\ B {\bf 563} (1999) 477 
[hep-ph/9903515];\\
Z.\ Bern, V.\ Del Duca and C.R.\ Schmidt, Phys.\ Lett.\ B {\bf 445}
(1998) 168 [hep-ph/9810409];\\
Z.\ Bern, V.\ Del Duca, W.B.\ Kilgore and C.R.\ Schmidt, Phys.\ Rev.\ D
{\bf 60} (1999) 116001 [hep-ph/9903516].

\bibitem{oneloopsoft}
S.\ Catani and M.\ Grazzini, Nucl.\ Phys.\  B {\bf 591} (2000) 435 
[hep-ph/0007142];\\


\bibitem{onelstr1}
D.A.~Kosower,
Phys.\ Rev.\ Lett.\  {\bf 91} (2003) 061602
[hep-ph/0301069].
\bibitem{onelstr2}
S.~Weinzierl,
JHEP {\bf 0307} (2003) 052
[hep-ph/0306248].

\bibitem{twolstr}
C.~Anastasiou, Z.~Bern, L.J.~Dixon and D.A.~Kosower,
Phys.\ Rev.\ Lett.\  {\bf 91}, 251602 (2003)
[hep-th/0309040];\\
Z.~Bern, L.J.~Dixon and D.A.~Kosower,
JHEP {\bf 0408} (2004) 012
[hep-ph/0404293];
S.D.~Badger and E.W.N.~Glover,
  JHEP {\bf 0407} (2004) 040
  [hep-ph/0405236].

\bibitem{babisdy}
C.~Anastasiou, L.J.~Dixon, K.~Melnikov and F.~Petriello,
  Phys.\ Rev.\ Lett.\  {\bf 91} (2003) 182002
  [hep-ph/0306192];\
Phys.\ Rev.\ D {\bf 69} (2004) 094008
[hep-ph/0312266].



\bibitem{audenigel}
A.~Gehrmann-De Ridder and E.W.N.~Glover,
Nucl.\ Phys.\ B {\bf 517} (1998) 269
[hep-ph/9707224];\\

\bibitem{campbell}
J.\ Campbell and E.W.N.\ Glover,
Nucl.\ Phys.\ B {\bf 527} (1998) 264 [hep-ph/9710255].

\bibitem{cg}
S.\ Catani and M.\ Grazzini, Phys.\ Lett.\ B {\bf 446} (1999) 143
[hep-ph/9810389];
Nucl.\ Phys.\ B {\bf 570} (2000) 287 [hep-ph/9908523].

\bibitem{campbellandother}
F.A.\ Berends and W.T.\ Giele, Nucl.\ Phys.\ B {\bf 313} (1989) 595;\\
V.~Del Duca, A.~Frizzo and F.~Maltoni,
Nucl.\ Phys.\ B {\bf 568} (2000) 211
[hep-ph/9909464];\\
T.G.~Birthwright, E.W.N.~Glover, V.V.~Khoze and P.~Marquard,
JHEP {\bf 0505} (2005) 013 [hep-ph/0503063].

\bibitem{uwer}
D.A.~Kosower and P.~Uwer,
Nucl.\ Phys.\ B {\bf 674} (2003) 365
[hep-ph/0307031].

\bibitem{fg}
D.~de Florian and M.~Grazzini,
Nucl.\ Phys.\ B {\bf 616} (2001) 247
[hep-ph/0108273];
 {\bf 704} (2005) 387
  [hep-ph/0407241].

\bibitem{daleo}
A.~Daleo, C.A.~Garcia Canal and R.~Sassot,
Nucl.\ Phys.\ B {\bf 662} (2003) 334
[hep-ph/0303199];\\
  A.~Daleo, D.~de Florian and R.~Sassot,
  Phys.\ Rev.\ D {\bf 71} (2005) 034013
  [hep-ph/0411212].

\bibitem{adamson}
  K.L.~Adamson, D.~de Florian and A.~Signer,
  Phys.\ Rev.\ D {\bf 65} (2002) 094041
  [hep-ph/0202132];
  Phys.\ Rev.\ D {\bf 67} (2003) 034016
  [hep-ph/0211295].


\bibitem{nnlosub1}
D.A.~Kosower, Phys.\ Rev.\ D {\bf 67} (2003) 116003
[hep-ph/0212097].
\bibitem{nnlosub2}
S.~Weinzierl,
JHEP {\bf 0303} (2003) 062
[hep-ph/0302180].
\bibitem{nnlosub3}
W.B.~Kilgore,
Phys.\ Rev.\ D {\bf 70} (2004) 031501
[hep-ph/0403128].
\bibitem{nnlosub4}
M.\ Grazzini and S.\ Frixione,
JHEP {\bf 0506} (2005) 010
[hep-ph/0411399].
\bibitem{nnlosub5}
G.~Somogyi, Z.~Trocsanyi and V.~Del Duca,
JHEP {\bf 0506} (2005) 024
[hep-ph/0502226].


\bibitem{secdec}
G.~Heinrich,
Nucl.\ Phys.\ Proc.\ Suppl.\  {\bf 116} (2003) 368
[hep-ph/0211144];\\
C.~Anastasiou, K.~Melnikov and F.~Petriello,
Phys.\ Rev.\ D {\bf 69} (2004) 076010
[hep-ph/0311311];\\
T.~Binoth and G.~Heinrich,
Nucl.\ Phys.\ B {\bf 693} (2004) 134
[hep-ph/0402265];\\
G.~Heinrich,
Nucl.\ Phys.\ Proc.\ Suppl.\  {\bf 135} (2004) 290
[hep-ph/0406332].

\bibitem{ggh}
A.~Gehrmann-De Ridder, T.~Gehrmann and G.~Heinrich,
Nucl.\ Phys.\ B {\bf 682} (2004) 265 [hep-ph/0311276].


\bibitem{babis2j}
C.~Anastasiou, K.~Melnikov and F.~Petriello,
Phys.\ Rev.\ Lett.\  {\bf 93} (2004) 032002
[hep-ph/0402280].

\bibitem{babishiggs}
C.~Anastasiou, K.~Melnikov and F.~Petriello,
Phys. Rev. Lett. {\bf 93} (2004) 262002
[hep-ph/0409088]; 
Nucl.\ Phys.\ B {\bf 724} (2005) 197
[hep-ph/0501130].


\bibitem{babismu}
  C.~Anastasiou, K.~Melnikov and F.~Petriello,
  hep-ph/0505069.


\bibitem{ERT}
R.K.~Ellis, D.A.~Ross and A.E.~Terrano,
Nucl.\ Phys.\ B {\bf 178} (1981) 421.



\bibitem{ert2}
K.~Fabricius, I.~Schmitt, G.~Kramer and G.~Schierholz, Z.~Phys.~C {\bf
  11} (1981) 315.

\bibitem{kn}
Z.~Kunszt and P.~Nason, in {\it Z Physics at LEP 1}, CERN Yellow Report
89-08, Vol.~1, p.~373.



\bibitem{singleun}
W.T.~Giele and E.W.N.~Glover,
Phys.\ Rev.\ D {\bf 46} (1992) 1980;\\
Z.~Kunszt and D.E.~Soper,
Phys.\ Rev.\ D {\bf 46} (1992) 192;\\
S.~Frixione, Z.~Kunszt and A.~Signer,
Nucl.\ Phys.\ B {\bf 467}, 399 (1996)
[hep-ph/9512328].


\bibitem{irsafe}
  G.~Sterman and S.~Weinberg,
  Phys.\ Rev.\ Lett.\  {\bf 39} (1977) 1436;\\
  G.~Sterman,
  Phys.\ Rev.\ D {\bf 17} (1978) 2773;
 {\bf 17} (1978) 2789.


\bibitem{cs}
S.~Catani and M.H.~Seymour,
Nucl.\ Phys.\ B {\bf 485} (1997) 291; {\bf 510} (1997) 503(E)
[hep-ph/9605323].



\bibitem{ant}
D.A.~Kosower,
Phys.\ Rev.\ D {\bf 57} (1998) 5410
[hep-ph/9710213];
Phys.\ Rev.\ D {\bf 71} (2005) 045016
[hep-ph/0311272].

\bibitem{colord}
F.A.\ Berends and W.T. Giele, Nucl.\ Phys.\ B {\bf 294} (1987) 700;\\
M. Mangano, S. Parke and Z. Xu, Nucl.\ Phys.\ B {\bf 298} (1988) 653;\\
M. Mangano, Nucl. Phys. B {\bf 309} (1988) 461;\\
L.J.~Dixon,
Proceedings of ``Theoretical Advanced Study Institute in Elementary Particle
Physics (TASI '95): QCD and Beyond'', ed.\ D.\ Soper, World Scientific
(Singapore, 1996),
p.539 [hep-ph/9601359].

\bibitem{ddm}
V.~Del Duca, L.J.~Dixon and F.~Maltoni,
Nucl.\ Phys.\ B {\bf 571} (2000) 51
[hep-ph/9910563];\\
F.~Maltoni, K.~Paul, T.~Stelzer and S.~Willenbrock,
Phys.\ Rev.\ D {\bf 67} (2003) 014026
[hep-ph/0209271].


\bibitem{our2j}
A.~Gehrmann-De Ridder, T.\ Gehrmann and E.W.N.\ Glover, 
Nucl.\ Phys.\ B {\bf 691} (2004) 195
[hep-ph/0403057].


\bibitem{our3j}
A.~Gehrmann-De Ridder, T.~Gehrmann and E.W.N.~Glover,
Nucl.\ Phys.\ Proc.\ Suppl.\  {\bf 135} (2004) 97
[hep-ph/0407023].


\bibitem{chi}
A.~Gehrmann-De Ridder, T.~Gehrmann and E.W.N.~Glover,
Phys.\ Lett.\ B {\bf 612} (2005) 36 [hep-ph/0501291].


\bibitem{h}
A.~Gehrmann-De Ridder, T.~Gehrmann and E.W.N.~Glover,
Phys.\ Lett.\ B {\bf 612} (2005) 49 [hep-ph/0502110].


\bibitem{catani}
S.\ Catani, Phys.\ Lett.\ B {\bf 427} (1998) 161
[hep-ph/9802439];\\ 
G.~Sterman and M.E.~Tejeda-Yeomans,
Phys.\ Lett.\ B {\bf 552} (2003) 48
[hep-ph/0210130].


\bibitem{kln}
T.~Kinoshita,
J.\ Math.\ Phys.\  {\bf 3} (1962) 650;\\
T.D.~Lee and M.~Nauenberg,
Phys.\ Rev.\  {\bf 133} (1964) B1549.

\bibitem{hpl}
E.\ Remiddi and J.A.M.\ Vermaseren, Int.\ J.\ Mod.\ Phys.\ A {\bf 15}
(2000) 725 [hep-ph/9905237].

\bibitem{3jmi}
T.\ Gehrmann and E.\ Remiddi, Nucl.~Phys.~B {\bf 601} (2001) 248
[hep-ph/0008287];
{\bf 601} (2001) 287 [hep-ph/0101124].


\bibitem{nielsen}
N.~Nielsen, Nova Acta Leopoldiana (Halle) {\bf 90} (1909) 123;\\
K.S.\ K\"olbig, J.A.\ Mignaco and E.\ Remiddi, BIT {\bf 10} (1970) 38.





\bibitem{hplnum}
T.~Gehrmann and E.~Remiddi,
Comput.\ Phys.\ Commun.\ {\bf 141} (2001) 296 [hep-ph/0107173];
  Comput.\ Phys.\ Commun.\  {\bf 144} (2002) 200
  [hep-ph/0111255].



\bibitem{vegas}
 G.P.~Lepage,
  J.\ Comput.\ Phys.\  {\bf 27} (1978) 192.

\bibitem{rambo}
 R.~Kleiss, W.J.~Stirling and S.D.~Ellis,
  Comput.\ Phys.\ Commun.\  {\bf 40} (1986) 359.

\bibitem{new3j}
A.~Gehrmann-De Ridder, T.~Gehrmann, E.W.N.~Glover and G.\ Heinrich,
work in progress.


\bibitem[A1]{weinzierlnew}
  S.~Weinzierl,
  Phys.\ Rev.\ D {\bf 74} (2006) 014020
  [hep-ph/0606008].


\end{thebibliography}
 
\end{document}